\documentclass[twocolumn]{aa} 
\usepackage{graphicx}
\usepackage{amsmath}
\usepackage{txfonts}
\usepackage{natbib}
\usepackage{hyperref}

\usepackage{lipsum}
\usepackage{mathtools}
\usepackage{cuted}

\bibpunct{(}{)}{;}{a}{}{}
\newcommand\bx{\textbf{x}}

\newcommand{\sub}[1]{_{\rm #1}}
\newcommand{\emm}[1]{\ensuremath{#1}}   
\newcommand{\emr}[1]{\emm{\mathrm{#1}}} 
\newcommand{\unit}[1]{\emm{\, \emr{#1}}}

\newcommand{\pscm}{\unit{cm^{-2}}}

\renewcommand{\deg}{\emm{^\circ}}
\newcommand{\chr}{^{\rm h}}
\newcommand{\cmin}{^{\rm m}}
\newcommand{\csec}{^{\rm s}}
\newcommand{\changed}[1]{#1}

\begin{document}
\title{Measuring the filamentary structure of interstellar clouds through wavelets}

\author{V. Ossenkopf-Okada\inst{1} \and R. Stepanov\inst{2}}

\offprints{V. Ossenkopf-Okada}

\institute{
I. Physikalisches Institut, Universit\"at zu K\"oln, Z\"ulpicher Strasse 77, D-50937 K\"oln, Germany\\
\email{ossk@ph1.uni-koeln.de}
\and
Institute of Continuous Media Mechanics of the Russian Academy of Science,
Ak. Korolev Str. 1, Perm 614013, Russia \email{rodion@icmm.ru}
             }

   \date{Received 2017/07/19; accepted 2018/11/02}


  \abstract
   {The ubiquitous presence of filamentary structures in the interstellar medium asks for
an unbiased characterization of their properties including a stability analysis.}
   {We propose a novel technique to measure the spectrum of filaments in any two-dimensional
data set. By comparing the power in isotropic and anisotropic structures
we can measure the relative importance of spherical and cylindrical collapse modes.
}
   {Using anisotropic wavelets we can quantify and distinguish local and global anisotropies
and measure the size distribution of filaments. The wavelet analysis does not need any
assumptions on the alignment or shape of filaments in the maps, but directly measures their typical spatial
dimensions.  In a rigorous test program, we calibrate
the scale-dependence of the method and test the angular and spatial sensitivity. We apply the method
to molecular line maps from magneto-hydrodynamic (MHD) simulations and observed column density maps from
\textit{Herschel} observations.
  }
   {When applying the anisotropic wavelet analysis to the MHD data, we find that the
observed filament sizes depend on the combination of magnetic-field dominated density-velocity
correlations with radiative transfer effects. This can be exploited by observing tracers
with different optical depth to measure the transition from a globally ordered large-scale structure
to small-scale filaments with entangled field lines. The unbiased view to Herschel column density
maps does not confirm a universal characteristic filament width. The map of the Polaris Flare
shows an almost scale-free filamentary spectrum up to the size of the dominating filament of
about 0.4~pc. For the Aquila molecular cloud the range of filament widths is limited to 
0.05--0.2~pc. The filaments in Polaris show no preferential direction in contrast to the global alignment
that we trace in Aquila.
   }
   {By comparing the power in isotropic and anisotropic structures we can measure the relative importance of spherical and cylindrical collapse modes and their spatial distribution.
   }


\maketitle

%

\section{Introduction}

Recent mapping observations of the interstellar medium reveal more
and more filamentary structures. \changed{In particular \textit{Herschel}\footnote{\textit{Herschel}
is an ESA space observatory with science instruments provided by
European-led Principal Investigator consortia and with important participation from NASA.}
continuum maps show a network of filamentary structures in many
interstellar clouds} \citep[e.g.][]{Andre2010, Schneider2010, MivilleDeschenes2010}.
\changed{They seem to dominate the structure both in star-forming and starless
clouds.}
However, it is still unclear to what degree the filaments dominate the
dynamical evolution of the clouds, whether there is a clear hierarchy
of filaments and even whether the filaments
showing up in the column density maps actually reflect true filaments
in the underlying three-dimensional (3-D) structure. In many models
'filaments' show up as projections of crossing sheets from interacting
shocks or from a 'stretch' induced by turbulence \citep[e.g.][]{Hennebelle2013}.
In case of a global 3-D isotropy, the projection effects have been estimated
\citep[e.g.][]{VazquezSemadeniGarcia2001, BurkhartLazarian2012}, but
\changed{this assumption is obviously questionable when highly anisotropic
structures, such as filaments, dominate the interstellar clouds. Therefore
a global statistics of the anisotropy of the interstellar medium is highly
demanded. Here, we try to contribute towards this general goal.

As most prestellar cores are found within dense filaments the question arises
whether the filaments channel the flow of material towards the cores and
consequently towards star formation or whether the cores form from the direct
fragmentation of filaments \citep{Federrath2016,Andre2017}.
Estimates on the gravitational stability of the filaments indicate that
many of them are unstable against fragmentation \citep[e.g.][]{FischeraMartin2012}
but this requires an accurate determination of their physical properties.
Measuring the width of the filaments in a network of structures including
background emission is not trivial.

Frequently used methods} for the identification of the filaments from the
morphological structure in the maps, such as
{\it DisPerSe} \citep{Sousbie2011},  {\it getFilaments} \citep{Menshchikov2010},
and {\it FilFinder} \changed{\citep{KochRosolowsky2015} apply some map
filtering and filament truncation that may affect} the measured filament
properties such as their typical length, width,
and transversal column density distribution \changed{\citep{Panopoulou2017}.
The existence of a universal filament width across different clouds is
highly debated \citep{Andre2017}. We try to contribute to the discussion here
by suggesting an unbiased method to measure the geometrical parameters of
filaments from any astrophysical map.}


\changed{The detection of characteristic scales describing length and
width of filaments can be regarded as an analysis of the spectral properties
of the maps. They are most directly measured by the power spectrum obtained
from a discrete Fourier transform. However, the power spectrum does not
retain any information on the localization of individual structures
in normal space, describing global properties only. To overcome this problem \citet{Gabor1946}
already introduced a combination of the Fourier transform with a windowing
technique that allowed him to localize individual frequency components
in a signal. This was extended by \citet{GrossmanMorlet1984} to a theory of
wavelet transforms using constant shape wavelets that fulfil an ``admissibility condition'' allowing
for an isometric representation of any data in combined frequency-localization
space. Wavelets are quadratically integrable in normal and
Fourier space and have a zero mean. This asks for smoothly tapered oscillatory
functions. \citet{Walker2008} provides a comprehensive introduction.
A first application to astronomical maps was presented by
\citet{GillHenriksen1990}. \citet{Robitaille2014} present a broader
overview on the application of wavelet analyses for interstellar clouds
maps.}

We propose a novel technique to quantify
the general structure of anisotropic structures in clouds
allowing to trace the degree of anisotropy as a function of spatial scale.
We use the anisotropic wavelet transform to characterize the spectral
energy distribution depending on scale and orientation of cloud
structures \citep[see e.g.,][]{Patrikeev2006,frick16}. This allows us to
detect scales and directions of anisotropies in the observed structure in
any 2-D maps. Extension to full 3-D structures is planned.
Compared to existing techniques, it does not only characterize the
global anisotropy, but also quantifies local anisotropies that trace
structures with changing directions and warps, produced for example by
entangled magnetic fields.
We can also use it to quantify the dominant modes of gravitational collapse in
molecular clouds to estimate the importance of filamentary structures
relative to spherical clumps for the global star formation process.

In Sect.~\ref{sec:theory} we introduce the formalism of the anisotropic wavelet analysis.
Sect.~\ref{sec:simpletests} shows the application of the method to well-controlled test
structures to determine which parameters are best suited for a
local and global analysis and to demonstrate the resolving power of the
anisotropic wavelet analysis. In Sect.~\ref{sect:MHD} we analyze the anisotropic
structure created in MHD turbulence simulations, in Sect.~\ref{sect:obs}
we apply the method to observed column density maps of interstellar clouds,
and in Sect.~\ref{sec:discussion} we discuss the
implications for the characterization of the role of filaments and
magnetic fields.

\section{Anisotropic wavelets}
\label{sec:theory}


The continuous wavelet transform of a 2-D map $f(\bx)=f(x,y)$ is
defined by
\begin{equation}
W(s,\varphi,\bx) = {1 \over {s^{3/2}}}\int_{-\infty}^{+\infty}
\int_{-\infty}^{+\infty} f(\bx') \mathrm{\psi}_{\varphi}^* \left(
{{\bx'-\bx}\over s}\right) d \bx', \label{cwt}
\end{equation}
\changed{where $s$ is the spatial scale and} the wavelet $\mathrm{\psi}_{\varphi}(x,y)$ obtained by
rotating an asymmetric wavelet $\mathrm{\psi}(x,y)$
by the position angle $\varphi$
\begin{equation}
\mathrm{\psi}_{\varphi}(\bx)=\mathrm{\psi}(x \cos\varphi - y
\sin\varphi,x \sin\varphi + y \cos\varphi).
\end{equation}
\changed{This definition implies that $\varphi$ is zero in positive
$x$ direction and increases counterclockwise.}
The scaling by $s^{-3/2}$ \changed{provides a wavelet energy spectrum that
corresponds to the energy density spectrum \cite{frick01}, allowing for an
easy comparison with the Fourier transform.}

\begin{figure}
\centerline{\includegraphics[width=0.9\columnwidth]{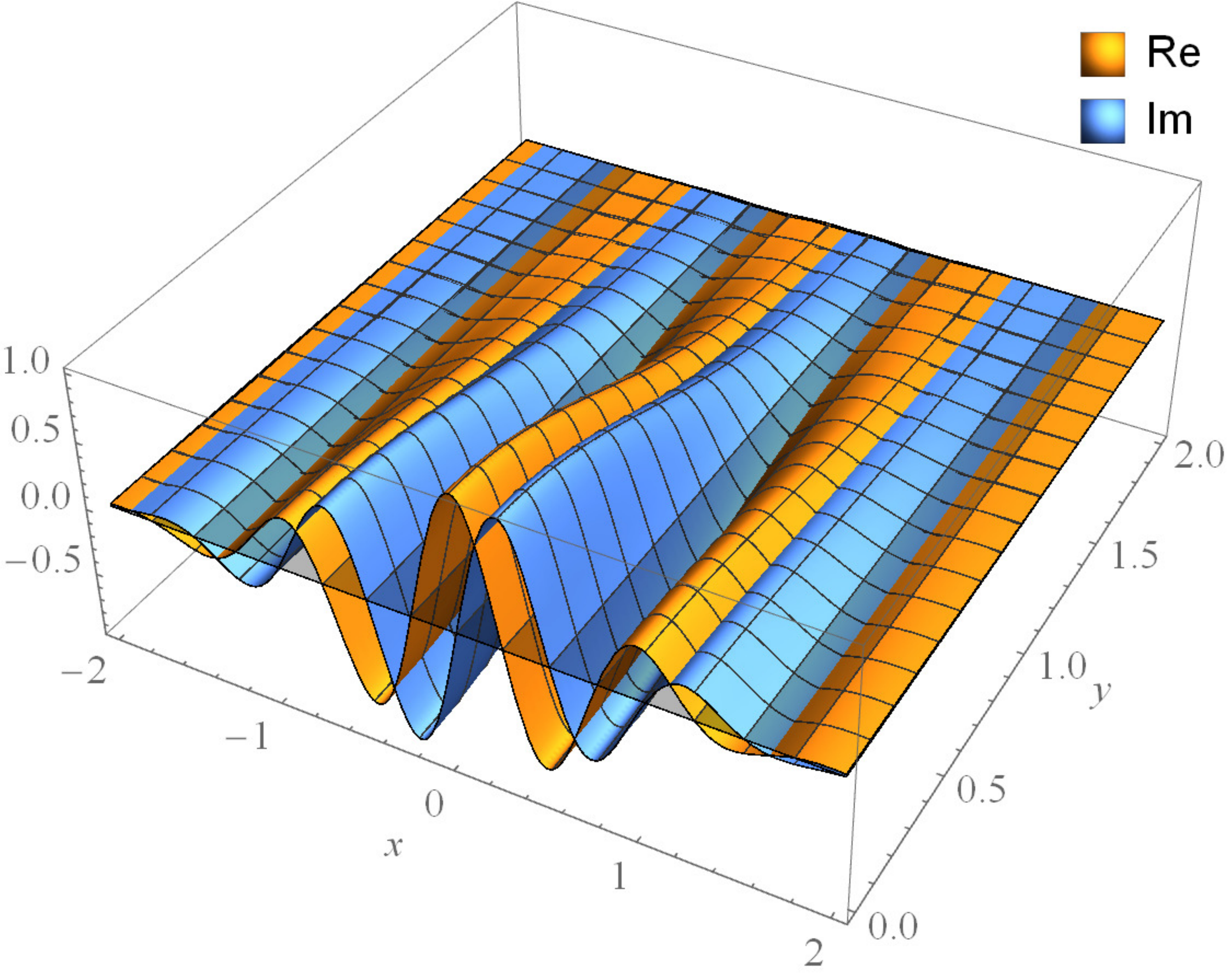}}
\caption{\changed{Cut through the} basic two-dimensional Morlet wavelet
used for the anisotropic
analysis (Eq.~\ref{eq:morlet} for $b=1$). The yellow surface represents the
real part, the blue surface the imaginary part of the wavelet
function. For better visibility of both parts, the range $y<0$ is
omitted in the display. For the full analysis, the wavelet is stretched by a factor $s$ and
rotated by an angle $\varphi$.
\label{fig_wavelet}}
\end{figure}

We use a complex Morlet mother wavelet function in the $x$-direction
(see Fig.~\ref{fig_wavelet}, \citet{Walker2008})
\begin{equation}
\mathrm{\psi}(x,y) =
{1 \over b} \left[\exp(2\pi\imath x)
-\exp(-\pi^2 b^2) \right] \exp\left(\frac{-x^2-y^2}{b^2}\right),
\label{eq:morlet}
\end{equation}
where the localization parameter $b$ allows for a scaling of the
isotropic Gaussian filter relative to the anisotropic sinusoidal part.
In Sect.~\ref{sect_filtershape} we show how it affects the
resolution of the wavelet in terms of scales and in the position in space.
\changed{The  small constant
$-\exp(-\pi^2 b^2)$ and the localization prefactor $1/b$  are required to fulfill, correspondingly,  the wavelet conditions
of a zero mean and a square norm of unity.}  A similar wavelet
was already proposed  by \citet{Robitaille2014} within a larger, more general
set of wavelets. However, they did not allow for a variation of the
localization parameter to avoid the variable offset in the normalization in Eq.~\ref{eq:morlet}.
We test the impact of \changed{localization parameters $b$ between $1/\sqrt{2}$ and $\sqrt{2}$
in Sect.~\ref{sect_filtershape}.}

The 2-D wavelet spectrum is defined as
\begin{equation}
M(s,\varphi) = \left\langle |W(s,\varphi,\bx)|^2 \right\rangle_\bx,  \label{w_spec}
\end{equation}
where $\langle \cdot\rangle_{\vec{x}}$ denotes the average over the map.
The wavelet spectrum characterizes the distribution of energy as a function of scale $s$ and
orientation $\varphi$. Isotropic $m^i$ and  anisotropic $m^a$ parts of
energy can be defined as the amplitude of the $n=0$ and $n=2$
complex rotational modes of $|W(s,\varphi,\bx)|^2$ at the
given position and scale, namely
\begin{eqnarray}
m^i(s,\bx) &=& (2\pi)^{-1} \int_{-\pi}^{+\pi} |W(s,\varphi,\bx)|^2 d \varphi , \\
m^a(s,\bx) &=& (2\pi)^{-1} \int_{-\pi}^{+\pi} |W(s,\varphi,\bx)|^2
e^{ 2 \imath \varphi}d \varphi. \label{w_part}
\end{eqnarray}
The $n=0$ mode $m^i(s,\bx)$ is real. It can be interpreted as the
local isotropic energy.  The mode $n=1$ is zero because of the
symmetry $W(s,\varphi,\bx)=W^*(s,\varphi+\pi,\bx)$.
The complex $n=2$ mode $m^a(s,\bx)$ quantifies the anisotropy in the
structure, where the amplitude $|m^a(s,\bx)|$ characterizes the
energy.\footnote{The restriction to the $n=2$ mode
excludes higher order anisotropies here, i.e. our definition is not
sensitive to cross- or asterisk-like structures.}
The phase $\varphi_0(s,\bx)=\arg (m^a(s,\bx))$ gives the preferred
direction of the local anisotropy.\footnote{Note that
position angle $\varphi_0$ can be
different from the maximum energy position angle $\varphi_m$
defined as
\begin{equation}
|W(s,\varphi_m,\bx)| =
\max_{\varphi\in[0,\pi]}{|W(s,\varphi,\bx)|}. \label{wtmax}
\end{equation}}

Using these distributions we can introduce the isotropic and anisotropic spectra
\begin{eqnarray}
  M^i(s) &=& \left\langle m^i(s,\bx) \right\rangle_\bx, \\
  M^a(s) &=& \left\langle |m^a(s,\bx)| \right\rangle_\bx.
\label{w_parts}
\end{eqnarray}
The spectrum $M^i(s)$ measures the energy as a function of scale of the isotropic
wavelet transform. \changed{Analytic expressions for the wavelet
spectrum can be given for idealized test structures. Appx.~\ref{appx_an}
covers the case of elliptical Gaussians.}
We can compare  $M^i(s)$ to the spectrum obtained from
filtering with isotropic wavelets such as the Mexican hat filtering
\changed{by \citet{frick01} and \citet{Arevalo2012} also used in} the
$\Delta$-variance spectrum, $\sigma_\Delta^2(s)$,
\citep{Stutzki1998, Ossenkopf2008a}. We discuss the comparison
for selected examples in Appx.~\ref{sect_comp_deltavar}.

It is now natural to introduce the degree of anisotropy as
\begin{equation}
d^w_{loc}(s) = M^a(s)/M^i(s).
\label{eq:dwloc}
\end{equation}
This characterizes the degree of the local anisotropies
throughout the map. \changed{The spectrum of the degree of anisotropy
then quantifies the scale dependence of the ratio of the energy of
anisotropic variations compared to that of isotropic variations.
It does not directly identify the size of individual structures
but the size where anisotropic variations dominate (Sect.~\ref{sec:simpleGaussian}).}

%

If one is only interested in global anisotropies, one can use the average
of the complex wavelet coefficients $m^a(s,\bx)$  instead of their
amplitudes so that different phases cancel each other. Then we
obtain a global degree of anisotropy from the wavelet transform,
\begin{equation}
d^w_{glob}(s) = \frac{\left| \langle m^a(s,\bx) \rangle_\bx
\right|}{M^i(s)}.
\label{eq:dwglob}
\end{equation}

In the case of a significant global anisotropy one can calculate
the preferred direction as the angle
\begin{equation}
\varphi^w(s) = \frac{1}{2}\arg \left(\langle m^a(s,\bx) \rangle_\bx\right).
\end{equation}
Averaging $\varphi_0$ over the map instead does not make sense
due to the periodicity of the angles.

To combine the spatial and angular information we can study the distribution
 of angles $\varphi_0(s,\bx)$ of the anisotropic modes $m^a(s,\vec{x})$.
We define the two-dimensional anisotropic mode spectrum as
\begin{equation}
A(s,\varphi) = \left\langle \left| m^a(s,\bx)|_{\varphi_0=\varphi}\right| \right\rangle_\bx,
\label{2dspec}
\end{equation}
where the $|_{\varphi_0=\varphi}$ symbol indicates that the averaging
is to be taken over the coefficients with ${\arg (m^a(s,\bx))=\varphi}$.

In the same way, we can define a two-dimensional spectrum of the degree
of anisotropy by normalizing the distribution of angles $A(s,\varphi)$
to the spectrum of isotropic modes
\begin{equation}
\tilde{A}(s,\varphi) = A(s,\varphi)\large/{M^i(s)}.
\label{2dnormspec}
\end{equation}
The two-dimensional spectra then provide information on the scale and the
angle dependence of the anisotropic wavelet coefficients. The inspection of
$A(s,\varphi)$ and $\tilde{A}(s,\varphi)$ provides a complementary view
on the ratio between local and global anisotropies that will help us
to understand the relation between $d^w_{loc}(s)$ and $d^w_{glob}(s)$.


The global anisotropy measured
through the wavelet spectrum can be compared with the outcome of
existing methods. In Appx.~\ref{sect_comp_sf} we show a
comparison with the structure-function-based anisotropy definition
from \citet{EsquivelLazarian2011}.
\changed{The wavelet-based anisotropy is independent of the choice of scaling normalization in Eq.~(\ref{cwt}).}
As shown for the $\Delta$-variance as an
isotropic wavelet analysis, the wavelet spectrum can also be related
to the Fourier
spectrum of the distribution \citep{Stutzki1998,Ossenkopf2008a}.
Scale $s$ and wave number $k$ are related as $s=X\sub{tot}/k$
where $X\sub{tot}$ is the total size of the map.
If $\hat{f}(k_x,k_y)=\hat{f}(k \cos \varphi,k \sin \varphi)$ is
the Fourier transform of the map $f(\vec{x})$ and
\begin{equation}
F(k)=(2\pi)^{-1} \int_{-\pi}^{+\pi}
|\hat{f}(k \cos \varphi,k \sin \varphi)|^2 k d \varphi
\label{eq:fourier_i}
\end{equation}
is the energy density of the map in $k$-space
we can decompose them into angular modes. The isotropic
contribution is just given by the energy density $F^i(k)=F(k)$
and the anisotropic contribution is
\begin{equation}
F^a(k) = (2\pi)^{-1} \int_{-\pi}^{+\pi} |\hat{f}(k \cos
\varphi,k \sin \varphi)|^2 e^{ 2 \imath \varphi} k d \varphi.
\label{f_part}
\end{equation}
However, as \changed{this definition implies a global integral of the
phases}, $F^a(k)$ can only characterize
global anisotropies. We obtain the degree of anisotropy as
\begin{equation}
d^F_{glob}(k) = |F^a(k)|/F^i(k),
\label{eq_da}
\end{equation}
which should be equivalent to the wavelet-based global
degree of anisotropy $d^w_{glob}(k)$. By testing that
both methods provide equivalent results for the global
anisotropy we can obtain a feeling for the general
reliability of the wavelet-based method that holds then
as well for local anisotropies.
%
%
%
%
%
%
For an easy comparison of the energy density spectra and the wavelet
coefficients, we express all scales in terms of the corresponding filter
size by $s = X\sub{tot}/k$.

\changed{
The scaling normalization in Eq.~(\ref{cwt}) means that structures of
similar shape but different size have the same contribution to $M^i(s)$
if their total energies per scale are equal. We can use these wavelet
spectra for a direct measurement of the energy density, equivalent to the
Fourier spectra. Our wavelet and Fourier spectra characterize
the energy density of the structures as a function of their scale in units of
the square of the amplitude in $f$ multiplied by the pixel scale.

However, for very shallow
structures like the Plummer profiles with $p=2$ discussed in
Sect.~\ref{sect_plummer} the energy normalization leads to a
peak of the spectrum at infinite scales.
For those structures the $M^i(s)$ spectrum cannot be directly used to
measure their size. This can be overcome by rescaling the spectrum by
$s^{-2}$. The rescaled spectra $M^i(s)/s^2$ measure amplitude per scale equality
of structures with similar shape and different size. The additional factor
$1/s^2$ always guarantees an appearance of a peak related to size of structure.
On top of the slope change we
can additionally normalize the spectra by the variance of the original maps, $\sigma_f^2$,
to separate the effects of the signal amplitude from the spatial structure.
For all applications where we want to measure the size of individual
structures we therefore switch to this combined rescaling.
The units of the rescaled spectra are consequently given
by the inverse pixel scale.}


\section{Test of the technique with artificial data sets}
\label{sec:simpletests}

To \changed{test} how the anisotropic wavelet analysis describes
inherent and coincidental anisotropies in different data sets
we performed a large set of tests using maps with simple
geometrical structures and computed the different coefficients
defined in Sect.~\ref{sec:theory}.
\changed{We start with a kernel width of $b=1$. A more extended
set of tests including other kernels is presented in Appx.~\ref{appx_tests}.}

\subsection{Individual structures}
\label{sec:single}

\subsubsection{Sinusoidal pattern}
\label{sec:sinus}

\begin{figure}
\centering
\includegraphics[angle=90,width=0.88\columnwidth]{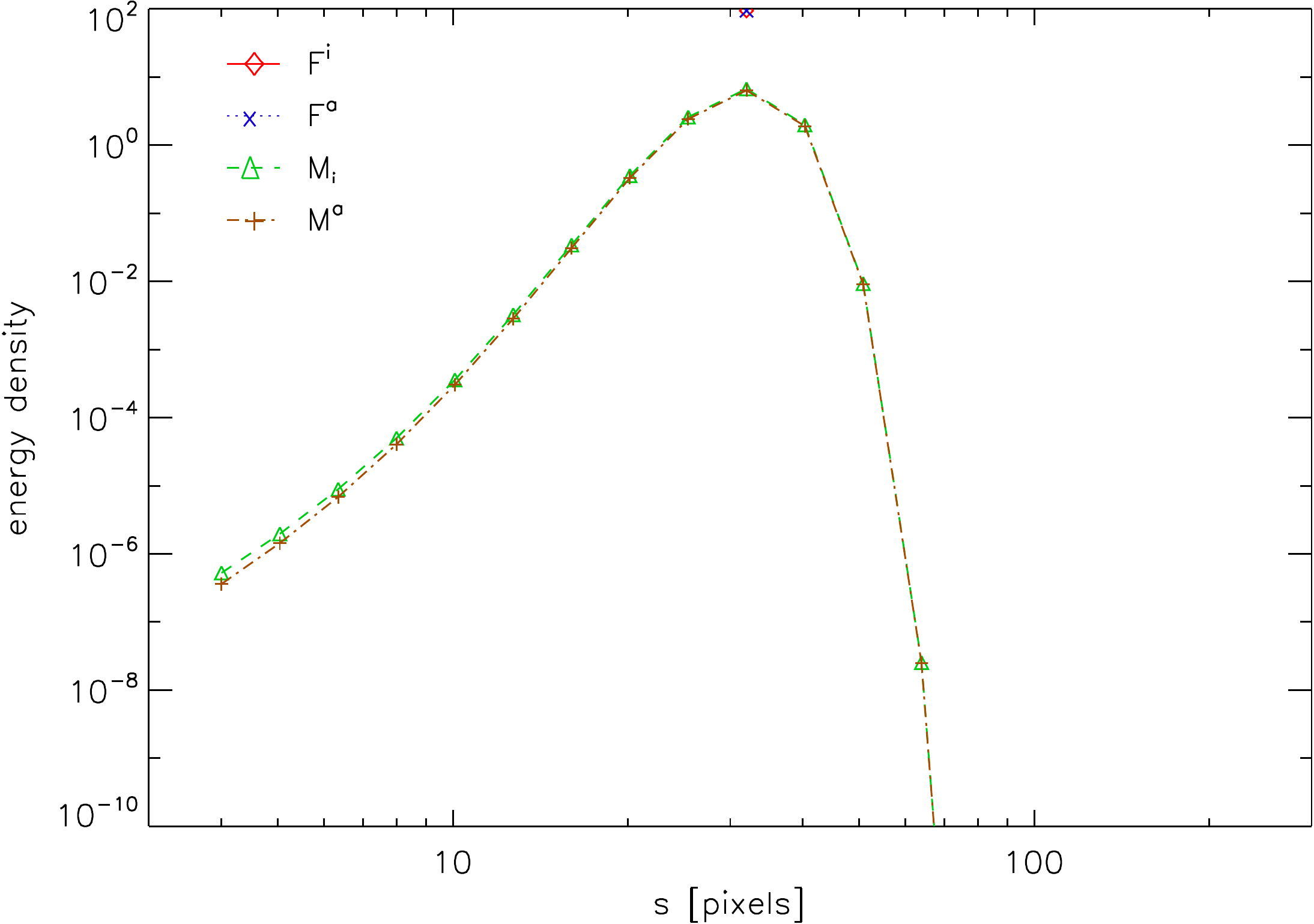}
\caption{Spectra of isotropic and anisotropic wavelet and Fourier coefficients
for a map containing a sinusoidal stripe pattern with $k_y=16$, i.e. a period
of $p=32$~pixels in a $512^2$ pixels image. The two Fourier spectra
consist only of a single point at $s=32$~pixels and are otherwise zero.
{\changed In all figures} we plot the absolute value for the anisotropic
Fourier spectrum. All other coefficients are real. The wavelet analysis used
a localization parameter $b=1$.}
\label{fig_stripes}
\end{figure}

\changed{
As the most simple anisotropic structure in Fourier space, we start
with a sinusoidal pattern.}
In Fig.~\ref{fig_stripes} we test a sine pattern in $y$
direction with a wave number of $k=16$ providing a period of $32$ pixels
in a $512^2$ pixels image, representing an extremely anisotropic
structure. The Fourier spectra consist of a single
point at the characteristic structure length of $s=32$.
The wavelet spectra show a broader peak at the same scale
\changed{due to finite scale resolution of wavelets controlled by
the localization parameter $b$.
}
As the only structure
in the map is 100\,\% anisotropic, isotropic and anisotropic
coefficients fall basically on top of each other.
The resulting ratios
between anisotropic and isotropic coefficients, providing the
degrees of anisotropy, are therefore close to unity at all scales
where the structure is detected and drop to a value of $\approx 0.8$ at
small scales where no structure is detected any more.
There is no difference between the global and local degrees
of anisotropy as the only structure is globally anisotropic.

When combining stripes in orthogonal directions, the $\pi$
symmetry in our definition of the anisotropic coefficients
in Eqs.~\ref{w_part} and \ref{f_part} leads to a cancellation of
the contributions from structures with a relative angle of
90~degrees. Therefore we find vanishing anisotropies when
analyzing the sum of two orthogonal patterns like
Fig.~\ref{fig:appx_std2} or from star-like structures. In principle
this limitation can be overcome by including higher moments
in our definitions but for the practical purpose of quantifying
filamentary structures those cases should be irrelevant.

\subsubsection{Noise maps}
\label{sec:noisemaps}

\changed{A special type of isotropic structures, well defined
in Fourier space, is given by fractional
Brownian motion structures \citep[fBm's, see e.g.][and Fig.~\ref{fig:appx_fbm}]{Stutzki1998}.
They are given by a power-law power spectrum $F(k) \propto k^{1-\beta}$
and random phases. One example is given by white noise maps having an
energy density spectrum $F(k) \propto k$. Other spectral
indices $\beta \ne 0$ correspond to maps of colored noise. The wavelet analysis
for two fBm's is shown in Appx.~\ref{appx_tests}.  We find
the same spectral index measured by the slope of the Fourier spectra and the
wavelet spectra. The spectra of anisotropic wavelet coefficients
are proportional to those of the isotropic coefficients
providing an almost constant local degree of anisotropy of about 0.3
independent of the spectral index. This indicates a significant
fraction of accidental local anisotropies that are caught by
the wavelets when considering all orientations.
When only considering global anisotropies, as measured by the
Fourier coefficients or the sum of anisotropic wavelet coefficients
including their phase (Eq.~\ref{eq:dwglob} and \ref{eq_da}),
we obtain vanishing global degrees of anisotropy below $10^{-4}$.}

\subsubsection{Gaussian clumps}
\label{sec:simpleGaussian}

\changed{
Gaussians represent a type
of clumps that is sharply confined and numerically well-behaving.
In an elliptic representation having different widths along the two main axes
$a$ and $c$ Gaussians are given by
\begin{equation}\label{eq_gaussian}
  g(x,y)=\hat{g} \times \exp\left({- \frac{(x \cos \phi + y \sin \phi)^2}{2\sigma_a^2}
-\frac{(x \sin \phi - y \cos \phi)^2}{2\sigma_c^2}}\right),
\end{equation}
where $\hat{g}$ specifies the amplitude, $\phi$ the angle of the $c$ axis in the $x, y$ plane
and $\sigma_a$ and $\sigma_c$ are the standard deviations of the
ellipse in both directions.}

\changed{For the most simple case of a single Gaussian clump,
analytic formulae for the wavelet spectra are derived in Appx.~\ref{appx_an}.
Here we only summarize the most important parameters and show the
results for localization parameters $b$ between $1/\sqrt{2}$ and $\sqrt{2}$.}

\begin{figure}[ht]
\centering
\includegraphics[angle=90,width=0.88\columnwidth]{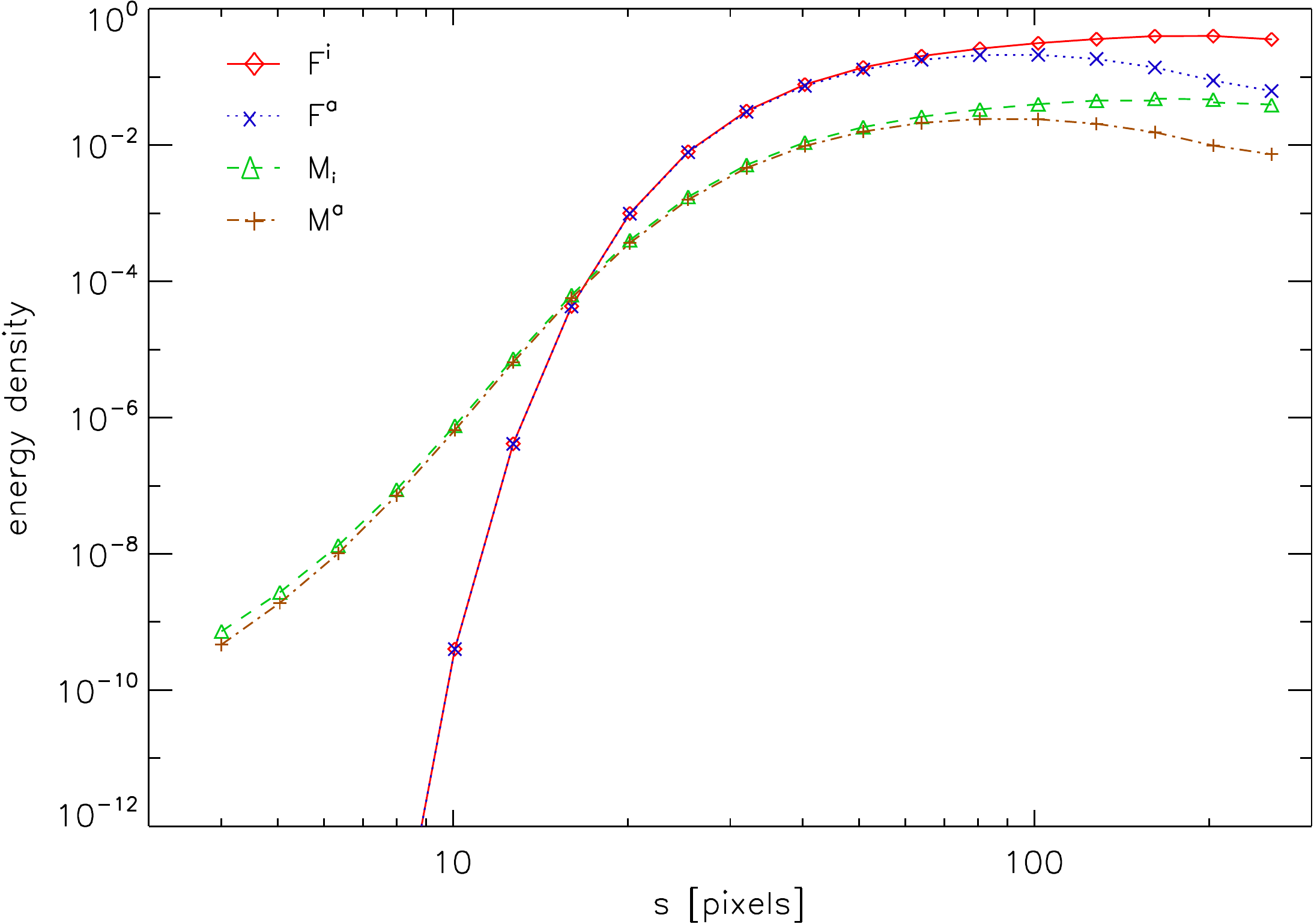}\vspace{0.3mm}
\includegraphics[angle=90,width=0.87\columnwidth]{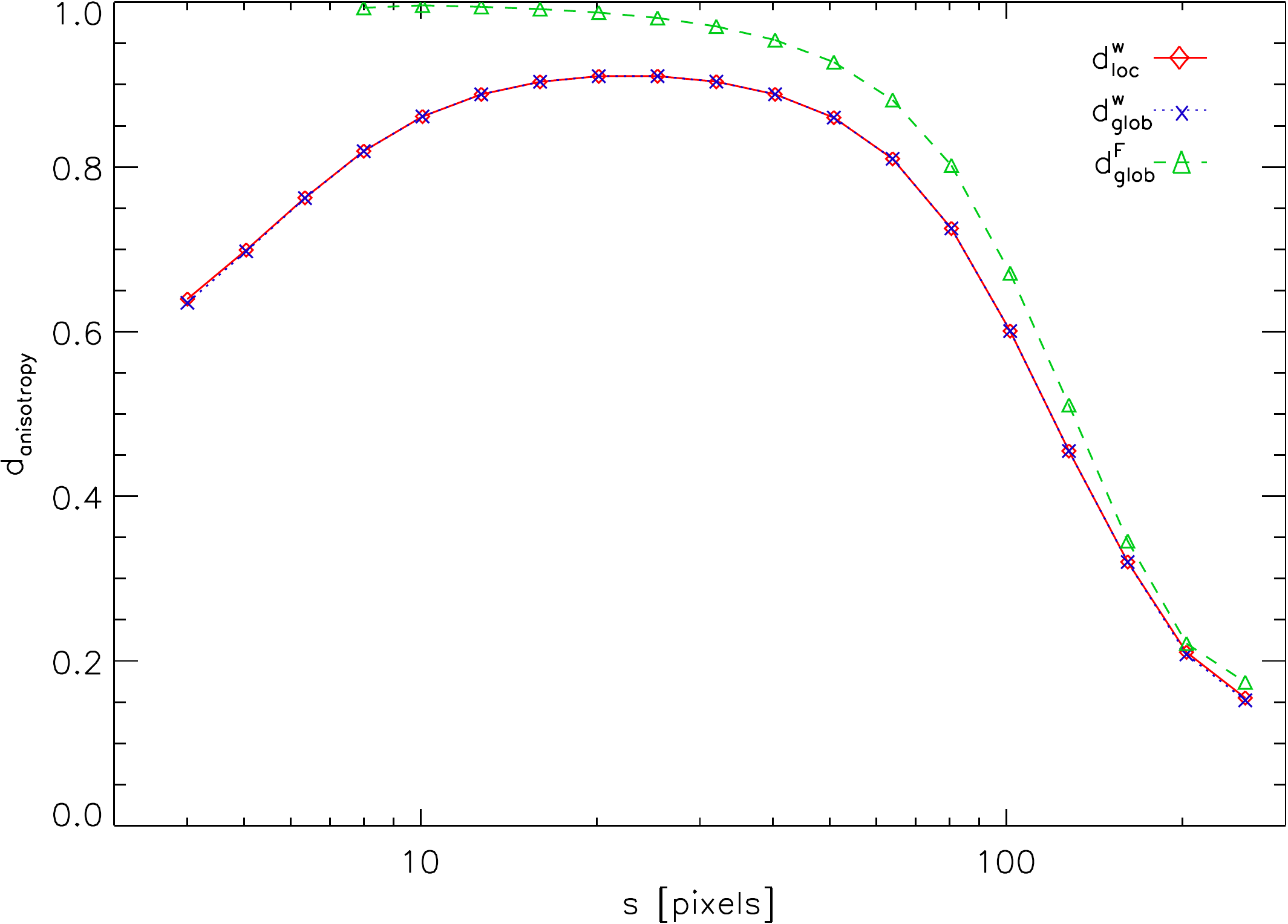}
\caption{Anisotropic wavelet analysis for a map containing
an anisotropic Gaussian clump with standard deviations of $\sigma_a=8$~pixels,
$\sigma_c=32$~pixels. The upper panel shows the isotropic and anisotropic
wavelet and Fourier spectra and the lower panel shows the resulting
local and global degrees of anisotropy. }
\label{fig:singleGaussian}
\end{figure}

\changed{
Figure~\ref{fig:singleGaussian} shows the results from the analysis of a
Gaussian clump with axes of $\sigma=8 \times 32$~pixels when using a filter parameter $b=1$.
It contains the isotropic and anisotropic wavelet and
Fourier spectra (Eqs.~\ref{w_parts}, \ref{eq:fourier_i}, \ref{f_part}) in the upper
panel and the local and global degree of anisotropy measured through wavelet and
Fourier coefficients (Eqs.~\ref{eq:dwloc}, \ref{eq:dwglob}, \ref{eq_da}) in the
lower panel.
The shape of the spectra of wavelet and Fourier coefficients reflects
the size of the Gaussians. The isotropic spectra, $M^i$  and $F^i$,
show the same behavior at scales above about 30~pixels. Their maxima appear at
about 180~pixels in a broad peak.} The Fourier spectra, $F^i$ and $F^a$,
show a steep increase from about 8 to 30 pixels. The spectra
of wavelet coefficients, $M^i$ and $M^a$, have a shallower increase
but anisotropic and isotropic coefficients \changed{behave very similarly for both quantities.}
The anisotropic spectra peak at smaller scales \changed{so that
the degrees of anisotropy, given by the ratio of the spectra, peak at
even lower scales. The wavelet-based degrees of anisotropy,
$d^w\sub{loc}$ and $d^w_{glob}$, have their maximum at $s^*\equiv s(d^w_{loc})\sub{max} \approx 23$~pixels; 
the Fourier-transform-based degree of anisotropy, $d^F\sub{glob}$, continues to rise up
to unity for the vanishing Fourier coefficients towards the small scale limit.
As there is only one anisotropic structure in the map, the local degree of
anisotropy is identical to the global degree of anisotropy.
The match between the peaks of the wavelet-based and the Fourier-transform-based
isotropic and anisotropic spectra indicate, as for the sinusoidal pattern, that
anisotropic wavelet and Fourier analysis measure the same scale of the maximum
energy density for isotropic and anisotropic variations.
}

\begin{figure}
\centering
\includegraphics[angle=90,width=0.88\columnwidth]{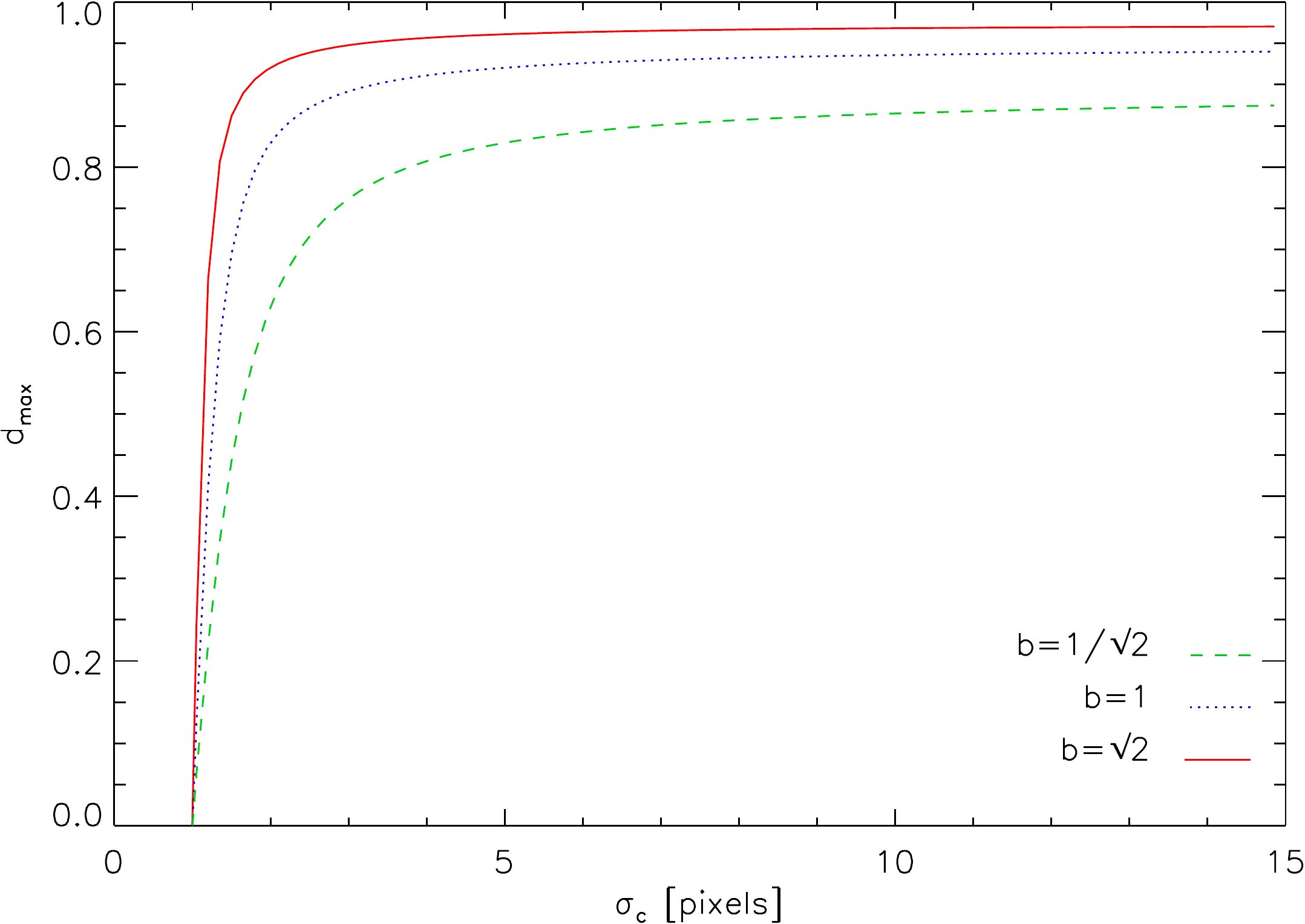}
\caption{Peak value of the degree of local and global anisotropy from
the wavelet analysis of Gaussians as a function of their aspect ratio.
\label{fig:gaussian-dmax}}
\end{figure}

\changed{The general formulae from Appx.~\ref{appx_an} show that
the peak of the local degree of anisotropy $d^w_{loc}(s^*)$ measures
the aspect ratio of the Gaussian clumps $\sigma_c/\sigma_a$. This is demonstrated in
Fig.~\ref{fig:gaussian-dmax} giving the maximum degree of anisotropy as
a function of the width of the major axis of a Gaussian when the minor
axis is fixed to $\sigma_a=1$~pixel. For aspect ratios below three we find a steep
dependence. For larger aspect ratios the maximum degree of anisotropy saturates.}

The location of the peak of the degree of anisotropy falls at
$s^*=\sqrt{2 \sigma_a \sigma_c}/b$.
However, with the extended plateau around the peak, the exact position of the peak
is hard to determine when applying the method to observational data
that may be affected by noise and other observational errors.
It is better to parameterize the whole plateau, by measuring
its edges. In agreement with typical observational errors in the order
of 10\,\% we define the plateau here as the region where the spectrum
falls above 90\,\% of the peak value.

\begin{figure}
\centering
\includegraphics[angle=90,width=0.88\columnwidth]{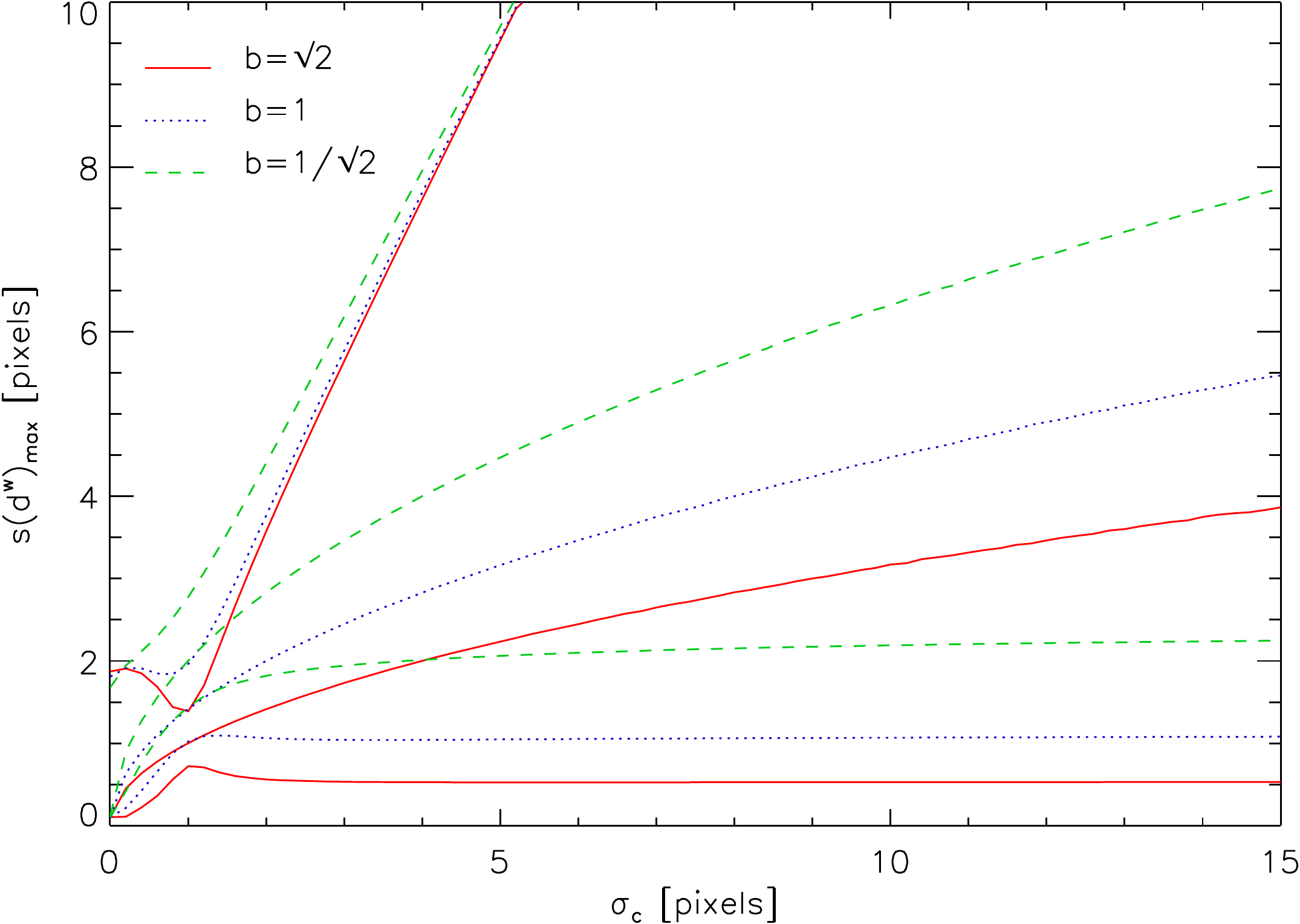}
\includegraphics[angle=90,width=0.88\columnwidth]{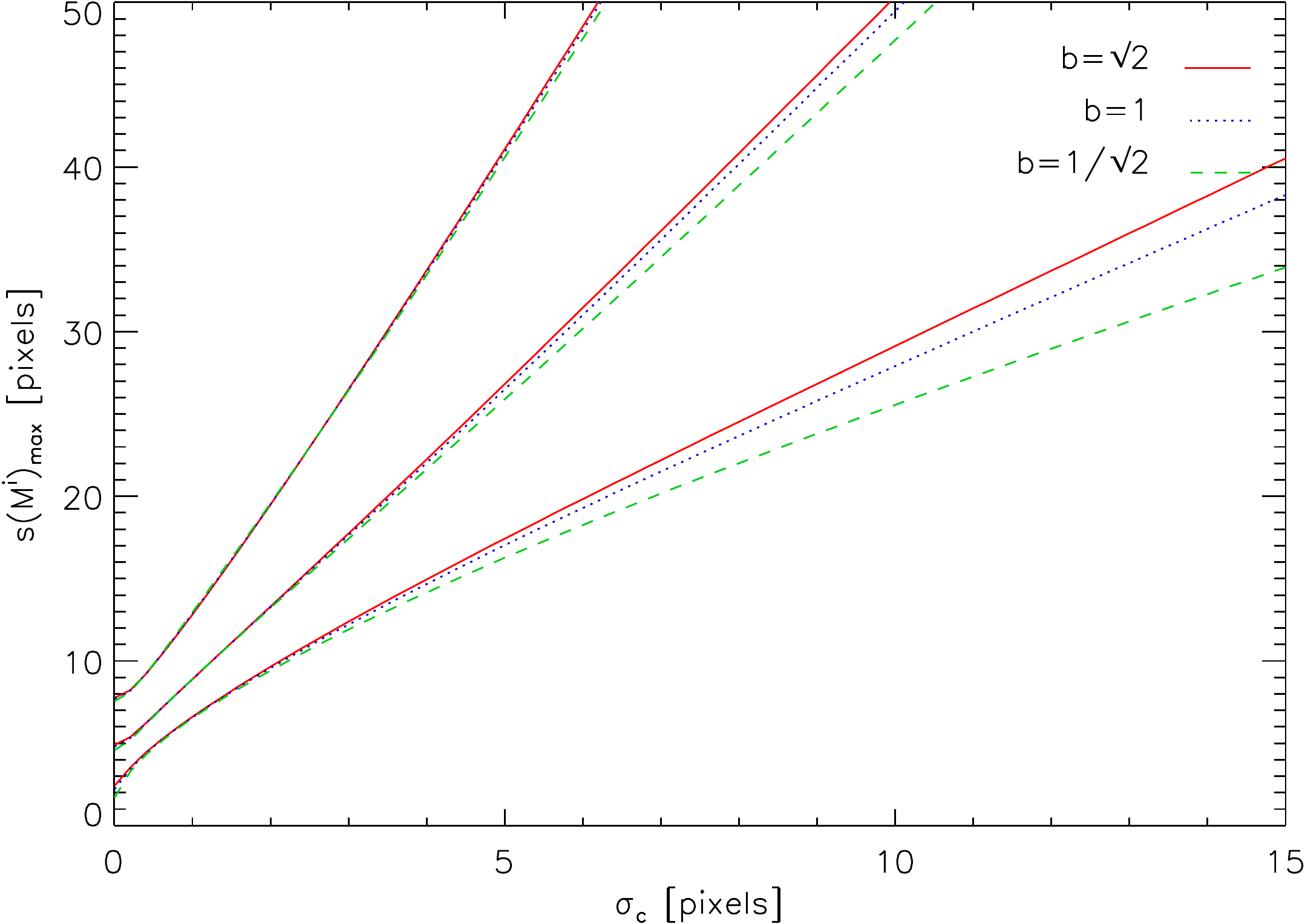}
\includegraphics[angle=90,width=0.88\columnwidth]{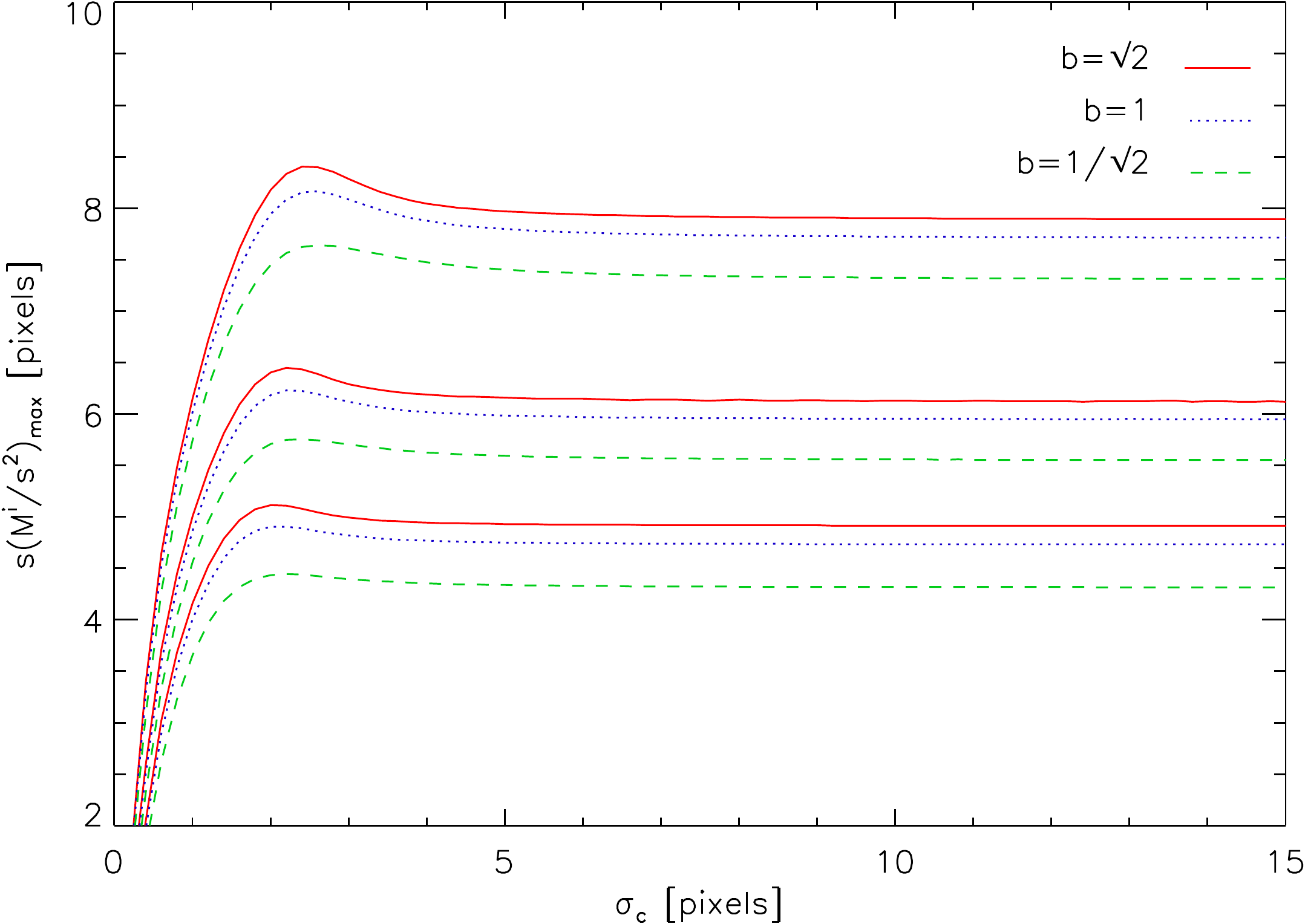}
\caption{Dependence of the peak region of the degree of anisotropy (upper plot),
the spectra of isotropic wavelet coefficients, $M^i$ (central plot),
and the \changed{quadratic-}scale-normalized spectra of isotropic wavelet coefficients,
$M^i/s^2$ (lower plot) on the \changed{aspect ratio} of the Gaussian ellipses. We changed the
\changed{major axis $\sigma_c$ of the ellipses while keeping
the minor axis $\sigma_a=1$~pixel} constant. For every filter,
three curves are plotted giving the location of the peak and the two
plateau edges at 90\,\% of the peak.
\label{fig:gaussian-plateau}}
\end{figure}

In Fig.~\ref{fig:gaussian-plateau} we show the dependence of the plateau
parameters for the degree of anisotropy, $d^w$, and the spectra of isotropic wavelet
coefficients, $M^i$, on the \changed{aspect ratio} of the Gaussian ellipses.
As discussed in \changed{Sect.~\ref{sec:theory}, a better sensitivity
to prominent scales is achieved when normalizing the isotropic spectrum
by the quadratic spatial scales $M^i/s^2$.} Therefore we also show the peak parameters
for the $M^i/s^2$ spectra in the figure.
For every \changed{localization} parameter $b$ we plot the three curves that provide
the location of the peak and the two plateau edges at 90\,\% of the peak
value.

For the degree of anisotropy \changed{(upper panel)} the upper end of the
plateau hardly changes with the $b$ parameter. The edge can be \changed{approximated}
by a simple linear function $\hat{s}_{90\,\%} \approx 2\sigma_c-1/2\sigma_a$ in the
range $\sigma_c>\sigma_a$. The lower edge does not depend on $\sigma_c$, but $b$ with
$\check{s}_{90\,\%} \approx 1.1\sigma_a/b^2$. The square root $\sigma_c$-dependence of
the peak, $s^*=\sqrt{2 \sigma_a \sigma_c}/b$, thus falls between the constant and the linear behavior of the two
edges. \changed{The plateau gets wider with increasing aspect ratio.}

The peak and edges of the isotropic wavelet spectrum \changed{(central panel)}
hardly depend on $b$ and are linear in $\sigma_c$.
The peak position falls at about $s(M^i)\sub{max} \approx 9/5\sigma_c+5\sigma_a$ for $\sigma_c>\sigma_a$.
\changed{This means that the peak of the isotropic wavelet spectrum, $M^i$,
measures mainly the length of the filaments, not their width.}
\changed{In contrast, the peak of the renormalized spectra $M^i/s^2$ (lower panel) mainly
depends on the filament width for aspect ratios above 1.5. We can describe the plateau
by a square root $\sigma_c$ dependence at small \changed{aspect ratios} and a constant
for larger aspect ratios $\sigma_c/\sigma_a$: $(\check{s}_{90\,\%},s(M^i/s^2)\sub{max},\hat{s}_{90\,\%})
 \approx (3.9,4.8,6.2)\times \sqrt{\sigma_a \sigma_c}$ for $\sigma_c < 1.5\sigma_a$; $(4.8,6.0,7.7) \times \sigma_a $
for $\sigma_c>1.5\sigma_a$ when using the $b=1$ filter. There is a weak $b$-dependence
so that all values are higher by 2\,\% for the $b=\sqrt{2}$ filter and
lower by 5\,\% for the $b=1/\sqrt{2}$ filter.}
For circular structures, the peak falls at 4.8 times the radius, $\sigma_a$, consistent with
the comparison to the $\Delta$-variance in Appx.~\ref{sect_comp_deltavar}.

The results show that for elliptic structures with an axes ratio of \changed{1.5} or more,
the peak of the spectrum of isotropic wavelet coefficients is determined
by the length of the major axis \changed{(central panel)}, the peak of the spectrum
of scale-normalized wavelet coefficients is determined by the minor axis \changed{(lower panel)},
while the plateau of the degree of anisotropy extends from $\check{s}_{90\,\%}$, determined
by the minor axis, to $\hat{s}_{90\,\%}$, determined by the major axis \changed{(upper panel)}.

\subsubsection{Plummer profiles}
\label{sect_plummer}

\changed{
Column-density maps of interstellar clouds rarely show
boundaries as steep as Gaussian clumps. In particular filamentary structures
are rather described by Plummer profiles \citep[][]{Plummer1911}
\begin{equation}\label{eq_plummer}
  p(r)=\hat{p} \times \left(1+\frac{r^2}{R_{a}^2}\right)^{-({p-1})/{2}} \;,
\end{equation}
where $R_{a}$ describes the radius of an inner core with a flat density
distribution and the exponent $p$ the outer radial decay
\citep[see e.g.][]{Ostriker1964}.
Therefore we implement filaments with a Plummer-type radial structure as
another test data set.

The value of $p=2$ found e.g. by
\citet{Arzoumanian2011,Malinen2012,Juvela2012} for many interstellar filaments
represents another extreme compared to the Gaussians, as $p=2$ Plummer profiles
are spatially ill confined having a diverging total mass due to the
$1/r$ dependence at large radii. The divergence prevents us from
providing analytical solutions but numerical computations on a finite
domain still provide us with significant numbers.
For the major axis of the clumps we stick to a Gaussian profile for which
we already know the imprint on the wavelet spectra from
Sect.~\ref{sec:simpleGaussian}.
The full Plummer test clump structure used here is thus given as
\begin{eqnarray}
  p(x,y)=\hat{p} \times \left(1+\frac{(x \cos \phi + y \sin \phi)^2}{R_{a}^2}\right)^{-({p-1})/{2}}\nonumber  \\
	\times \exp\left({- \frac{(x \sin \phi - y \cos \phi)^2}{2\sigma_c^2}}\right)
\;,
\label{eq_plummer2D}
\end{eqnarray}
where $R_{a}$ is the inner core radius of the Plummer profile in the direction of the
minor axis and $\sigma_c$ is the standard deviations of the Gaussian profile in the
direction of the major axis. Consequently, we only consider $\sigma_c> R_{a}$.
}

\begin{figure}
\centering
\includegraphics[angle=90,width=0.88\columnwidth]{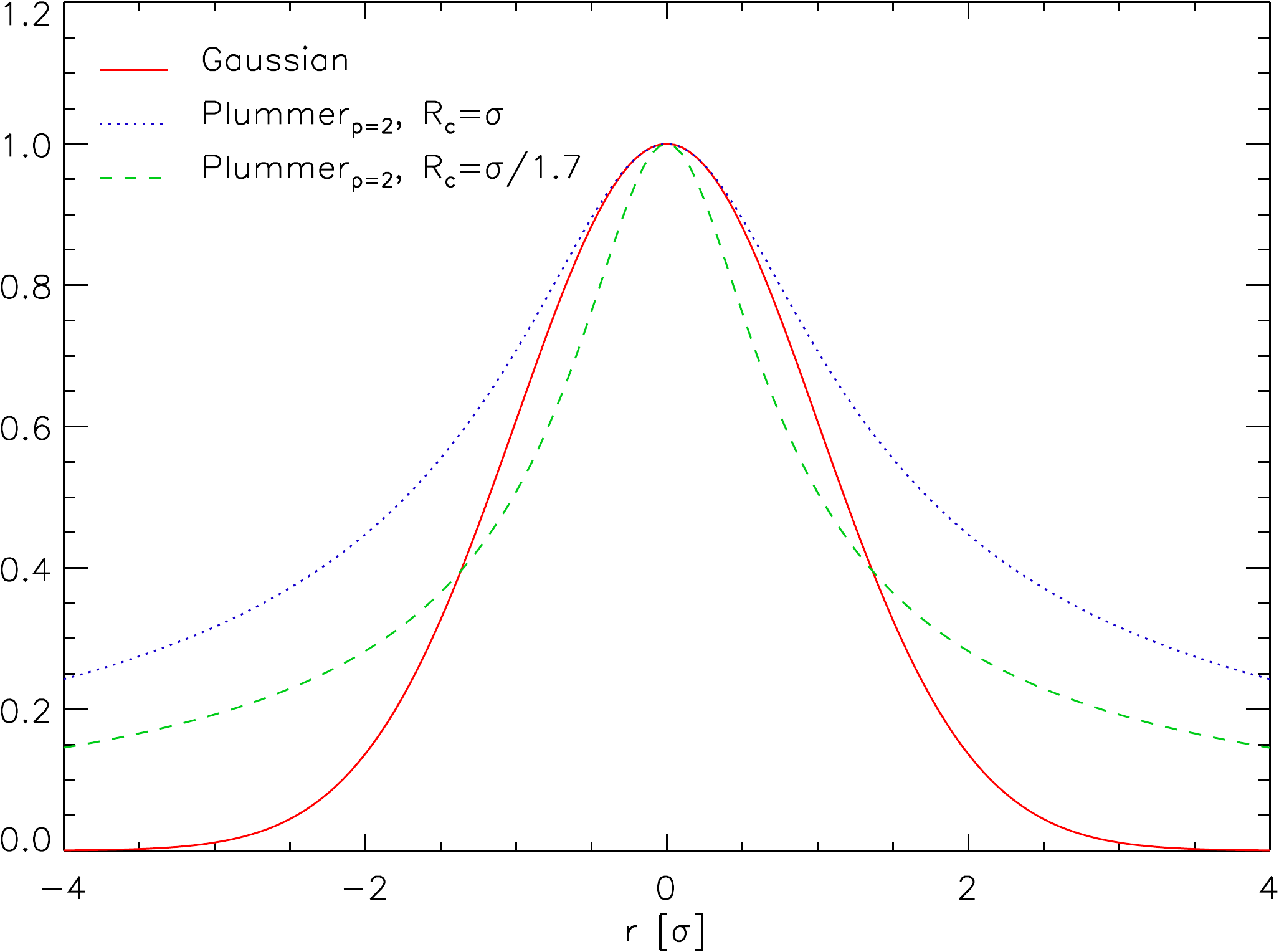}
\caption{Comparison of a Gaussian profile with a Plummer profile for $p=2$
when using the Gaussian standard deviation for the plateau radius $R_c$ or
using a 1.7 times smaller plateau radius leading to the same 40\,\% width.
}
\label{fig:plummer}
\end{figure}

\begin{figure}
\centering
\includegraphics[angle=90,width=0.88\columnwidth]{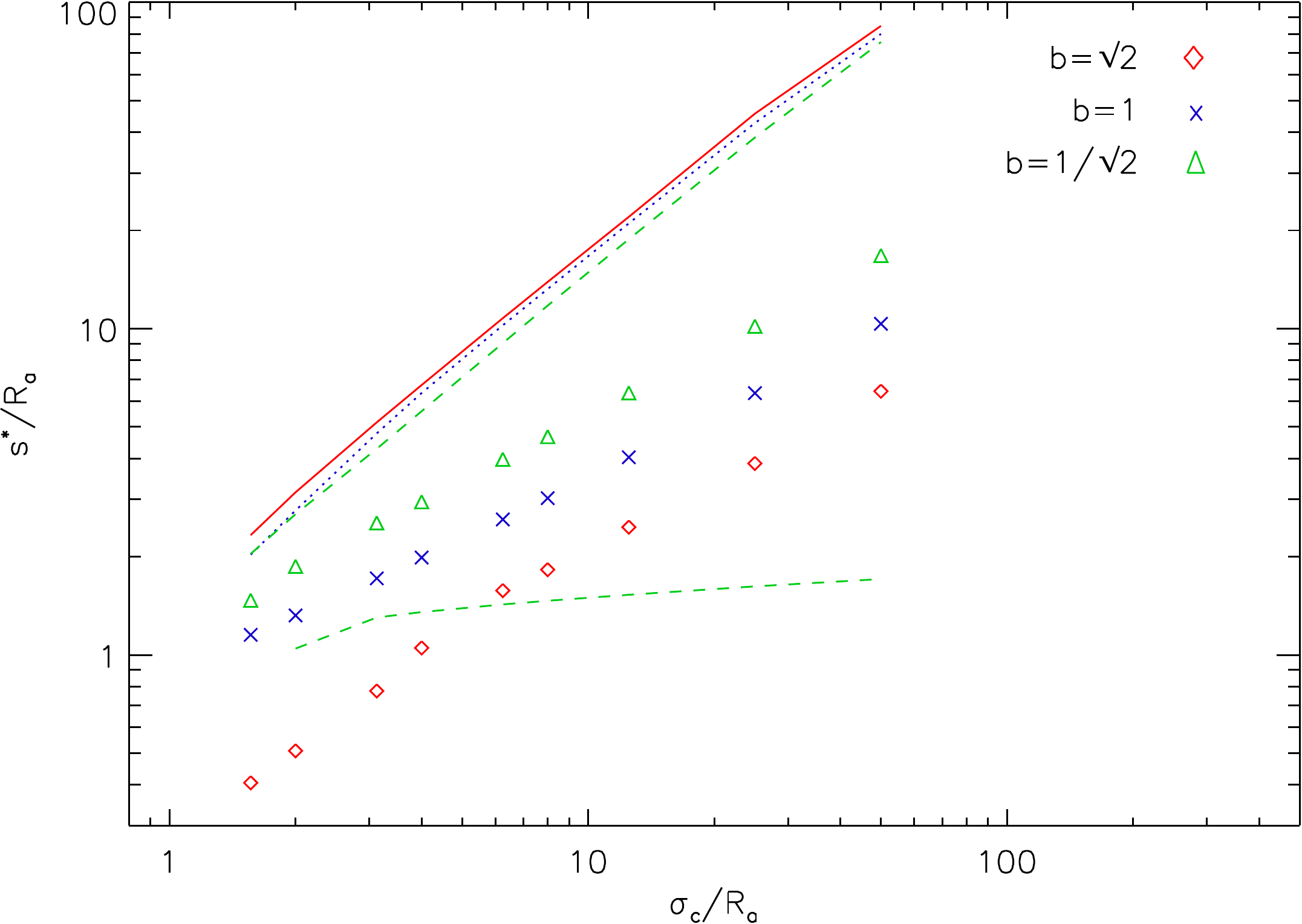}\vspace{3mm}
\includegraphics[angle=90,width=0.88\columnwidth]{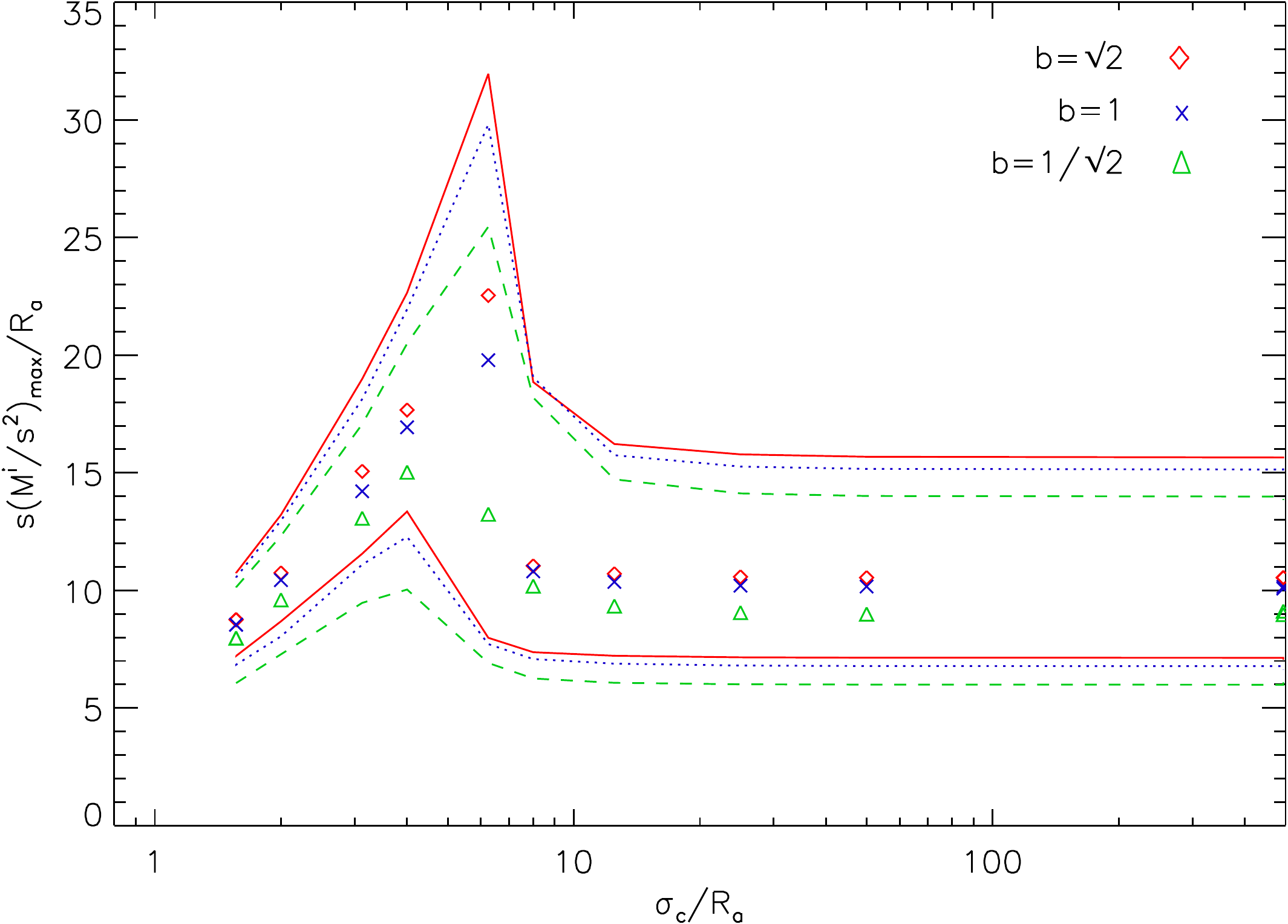}
\caption{Parameters of the peak plateau regions for the degree of anisotropy (upper plot),
and the \changed{rescaled spectra of isotropic wavelet coefficients,
$M^i/s^2$ (lower plot),} as a function of the aspect ratio of the Plummer ellipses.
We labeled the $y$-axes by the variables for the peak of the two quantities.
For every filter,
the central dots give the location of the peak and the two outer lines show the
plateau edges, at 90\,\% of the peak. For the wavelet coefficients, the dots at the
right edge of the plotted range represent the values for an infinite \changed{aspect ratio},
i.e. a one-dimensional Plummer profile. In the degree of anisotropy we were not
able to measure the lower edge of the plateau except for the $b=1/\sqrt{2}$
filter as the peak is extended and close to the minimum scale of the maps.
\label{fig:plummer-plateau}}
\end{figure}

\changed{
In reality all clumps should
show an intermediate behavior between the Gaussians with the steep
exponential boundaries and the Plummer profiles with $p=2$ having
very shallow boundaries.
Figure~\ref{fig:plummer} compares the radial profiles for the Gaussian case with
the $p=2$ Plummer distribution. For $R_a=\sigma$ the inner part of both profiles
agree but the Plummer profile has much shallower wings. We also show a Plummer
profile that is narrower by a factor 1.7. Below we demonstrate that this one has
the same peak of the renormalized wavelet spectra as the Gaussians (Fig.~\ref{fig:plummer-plateau}).
It corresponds to a match of the 40\,\% of the maximum levels between Gaussian and Plummer profile.
}

\begin{figure}[ht]
\centering
\includegraphics[angle=90,width=0.88\columnwidth]{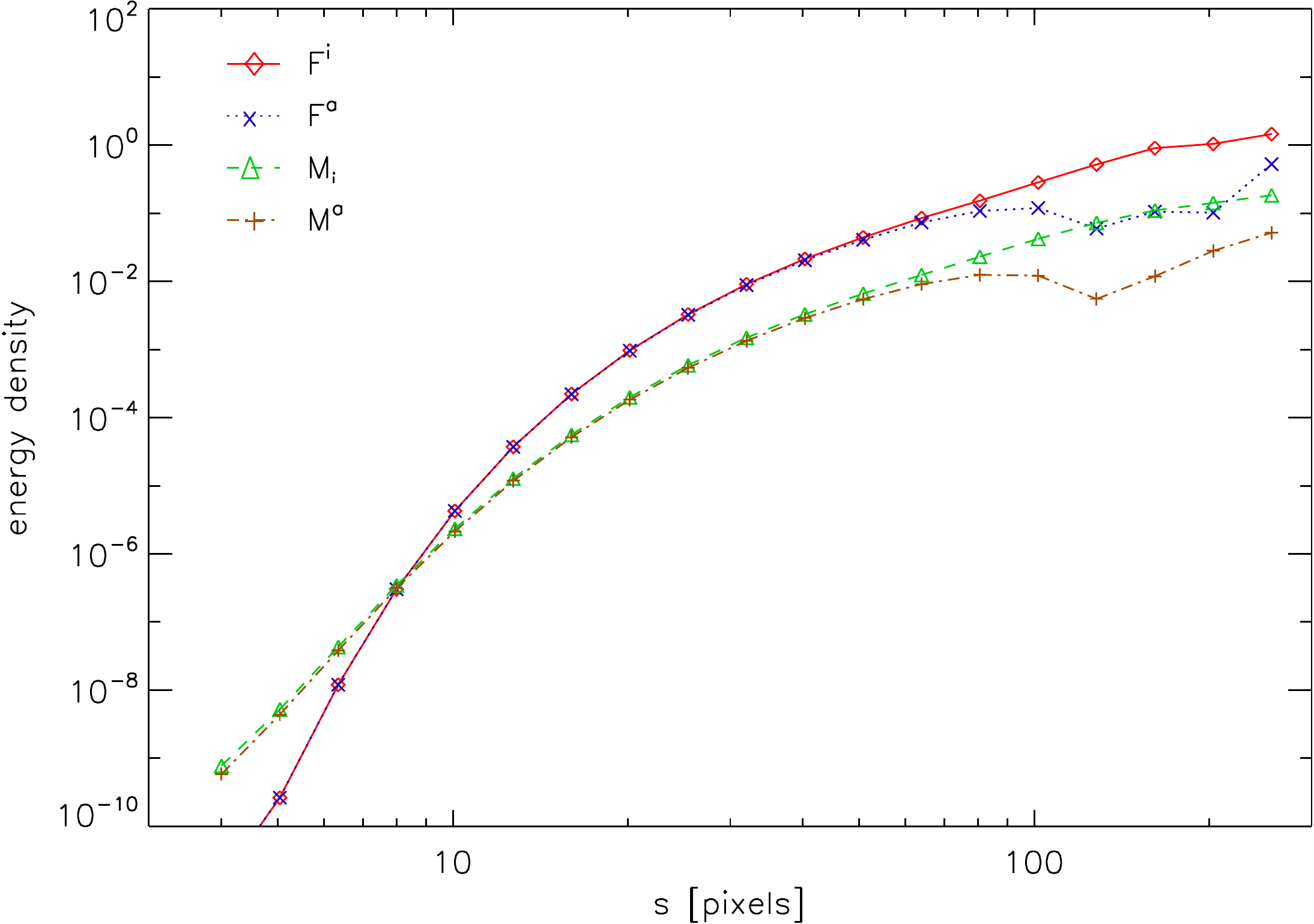}\vspace{0.3mm}
\includegraphics[angle=90,width=0.87\columnwidth]{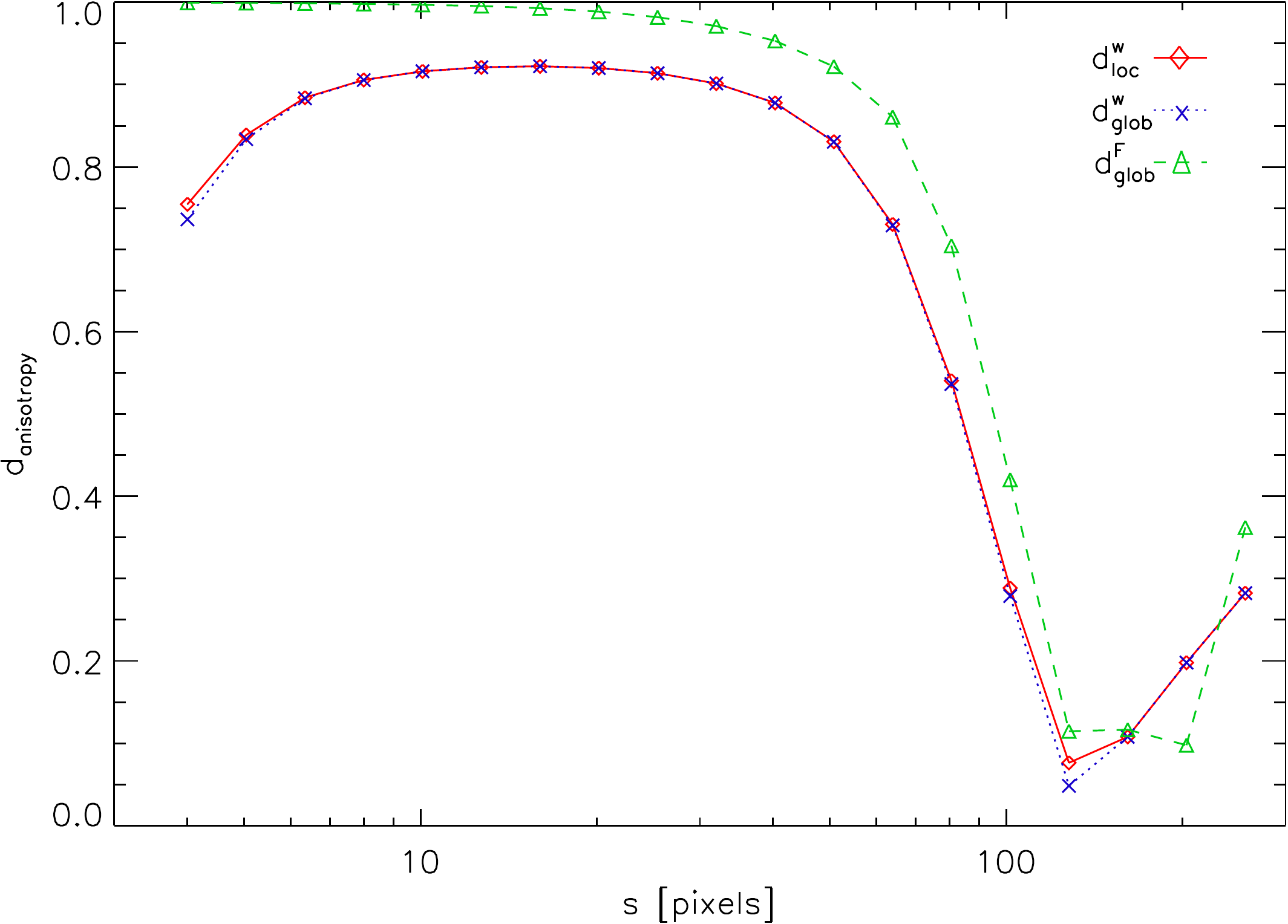}
\caption{Anisotropic wavelet analysis for a map containing
an anisotropic Plummer profile with a core radius $R_{a}=8$~pixels and
$\sigma_{c}=32$~pixels. The panels and parameters are the same as in
Fig.~\ref{fig:singleGaussian}.}
\label{fig:singlePlummer}
\end{figure}

\changed{Figure~\ref{fig:singlePlummer} shows the results from the wavelet
and Fourier analysis for a $p=2$ Plummer profile with a core radius of
$R_{a}=8$~pixels and a Gaussian length $\sigma_c=32$~pixels, equivalent to
Fig.~\ref{fig:singleGaussian}. Again, Fourier and wavelet spectra behave very
similar at scales above $s=30$~pixels while the Fourier spectra show a steeper
slope at small scales. For the Plummer profiles the isotropic wavelet coefficients
grow monotonically with scale showing no local maximum but continuing to rise
towards infinite size scales with a slope of $s^{3/2}$. The anisotropic spectra
show a local maximum at about 90~pixels leading to a peak of the wavelet-based
degrees of anisotropy at $s^* \approx 15$~pixels while the
Fourier-transform-based degree of anisotropy approaches unity towards the
small scale limit.  As there is only one anisotropic structure in the map,
local and global degrees of anisotropy agree.
Due to the unboundedness of the shallow profiles the energy density
is no useful quantity to measure the characteristic size of the clumps.
This can be overcome by rescaling the spectra by $s^{-2}$, providing the
amplitude per scale as mentioned in Sect.~\ref{sec:theory}. For the rest of
the paper we will therefore compare rescaled wavelet spectra, $M/(s^2 \sigma_f^2)$.}

\changed{In Fig.~\ref{fig:plummer-plateau} we show the dependence of the plateau
parameters for the degree of anisotropy, $d^w$ (upper panel), and the renormalized spectra of
isotropic wavelet coefficients, $M^i/s^2$ (lower panel), on the aspect ratio of the
Plummer profile clumps, $\sigma_c/R_a$.
For the three localization parameters $b=\sqrt{2}$, $b=1$, and $b=1/\sqrt{2}$
and a number of aspect ratios we give the location of the peak as a central dot
and the 90\,\% plateau edges, $\check{s}_{90\,\%}$ and $\hat{s}_{90\,\%}$ as
outer lines. In the degree of anisotropy the lower edge of the plateau falls
at very small scales, close to the minimum scale of the map, so that we were
only able to reliably measure this edge with the $b=1/\sqrt{2}$ filter.
For the wavelet coefficients, the points at the right edge of the plotted range
indicate the values for an infinite aspect ratio, given by an extended
one-dimensional Plummer profile.

For the degree of anisotropy (upper panel) we find a behavior that is close to that of
the Gaussian clumps when multiplying the plateau scales by a factor 0.8.
The upper end of the plateau is again independent of $b$ and well approximated
by $\hat{s}_{90\,\%} \approx 0.8\times (2\sigma_c-1/2 R_a)$. The dependence
of the peak location, $s^*\equiv s(d^w_{loc})\sub{max}$, on the aspect ratio is somewhat steeper than
for the Gaussians. Instead of an exponent of 0.5 we fit a common exponent of $0.63\pm 0.02$
when excluding the smallest aspect ratios for the $b=\sqrt{2}$ filter. However,
simply using the square-root scaling law from the Gaussian clumps, multiplied
by the factor 0.8, also gives a reasonable representation of the peak location
within 20\,\% when excluding aspect ratios above 40 and
the aspect ratios below 6 for the $b=\sqrt{2}$ filter.
The lower edge of the plateau, $\check{s}_{90\,\%}$, seems to fall at slightly
smaller scales than the 0.8-fold of the Gaussian value, only approaching it for
large aspect ratios, but we cannot give solid estimates here.
As a consequence we get an approximate match of the peak positions of the
degrees of anisotropy between Gaussians and Plummer profiles if we increase the
size of the Plummer profiles by the factor $1/0.8=1.25$ and exclude the $b=\sqrt{2}$
filter. In Sect.~\ref{sect_filtershape} we will see anyway that this filter is
less suited to study the degree of anisotropy.

For the normalized isotropic wavelet spectra, $M^i/s^2$ (lower panel), we also find a constant
position of the peak plateau, but only for large aspect ratios,
$\sigma_c/R_a \ga 8$. Like for the Gaussians, the peak position only depends
on the filament width in this range: $(\check{s}_{90\,\%},s(M^i/s^2)\sub{max},\hat{s}_{90\,\%})
 \approx (6.7,10.2,15.2) \times R_a$ when using the $b=1$ filter. For
the $b=\sqrt{2}$ filter, the values are higher by 4\,\%, for the
$b=1/\sqrt{2}$ filter lower by 10\,\%. Compared to the Gaussian ellipses
the scale of the lower edge of the plateau is 1.4 times higher, the
peak position 1.7 times higher and the upper edge 2.0 times higher. The
peak plateau is thus shifted by a factor 1.7 and widened by 40\,\%.
As seen in Fig.~\ref{fig:plummer}, a match between the peak scales of highly elongated
Gaussian and Plummer profiles is consequently reached when comparing the 40\,\% of the
maximum contours.

However, an excursion exists for aspect ratios $\sigma_c/R_a \approx 3 \dots 6$.
There, the location of the peak of the isotropic wavelet spectra would overestimate
the filament width by up to a factor of two. The deviation is small for the
$b=1/\sqrt{2}$ filter and largest for the $b=\sqrt{2}$ filter. Inspection of the
wavelet coefficient maps shows that this effect results from the varying match of the
filter shape to the wings of the clump. Due to the shallow decrease of the Plummer
profile in the direction of the minor axis, only the central contours of the
clump reflect the input aspect ratio. For an aspect ratio $\sigma_c/R_a= 6.0$,
the 40\,\% contour
only shows an aspect ratio of 3.5, and the 10\,\% contour is already almost round.
The filter with the wide localization parameter $b$ is more sensitive to these
broad structures resulting in an almost spherical distribution of the wavelet coefficients
at the scale of the peak of $M^i/s^2$. There it obviously does not trace the
filamentary structure any more. Instead of providing an additional fit for the
range of aspect ratios between $\sigma_c/R_a \approx 3 \dots 6$ we consider
this rather as an uncertainty of the method for the particular case of relatively
short filaments with a shallow radial profile. In all cases where the 40\,\%
contour has an aspect ratio of four or above, the fixed relation between the
peak of the $M^i/s^2$ spectra and the filament width can be used to reliably measure the
width of the 40\,\% contour independent of whether the filament profile is as steep
as a Gaussian or as shallow as a $p=$ Plummer profile. In terms of the 40\,\% of
the maximum width the peak of the $M^i/s^2$ spectra fall at $s(M^i/s^2)\sub{max} = 2.2\mathrm{FW}_{40\%}$.
When sticking to the commonly used full-width-half-maximum instead, the relation is
$s(M^i/s^2)\sub{max} \approx 2.7\mathrm{FWHM}$ but this introduces an uncertainty of 8\,\% due to
the unknown density structure falling somewhere in between the Gaussian and the
$p=2$ Plummer profile.}

\begin{figure}
\centering
\includegraphics[angle=90,width=0.88\columnwidth]{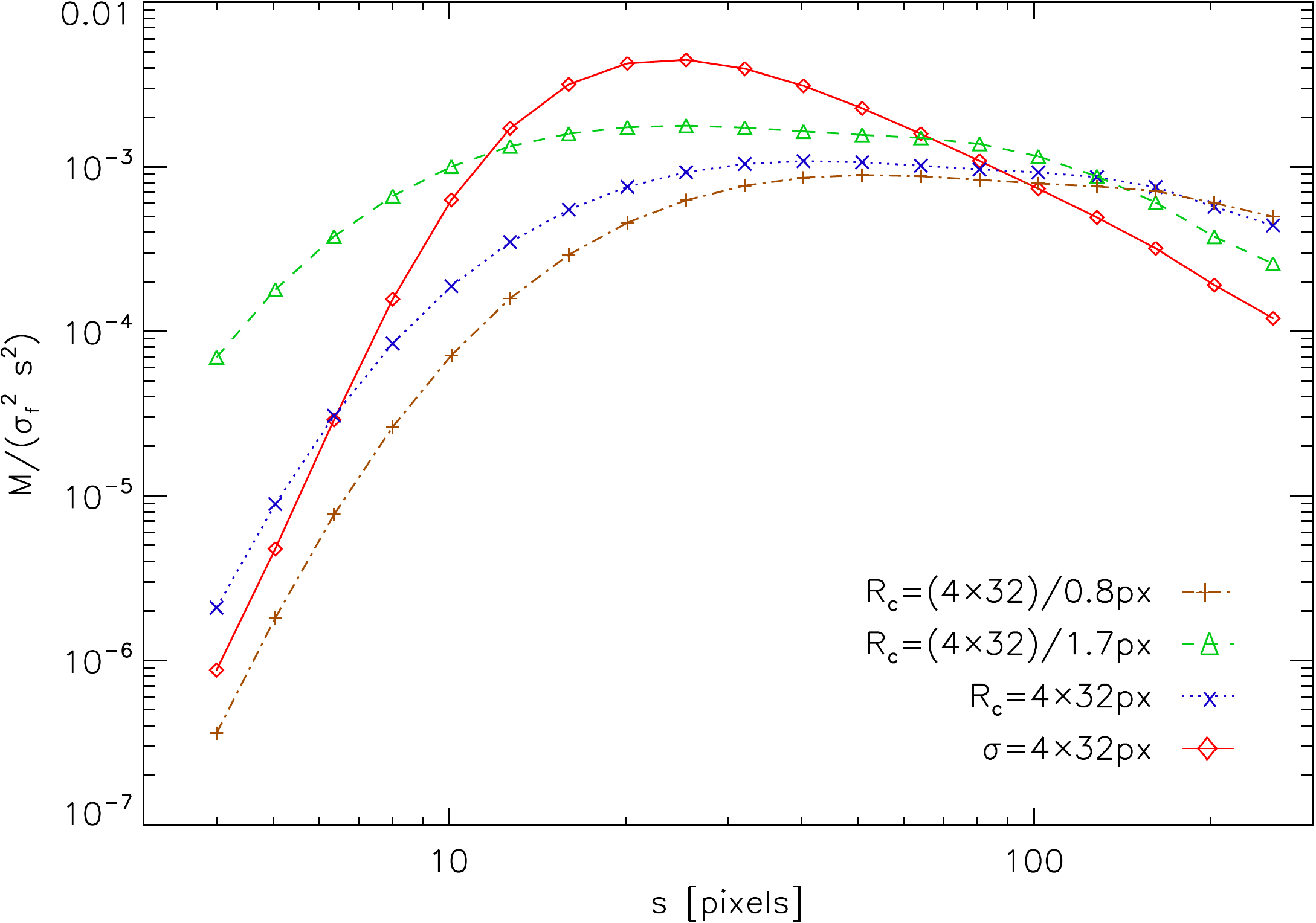}\vspace{3mm}
\includegraphics[angle=90,width=0.88\columnwidth]{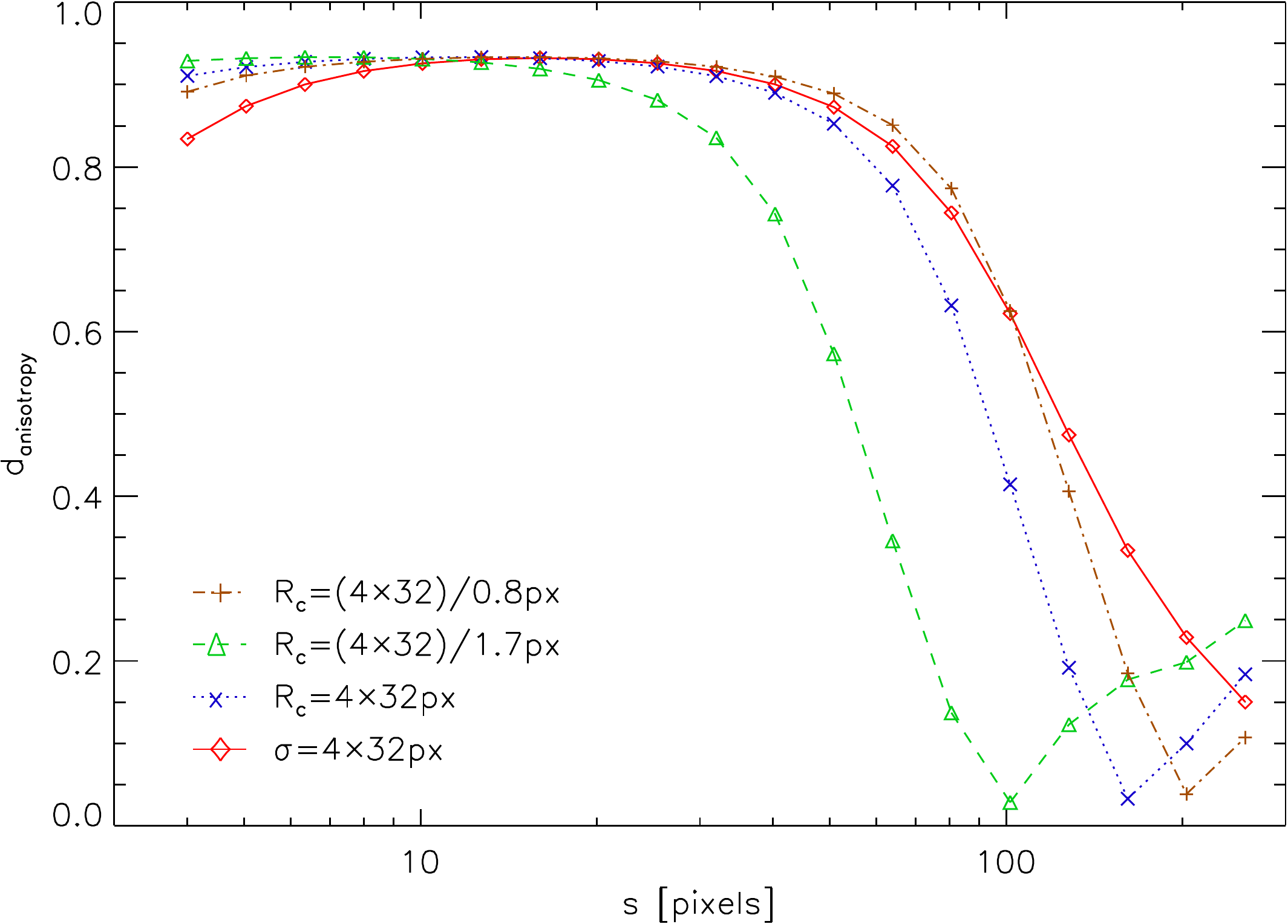}
\caption{Comparison of the rescaled spectra of the isotropic wavelet coefficients (top) and
the local degree of anisotropy (bottom) for elliptical Gaussian clumps with $\sigma_a=4$~pixel,
$\sigma_c=32~$~pixel and Plummer profile clumps with $R_a=4$~pixel and $\sigma_c=32$~pixel.
To test the fit relations for the peaks in the wavelet spectra and the degree of
anisotropy, we also show the results for Plummer profile clumps that are smaller by
a factor 1/1.7 and larger by the factor 1/0.8.}
\label{fig:plummer_spectra}
\end{figure}

\changed{To demonstrate this approach we compare the normalized isotropic spectra and the
local degree of anisotropy for Gaussian clumps and Plummer profiles with the same
axes ratios in Fig.~\ref{fig:plummer_spectra}.
The shallow density profile of the Plummer ellipses creates shallower
spectra in both quantities. When using the same axes $R_a=\sigma_a$ and $\sigma_c$
for both clumps, the isotropic wavelet spectra of the Plummer profiles
peak at much larger scales than for the Gaussian profiles and the degrees of
anisotropy peak at slightly smaller scales. A match of the peak position of the
wavelet spectra is achieved when reducing the size of the Plummer profile
by the factor of 1.7, a match in the peak position of the degree of anisotropy
when increasing the size of the Plummer profile by the factor 1.25.
Both quantities react in a different way to the shallower density profile.
For single-sized clumps with known aspect ratio one could therefore use the
comparison between the two to distinguish the density profile. Vice versa, if
the density profile is known one could derive the width and the aspect ratio.
For real maps consisting of multiple structures with different sizes, aspect
ratios, and density profiles this is however practically impossible so that
we concentrate here on a robust way to measure the width of the filaments
given by the 40\,\%-contours independent of the aspect ratio and the density
profile. Unfortunately, the scale sensitivity of the wavelet analysis is significantly reduced
when the analyzed structures are less pronounced being embedded in shallow density
halos. The plateau around the peaks is wider so that an accurate determination
of the peak location becomes more difficult than for sharply confined structures.
}


\subsubsection{Noise effects}

\changed{In observational data the analysis will be affected by observational
uncertainties, in particular noise. This may introduce an error in the
measurement of the characteristic scale of any structure. From Figs.~\ref{fig_stripes} and
\ref{fig:singleGaussian} it is
clear that in principle the maximum of the power spectrum measures the
characteristic scale most sharply, but only for infinitely extended structures.
For spatially constrained structures, the wavelet analysis provides a good
compromise.

\begin{figure}[ht]
\centering
\includegraphics[width=0.98\columnwidth]{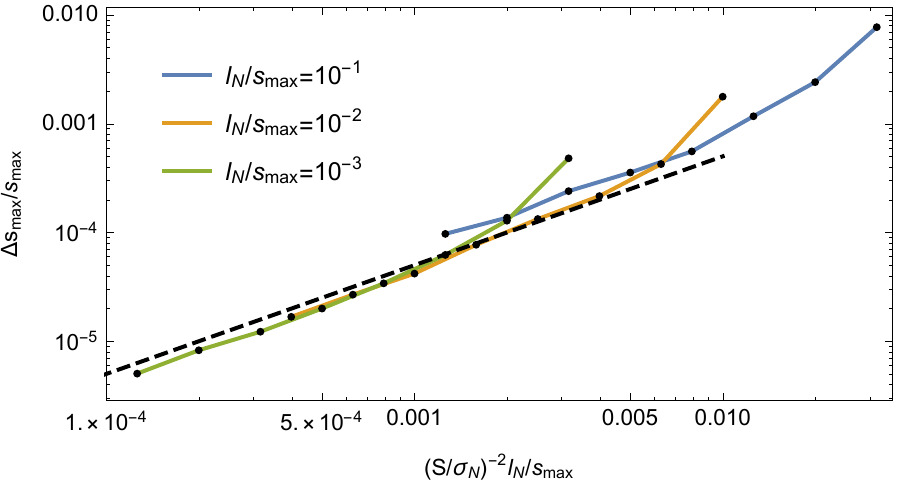}
\caption{Relative error of $s_{\rm max}$ simulated for noise with different values
of the signal-to-noise ratio, $S/\sigma_N$, and the relative noise correlation length,
$l_N/s_{\rm max}$, superimposed on Gaussian structures. The dashed line represents
the approximation for small noise correlation lengths, $\triangle s_{\rm max}/s_{\rm max}
= 0.05 (S/\sigma_N)^{-2} l_N/s_{\rm max}$.}
\label{fig:errors}
\end{figure}

The standard deviation of the wavelet coefficients $\sigma_w$ caused by
noise with a correlation length $l_N$ is approximately $\sigma_{w_s}^2\sim\sigma_N^2 l_N/s$.
This leads to the systematic variation of the wavelet spectra and a possible shift
of the peak of the rescaled isotropic spectra $s(M^i/s^2)_{\rm max}$ used for the
filament width measurement. We estimated the error for Gaussian profiles of
different widths but find qualitatively the same behaviour for all structures
discussed so far. Figure~\ref{fig:errors} shows the results of numerical tests,
providing the relative uncertainty of the peak position as a function of the
signal to noise ratio, $(S/\sigma_N)$, and the noise correlation length relative
to the characteristic structure scale, $l_N/s_{\rm max}$. Because the uncertainty
in $\sigma_{w_s}^2$ is determined by $\sigma_N^2 l_N$ the uncertainty in the
peak position also depends only on the product $(S/\sigma_N)^{-2}(l_N/s_{\rm max})$.
The dashed line represents a linear dependence, providing an approximate
relation for relative error of the peak position
$\triangle s_{\rm max}/s_{\rm max}=c (S/\sigma_N)^{-2} l_N/s_{\rm max}$ with $c\approx 0.05$
for $l_N<s$. Even for relatively large noise amplitudes, noise will thus only
affect our size estimates for size estimate at scales below
the noise correlation length. They are easily identified by increased variances
(see the discussion of the Polaris column density map in Sect.~\ref{sect:obs}).}

\subsection{Superposition effects}
\label{sect_superposition}

\begin{figure}[ht]
\centering
\includegraphics[angle=90,width=0.8\columnwidth]{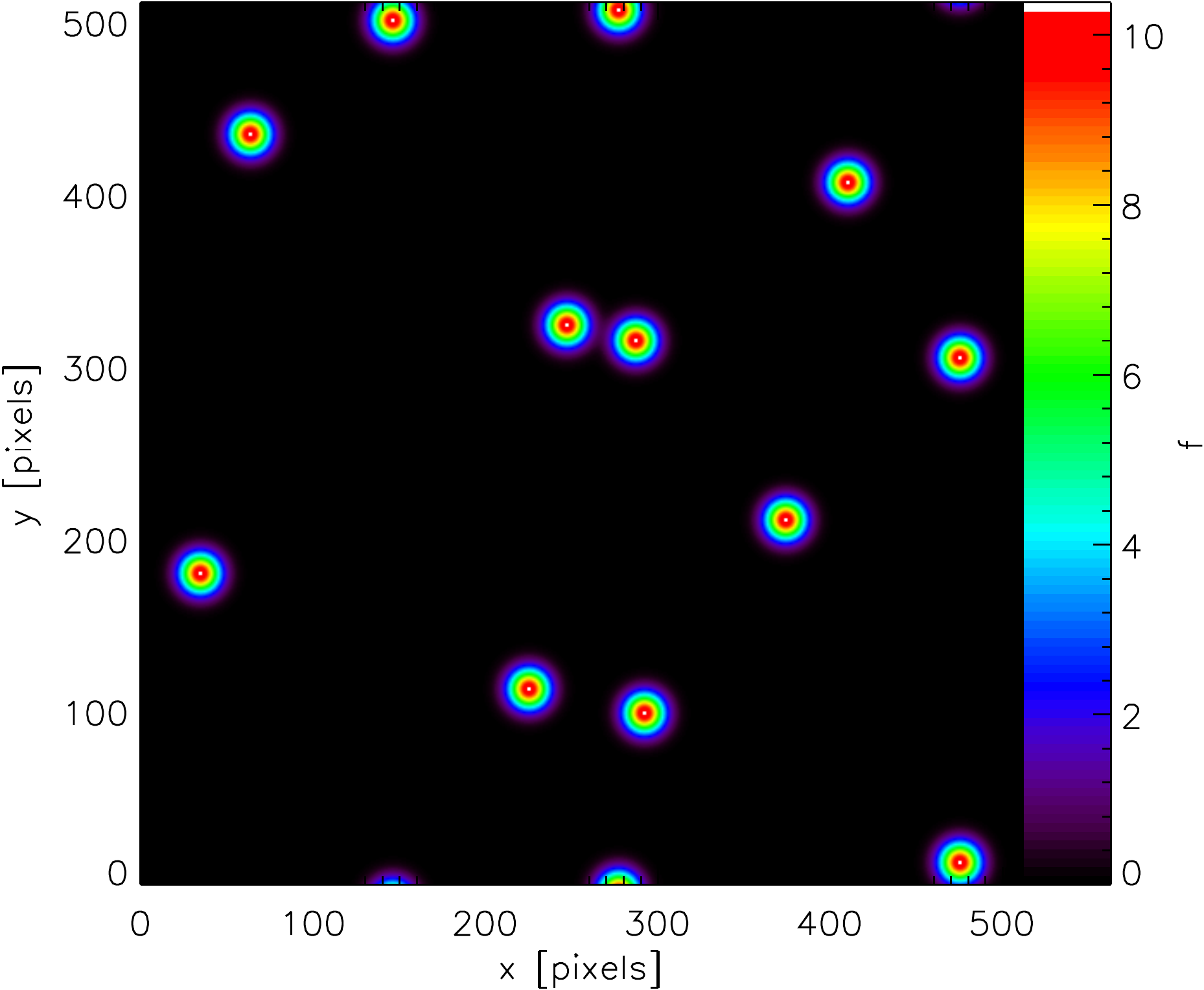}\vspace{0.3mm}
\includegraphics[angle=90,width=0.88\columnwidth]{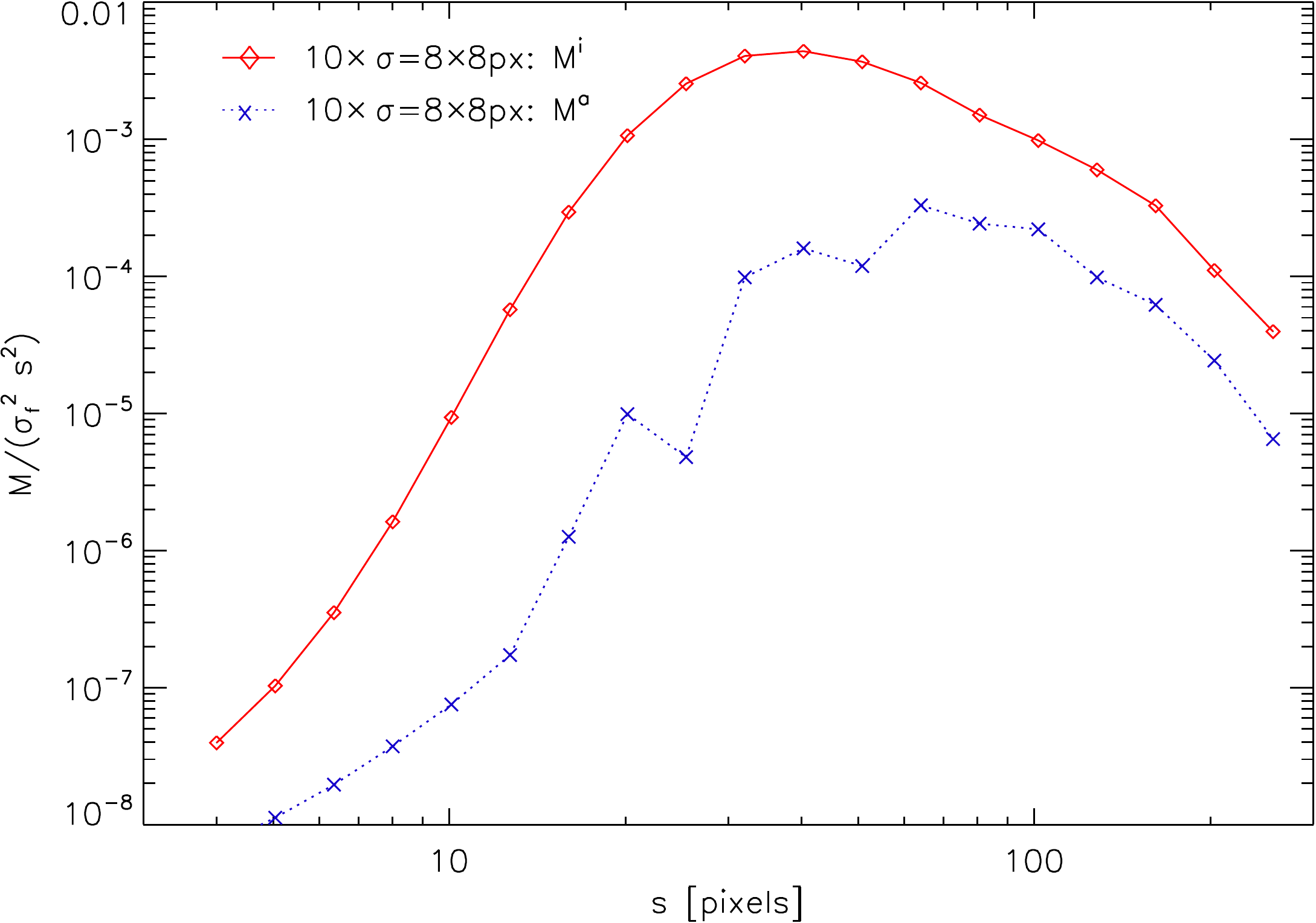}\vspace{0.3mm}
\includegraphics[angle=90,width=0.87\columnwidth]{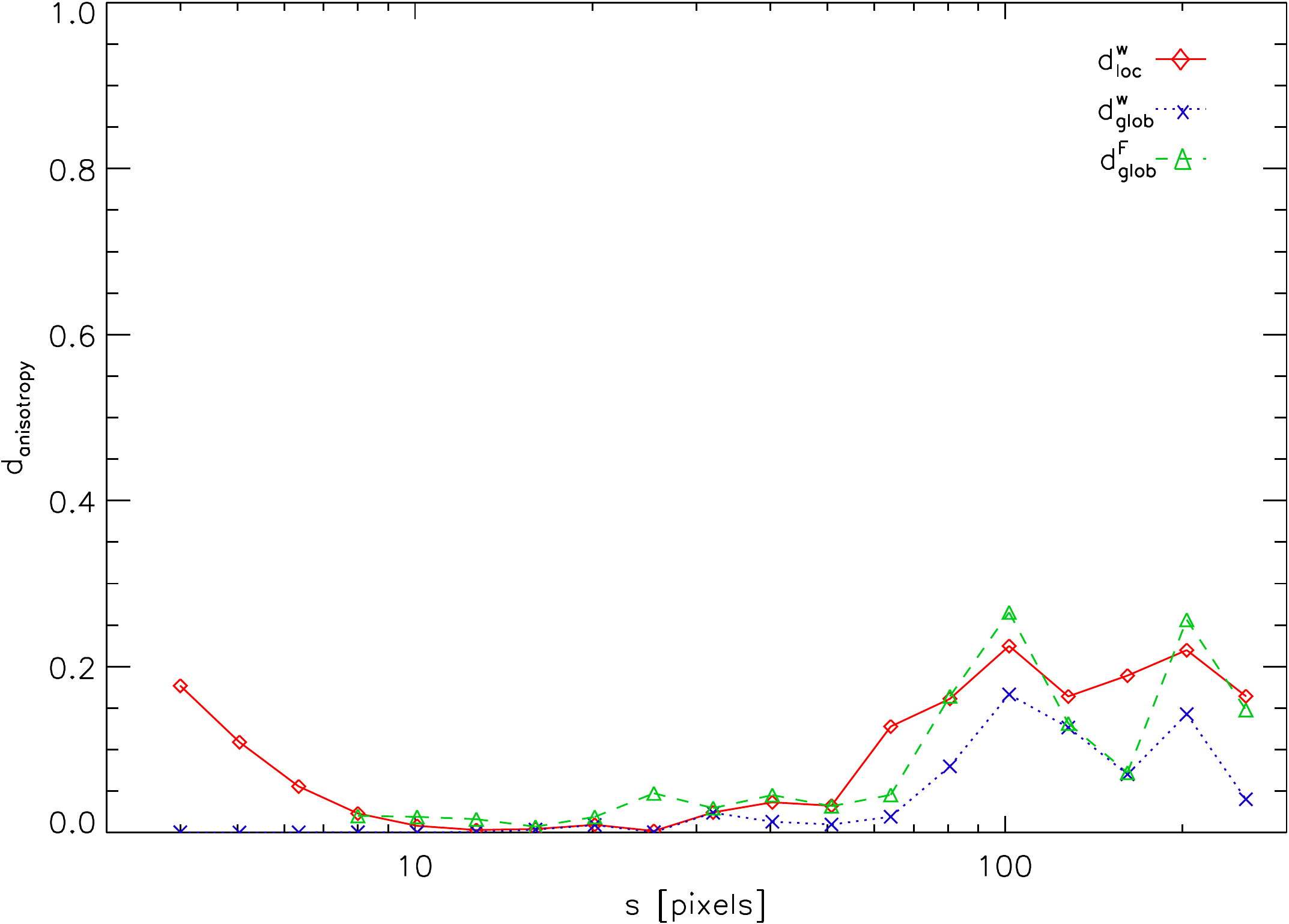}
\caption{Anisotropic wavelet analysis for a map containing
10 Gaussian clumps with a standard deviation of $\sigma=8$~pixels,
i.e. a full-width-half maximum (FWHM) of 19~pixels. The upper panel shows the original map, the
central panel shows the scale-normalized isotropic and anisotropic wavelet
spectra, and the lower panel shows the resulting
local and global degrees of anisotropy.}
\label{fig:clumps}
\end{figure}

\changed{Astronomical maps usually contain multiple structures.
As the wavelet analysis is a linear transformation, the wavelet spectra,
averaging the square of the transform, do not distinguish between a map
consisting of 10 separated clumps, like in Figs.~\ref{fig:clumps} and \ref{fig:ellipses},
and a map consisting of a single clump with an amplitude that is higher by the
factor $\sqrt{10}$. The distinction has to come from the spatial
distribution of the wavelet coefficients. The contribution of every
individual structure to the total wavelet spectrum is always determined
by its square-amplitude weighted fraction of the total map. The area
filling enters only as a simple scaling parameter that is eliminated
when we divide the spectra by the variance of the maps, $\sigma_f^2$,
as proposed in Sect.~\ref{sec:theory}. If there is more than one component
that contributes to a map it is therefore necessary to inspect the maps of
wavelet coefficients to judge the number and relative contribution
of different structures in the map.

However, the spatial correlation of multiple clumps
may create some effects that are not visible in the analysis of
the single structures considered above. To study superposition effects
we limit ourselves here to superpositions of Gaussians because they
are numerically well behaving and the general impact of the superposition effects
does not depend of the shape of the individual structures.}

\changed{
We start with the superposition of isotropic clumps. In
Fig.~\ref{fig:clumps} we use a random superposition of ten Gaussian clumps.
They} are inherently isotropic but may create some anisotropy due to their
random placement in the map.
\changed{The upper panel of the plot
shows the analyzed map, the central panel gives the scale-normalized isotropic
and anisotropic wavelet spectra, $M^i/(s^2 \sigma_f^2)$ and $M^a/(s^2 \sigma_f^2)$, and the
lower plot shows the local and global degree of anisotropy measured
through wavelet and Fourier coefficients.
The shape of the isotropic wavelet spectra roughly matches the behavior of
single Gaussians. In agreement with the
results from Sect.~\ref{sec:simpleGaussian} the peak of the renormalized
spectrum falls at at $s \approx 4.8\sigma_a$. The anisotropic wavelet spectrum
is non-negligible, roughly following the isotropic spectrum, but at a ten to hundred
times lower level.} The global
degrees of anisotropy rise to a noticeable level at scales
above 60~pixels where the random placement of the Gaussians creates
some larger structure. Small local anisotropies are also visible at scales
below 8~pixels where the filter picks up the edges of the individual
circles as some anisotropy, but obviously without any preferred
direction.

\changed{In the actual numerical implementation, we noticed that it is
very difficult to create perfectly isotropic structures. Placing isotropic
clumps randomly on a rectangular grid already creates small deviations due to the gridding.
These anisotropies are picked up by the wavelet coefficients. A 1\,\% deviation
from isotropy already creates a local degree of anisotropy of 2.5\,\% for the $b=1$
filter and of 5\,\% for the $b=\sqrt{2}$ filter. This is expressed in the very steep
rise of the maximum degree of anisotropy as a function of the aspect
ratio in Fig.~\ref{fig:gaussian-dmax}. We therefore interpret degrees of anisotropy
below 10--20\,\% as isotropic even if the anisotropic wavelet spectra do not vanish.}

\begin{figure}
\includegraphics[angle=90,width=0.87\columnwidth]{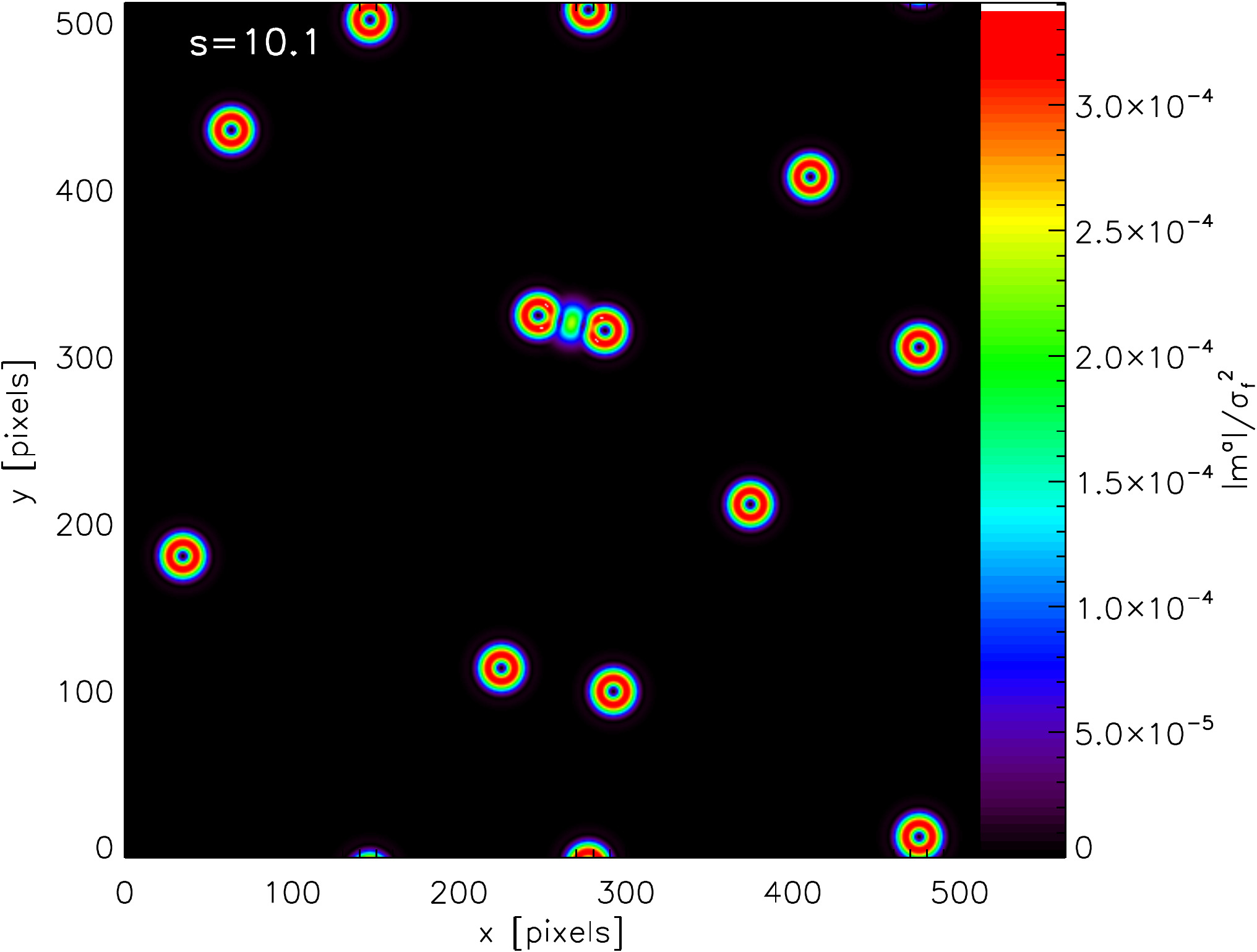}\vspace{3mm}
\includegraphics[angle=90,width=0.8\columnwidth]{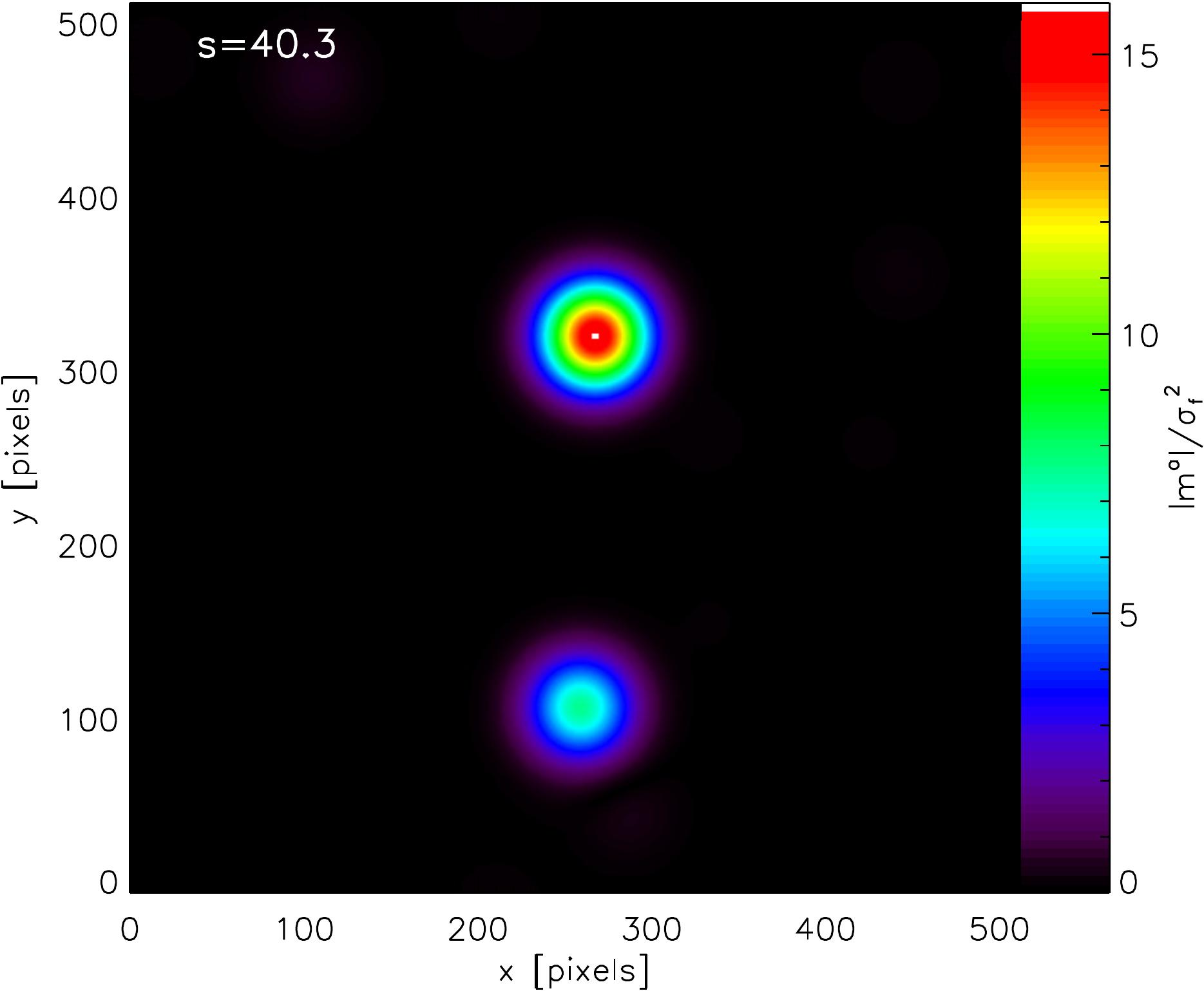}\vspace{3mm}
\includegraphics[angle=90,width=0.8\columnwidth]{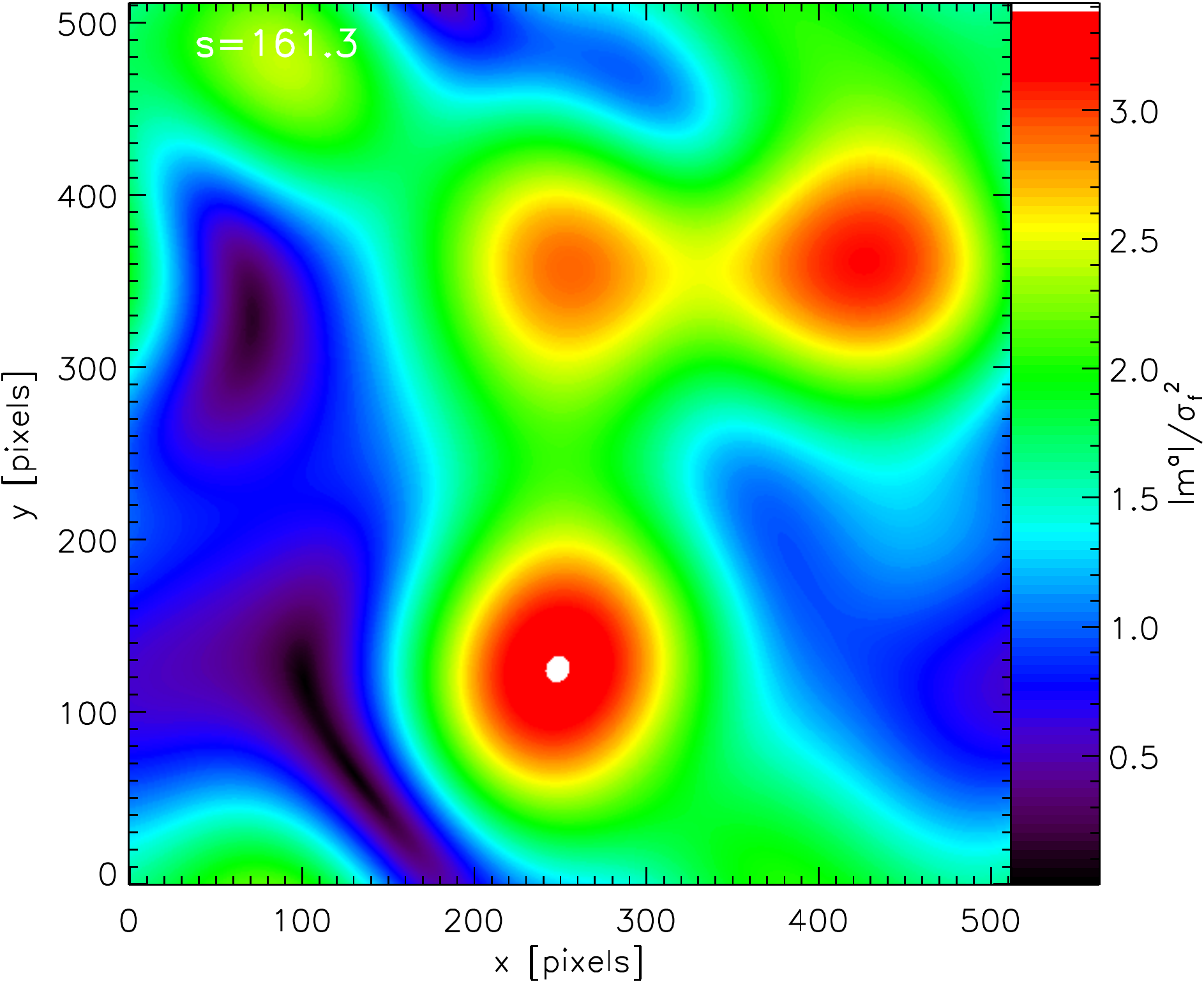}
\caption{Maps of anisotropic wavelet coefficients $|m^a(s,\vec{x})|$
for the field of 10 Gaussian clumps from Fig.~\ref{fig:clumps}.
The three panels show the coefficients for the filter sizes of
$s=10$ pixels, 40~pixels, and 160~pixels. \changed{To be independent of the
amplitude of the original signal we normalized the coefficients
by the map variance $\sigma_f^2$ as discussed in Sect.~\ref{sec:theory}.}
}
\label{fig_waveletmaps}
\end{figure}

To better understand the nature of the anisotropies
we show the maps of the \changed{individual anisotropic wavelet coefficients
$|m^a(s,\vec{x})|$, normalized by the total variance of the map, $\sigma_f^2$,
for some} filter sizes $s$ in
Fig.~\ref{fig_waveletmaps}. At small filter sizes, the curvature of the
edges of the Gaussians appears as anisotropy, but at a very low level.
If the filter size is in the order of the size of the individual
clumps, they are no longer visible as anisotropies, but the two closely
neighboring clumps in the upper central part of the map (Fig.~\ref{fig:clumps} top)
appear as main
anisotropic structure. With larger filter sizes, other groups of clumps
appear as dominant anisotropies, all at a similar level.
As those groups represent global structures, we find that
the local and global degrees of anisotropy in Fig.~\ref{fig:clumps}
grow from values close to zero to values of about 0.2 for scales above
60~pixels. Degrees of anisotropy of 0.2--0.3 are thus naturally
expected for random configurations of overall isotropic structures.
\changed{As seen in Sect.~\ref{sec:noisemaps}, degrees of anisotropy of 0.3 
are a robust limit for random noise signals.}
 
\changed{As mentioned above, the map of wavelet coefficients is moreover
a necessary tool to interpret the wavelet spectra in terms of the number of
contributions. At small scales, we can clearly count the ten clumps in the
map. At larger scales fewer structures contribute. The relatively low
number of structures at wavelet scales above the clump size confirms
that superposition effects are relatively small in this example so that
the wavelet spectra are still close to those of the individual clumps.}

\changed{
Equivalent tests for anisotropic structures, laid out in detail in
Appx.~\ref{appx_anisotropy}, show that the spectra of wavelet coefficients
and degrees of anisotropy for ensembles of anisotropic structures
are simply given by a combination of the behavior of individual
anisotropic clumps as discussed in Sect.~\ref{sec:simpleGaussian}
and the superposition effects discussed for the isotropic case above.
The isotropic wavelet spectra closely follow the spectra measured
for individual clumps and at scales up to about four times the major
axis of the clumps, $\sigma_c$, the local degree of anisotropy
also matches the curve of the individual clumps.  Differences only
occur at larger scales where the random superposition of the clumps
creates local and global anisotropies of about 20\,\%. The mutual
alignment of the individual clumps, however, creates a huge difference
in the global degree of anisotropy at small scales. For the random
angular placement of the clumps, the global anisotropy vanishes at
the scales below $4\sigma_c$ while for a parallel alignment the
global anisotropy is identical to the local anisotropy in this scale
range. Combining local and global degree of anisotropy as a function
of scale then allows us to characterize both the anisotropy of
individual structures and their mutual alignment.}

\begin{figure}
\centering
\includegraphics[angle=90,width=0.88\columnwidth]{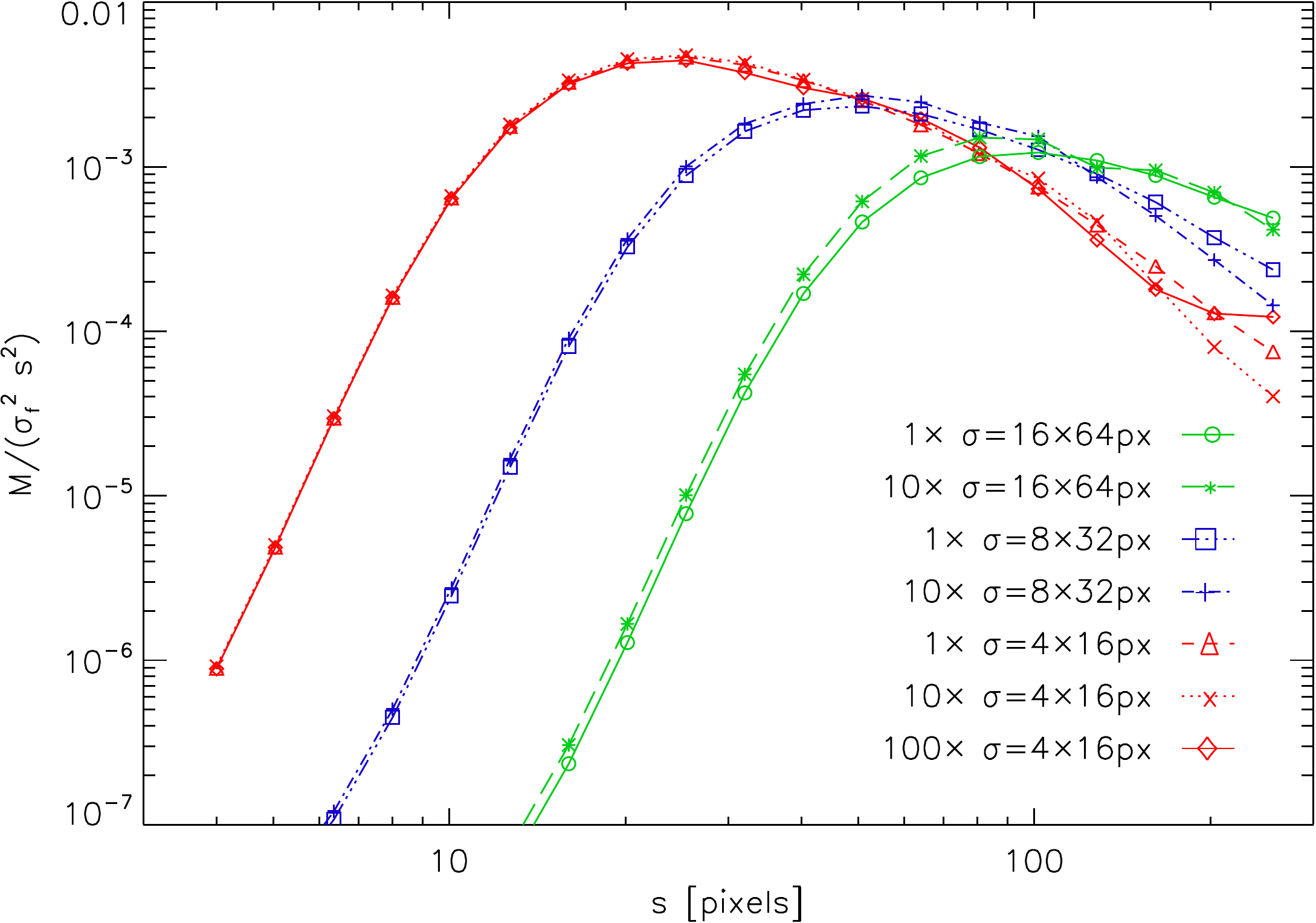}\vspace{3mm}
\includegraphics[angle=90,width=0.87\columnwidth]{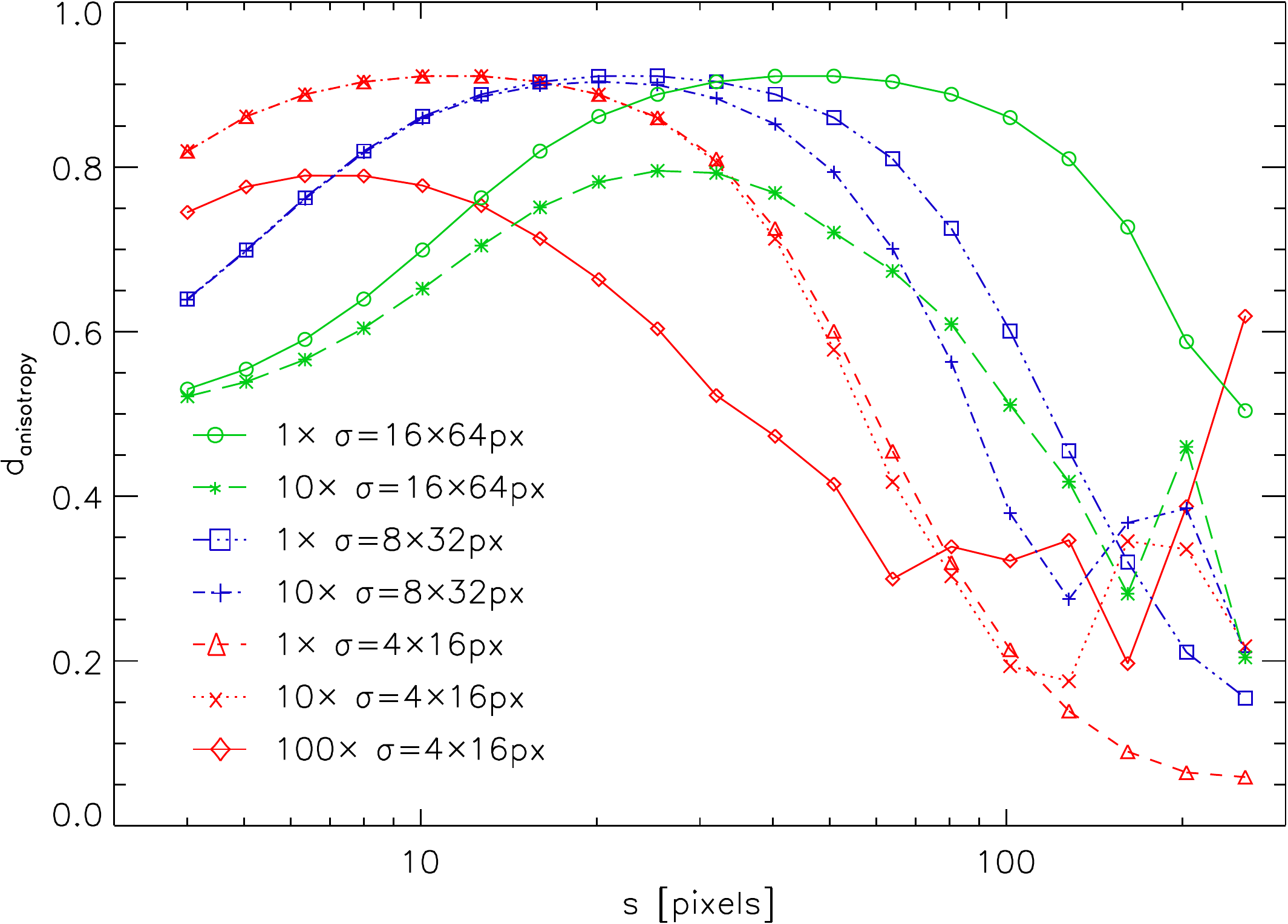}
\caption{Comparison of the wavelet spectra and local degrees of anisotropy measured  in
the anisotropic wavelet analysis for maps containing randomly placed and oriented
ellipses with different sizes,
all having an axes ratio of 4:1 ($\sigma=4\times 16$~pixels, $\sigma=8\times 32$~pixels,
$\sigma=16\times 64$~pixels). By changing the number of the clumps, we introduce
different degrees of anisotropy on the scale of the typical distance of the
clumps.
The wavelet spectra are normalized \changed{for amplitude per scale weighting}
by dividing them by $\sigma\sub{f}^2\times s^2$.
}
\label{fig:compare_sizes}
\end{figure}

\changed{The maps used so far still contained only relatively few
clumps, leading to small superposition effects. To combine all
effects discussed so far with a large number of clumps we compare
in Fig.~\ref{fig:compare_sizes} maps of individual clumps of different
sizes to maps of superpositions containing a variable number of clumps.}
We randomly placed and oriented elliptical
clumps with an aspect ratio of 4:1 using different sizes and numbers of clumps.
For the degree of anisotropy we only show the local degree $d\sub{loc}^w$
because we find only small accidental global anisotropies.
The spectra of wavelet coefficients show the expected broad peaks and very
little variation when changing the number of clumps. \changed{All peak
positions fall at six times the standard deviation of the clump minor axis
and the range above the 90\,\%-of-the-peak level extends from 20\,\% below
the peak position to 30\,\% above the peak position. At a scale of three times
the clump minor axis, the coefficients drop to about $1/\mathrm{e}$ and
at the scale of the minor axis to $2\cdot 10^{-4}$ of the peak values.}
The difference between 1 clump and 100 clumps \changed{can provide a variation
of the peak position by up to 15\,\% and some} deviation at large
scales due to some random superpositions.
The situation is very different when inspecting the degree of isotropy.
The simple size-peak relation for single elliptic clumps in the maps is
quickly distorted when adding more clumps. The anisotropy
at large scales is systematically reduced due to the random relation between
neighboring clumps. The affected scales are determined by the typical distance
between the clumps. We measured the average distance between the clumps through
Delaunay triangulation and obtained a broad distribution with a mean of 160~pixels
and a standard deviation of 70~pixels for ten clumps and a mean of 50~pixels and a standard
deviation of 40~pixels for hundred clumps. As a consequence, the maximum of the
smallest clumps between 4 and 30~pixels is not affected when combining only
10 clumps, but the large-scale wing is significantly reduced when combining
100 clumps. The resulting peak position is shifted to smaller scales.
The same effect is prominent for the largest clumps when combining 10 of them.
\changed{When combining many clumps the mutual alignment or misalignment from
the random placement always destroys the anisotropy in the map at scales of
the average clump distance, both globally and locally, but provides only a
minor modification to the isotropic wavelet coefficients.}

%

\subsection{Angular sensitivity}
\label{sect_ecc_calibration}

\begin{figure}
\centering
\includegraphics[angle=90,width=0.8\columnwidth]{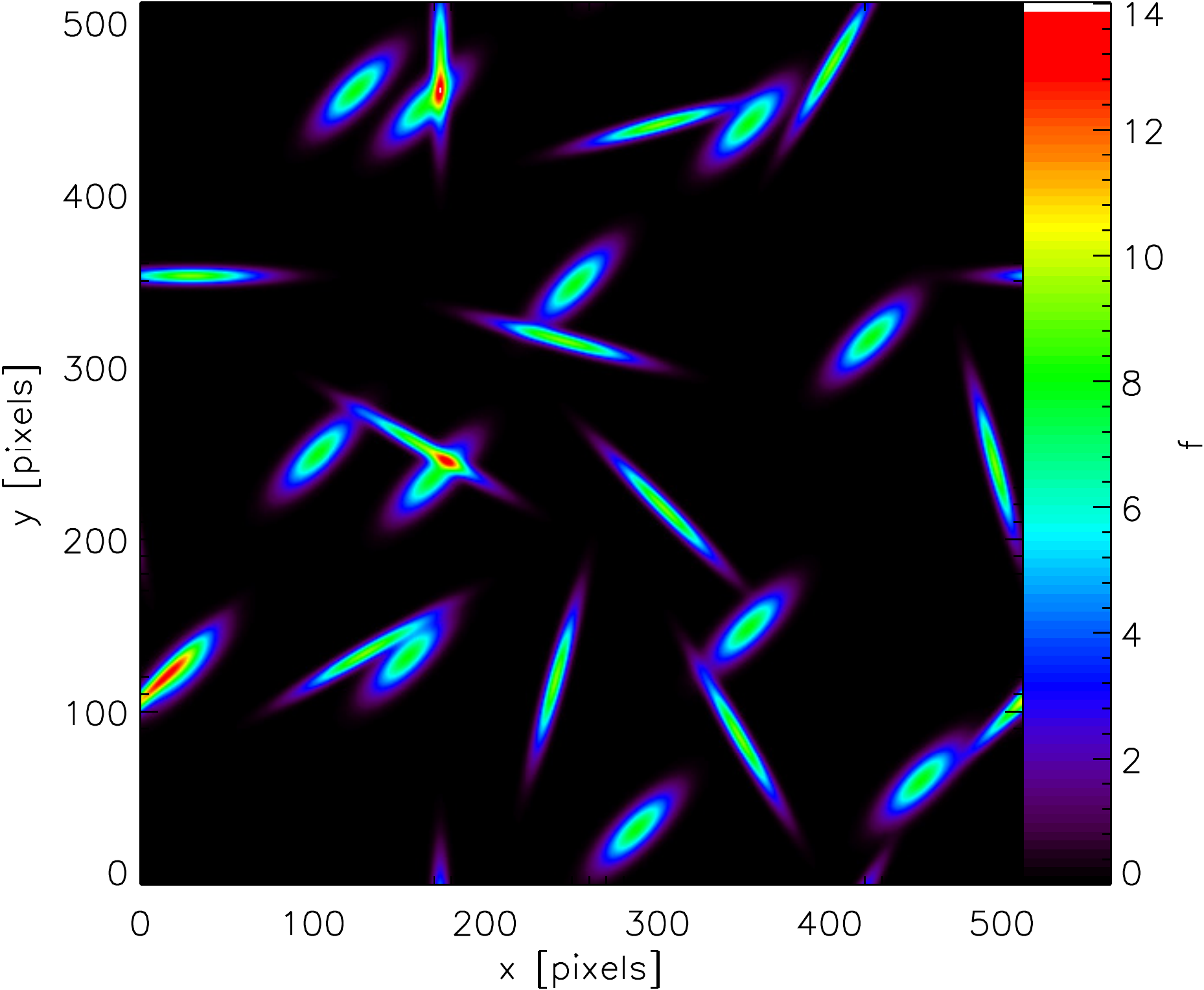}\vspace{3mm}
\includegraphics[angle=90,width=0.87\columnwidth]{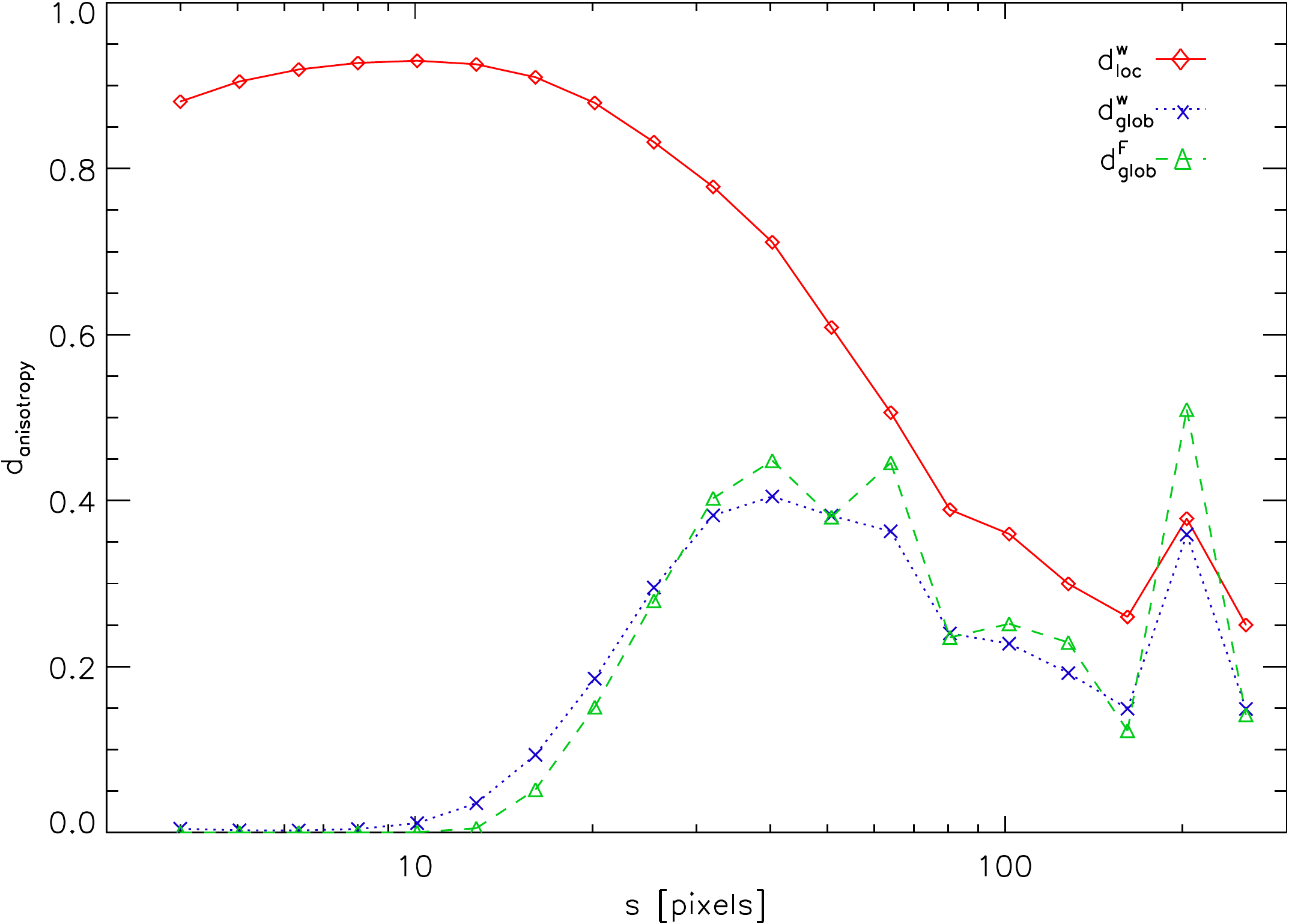}
\caption{\changed{Local and global degrees of anisotropy for the map containing
a superposition of two ensembles of elliptic clumps, shown on the top panel.}
All clumps with standard deviations of 6 and 18~pixels are oriented at 45 degrees;
ellipsoidal clumps with standard deviations of 3 and 27~pixels, i.e.
an aspect ratio of 9:1, are distributed at uniformly spaced angles
covering 360~degrees.}
\label{fig:superposition_of_ellipses}
\end{figure}

In Fig.~\ref{fig:superposition_of_ellipses} we combine two anisotropic structures
with similar sizes but different aspect ratios and
orientations. Comparing the local and global degree of anisotropy then
allows us to assess the sensitivity of the method to the \changed{aspect ratio} of the
structures. Ten Gaussian clumps with
standard deviations of 6 and 18~pixels for their main axes
are oriented at 45 degrees while ten clumps with a larger aspect ratio,
provided by standard deviations of 3 and 27~pixels for their main axes,
have a uniform angular distribution.
The corresponding isotropic power spectra and spectra of wavelet coefficients,
not plotted here, show a similar behavior to the \changed{individual clumps
in Fig.~\ref{fig:singleGaussian}. The same is true for the degree of local
anisotropy at scales $s<27$~pixels.} The global anisotropy introduced
by the 45~degrees alignment of the clumps with the 3:1 axes ratio
is only apparent at scales between about 30 and 60 pixels.

\begin{figure}
\centering
\includegraphics[angle=90,width=0.87\columnwidth]{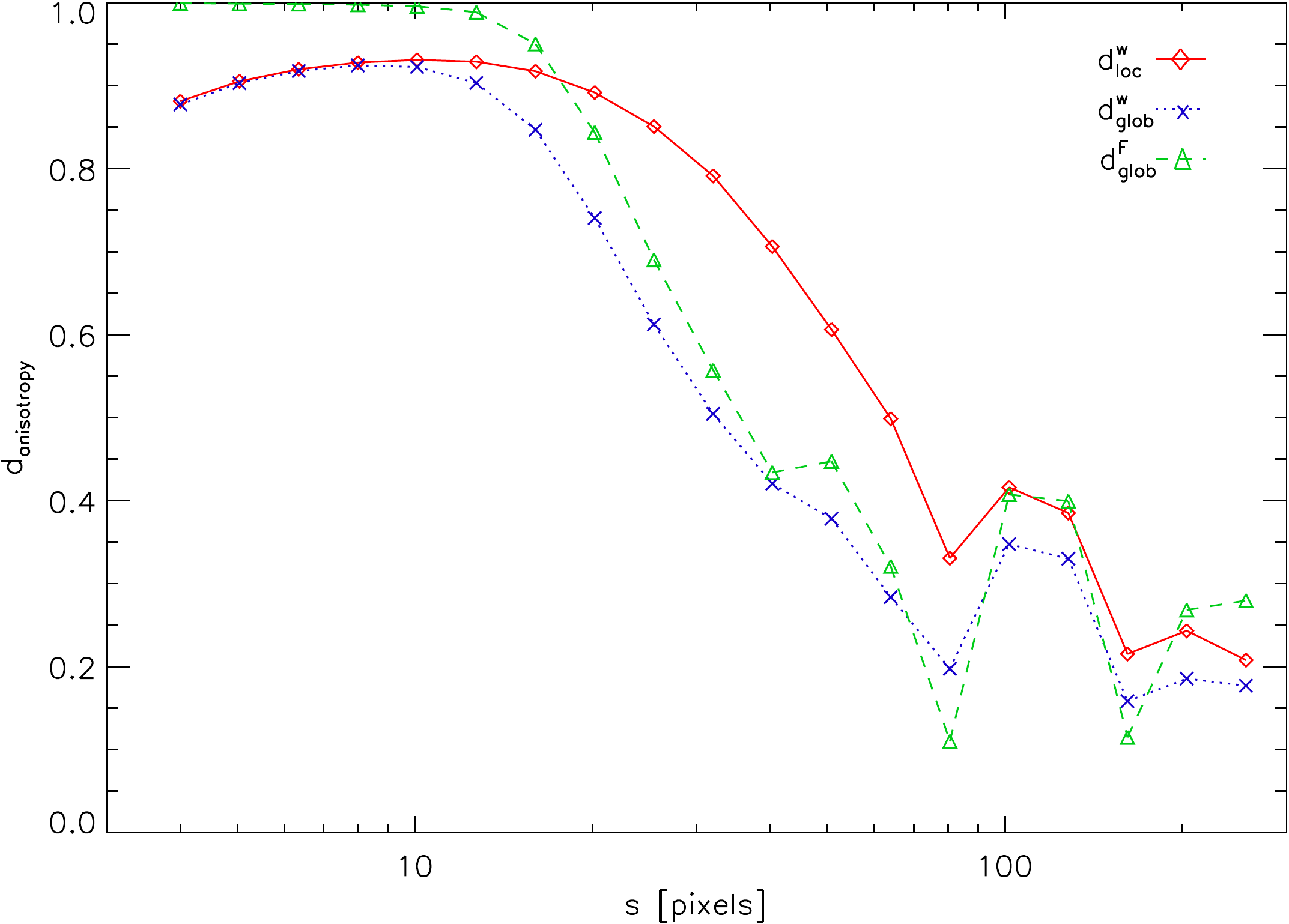}
\caption{\changed{Local and global degrees of anisotropy for} a map containing
a superposition of two ensembles of elliptic clumps. Ten clumps
with standard deviations of 3 and 27~pixels, i.e. an aspect ratio of 9:1
are aligned at 0 degrees, ten clumps with standard deviations of 6 and
18~pixels are distributed at uniformly spaced angles
covering 360~degrees.}
\label{fig:superposition_of_ellipses_reverse}
\end{figure}

In Fig.~\ref{fig:superposition_of_ellipses_reverse} we invert the situation in the
sense of aligning all ellipses with the high \changed{aspect ratio} at an angle of zero
while uniformly distributing the angles of the $\sigma=6\times 18$~pixels ellipses.
As the orientation of the ellipses is irrelevant for the local degree of
anisotropy, the curve for $d\sub{loc}^w$ agrees with the one from
Fig.~\ref{fig:superposition_of_ellipses}. However, the global degree of
anisotropy is much higher here at all scales below 30~pixels. Similar global degrees
as in Fig.~\ref{fig:superposition_of_ellipses} only occur for $s\ga 30$~pixels.

When inspecting the maps of anisotropic wavelet coefficients $|m^a(s,\bx)|$,
we find that at small scales the clumps with the higher \changed{aspect ratio} produce
nine times higher anisotropic wavelet coefficients than the $\sigma=6\times 18$~pixels
clumps. The relative contribution of the wavelet coefficients of the
clumps with lower \changed{aspect ratio} grows with scale, starting from the characteristic
anisotropy scale for the minor axis at $s=\sqrt{2}\times 6 \approx 8$~pixels,
until they show the same magnitude as the coefficients for the clumps with the high
\changed{aspect ratio} at a scale of about 30~pixels. Consequently, the map-averaged
anisotropic wavelet coefficients $M^a(s)$ are dominated by high \changed{aspect ratio}
clumps at all small scales while we find similar contributions at scales $s\ga 30$~pixels.
On very large scales, approaching the map size, only random anisotropies
appear from the mutual positioning of the individual clumps providing
degrees of anisotropy around 0.2--0.3.

\changed{To also inspect} the distribution of the directions of anisotropic
structures in the maps we need to look at the distribution of angles of the
anisotropic modes $m^a(s,\vec{x})$ measured by the two-dimensional
anisotropic mode spectra $A(s,\varphi)$ and $\tilde{A}(s,\varphi)$
(Eqs.~\ref{2dspec} and~\ref{2dnormspec}).

\begin{figure}
\centering
\includegraphics[angle=90, width=0.88\columnwidth]{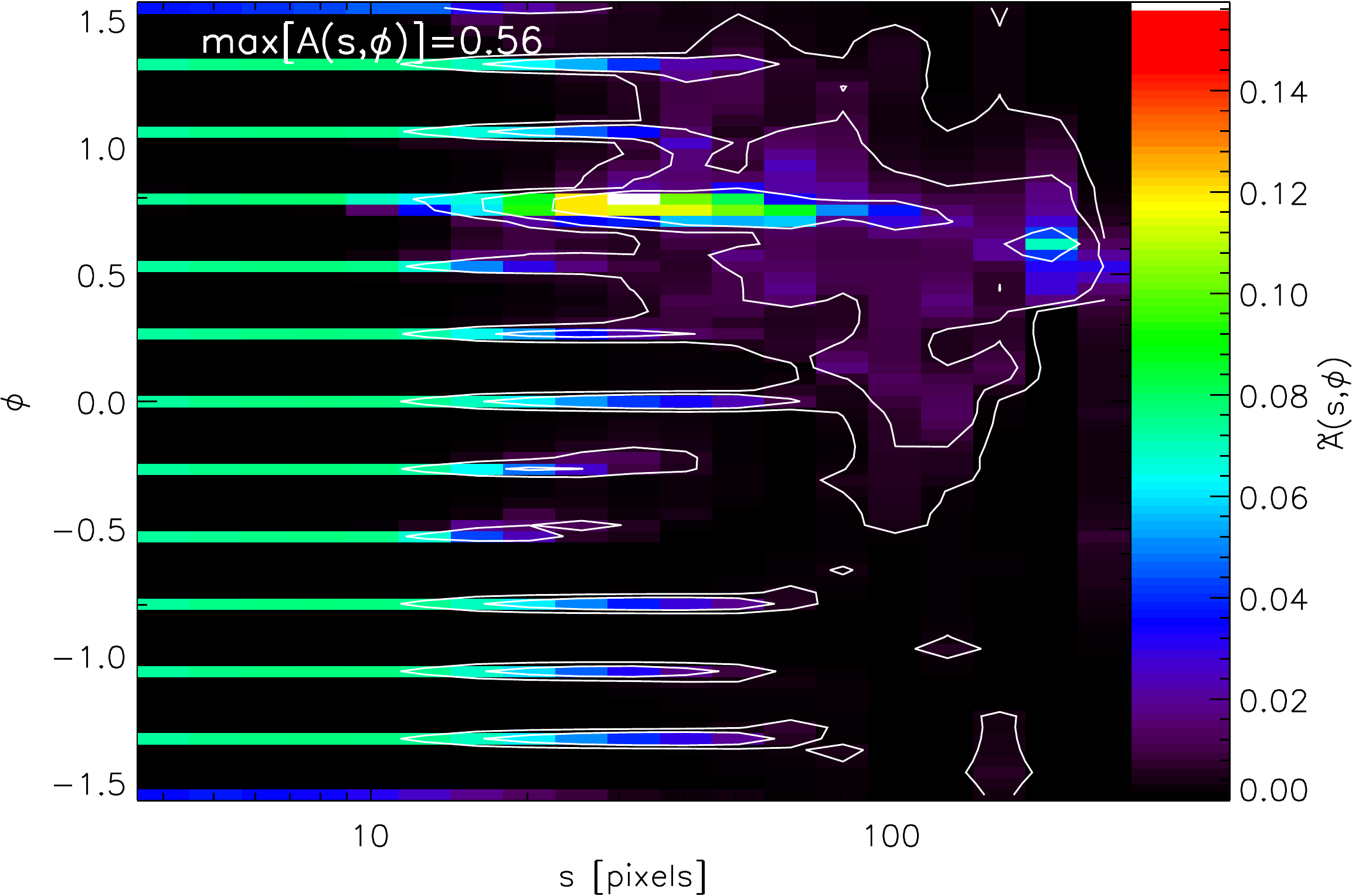}\vspace{3mm}
\includegraphics[angle=90,width=0.88\columnwidth]{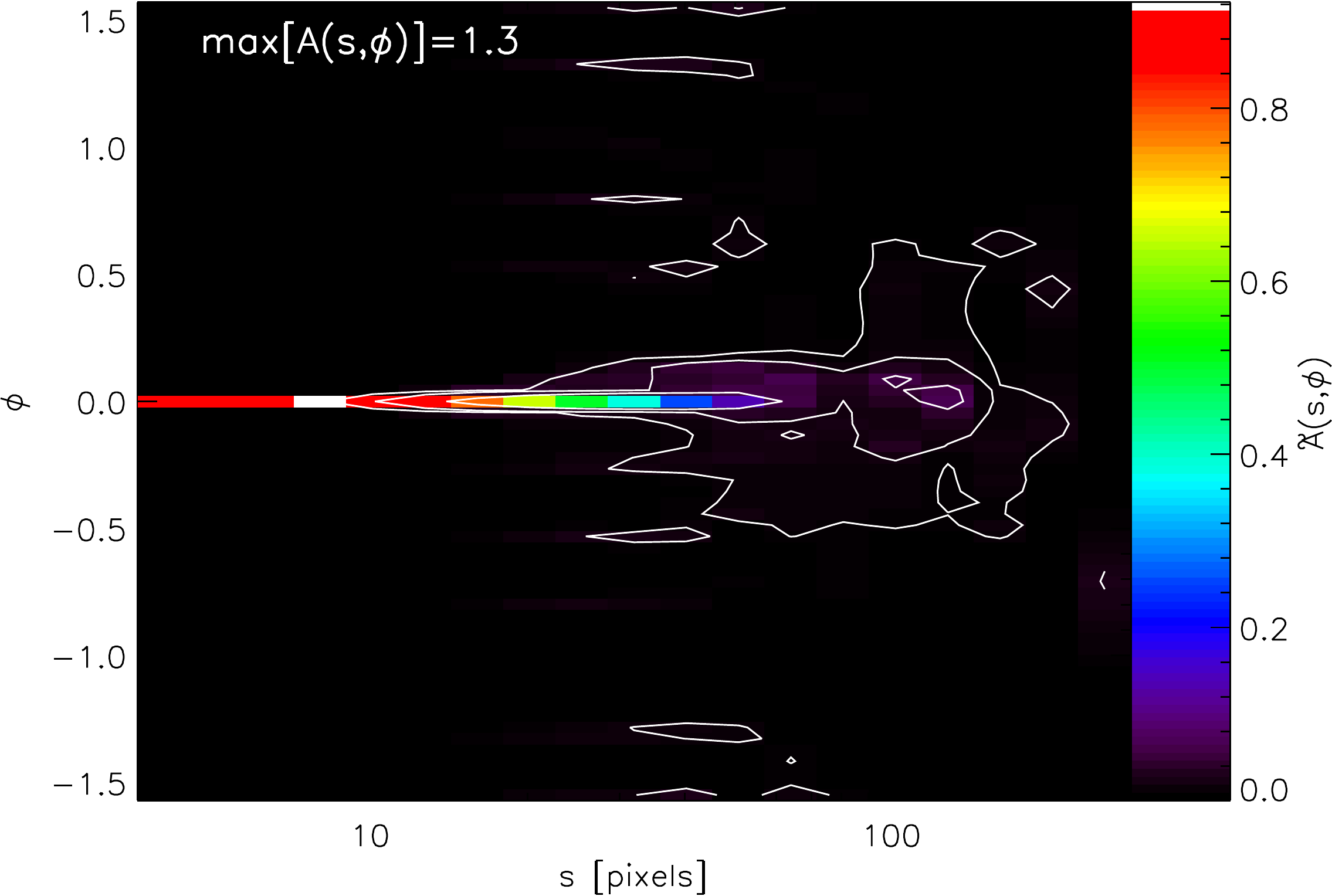}
\caption{Angular distribution of the anisotropic wavelet coefficients
for the maps containing the two ensembles of clumps from Figs.~\ref{fig:superposition_of_ellipses}
(top, $\sigma=6\times 18$~pixel clumps at 45~degrees,
$\sigma=3\times 27$~pixel clumps uniformly distributed) and
\ref{fig:superposition_of_ellipses_reverse} (bottom, $\sigma=6\times 18$~pixel clumps
uniformly distributed, $\sigma=3\times 27$~pixel clumps at 0 degrees). The contours show the
two-dimensional spectrum of wavelet coefficients $A(s,\varphi)$
at levels of 1/30, 1/10, and 1/3 of the peak value indicated in the top left
corner of the plot. The colors
represent the coefficients normalized by the spectrum of isotropic modes,
$\tilde{A}(s,\varphi)$ (Eq.~\ref{2dnormspec}), on a linear scale.}
\label{fig:angular_distribution}
\end{figure}

Figure~\ref{fig:angular_distribution} shows the angular distribution of the
anisotropic wavelet coefficients, in terms of $A(s,\varphi)$ (contours) and
$\tilde{A}(s,\varphi)$ (colors) for the examples from Figs.~\ref{fig:superposition_of_ellipses}
and \ref{fig:superposition_of_ellipses_reverse}. The upper plot refers to the
configuration where the low \changed{aspect ratio} clumps are aligned at 45~degrees; in the lower
plot the clumps with high aspect ratio are aligned at 0~degrees. In case of the uniform
angular distribution of the high \changed{aspect ratio} clumps,
we can see the contribution from every individual of these clumps
at small scales in the normalized coefficients $\tilde{A}(s,\varphi)$.
However, there the absolute magnitude of the wavelet
coefficients is small so that the clumps do not show up in the contours
of $A(s,\varphi)$. $A(s,\varphi)$ exceeds 3\,\% of its maximum only for
$s>10$~pixels. The peak is reached at $s\approx 30$~pixels where we also see
the contribution of the $\sigma=6\times 18$~pixels clumps concentrated at the
angle of $\pi/4=45$~degrees. Although the global anisotropy is not very
prominent in the spectra in Fig.~\ref{fig:superposition_of_ellipses}
it is \changed{very obvious} in the angular distribution. At large scales we find an
accidental anisotropy at an angle of about 30~degrees responsible for the
enhanced degree of anisotropy there.

In the lower plot where the high \changed{aspect ratio} clumps are aligned,
the whole angular distribution, both in terms of $A(s,\varphi)$ and $\tilde{A}(s,\varphi)$,
is dominated by the $\sigma=3\times 27$~pixels clumps.
One can only recognize the broad angular distribution of the low \changed{aspect ratio}
clumps in the $A(s,\varphi)$ contours at the level of $A(s,\varphi) \ge 3\,\%$
at scales between 30 and 60~pixels. This is hardly visible in the normalized
spectrum but is sufficient to lower the global degree of anisotropy at those
scales to the same level as measured for the aligned clumps with low aspect ratio
(see Fig.~\ref{fig:superposition_of_ellipses}).

Both angular distributions of wavelet coefficients, $A$ and $\tilde{A}$, are therefore useful
to judge the anisotropic structure in a map. The absolute coefficients,
$A(s,\varphi)$, add the angular information to the power of anisotropic
structural variations thereby providing a visual explanation for the measured
degree of global anisotropy in a map. Local and global anisotropy
are, however, better covered by the plot of $\tilde{A}(s,\varphi)$, combining
the information from $d\sub{loc}^w$ and $d\sub{glob}^w$ in a single two-dimensional
surface showing the full angular dependence. The color represents the
degree of local anisotropy as a function of size scale and angle; the
angular spread of the contributions provides a good assessment of the alignment
that can create some global anisotropy.

\changed{
\subsection{Three-dimensional structures}
\label{sect_3d}

\begin{figure}
\centering
\includegraphics[angle=90,width=0.88\columnwidth]{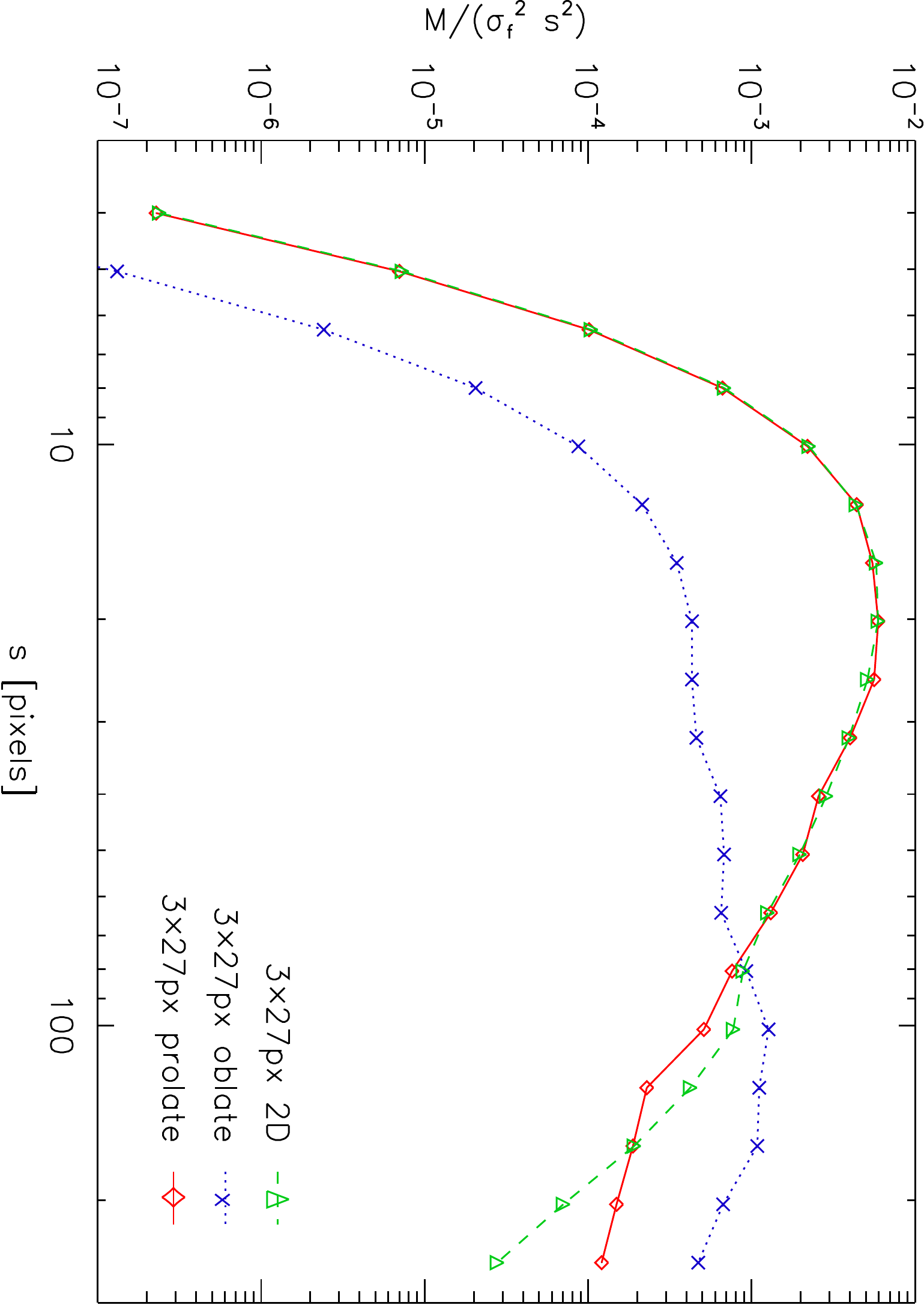}\vspace{3mm}
\includegraphics[angle=90,width=0.88\columnwidth]{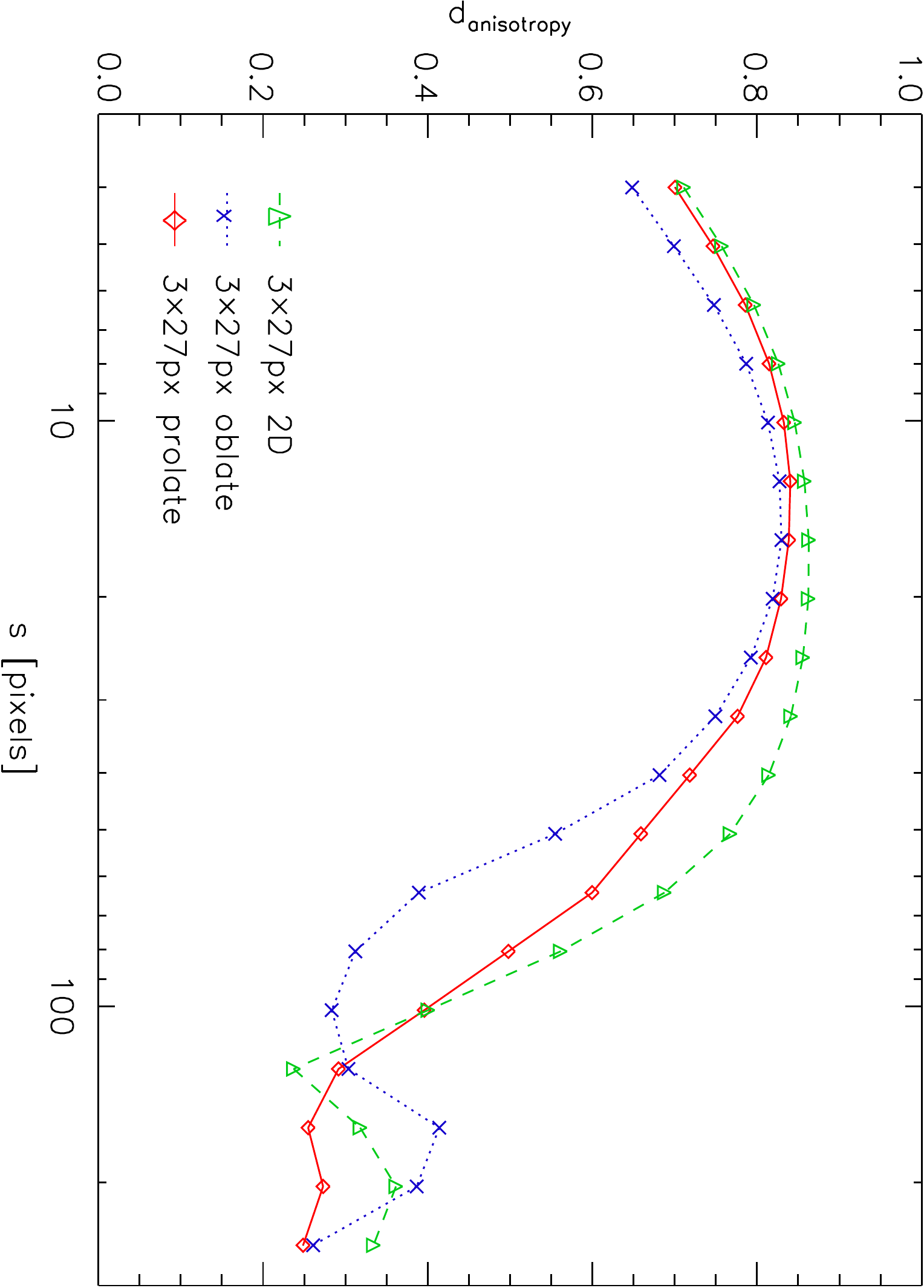}
\caption{Comparison of the normalized spectra of the isotropic wavelet coefficients (top) and
the local degree of anisotropy (bottom) for maps containing random projections of
ten three-dimensional Gaussian rotational ellipsoids with sizes of
$\sigma=3\times 3 \times 27$~pixels and $\sigma=3\times 27 \times 27$~pixels. The green line
shows the result for a map consisting of ten two-dimensional Gaussian ellipses
with  $\sigma=3\times 27$~pixels.}
\label{fig:3d}
\end{figure}

To address the question whether the anisotropy of the two-dimensional projection
eventually reflect elongated structures in the underlying three-dimensional
structure, or rather projections of sheets, we analyzed
projections of Gaussian rotational ellipsoids with main axes $a \times a \times c$.
For $a < c$ this corresponds to cigar-like, prolate
structures approximating the traditional picture of elongated filaments while
$a > c$ represents the oblate case of thin sheets. In the projection of a rotational ellipsoid
one axis corresponds to the length of the two common axes $a\sub{proj}=a$ while the
second axis is given by $c\sub{proj}^2={a \cos^2 \theta + c \sin^2 \theta}$, where
$\theta$ is the angle between the symmetry axis of the ellipsoid and the
direction of projection. Any
projection reduces the aspect ratio of the resulting two-dimensional ellipses
from the three-dimensional ellipsoids.

Figure~\ref{fig:3d} shows the normalized spectra of isotropic wavelet coefficients
and the degrees of local anisotropy for maps created from random projections of
10 prolate Gaussian clumps with $\sigma=3\times 3 \times 27$~pixels and 10 oblate clumps
($\sigma=3\times 27 \times 27$~pixels). For comparison we overplot the spectra for 10
randomly placed two-dimensional Gaussian ellipses with $\sigma=3\times 27$~pixels.
The spectra of isotropic wavelet coefficients show a close match between the
two-dimensional case and the projections of the prolate clumps.
The good match results from the combination of two effects already discussed in
Sect.~\ref{sec:simpleGaussian} and \ref{sect_ecc_calibration}. There we found
that for highly elongated structures the larger diameter hardly changes the wavelet
spectrum and that in a superposition of clumps those clumps with
the higher \changed{aspect ratio} dominate the wavelet spectrum. This means that for the
random projections of prolate structures the clumps with a projection angle
$\theta$ close to 90~degrees dominate the spectrum having the same shape as the
two-dimensional ellipses. The deviation between both curves at larger scales
is explained by the variable superposition effects (Sect.~\ref{sect_superposition}).
The wavelet spectrum for the projection of the oblate clumps is very shallow with
an onset of structures also at the
3~pixel scale, but a much wider distribution of scales because of the wide
distribution of the minor axes of the projected clumps covering the whole range
between 3 and 27~pixels. In contrast to the wavelet spectrum the degree of anisotropy
on its own does is not sufficient to distinguish between projections of oblate and prolate
spectra. As the average axes ratio of the projections is the same in both cases,
given by the mean projection angle, we find only small differences between the degrees
of anisotropy. The degree of anisotropy is always dominated by the few clumps that
appear very elongated in the projection close to the two-dimensional case that
represents the most elongated limit and thus has the broadest peak going up
to $\hat{s}_{90\,\%} \approx 2\sigma_c-1/2\sigma_a = 52$~pixels.

Therefore, the anisotropic wavelet analysis performed on projected maps can only
give some hint on the underlying three-dimensional structure. It reliably measures
the width of the narrowest filaments and the anisotropy of the most elongated
structures but a distinction between the oblate and the prolate case is only
possible if one has some a priori knowledge of the size distribution of the
clumps allowing for a detailed quantitative interpretation of the isotropic
wavelet spectrum. To what degree this can be performed in observed three-dimensional
position-position-velocity cubes will be the topic of a subsequent paper.}

\subsection{Spectra of clumps}
\label{sect_fractal}

\begin{figure}
\centering
\includegraphics[angle=90,width=0.8\columnwidth]{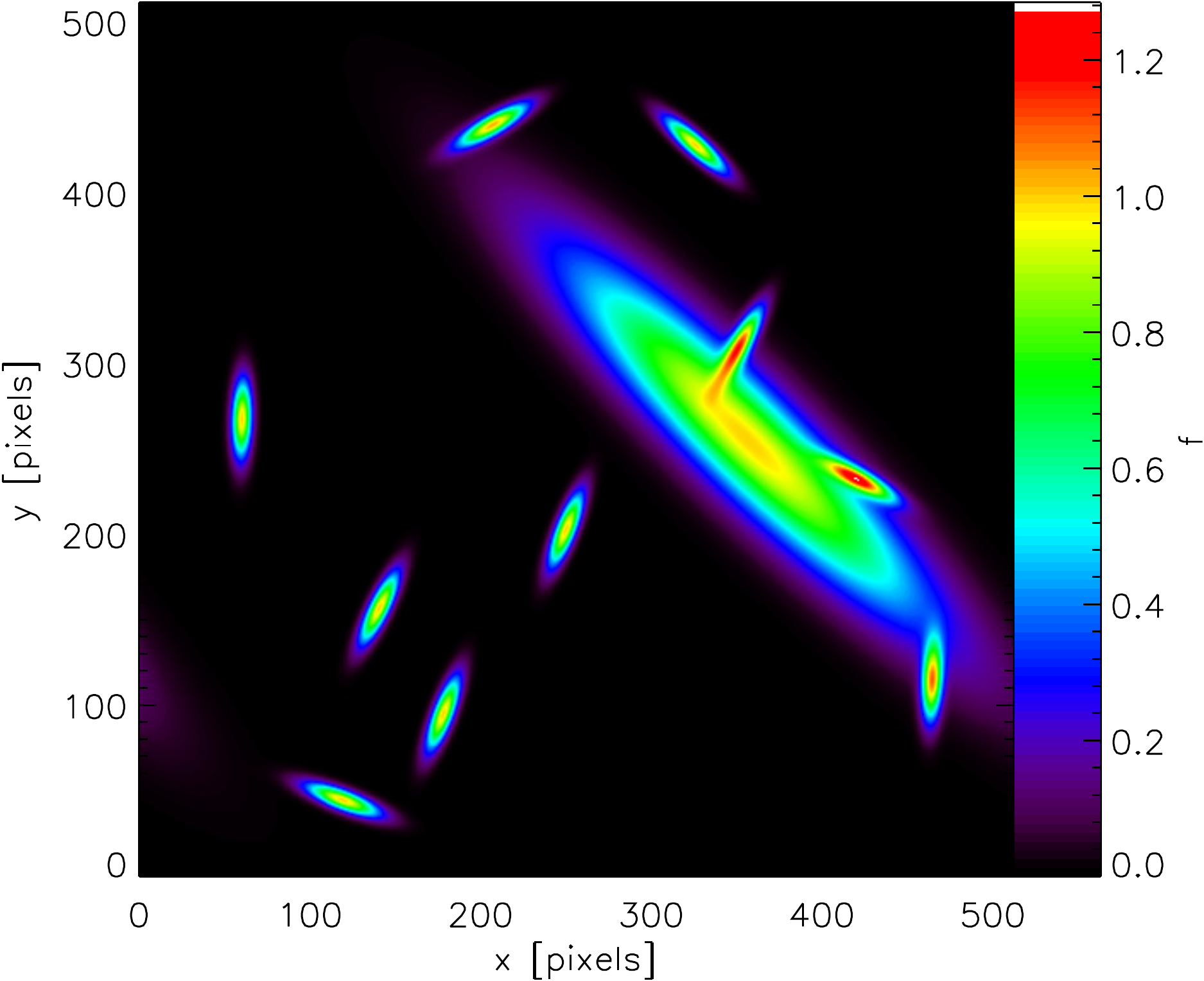}\vspace{3mm}
\includegraphics[angle=90,width=0.88\columnwidth]{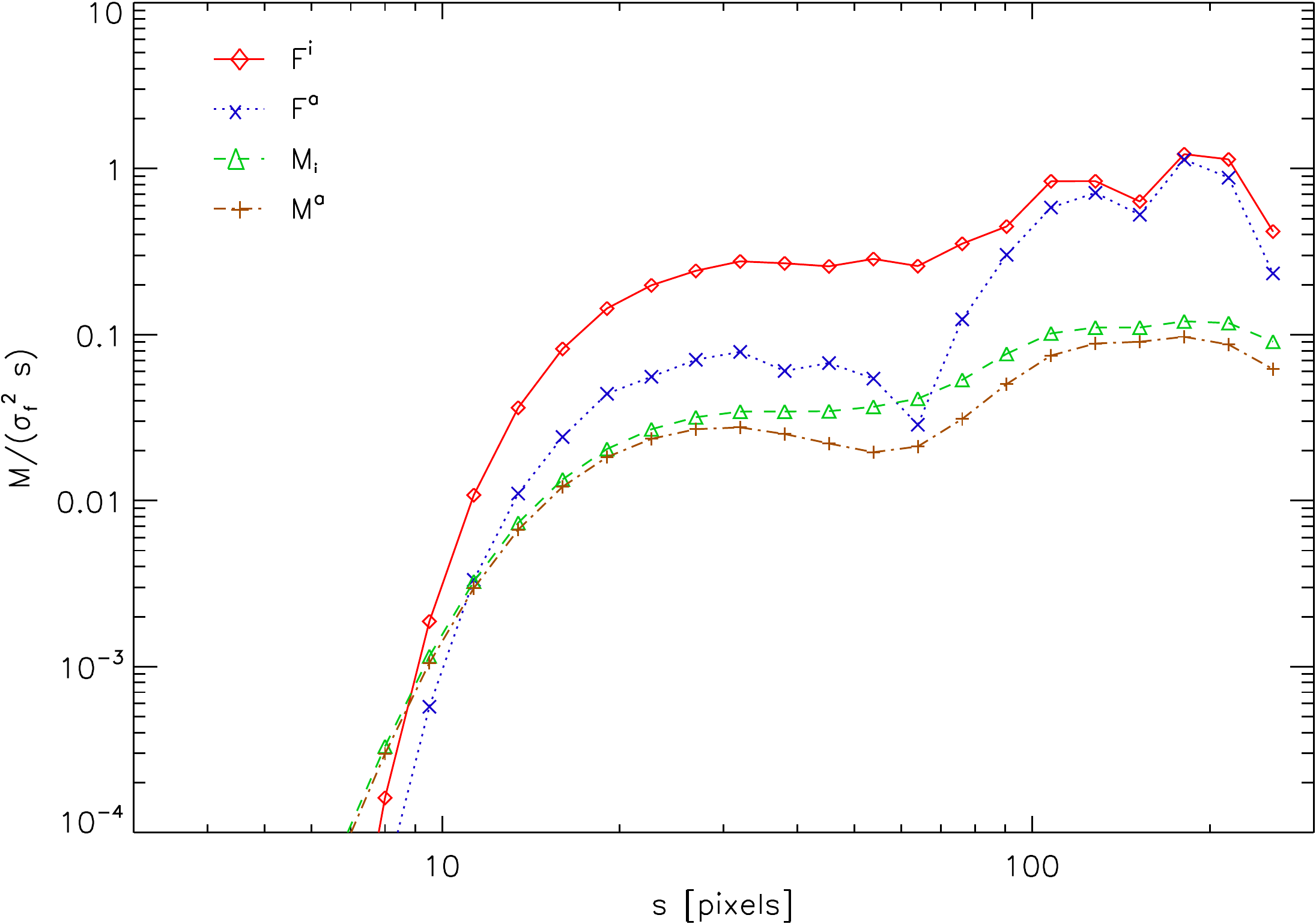}\vspace{3mm}
\includegraphics[angle=90,width=0.87\columnwidth]{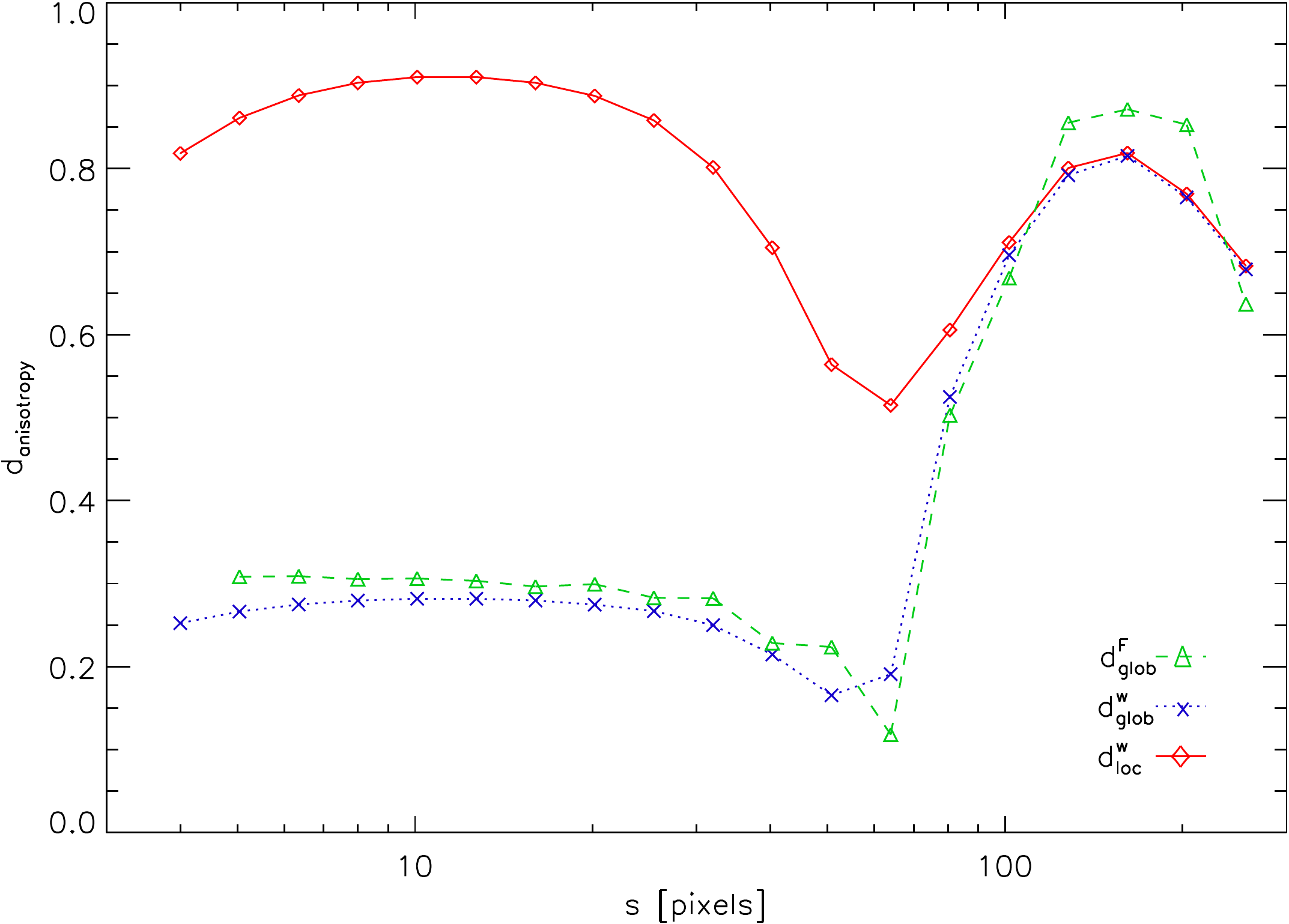}
\caption{Anisotropic wavelet analysis of a map with a superposition of a big elliptic
clump with $\sigma=20\times100$~pixels with 10 smaller ellipsoidal clumps with
$\sigma=4\times16$~pixels.}
\label{fig:small_anisotropic}
\end{figure}

In Fig.~\ref{fig:small_anisotropic} we combine anisotropic structures
on very different scales. A big ellipsoidal clump with $\sigma=20 \times
100$~pixels is oriented at -45 degrees, providing a global
anisotropy. Ten small ellipsoidal clumps with $\sigma=4\times 16$~pixels,
randomly oriented and positioned are superimposed.
The wavelet and Fourier spectra show the superposition of two broad maxima
from the two structures, having only a weak dip at the intermediate scales
\changed{of 60-70~pixels. In the degrees of anisotropy} we
find two well separated peaks.
The local degree of anisotropy shows the broad maximum between 4 and
30~pixels expected from the individual $\sigma=4\times 16$~pixel clumps, but
the maximum from the $\sigma=20\times 100$~ellipse does not start at 22~pixels,
\changed{as expected from the scaling relations for individual clumps},
but only  at about 120~pixels. \changed{This can be understood from the
effect seen in Fig.~\ref{fig:compare_sizes}.} The scale range below 100~pixels is
dominated by the arrangement of the individual smaller clumps. The mutual
alignment or misalignment from the random placement destroys the anisotropy
in the map at those scales, both globally and locally, while still providing
some noticeable contribution to the isotropic wavelet coefficients.

The global degrees of anisotropy are close to the
local ones at large scales where the map contains only one large anisotropic
structure. At small scales the global degrees also show a small enhancement due to
the accidental alignment of four of the clumps at an angle of about 60~degrees.

\begin{figure}
\centering
\includegraphics[angle=90,width=0.88\columnwidth]{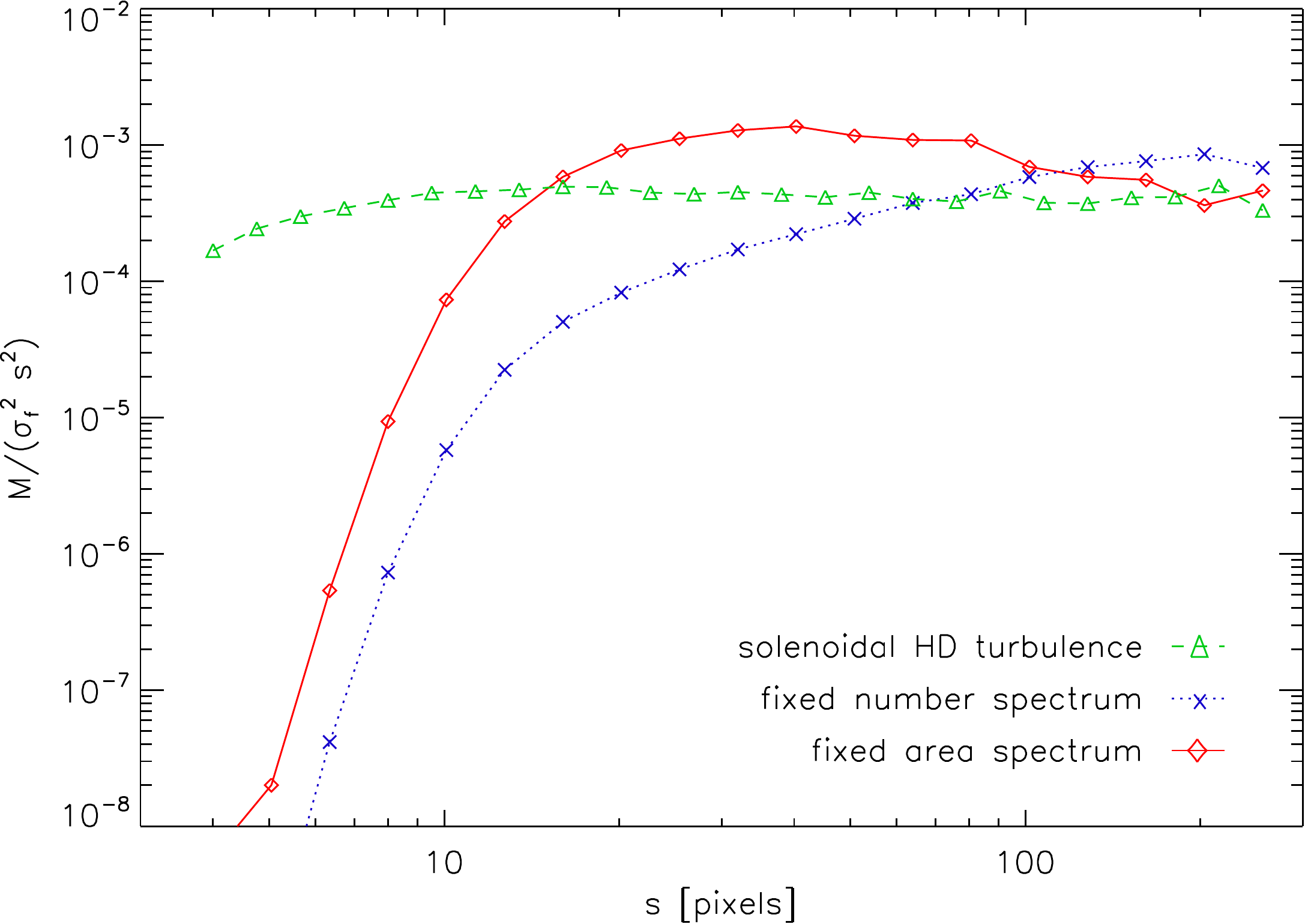}\vspace{3mm}
\includegraphics[angle=90,width=0.88\columnwidth]{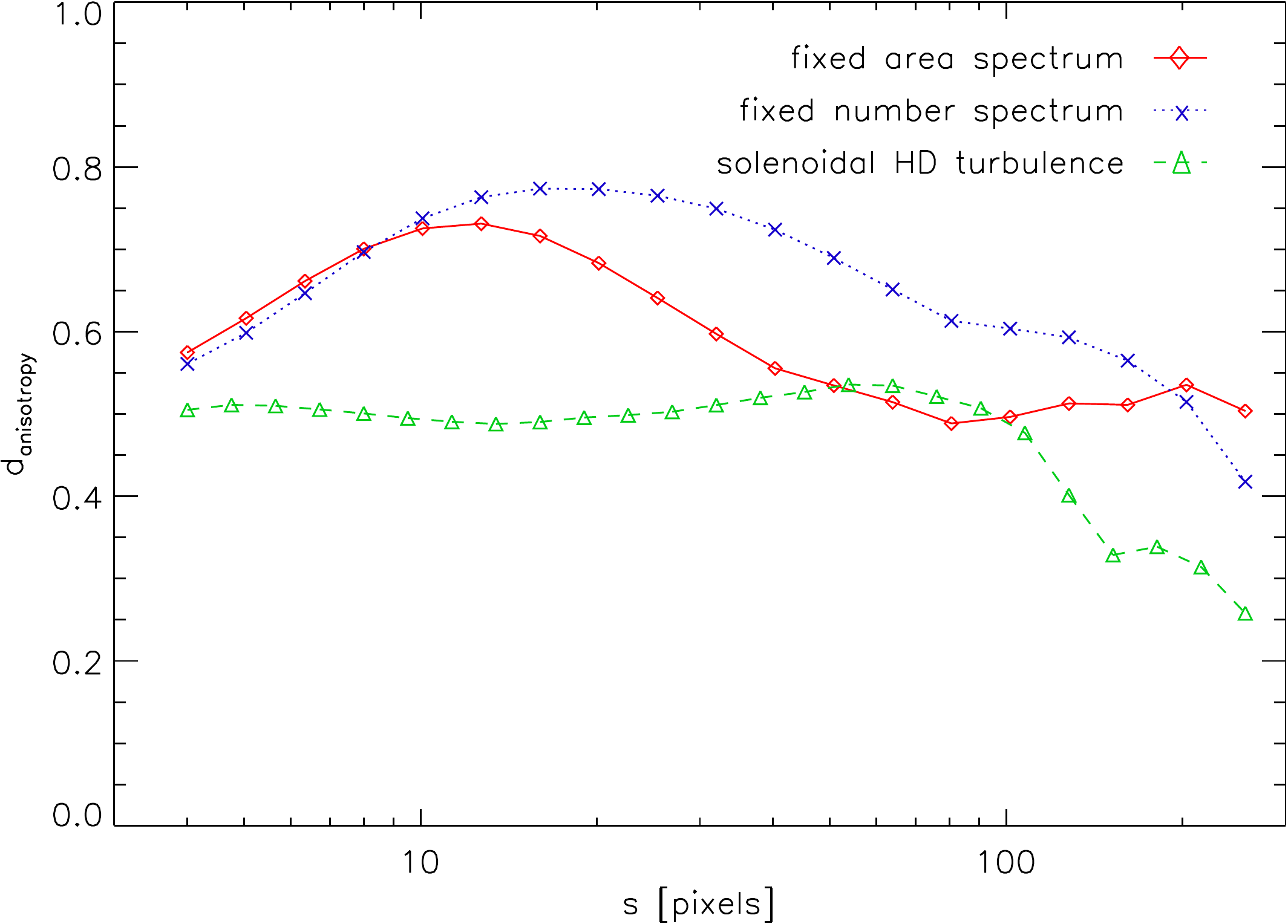}
\caption{Comparison of the normalized spectra of the isotropic wavelet coefficients (top) and
the local degree of anisotropy (bottom) for maps with a hierarchy of structures of different sizes.
The green line represents the projected density structure in a simulation of fully developed
hydrodynamic turbulence with a close-to-power-law power spectrum and no preferred direction
\citep{Federrath2010}. The blue and red curves show the results for maps composed of a hierarchy
of Gaussian clumps with sizes of $\sigma=4\times 16,\; 8\times 32,\; 16\times 64$,
and $32x128$~pixels. One map contains two clumps of each size, the other one a fixed
area for clumps of different sizes. i.e. 64~clumps of the smallest size and one clump of the
largest size.
}
\label{fig:clump_spectra}
\end{figure}

\changed{
All tests performed so far intentionally used structures with well defined sizes.
An opposite assumption is a continuous spectrum of sizes following an approximately
self-similar scaling. One case,
composed of discrete sizes, consists of a superposition of elliptical Gaussian clumps
with different sizes. We compare two different approaches: an equal number of
clumps independent of their size and an equal area filling, corresponding to a
quadratic decrease of the number of clumps as a function of their size. The smallest
clumps used here have a size of $\sigma=4\times 16$~pixels. As a
data set that is free of any discrete scales, we use a simulation of fully developed
hydrodynamic turbulence \citep{Federrath2010}. The column density structure
obtained in a FLASH3 simulation of isothermal turbulence in a periodic box driven by
solenoidal velocity perturbations provides a structure with a close-to-power-law
power spectrum and no preferred direction. After 10 autocorrelation times of the forcing,
structures of all sizes are created leading to a large inertial range of the power
spectrum of the maps.

Figure~\ref{fig:clump_spectra} shows the normalized spectra of isotropic wavelet coefficients and the degrees of local
anisotropy for the three data sets. The global degree of anisotropy is always vanishing.
The wavelet spectrum of the hydrodynamic simulations is flat over all scales
corresponding to a power spectrum with an exponent $\beta=3$. The spectra of the
clump ensembles show a steep decay below the predicted lower plateau edge at
$(\check{s}_{90\,\%} \approx 19$~pixels corresponding to the width of the smallest clumps
(see Sect.~\ref{sec:simpleGaussian}). Above that limit the spectra follow power laws.
The ensemble with an equal number of clumps of all sizes shows a
spectrum proportional to the scale $M/s^2 \propto s$.
This corresponds to the analytical relations given in Appx.~\ref{appx_an} in which an
individual clump contribution in a spectrum is proportional to its scale.
The spectrum with the same area of the clumps of different sizes shows a weakly decaying power-law spectrum
$M/s^2 \propto s^{-0.5}$. It somewhat shallower than the expected scaling $M/s^2 \propto s^{-1}$.
The large-scale wing of the individual clump spectrum decays $\propto s^{-1}$ but the
superposition of many small clumps with largest ones probably creates this deviation.

In the degrees of anisotropy we find a very flat spectrum for the hydrodynamic
turbulence simulation with a degree around 0.5 tracing the numerous filaments in
the structure. For the clump distribution at small scales the degree of anisotropy
matches the one of the individual clumps of $\sigma=4\times 16$~pixels
contained e.g. in Figs.~\ref{fig:compare_sizes} and \ref{fig:small_anisotropic}. It measures the structure
of the individual smallest clumps. However, the superposition with the larger
clumps leads to a higher anisotropy also at large scales. Because of the relatively
higher contribution of large clumps in the map with a fixed number of clumps per size,
the degree of anisotropy remains higher at large scales compared to the ensemble with
a fixed area of clumps. The noticeable decrease of the degree of anisotropy towards
large scales in both cases does not reflect the size of the largest clumps but is
rather due to the superposition of the different clumps destroying the local geometry
of the individual clumps in configurations with many clumps
as seen in Figs.~\ref{fig:compare_sizes} and \ref{fig:small_anisotropic}. This effect
does not seem to be present in the turbulence simulations indicating that it is an
artifact of our test setup of clump superpositions and not a general limitation of the method.
}

\subsection{Filter shape}
\label{sect_filtershape}

As shown in the previous section, for a superposition of
ellipses the spectrum of anisotropic wavelet coefficients tends to be
dominated by the clumps with the highest aspect ratio at any given scale.
The wavelet coefficients of the structures
with the higher anisotropy exceed those of structures with a lower anisotropy
by a large factor. This usually matches the visual
impression, by preferring the most filamentary structure in the spectrum,
but may hide relevant structures in a map.
To verify whether this effect is due to the accidental match of the
wavelet shape to the shape of the ellipses we have changed the
shape of the wavelet by varying the isotropic localization parameter $b$ in
Eq.~\ref{eq:morlet}.

\begin{figure}
\centering
\includegraphics[angle=90,width=0.88\columnwidth]{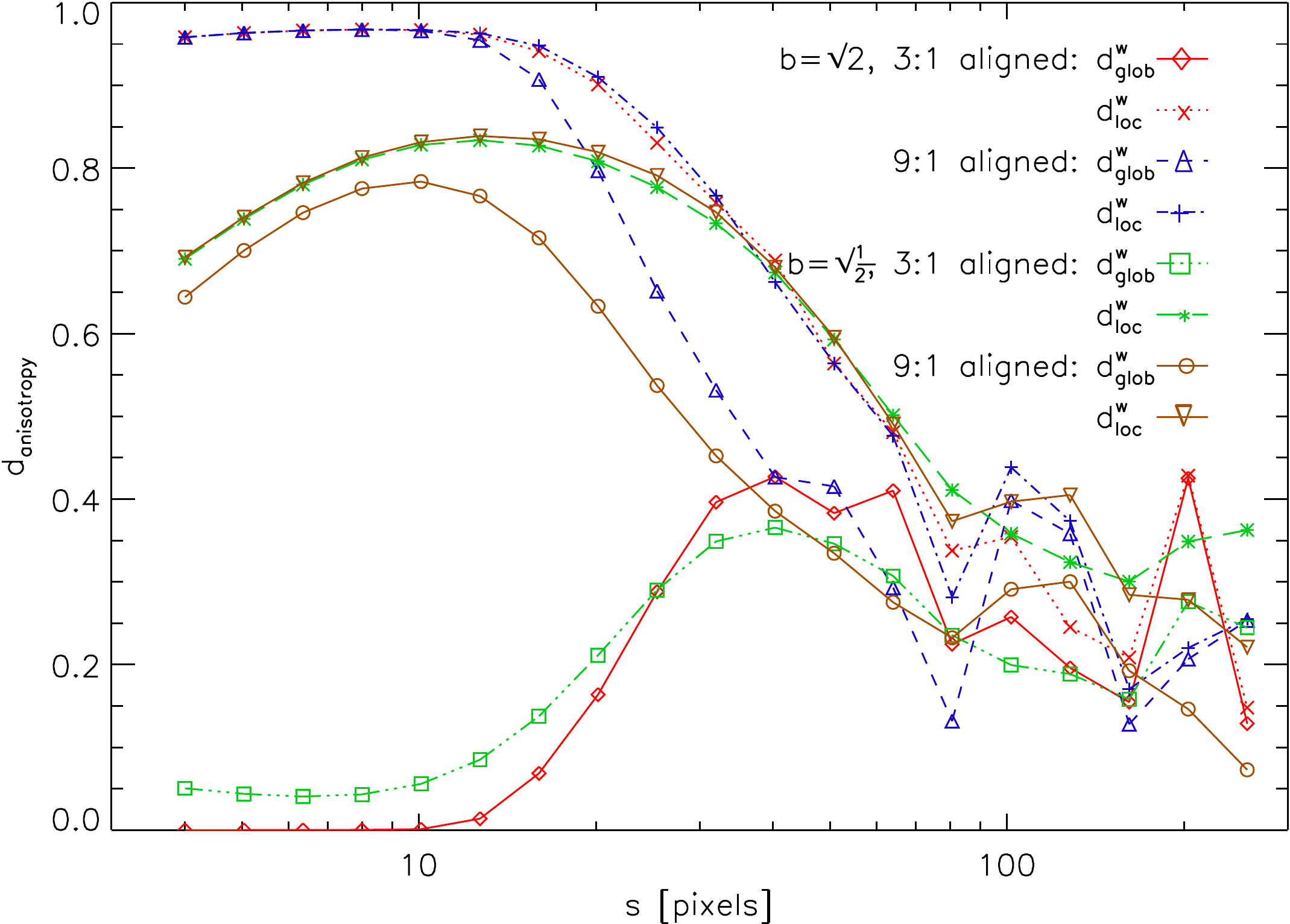}
\caption{Comparison of the degrees of anisotropy measured
by different filter shapes in the anisotropic wavelet analysis for
the ensembles from Figs.\ref{fig:superposition_of_ellipses} and
\ref{fig:superposition_of_ellipses_reverse} containing clumps
with $\sigma=6\times 18$~pixels and $\sigma=3\times 27$~pixels
where either of the two types is aligned.}
\label{fig:comparefilterellipses}
\end{figure}

Figure~\ref{fig:comparefilterellipses} presents the local and global degrees of
anisotropy for the two clump ensembles from Figs.~\ref{fig:superposition_of_ellipses} and
\ref{fig:superposition_of_ellipses_reverse} when using a $b$ parameter of
$1/\sqrt{2}$ and $\sqrt{2}$. The filter with $b=\sqrt{2}$ has a wider isotropic part
at the same scale
so that it can trace more elongated structures than the standard $b=1$ filter,
the filter with $b=1/\sqrt{2}$ is more localized leading to a lower effective
aspect ratio of the filter. Because a narrower isotropic filter corresponds
to the convolution with a broader Gaussian in Fourier space the peaks in the
isotropic spectra become broader. As the previous analysis used $b=1$, the curves
there show an intermediate behavior with respect to the two cases in
Fig.~\ref{fig:comparefilterellipses}.

The curves for the global degrees of anisotropy confirm the expected effect.
The filter with the lower aspect ratio given by the lower $b$ parameter
reduces the contributions of the 9:1 clumps relative to the 3:1 clumps
in the global degree. For the ensemble with the aligned 3:1 clumps, the
$b=1/\sqrt{2}$ filter reveals some global anisotropy also at low scales
that was not visible for higher $b$ values.
The value remains quite low, however. For the ensemble with the aligned 9:1 clumps
we see the corresponding opposite effect. The contribution of unaligned 3:1
clumps reduces the global degree of anisotropy drops from a value close to
unity to a peak below 0.8. A lower $b$ parameter thus dampens the high
sensitivity of the wavelet analysis to the most elongated structures. The
effect is, however, relatively small. The highest \changed{aspect ratio}
still provides the dominant wavelet coefficients.

As seen before already, the local degree of anisotropy does not depend
on the alignment of the different ensembles so that the two corresponding
curves for each filter always fall on top of each other. However the details of the
peak in the local degree of anisotropy are affected by the filter shape.
The change from the $b=\sqrt{2}$
filter to the $b=1/\sqrt{2}$ filter reduces the measured local degree of anisotropy
by 0.1-0.2 at small scales. In terms of the characterization of the structure
this seems undesired as the map consists of highly anisotropic clumps.
The peak is however more pronounced, allowing for a more reliable
measurement of the sizes of the anisotropic structures.

\begin{figure}
\centering
\includegraphics[angle=90,width=0.88\columnwidth]{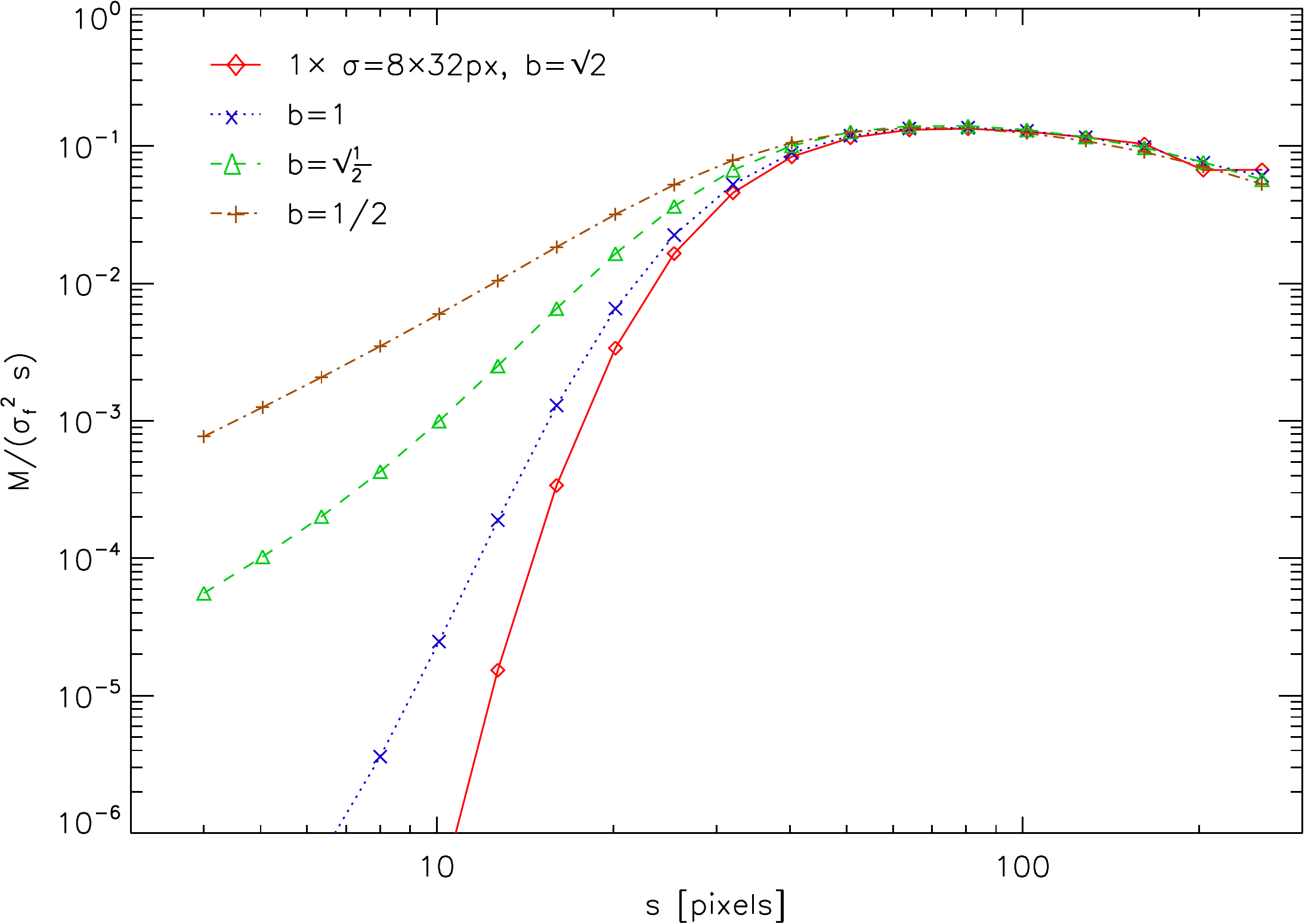}\vspace{3mm}
\includegraphics[angle=90,width=0.88\columnwidth]{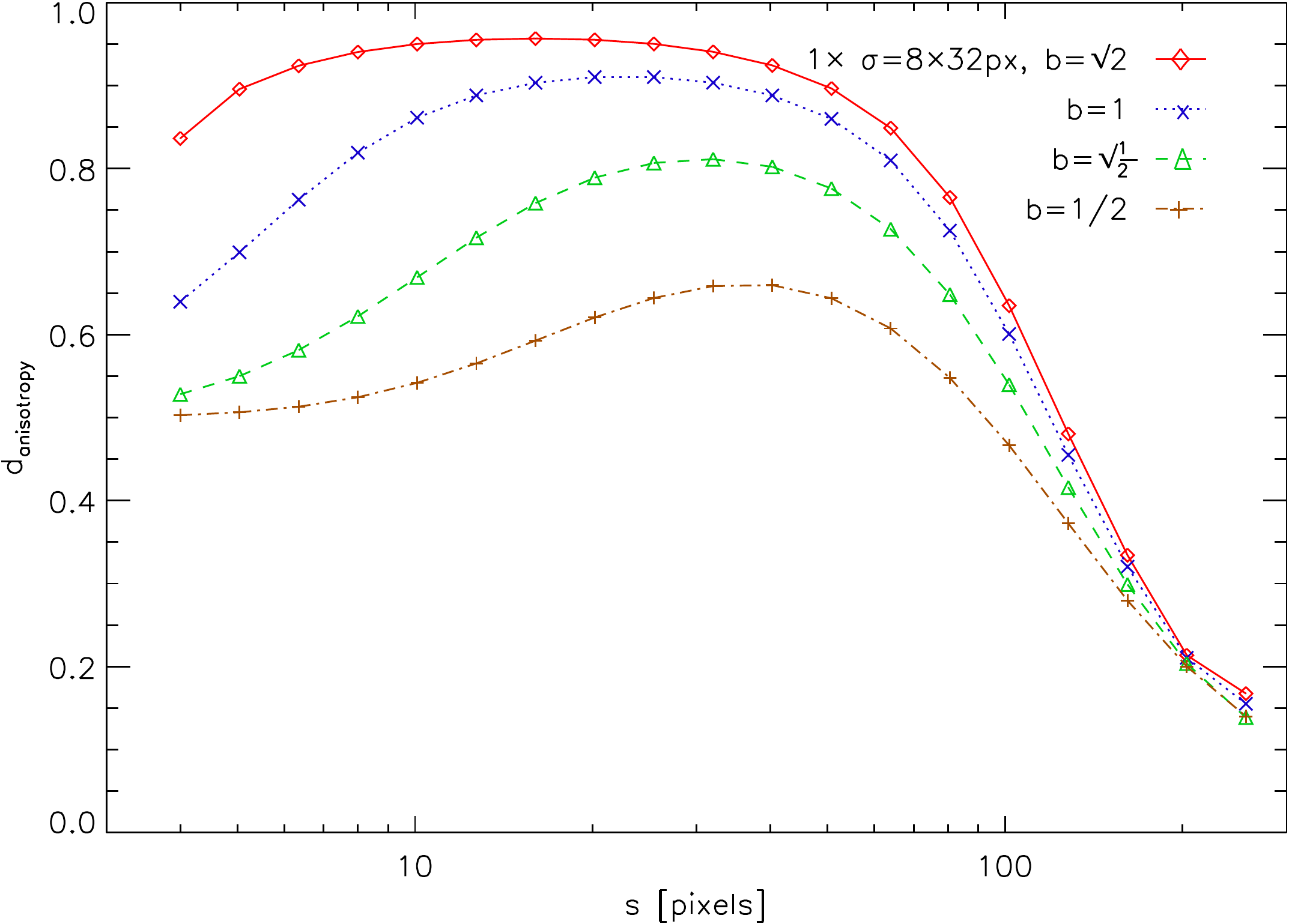}
\caption{Comparison of the wavelet spectra and local degrees of anisotropy
measured by different filter shapes for maps containing randomly placed and
oriented ellipses with a size of $\sigma=8\times 32$~pixels (see Fig.~\ref{fig:compare_sizes}).
The wavelet spectra are normalized relative to a slope of unity and the variance in the map.
}
\label{fig:comparefilterizes}
\end{figure}

\changed{The plateau parameters in Sect.~\ref{sec:simpleGaussian}) suggest that
at the 90\,\% level the peaks in the wavelet spectra hardly depend on the
localization parameter $b$ while the peak plateau in the degree of anisotropy
widens with $b^2$.}
To evaluate this scale sensitivity as a function of the filter shape
we compute the wavelet spectra and degrees
of anisotropy for the clumps with $\sigma=8\times 32$~pixels from
Fig.~\ref{fig:compare_sizes} when changing the filter shape.
The results are shown in Fig.~\ref{fig:comparefilterizes}. For a direct
comparison, we also include the curves for the $b=1$ filter already shown
in Fig.~\ref{fig:compare_sizes}, \changed{and to expand the parameter
range, we also add results for $b=1/2$. As expected,} the position
of the peaks of the wavelet spectra is hardly changed. \changed{However,
the decay at small scales changes significantly.}
The spectra turn shallower when lowering
the localization parameter $b$ approaching the spectra measured with the
$\Delta$-variance (see Appx.~\ref{sect_comp_deltavar}). A higher $b$ parameter
thus seems favorable for a sensitive size determination.
In the local degree of anisotropy we have, however, the opposite effect.
The peak becomes narrower when reducing the $b$ parameter
allowing for a better characterization of the size of anisotropic structures.
On the other hand, we find a decrease of the contrast of the peak that
prevents a good characterization of the degree of anisotropy when going to
too low $b$ parameters. A good compromise is given by $b=1/\sqrt{2}$
providing a pronounced peak and degrees of anisotropy up to about 0.8.

\begin{figure}
\centering
\includegraphics[angle=90,width=0.88\columnwidth]{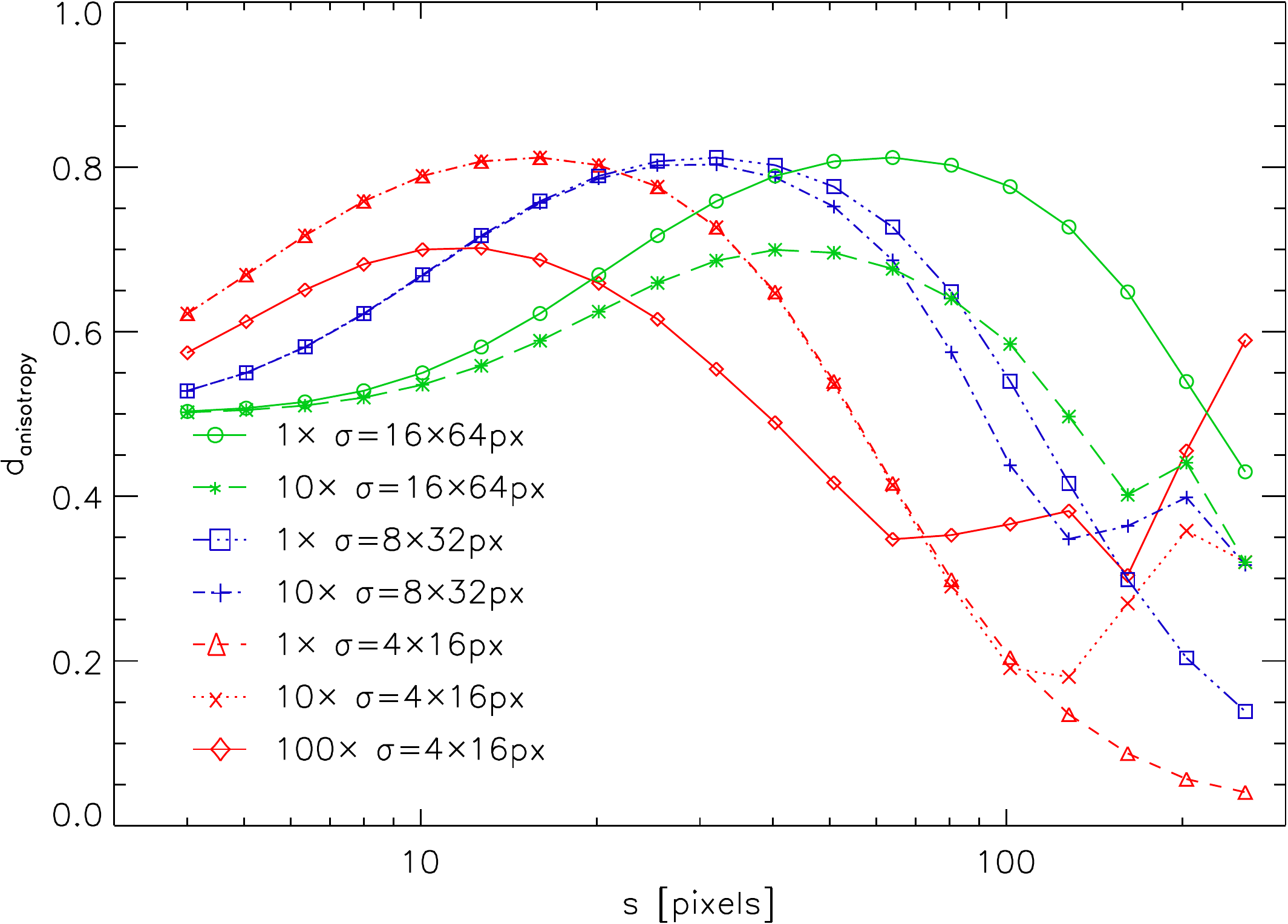}
\caption{Comparison of the local degrees of anisotropy measured by the filter with
$b=1/\sqrt{2}$ for the elliptic clumps from Fig.~\ref{fig:compare_sizes}.
}
\label{fig:comparesizes_filterE}
\end{figure}

When inspecting the maps of wavelet coefficients we also find that the
narrow $b=1/\sqrt{2}$ filter leads to an improved localization
of the coefficients that follow the individual clumps in the map more
closely being less extended along their edges. This has the positive side
effect that the analysis turns less sensitive to superposition effects,
better measuring the properties of individual clumps instead of the
impact of their mutual alignment or misalignment. This is demonstrated in
Fig.~\ref{fig:comparesizes_filterE} that shows the local degree of anisotropy
for the same clump configurations already analyzed in Fig.~\ref{fig:compare_sizes}.
Compared to Fig.~\ref{fig:compare_sizes}, the distortion of the anisotropy spectra
of the individual clumps due to the superposition of 10 or 100 of them
is noticeably reduced, but still significant.

The selection of the best localization parameter $b$ of the filter
requires some compromise. A large $b$ value provides a bad localization in
the map space, but an accurate localization in terms of the angular spectra
$\tilde{A}(s,\varphi)$ \changed{and steeper wavelet spectra, allowing
for a more accurate measurement of the minimum structure size.
However,} when looking for the location of spines of filamentary
structures in maps of wavelet coefficients, a smaller isotropic filter is
preferable. The better localization of the filter with a smaller $b$
also decreases the impact of superposition effects.
Moreover, a smaller extent of the filter in $y$ direction reduces the bias of the method towards the
most elongated structures better identifying contributions from structures
with moderate anisotropies. It also produces sharper spectra of the degrees
of anisotropy that give access to a more accurate determination of the filament
aspect ratio. On the other side it lowers the contrast in the measured degree of
anisotropy and provides smoother spectra of isotropic wavelet coefficients
turning them as broad as the $\Delta$-variance spectra providing the limit of fully
isotropic wavelets.
To allow for a systematic comparison of all these effects for individual
cases we provide the results from two filters with $b=\sqrt{2}$ and
$b=1/\sqrt{2}$ for a set of our test structures in Appx.~\ref{appx_tests}.

\changed{
To exploit the full strength of the method we combine the analysis with
different filter shapes depending on the quantity to be measured.
In the following we use the $b=\sqrt{2}$-filter when showing spectra
of isotropic or anisotropic wavelet coefficients so that we are most sensitive
to the size of the structures. This provides a fixed relation between the
peak of the isotropic wavelet spectra and the width of elongated filaments,
$s(M^i/s^2)\sub{max} \approx 2.7\mathrm{FWHM}$ with an 8\,\% error due to the unknown density
structure (see Sect.~\ref{sect_plummer}).
This filter is also used for the two-dimensional angular spectra of anisotropic
wavelet coefficients $\tilde{A}(s,\varphi)$ to obtain the best resolution in
scales and angles. To best measure the anisotropy of all filaments we
use the $b=1/\sqrt{2}$ filter for the spectra of the degrees of anisotropy.
It is also used to identify show the spines of
individual filaments in maps as it localizes their positions more
accurately. To be sure that we did not overlook any characteristic scales or structures
in the maps we always applied both filters and compared their results, but
we will only show the above mentioned combination of results in the following
unless explicitly stated.
}

\begin{figure*}
\centering
\includegraphics[angle=90,height=0.26\textwidth]{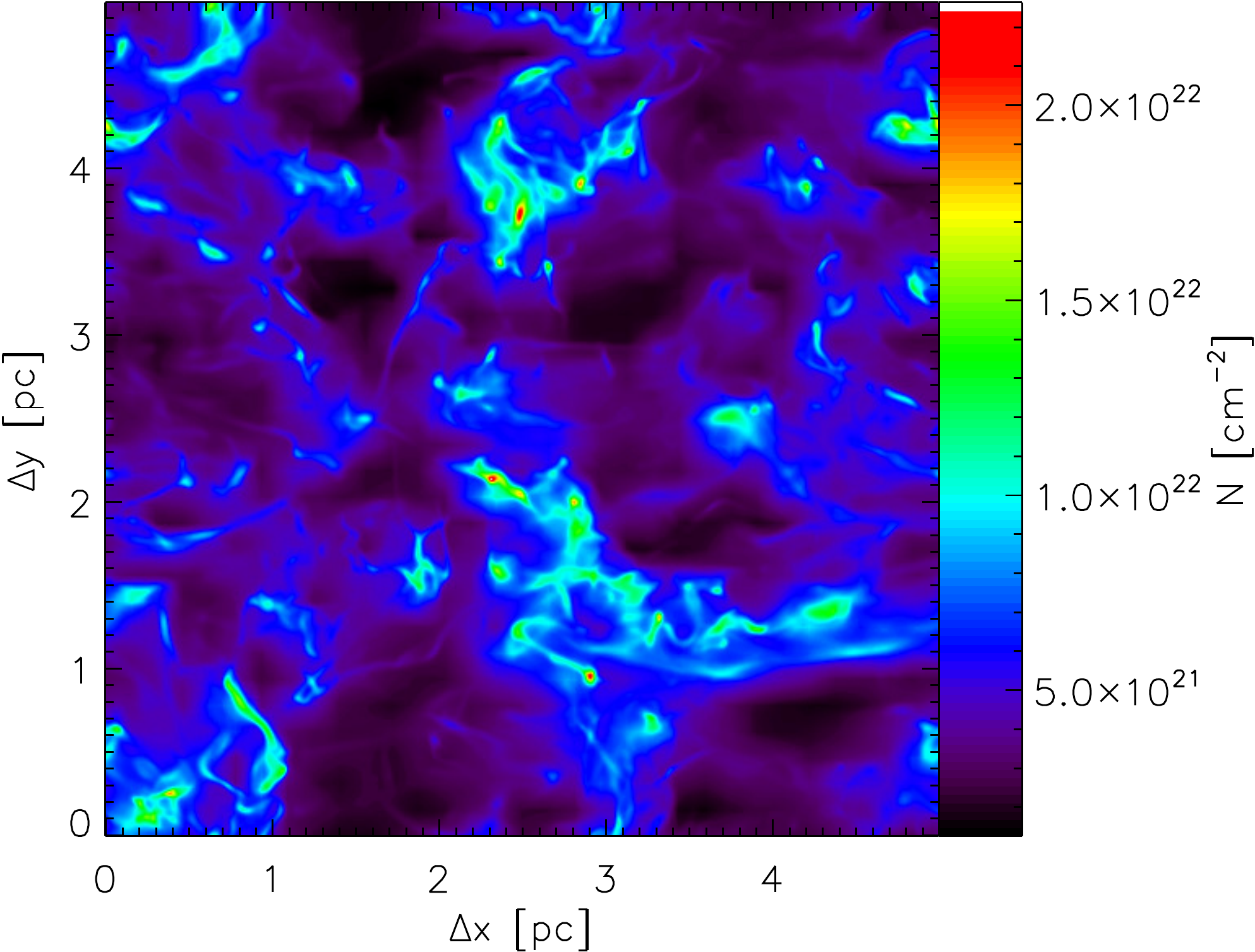}\hspace*{0.3cm}
\includegraphics[angle=90,height=0.26\textwidth]{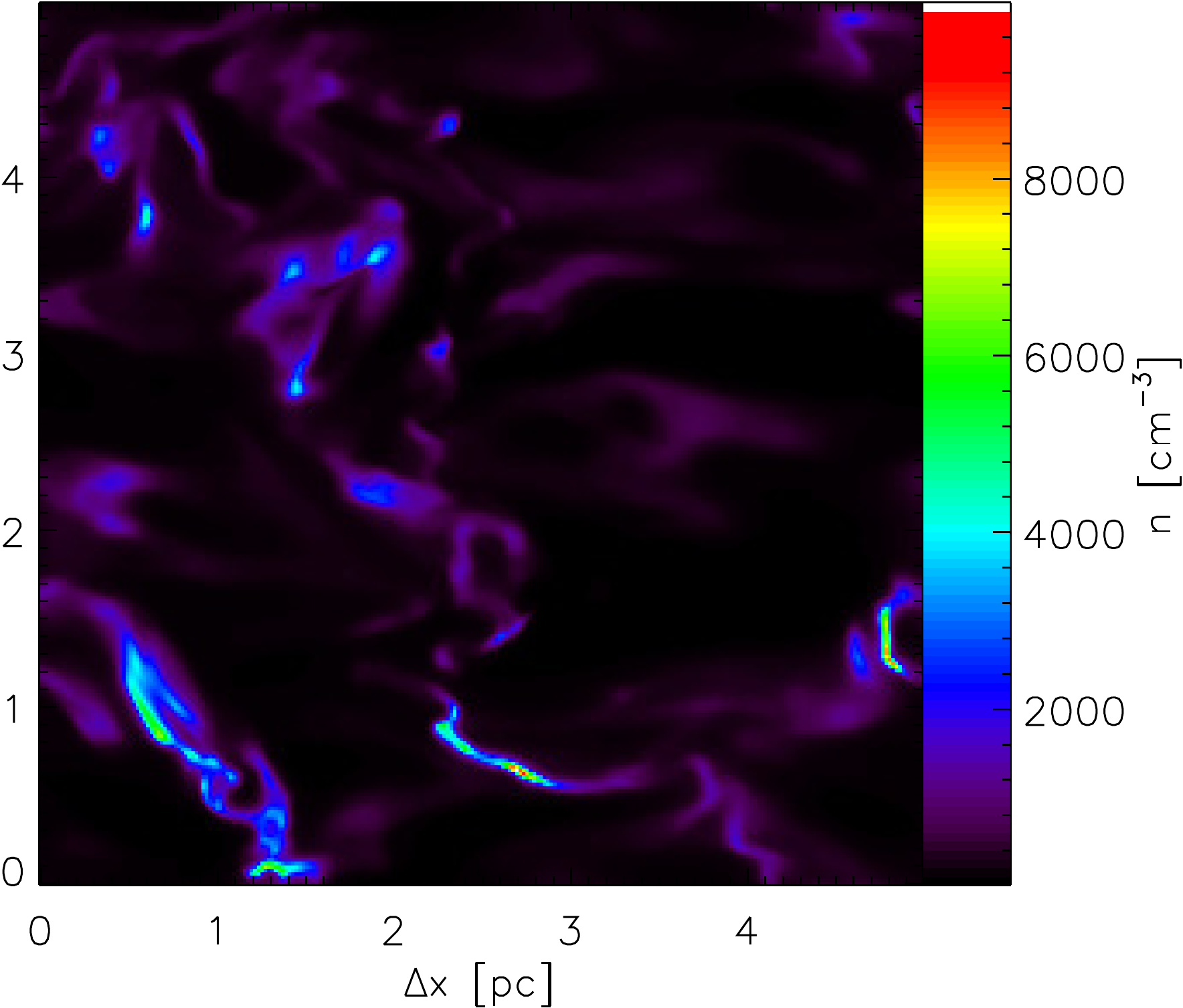}\hspace*{0.3cm}
\includegraphics[angle=90,height=0.26\textwidth]{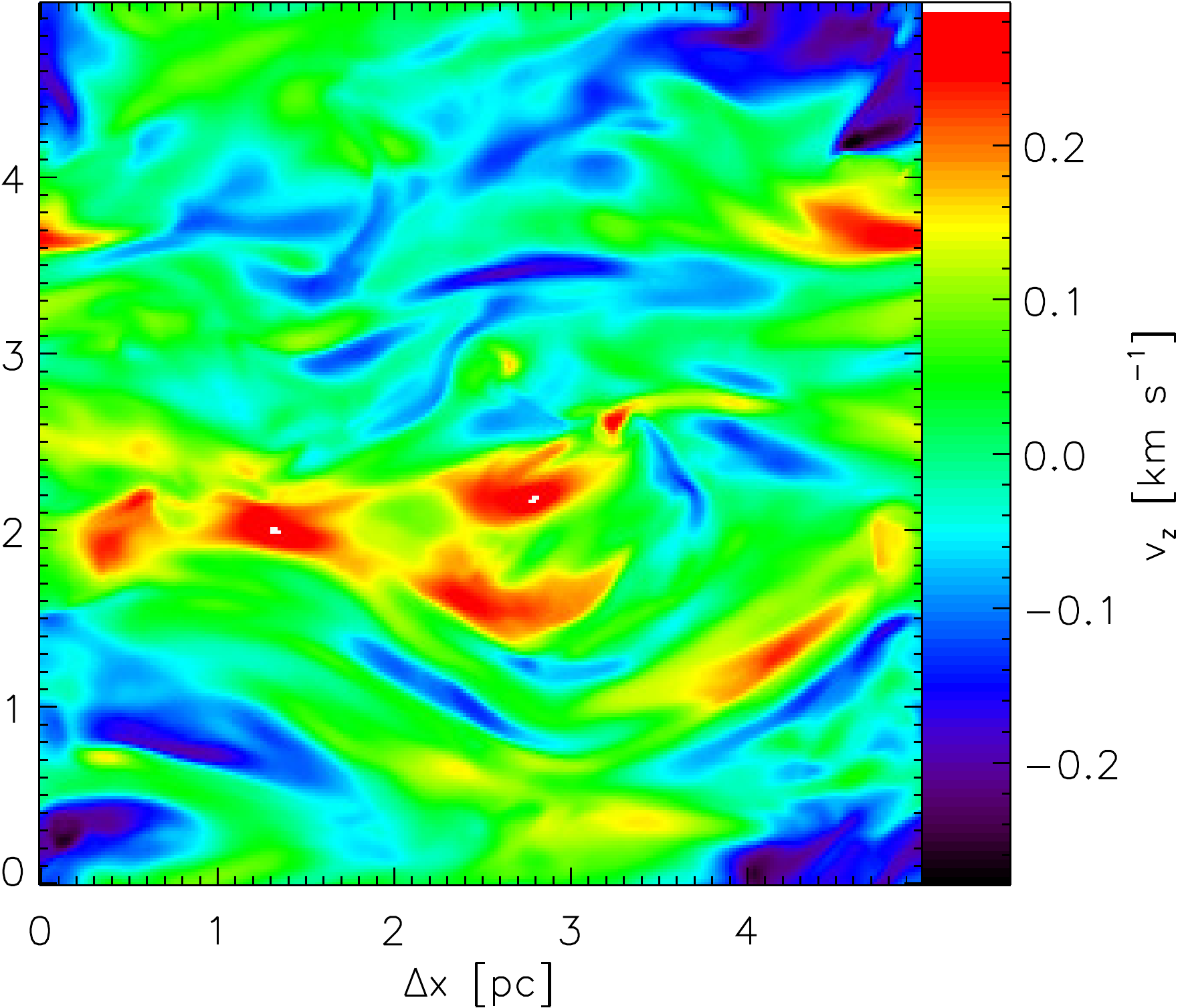}
\caption{Total column density and central slice through the density and the line-of-sight velocity
structure of the supersonic, sub-Alfv\'enic MHD turbulence simulation analyzed \citep{Burkhart2013}.}
\label{fig:mhd-simulation}
\end{figure*}

\begin{figure*}
\centering
\includegraphics[angle=90,height=0.26\textwidth]{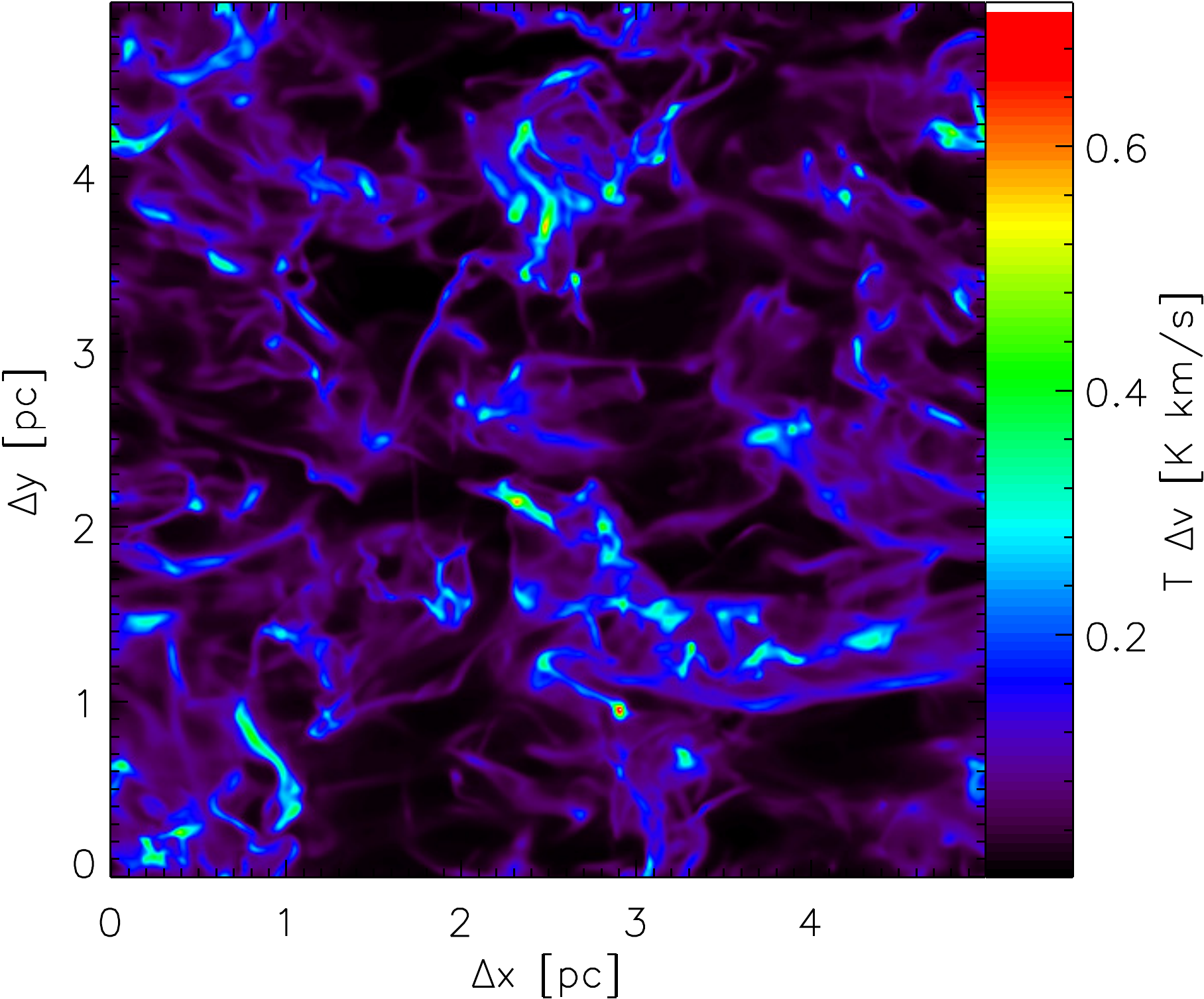}\hspace*{0.6cm}
\includegraphics[angle=90,height=0.26\textwidth]{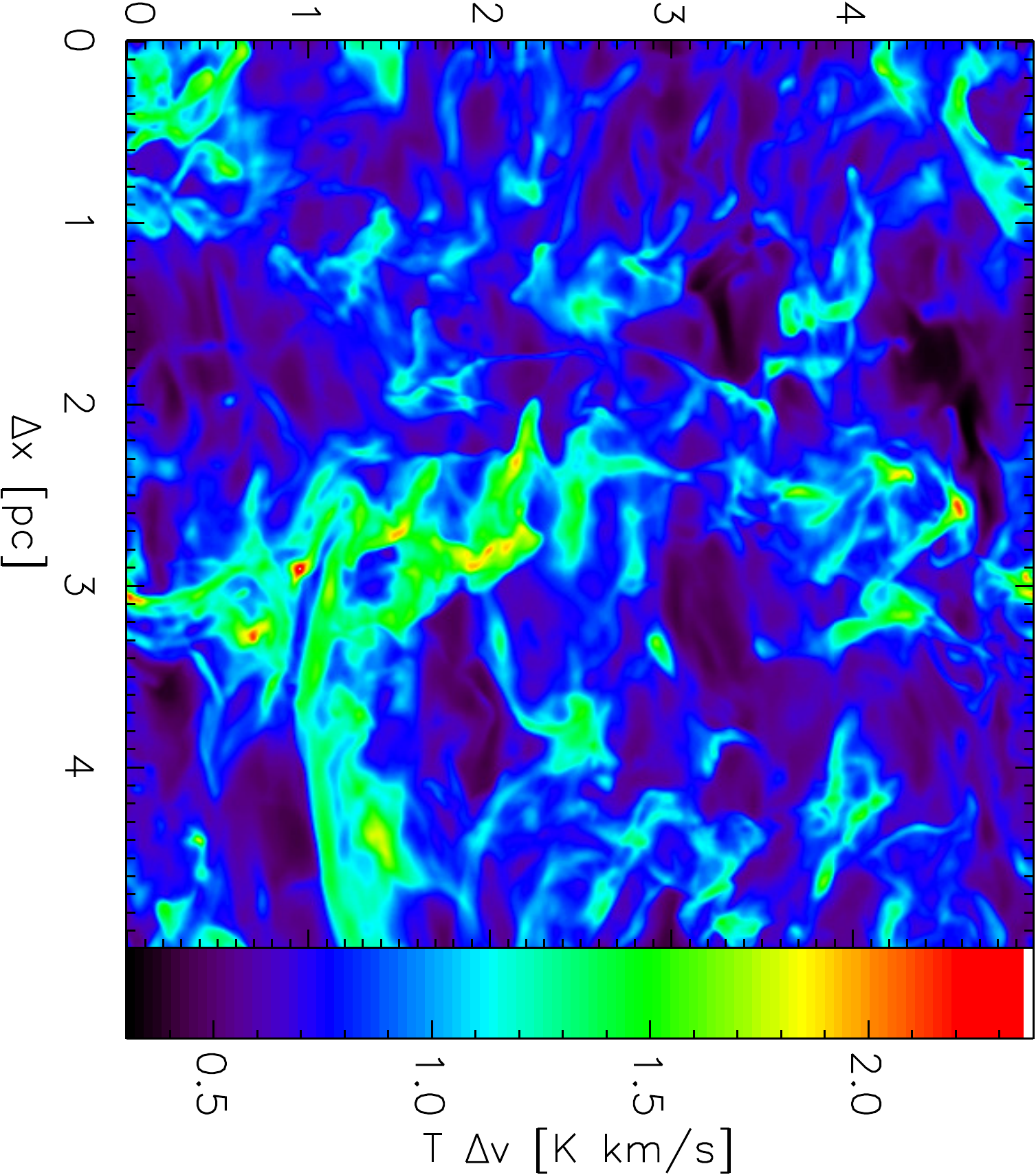}\hspace*{0.6cm}
\includegraphics[angle=90,height=0.26\textwidth]{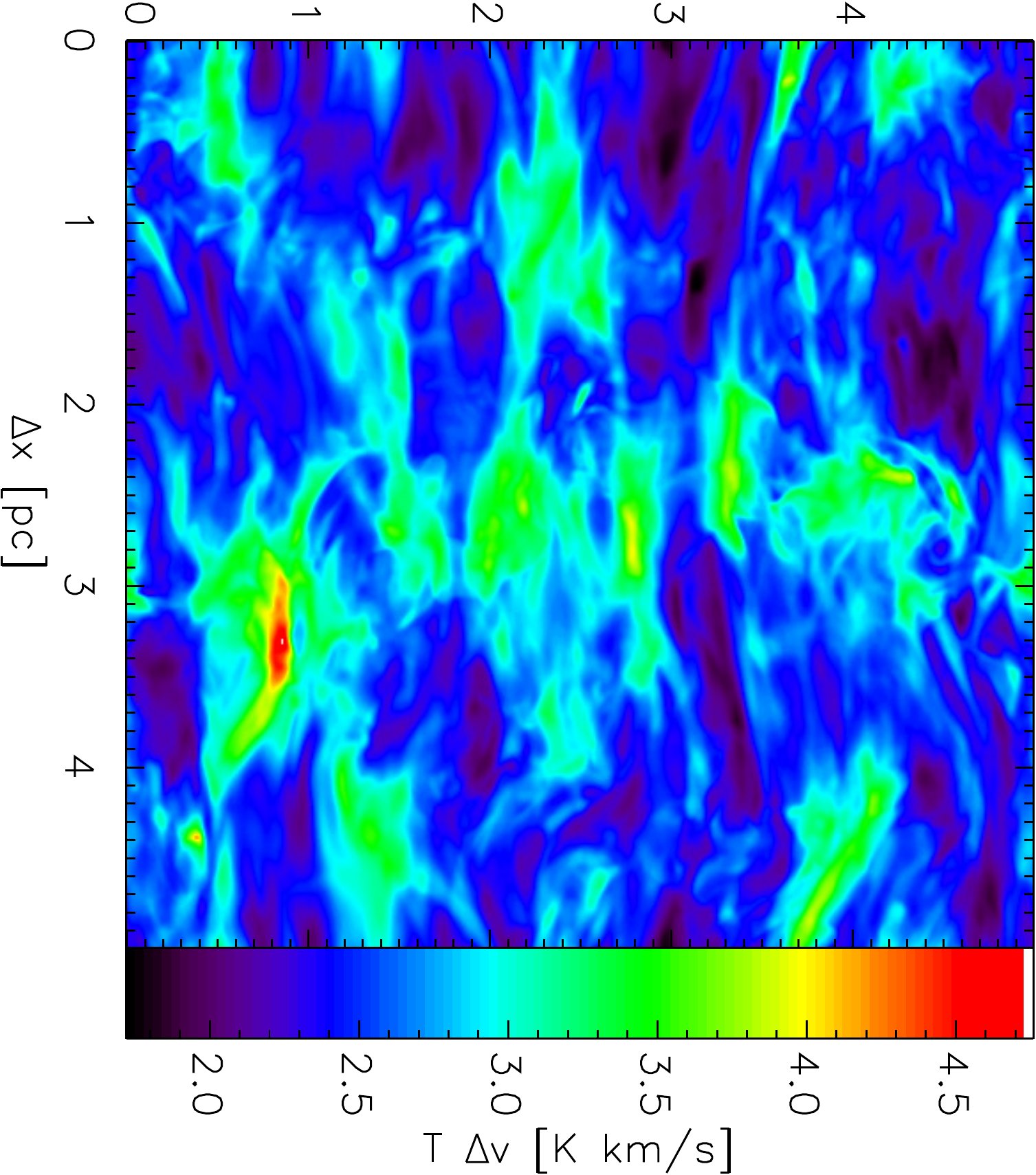}
\caption{Maps of line-integrated intensities of the $J=2-1$ transition of CO isotopologues
computed for the MHD simulation from Fig.~\ref{fig:mhd-simulation}. The
molecule abundances in the three panels are $X=5\times10^{-8}, 1.5\times 10^{-6}$, and
$4.5\times 10^{-5}$.}
\label{fig:rt-simulation}
\end{figure*}

\begin{figure*}
\centering
\includegraphics[angle=90,height=0.26\textwidth]{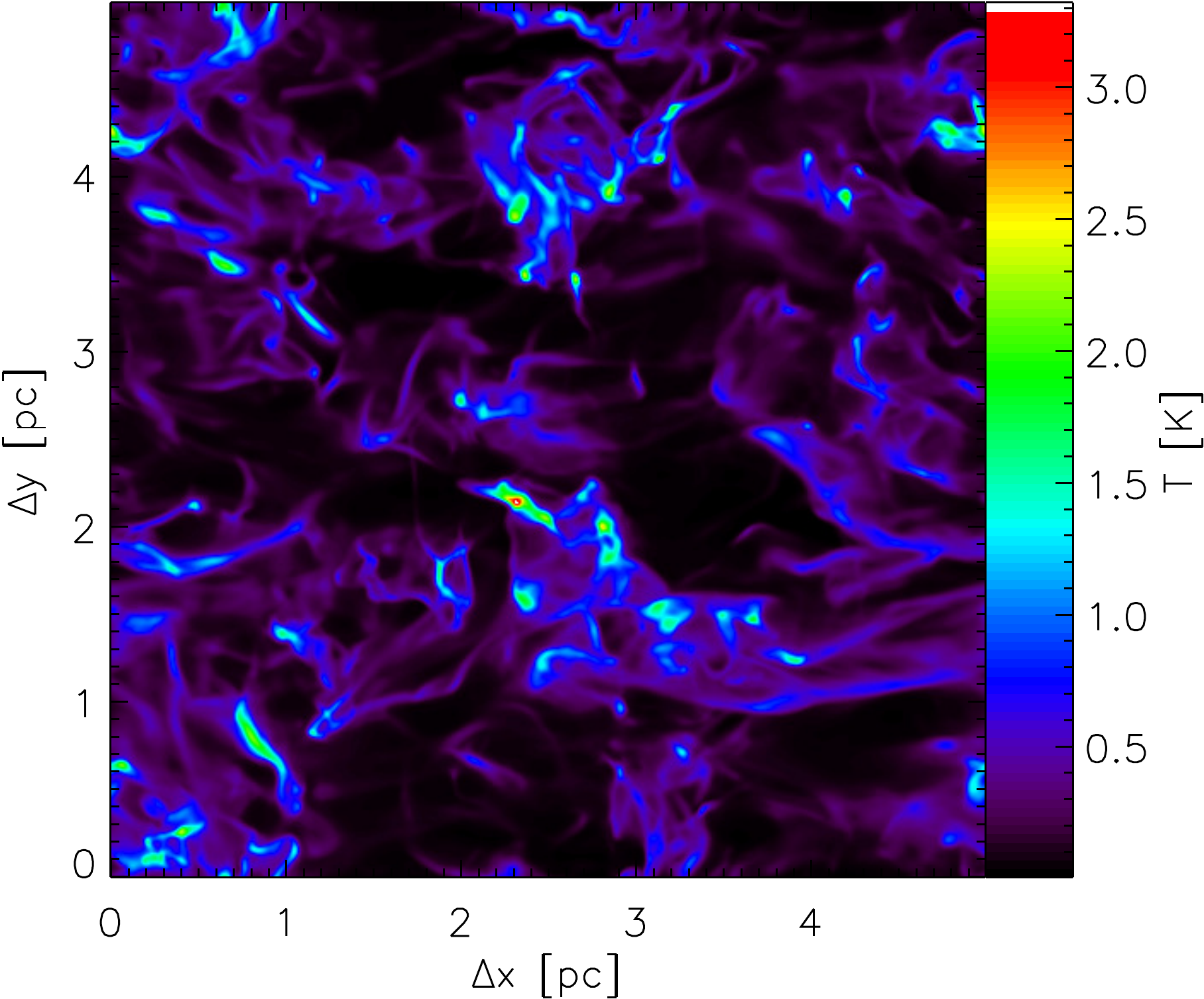}\hspace*{0.6cm}
\includegraphics[angle=90,height=0.26\textwidth]{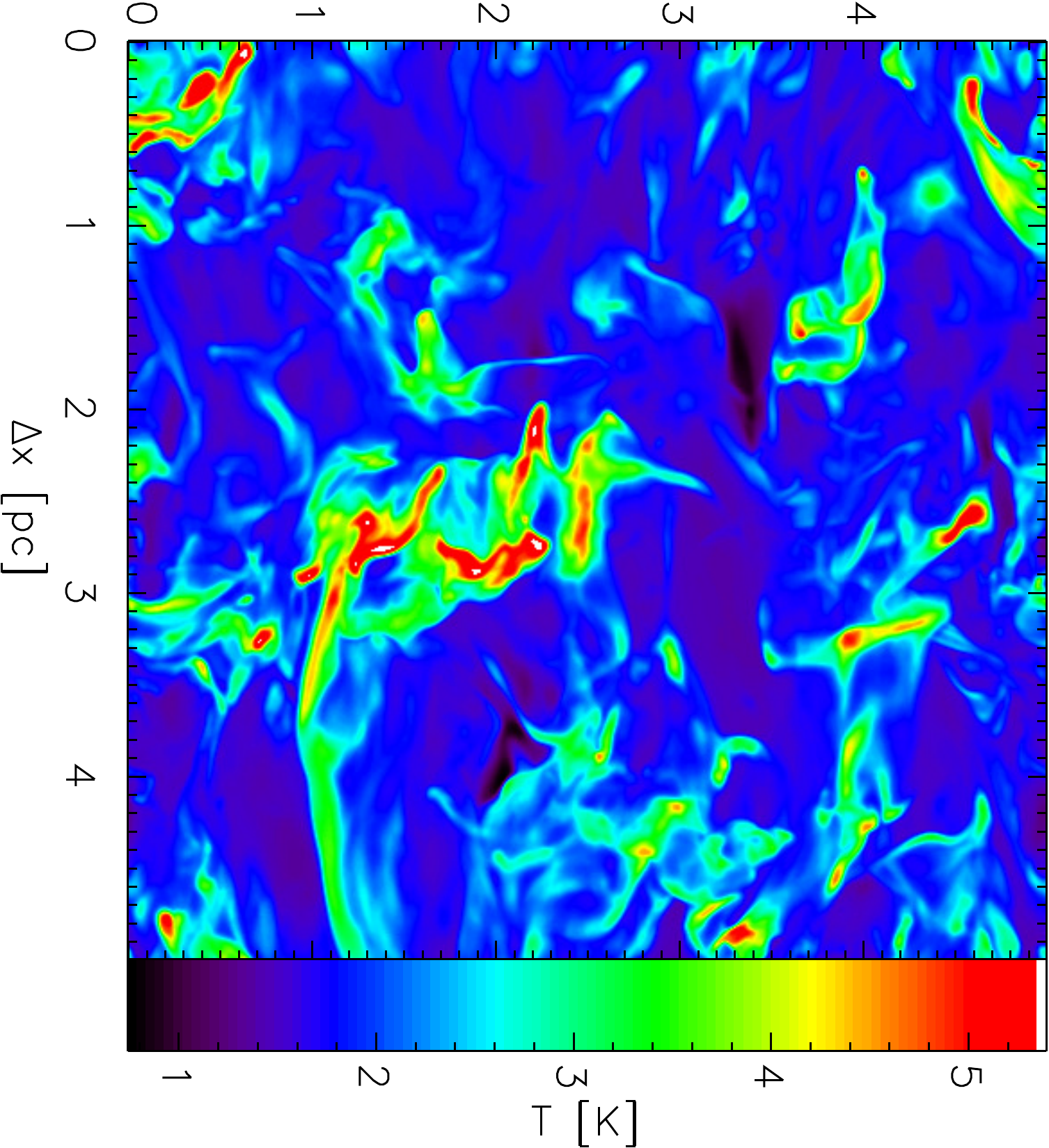}\hspace*{0.6cm}
\includegraphics[angle=90,height=0.26\textwidth]{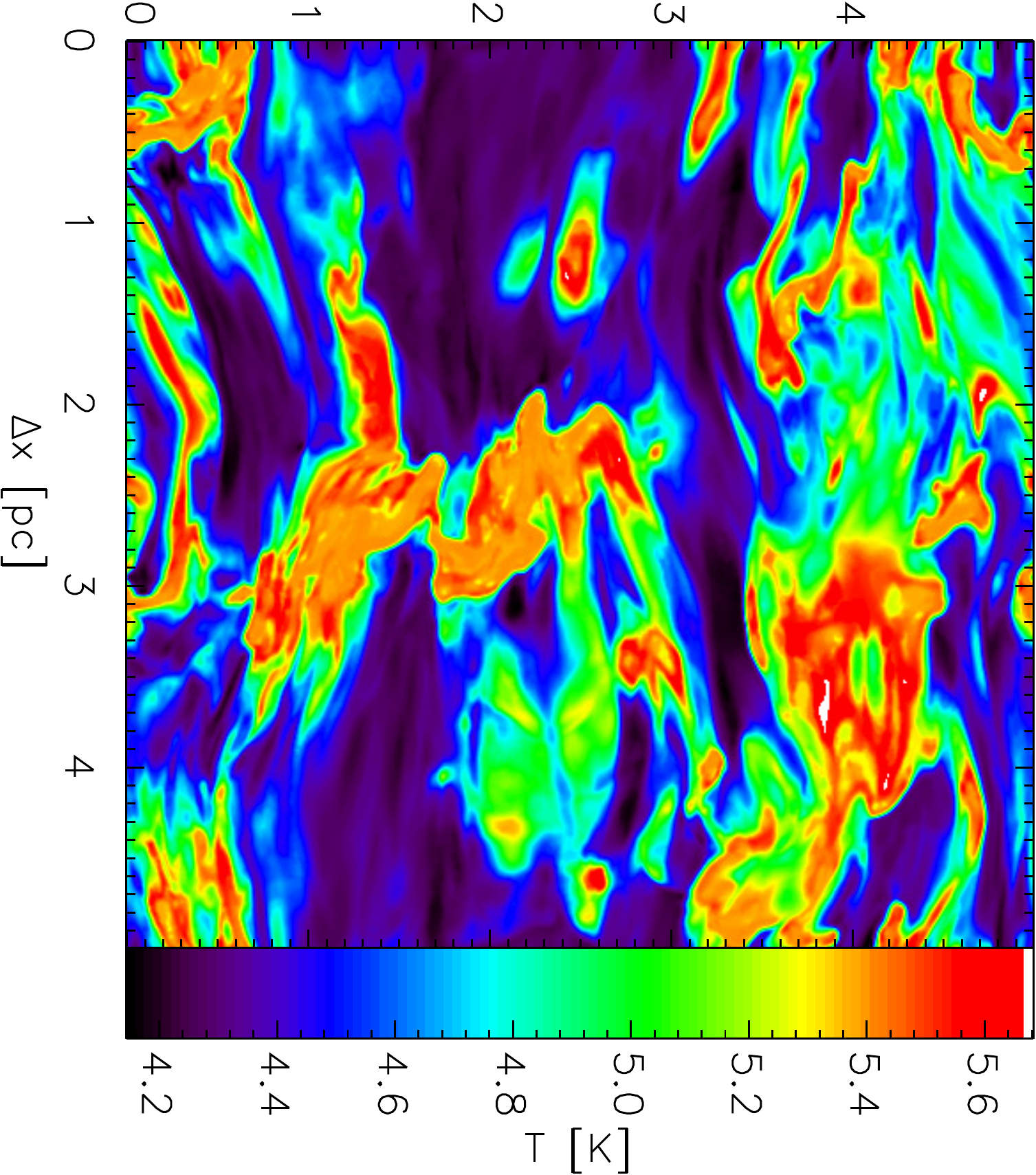}
\caption{Channel maps for the $v=0$ velocity channel of the line from
the radiative transfer simulations shown in Fig.~\ref{fig:rt-simulation}. }
\label{fig:v0-simulation}
\end{figure*}

\section{Application to MHD simulations}
\label{sect:MHD}

\subsection{Model setup}
To investigate whether the method can be used to quantify the role
of magnetic fields in the formation of anisotropic density structures
in interstellar clouds, we apply the method to a simulation
of isothermal compressible magneto-hydrodynamic (MHD) turbulence
with well controlled properties introduced by \citet{Burkhart2013}.
As a test set, we selected the model of supersonic ($M\sub{s}=7$) and
sub-Alfv\'enic ($M\sub{A}=0.7$) turbulence \citep[model 5 from Table 1 in][]{Burkhart2013}.
It is characterized by
a very filamentary density structure without any obvious preferential
direction, but a relatively anisotropic velocity structure that
is due to the combination of the magnetic field and the
large-scale turbulent driving. Fig.\ref{fig:mhd-simulation} shows the
column-density together with the density and line-of-sight
velocity in a slice through the simulation. The initial magnetic field
in the simulation was set up parallel to the $x$ direction.
The velocity structure traces this anisotropy by forming regions
of constant velocity flows parallel to the magnetic field. In contrast,
the density is dominated by individual narrow compression regions.

%
%
%
%
%

%
%

The radiative transfer simulations in \citet{Burkhart2013} and
\citet{Burkhart2013b} showed that in maps of integrated line intensities
optically thin tracers follow the highly entangled filamentary
structure of the density field while optically thick lines rather
reflect the anisotropic velocity field resulting
in thick filaments parallel to the magnetic field direction.

Following the approach described in \citet{Burkhart2013} we
generate spectral cubes of the 2--1 transition of different CO
isotopologues from the simulation using the SimLine-3D radiative transfer
algorithm \citep{Ossenkopf2002}. The code computes the excitation of the
molecules from collisions with the surrounding gas and
from the radiative excitation at the frequencies of the molecular
transitions through line and continuum radiation impinging from the
environment. In a second step it solves the ray-tracing problem
computing the observable intensities as a function of the line
velocity so that we obtain position-position-velocity cubes of the
molecular lines. The simulation is scaled to a total size of 5~pc and
a uniform gas temperature of 10~K is assumed.

To test the impact of the radiative transfer on the observable structure
we simulate the same molecular line, using the parameters of the
$^{13}$CO 2--1 transition, but change the molecular abundance to
represent different optical depths. Molecular abundances
of $X=N\sub{mol}/N\sub{H_2}=5\times10^{-8}, 1.5\times 10^{-6}$, and
$4.5\times 10^{-5}$ provide results that are \changed{roughly} representative
for observations of the C$^{18}$O, $^{13}$CO, and $^{12}$CO
isotopologues \changed{\citep[see e.g.][]{LangerGraedel1989,Bolatto2013}}.
Mean line-center optical depths in the simulations
are $\tau=0.1$, 3.4 and 74 for the three abundances.
The resulting velocity integrated intensity maps
are shown in Fig.~\ref{fig:rt-simulation}. Channel maps for the
component with zero line-of-sight velocity, $v=0$, are
shown in Fig.~\ref{fig:v0-simulation}.

Eye inspection of the maps of line-integrated intensities in
Fig.~\ref{fig:rt-simulation} confirms the findings from
\cite{LazarianPogosyan2004} and \cite{Burkhart2013b} on the impact
of the optical depth on the size spectrum of the observable structures.
When integrated over a broad velocity range, the intensity maps
show less small scale structures at high optical depths.
Larger structures become more prominent when going to
the high optical depths. This means that their power spectrum
turns shallower. On top of this known behavior we see, however,
that there is also an alignment effect. All maps look very
filamentary but while the filaments at low optical depths
($X=5\times10^{-8}$) show no preferential direction those at
high optical depths ($X=4.5\times10^{-5}$) are mainly aligned
parallel to the $x$ axis. They trace the velocity and magnetic
field structure.

The channel maps do not integrate over the velocity structure
therefore providing a different combination of density and
velocity structure \citep[see e.g.][]{LazarianPogosyan2000}.
The optically thin channel map in Fig.~\ref{fig:v0-simulation}
is very similar to the structure also seen in the line-integrated map.
For the map with an optical depth of 3.4 there is still some
resemblance but in the high optical depth case the $v=0$ channel
map looks completely different from the integrated line map.
The two cases with a significant optical depth also show some
increase of the typical structure size in the channel maps, but less
prominent than for the integrated line maps. This matches
the predictions from \cite{LazarianPogosyan2004} that thin slices
show a steeper power spectrum than integrated maps for our parameters.
In contrast to the integrated line maps, the high optical depth
channel map does not show a clear global alignment of the
structures. At the
highest optical depths there even seems to appear some alignment
perpendicular to the magnetic field direction.
For a quantification of these effects we need to apply the
anisotropic wavelet analysis to all six maps.

\subsection{Wavelet analysis}

\begin{figure}
\centering
\includegraphics[angle=90,width=0.88\columnwidth]{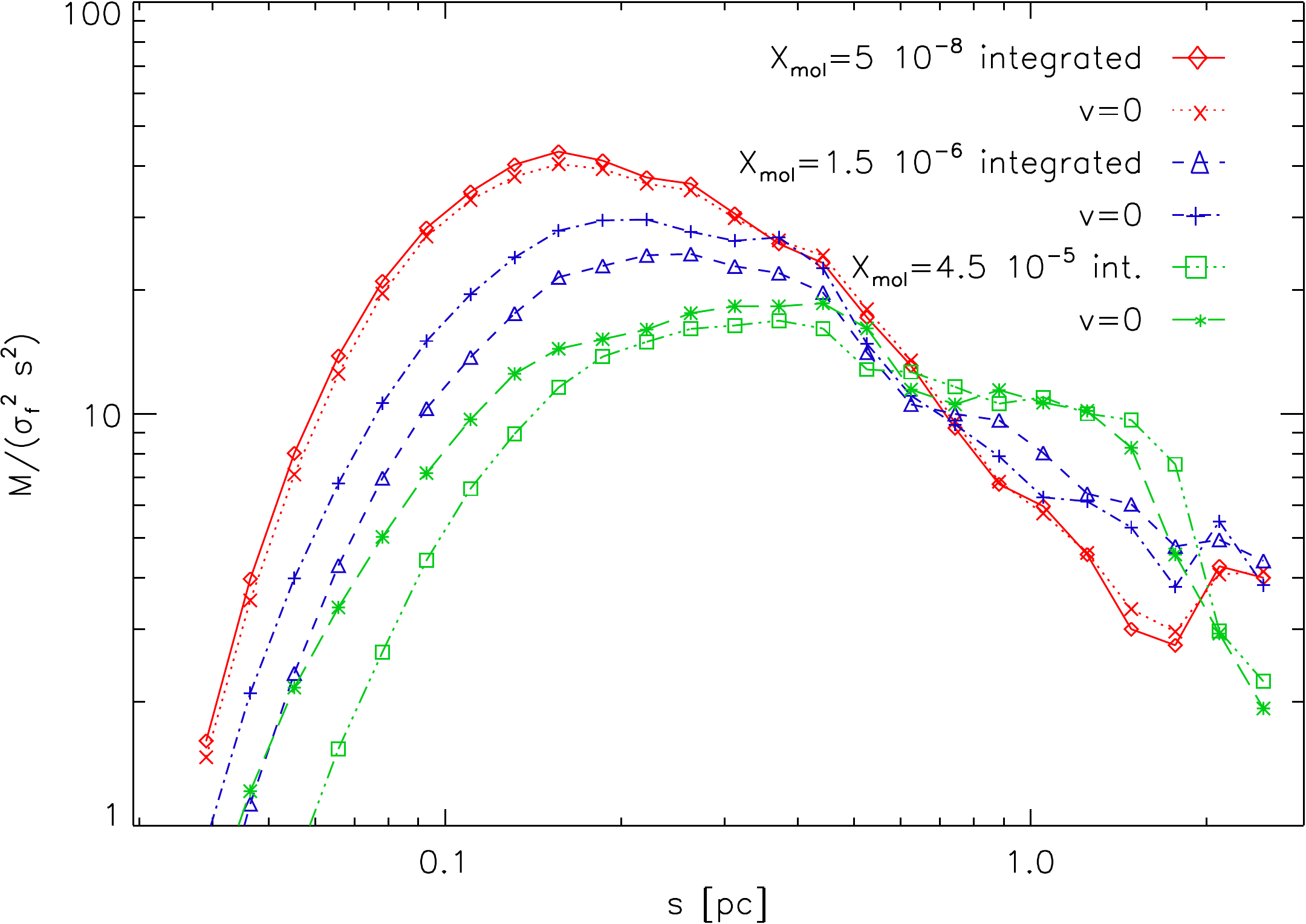}
\caption{Spectra of normalized isotropic wavelet coefficients
$M^i(s)/(\sigma_T^2 s^2)$ as a function of the filter
size for the six maps from Figs.~\ref{fig:rt-simulation} and \ref{fig:v0-simulation}.
\changed{Following Sect.~\ref{sect_filtershape} the $b=\sqrt{2}$ filter was used here.}
}
\label{fig:mhd-spectra}
\end{figure}

Figure~\ref{fig:mhd-spectra} shows the normalized spectra of isotropic wavelet coefficients
for the integrated line maps (Fig.~\ref{fig:rt-simulation}) and the maps of the
$v=0$ velocity slices (Fig.~\ref{fig:v0-simulation}) in the three simulations.
In the optically thin case, the spectra of the integrated line map and the
channel map agree. They show a relative surplus of small scale structures and
a lack of structures with sizes above 1~pc \changed{compared to the other maps}. The maps with the larger optical depth
are characterized by wavelet spectra \changed{with a higher contribution from larger
scales and a lower contribution from small-scale structures, corresponding to a higher slope
between 0.1 and 1.5~pc}. In agreement with the discussion by
\cite{Burkhart2013b} larger optical depths blur out all small filaments and merge
them into large systematic structures. The suppression of small scale structures
\changed{shifts the peak of the normalized spectra from 0.15~pc to 0.35~pc.}
The line-integrated intensity maps are more affected
than channel maps. At the largest optical depths the isotropic wavelet coefficients
\changed{form a plateau that extends up to $s \approx 1.5$~pc. Using the
calibration of the plateau edge for highly elongated Gaussians from Sect.~\ref{sec:simpleGaussian}
this corresponds to a maximum filament width of $\sigma\sub{minor} \approx 0.2$~pc or a
FWHM$\approx 0.45$~pc}.
This number is in rough agreement with the visual impression for the width of the
large filaments in Figs.~\ref{fig:rt-simulation} and \ref{fig:v0-simulation}.
In the
optically thinner maps we rather see a continuous hierarchy of smaller and smaller
filaments.

\begin{figure}
\centering
\includegraphics[angle=90,width=0.88\columnwidth]{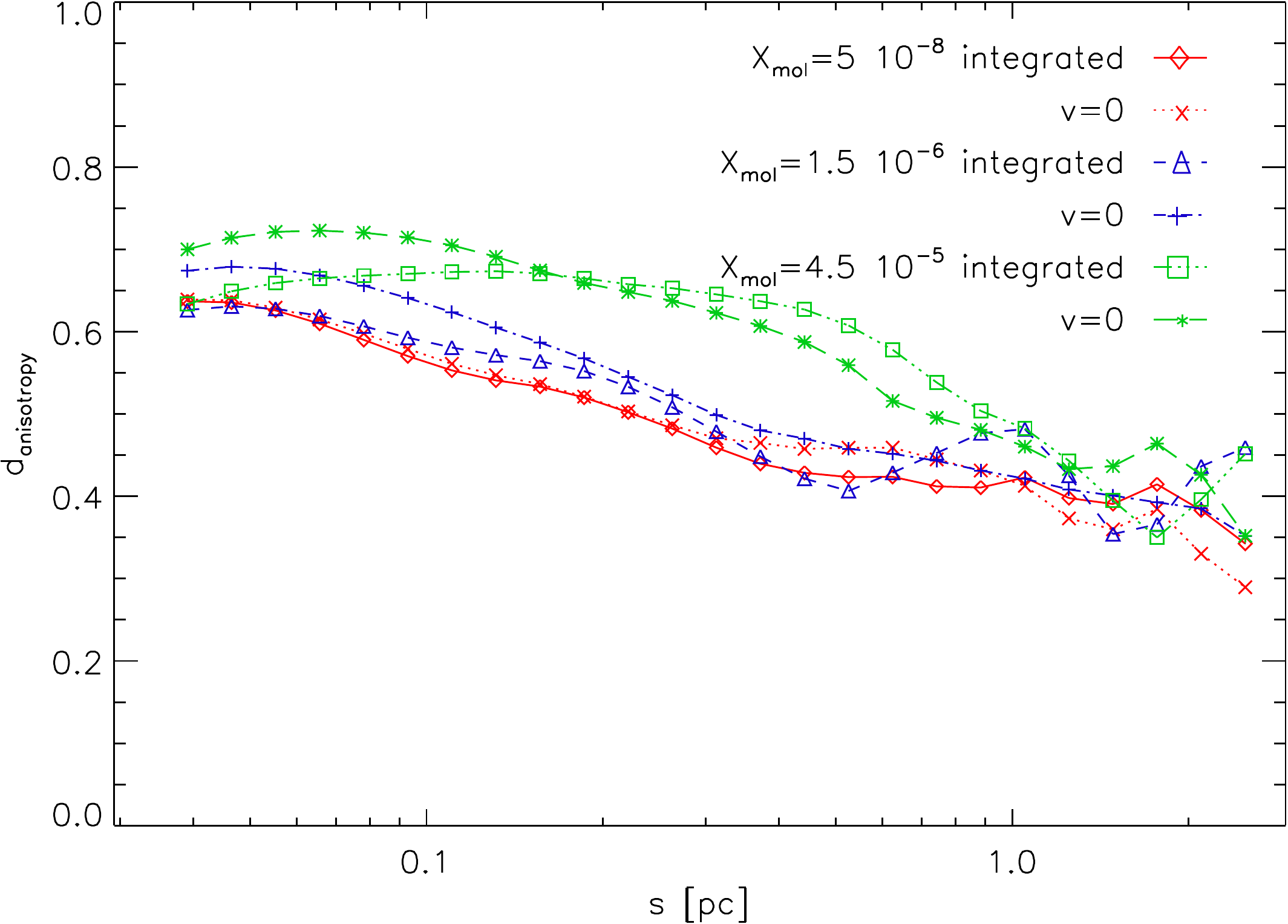}\vspace{3mm}
\includegraphics[angle=90,width=0.88\columnwidth]{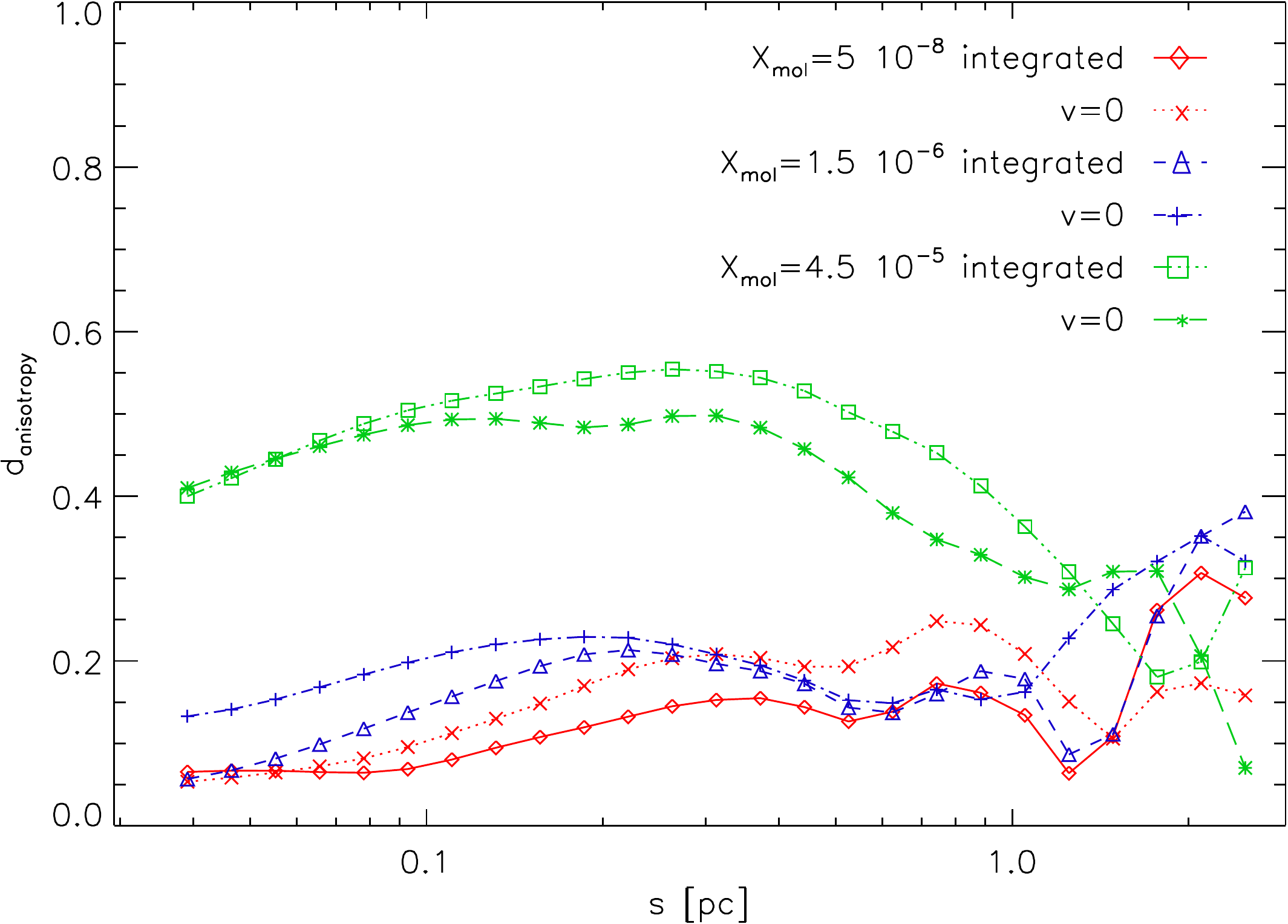}
\caption{Spectra of the local degree of anisotropy (top panel) and the global
degree of anisotropy (bottom panel) for the six maps from Figs.~\ref{fig:rt-simulation} and \ref{fig:v0-simulation}.
The filamentary density structure seen at low optical depths provides a high degree
of local anisotropy at small scales. But those filaments appear unaligned.
For large optical depths (brown and purple curves) the velocity filaments
dominate and the global anisotropy in $x$-direction becomes significant.
\changed{Following Sect.~\ref{sect_filtershape} the $b=1/\sqrt{2}$ filter was used for
the degrees.}}
\label{fig:mhd-degrees}
\end{figure}

Figure~\ref{fig:mhd-degrees} shows the spectra of the local (top panel) and global
degrees (bottom panel) of
anisotropy for the six maps. At small scales the local anisotropy is high in all maps,
the highest degree is seen in the channel map for the high optical depth case.
The degree of local anisotropy slightly decreases with scale. For the optically
thick maps it remains high up to scales of $s\approx 0.7$~pc, suggesting
an increase of the filament length with the optical depth. However, even the
optically thin maps still show a significant local anisotropy up to the largest
scales indicating that the individual small filaments visible in the maps create
a hierarchy of filamentary structures at larger and larger scales.

The biggest difference between the maps at high optical depths and those at low
and moderate optical depths shows up in the global degree of anisotropy. The
global alignment visible in the maps creates a degree of global anisotropy of
about 0.5 up to scales of  $s\approx 0.7$~pc for the optically thick maps
while there is negligible global anisotropy in the other maps at those scales.
The difference between the curves of the optically thick channel map and
the corresponding line-integrated map indicates that the globally aligned
structures are somewhat longer for the integrated map.
Some accidental global anisotropies show up in all maps at the largest
scales when the individual filaments merge in all cases.

Altogether, this approach helps to quantify the imprint of the magnetic field
on the structure formation as a function of the spatial scale. Small-scale
filaments are entangled with the field lines while the large-scale magnetic
field preserves the global anisotropy. As the density structure is dominated
by the small scale fluctuations while the velocity structure inherits more
of the global anisotropy, we can exploit the different selectivity of optically thin
and thick maps to the two aspects to characterize the anisotropy in both
structures. Higher optical depths emphasize the global anisotropy. They
produce maps with wider filaments aligned with the global magnetic field
as they trace the
broader velocity dispersion on larger scales and the connection of the
velocity field to the magnetic field structure. Optically thin lines trace
the individual small-scale shocks dominating the density structure without
any preferred direction. Both
structures are filamentary, but the ratio between local and global anisotropy
clearly separates them. The spectra of wavelet coefficients allows us
to quantify the distribution of the filament widths.

\section{Observed maps}
\label{sect:obs}

\begin{figure}
\centering
\includegraphics[angle=90,width=0.86\columnwidth]{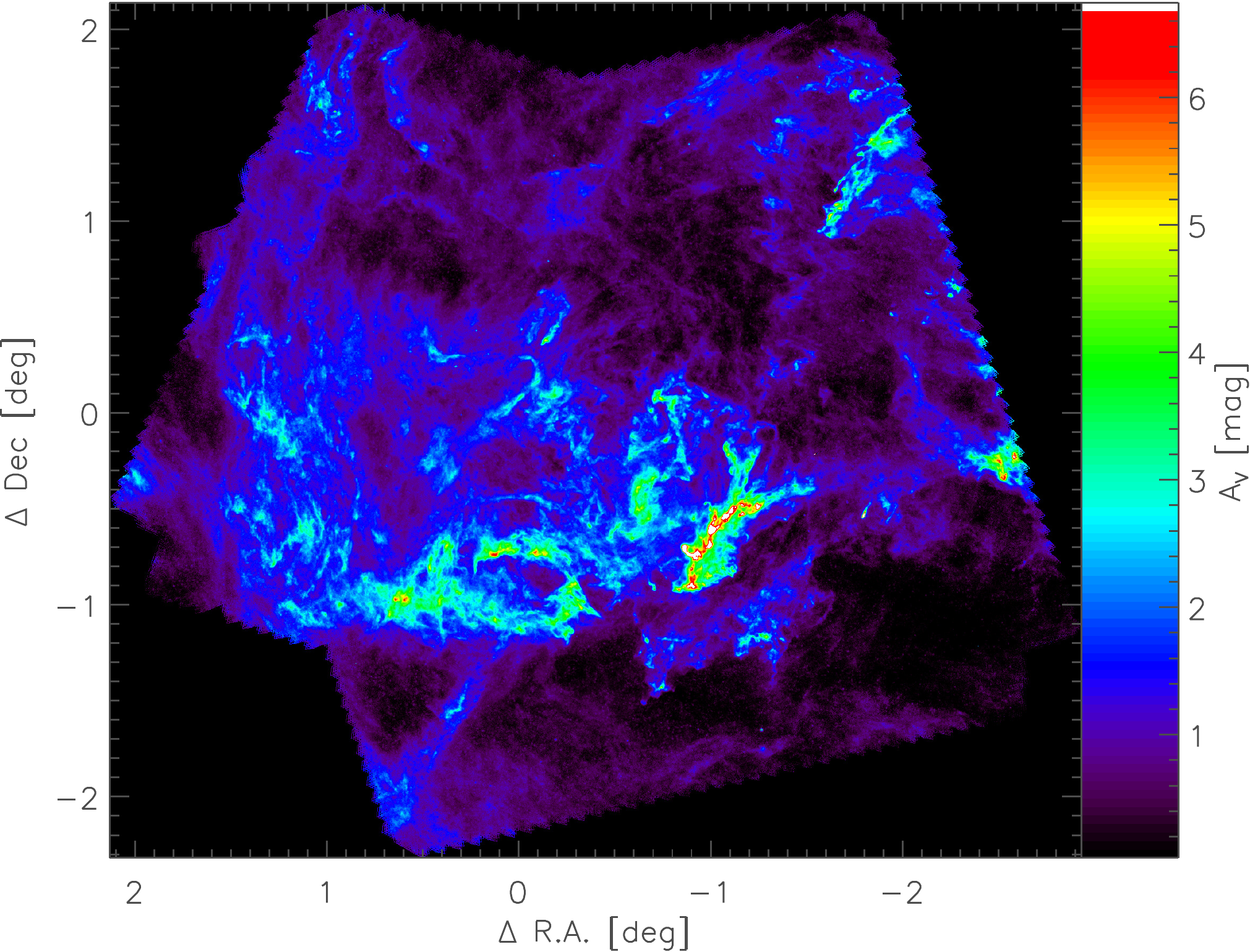}\vspace{3mm}
\includegraphics[angle=90,width=0.92\columnwidth]{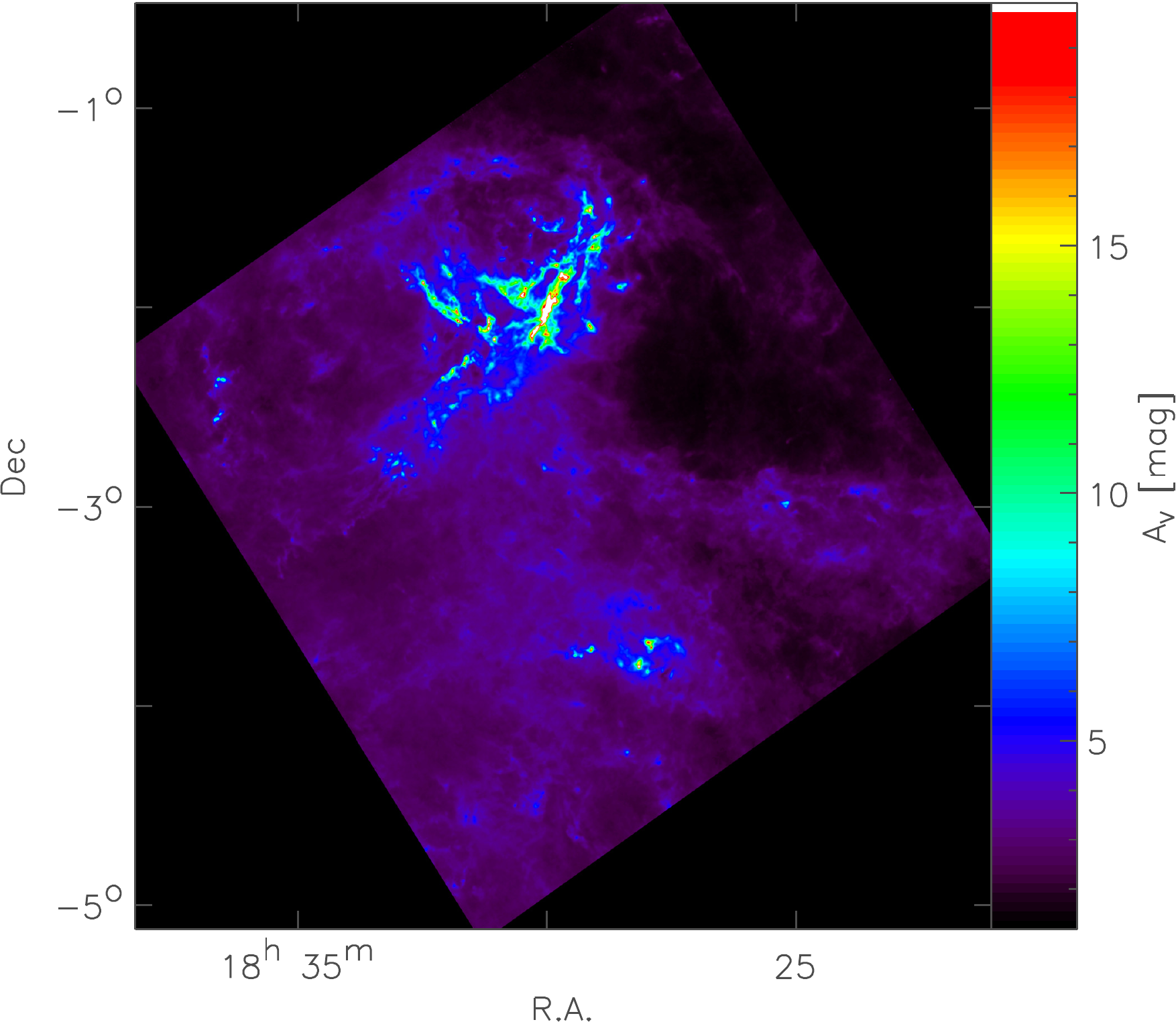}
\caption{\changed{Column density maps of the Polaris and Aquila regions derived from
Herschel dust emission data} previously
analyzed by \citet{Andre2010}, \citet{Schneider2013}, and \citet{Konyves2015}. As the map of Polaris includes
the celestial pole
coordinates are given relative to $3\chr28\cmin16.0\csec$, $88\deg34'32.0''$ (J2000).
One degree in the map corresponds to 2.6~pc in Polaris and 4.5~pc in Aquila.
\changed{We specify the column density in terms of the visual extinction using}
the standard factor $N\sub{H}/A\sub{V}=1.87\times 10^{21}$~\pscm{} \citep{Bohlin1978}.}
\label{fig:maps-aquila-polaris}
\end{figure}

To verify whether our method confirms the findings of \citet{Andre2014} and \citet{Arzoumanian2011}
that most filaments in different clouds have a characteristic \changed{FWHM} of about
0.1~pc, we apply our analysis to two different column density maps obtained
from observations of the ESA Herschel Space Observatory \citep{Pilbratt2010}.
The Aquila rift and Polaris Flare are very different
regions and were already compared by \citet{Andre2010} and \citet{Schneider2013}.
The  Aquila rift is an active high-mass star-forming region, the Polaris Flare
is still quiescent at a much lower density. Because of the different conditions
a common property such as a uniform filament width would give an important
clue on the underlying physics governing the ISM structure independent of the
evolutionary state.
The two column density maps are shown in Fig.~\ref{fig:maps-aquila-polaris}.
\changed{The column densities were taken from  \citet{Andre2010} and \citet{Konyves2015}.
They were obtained by a gray-body fit to the {\it Herschel} PACS and SPIRE continuum
maps and have a resolution of 36~arcsec.}
\changed{For practical reasons, we give the column density in units of the visual
extinction $A\sub{V}$ using the standard extinction factor
$N\sub{H}/A\sub{V} = 1.87 \times 10^{21}$~cm$^{-2}$ mag$^{-1}$
\citep{Bohlin1978}. This avoids wavelet coefficients of order $10^{45}$.}

\begin{figure}
\centering
\includegraphics[angle=90,width=0.86\columnwidth]{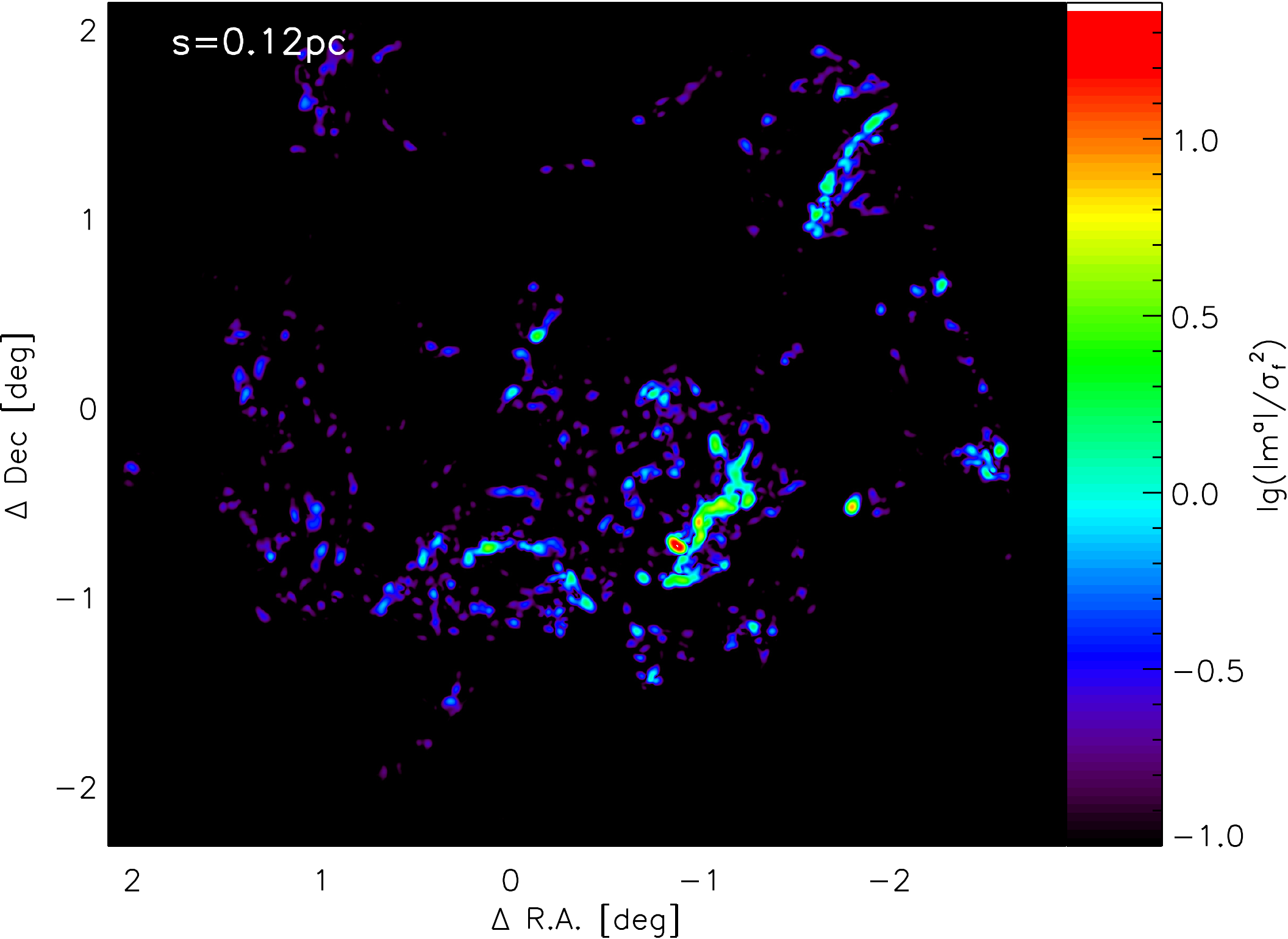}\vspace{3mm}
\includegraphics[angle=90,width=0.86\columnwidth]{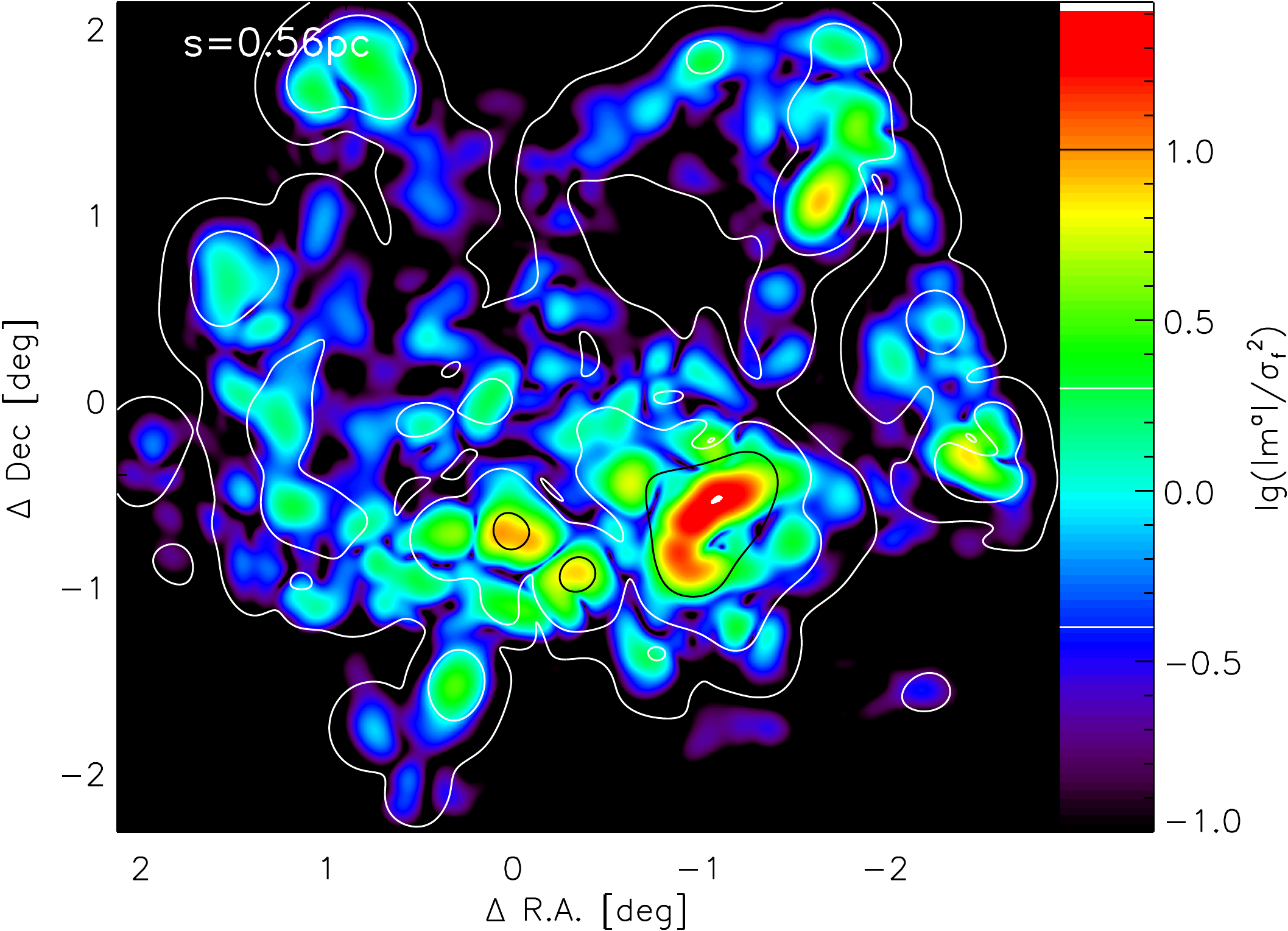}
\caption{Maps of the amplitude of the anisotropic wavelet coefficients
$|m^a(s,\vec{x})|$ in logarithmic scaling for the Polaris column density map. The upper panel
shows the coefficients for the filter sizes of $s=0.12~\mathrm{pc}=0.044\deg$, the lower panel for a
filter size of 0.56~pc=0.21\deg. In both maps the colors show the results for the $b=1/\sqrt{2}$-filter.
The contours in the large scale map represent the coefficients from the $b=\sqrt{2}$-filter.
We omit the corresponding contours in the small-scale map as they hide too many
details and only confirm the behavior also seen in the large-scale map.}
\label{fig:ma_polaris}
\end{figure}

To discuss the properties of individual filaments and to compare our results with
filament finders such as {\it DisPerSe} \citep{Sousbie2011} or {\it getFilaments}
\citep{Menshchikov2010} it is useful to directly inspect
the maps of anisotropic wavelet coefficients localizing the individual
filaments (see Fig.~\ref{fig_waveletmaps}).  In Fig.~\ref{fig:ma_polaris} we show
the coefficient maps for two filter sizes applied to the Polaris column density map.
The colors represent the results for the \changed{localization parameter $b=1/\sqrt{2}$
that should best trace the places of the filaments}. In the maps
from both filter sizes the main filament, often called the `saxophone', dominates
the wavelet coefficients. In the map for the small filter size chains of peaks follow the
spines of the individual filaments. The widespread emission and more isotropic
structures seen e.g. in the eastern part of the map are filtered out. Comparing
the map of wavelet coefficients with the original map provides a good impression
which structures provide the main anisotropies.
For the larger filter the somewhat broader structures east of the saxophone also
provide a significant contribution. More extended low column density filaments
become visible. In this way we can identify different types of filaments.
The coefficients for very
small filter sizes follow the filament spines also traced by the filament finders.
The coefficients for larger filters trace larger and larger filamentary structures
that are not necessarily correlated with those at small scales. By providing the
full spectrum of wavelet coefficients our analysis is not biased towards structures
of a particular size or width.

In the plot for the larger filter size, we include the equivalent results from the
$b=\sqrt{2}$-filter as contours \changed{having a better angular resolution but a lower
spatial sensitivity}. They show the same peaks but do not trace the shape of
the filaments in the same way as the coefficients from the $b=1/\sqrt{2}$-filter.
All contours are more roundish, not following all the spatial variations that
are seen in the color maps. We omitted to overplot the contours for the small scale
as they only show the same behavior on the smaller scale, not adding new
information, but making the picture less readable.
Equivalent plots for the Aquila map mainly trace the
main ridge and the associated filaments visible in the central North part of the map.

\begin{figure}
\hspace*{1mm}\includegraphics[angle=90,width=0.88\columnwidth]{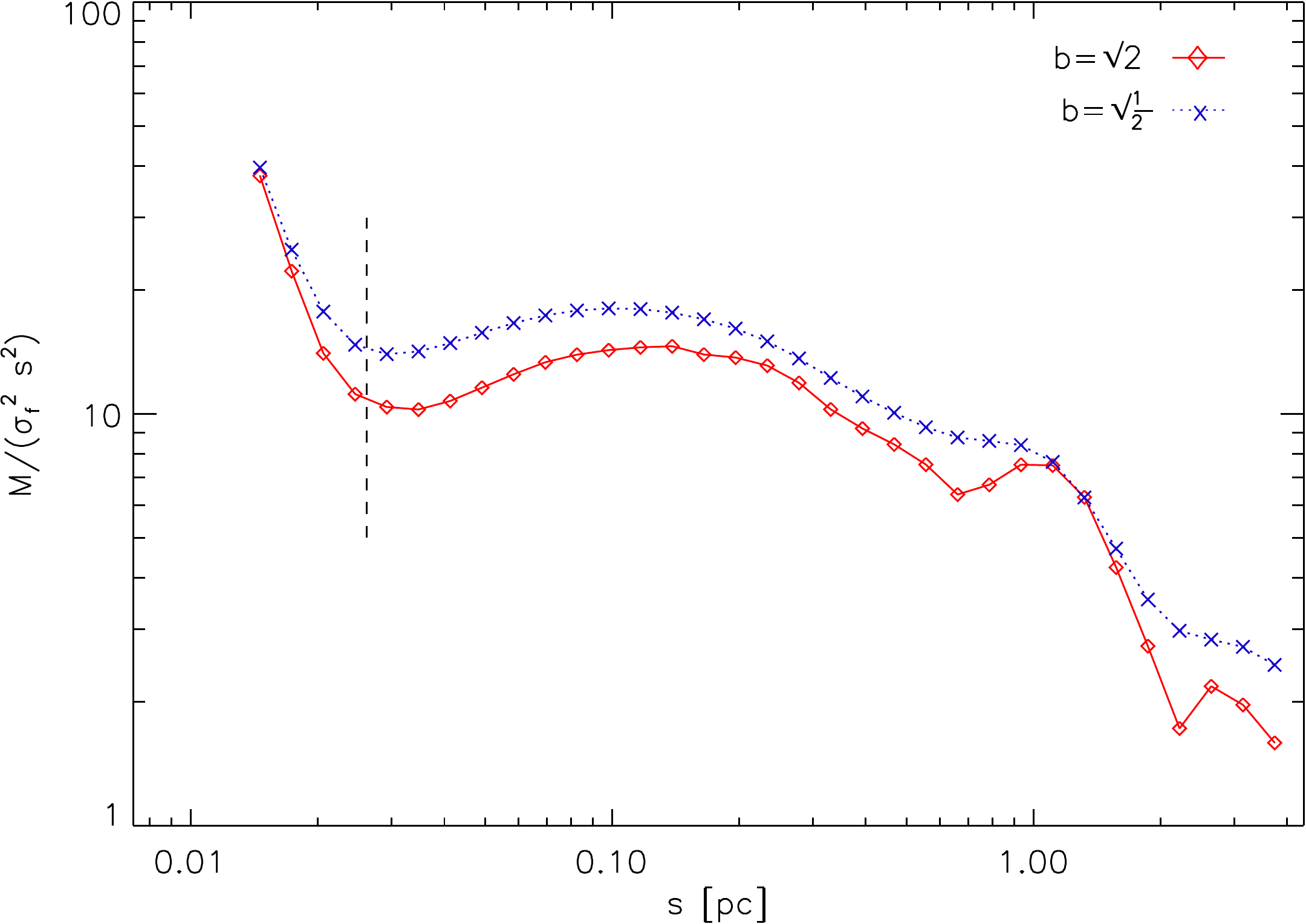}\vspace{3mm}
\includegraphics[angle=90,width=0.95\columnwidth]{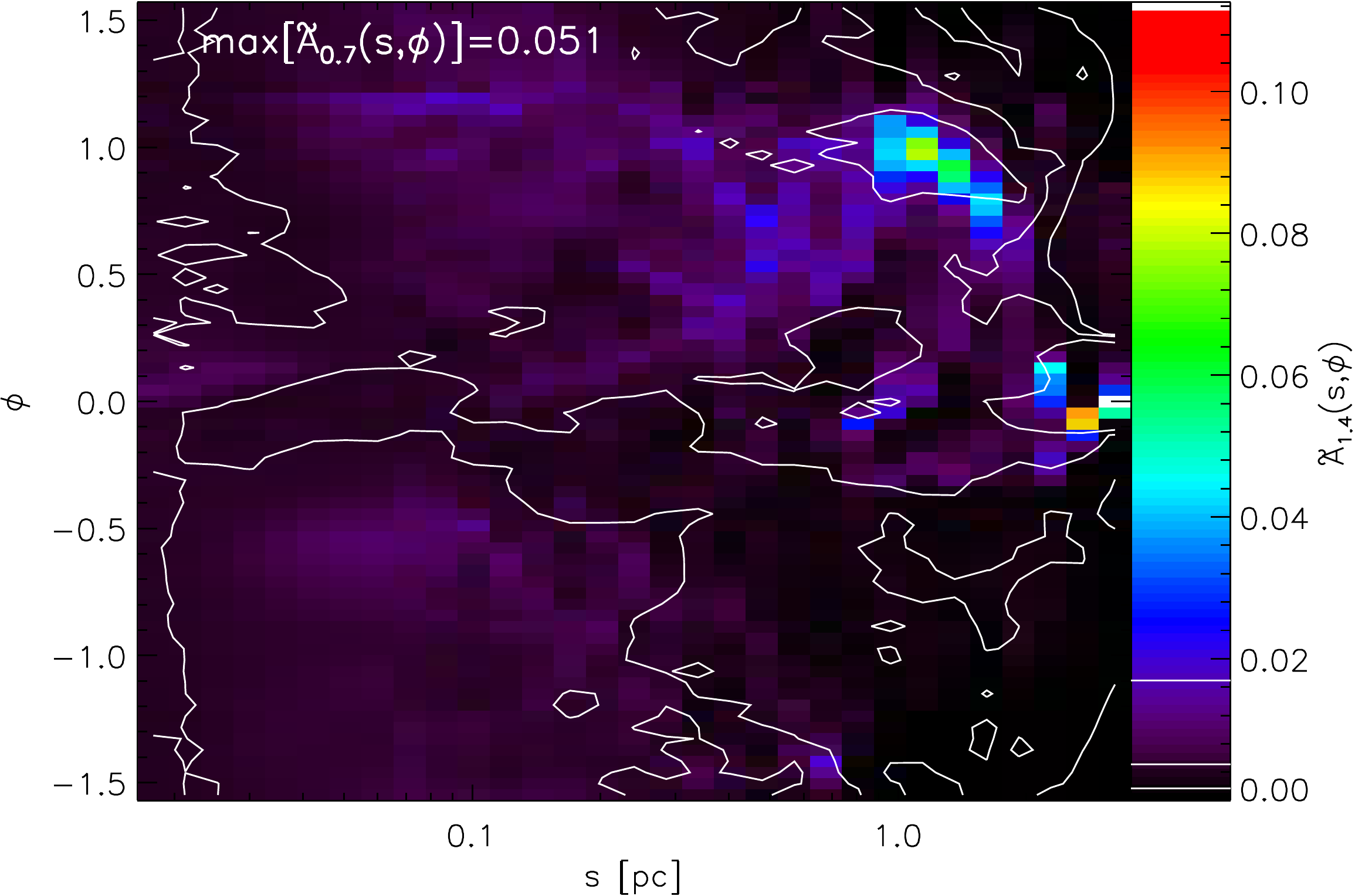}\vspace{3mm}
\hspace*{2mm}\includegraphics[angle=90,width=0.88\columnwidth]{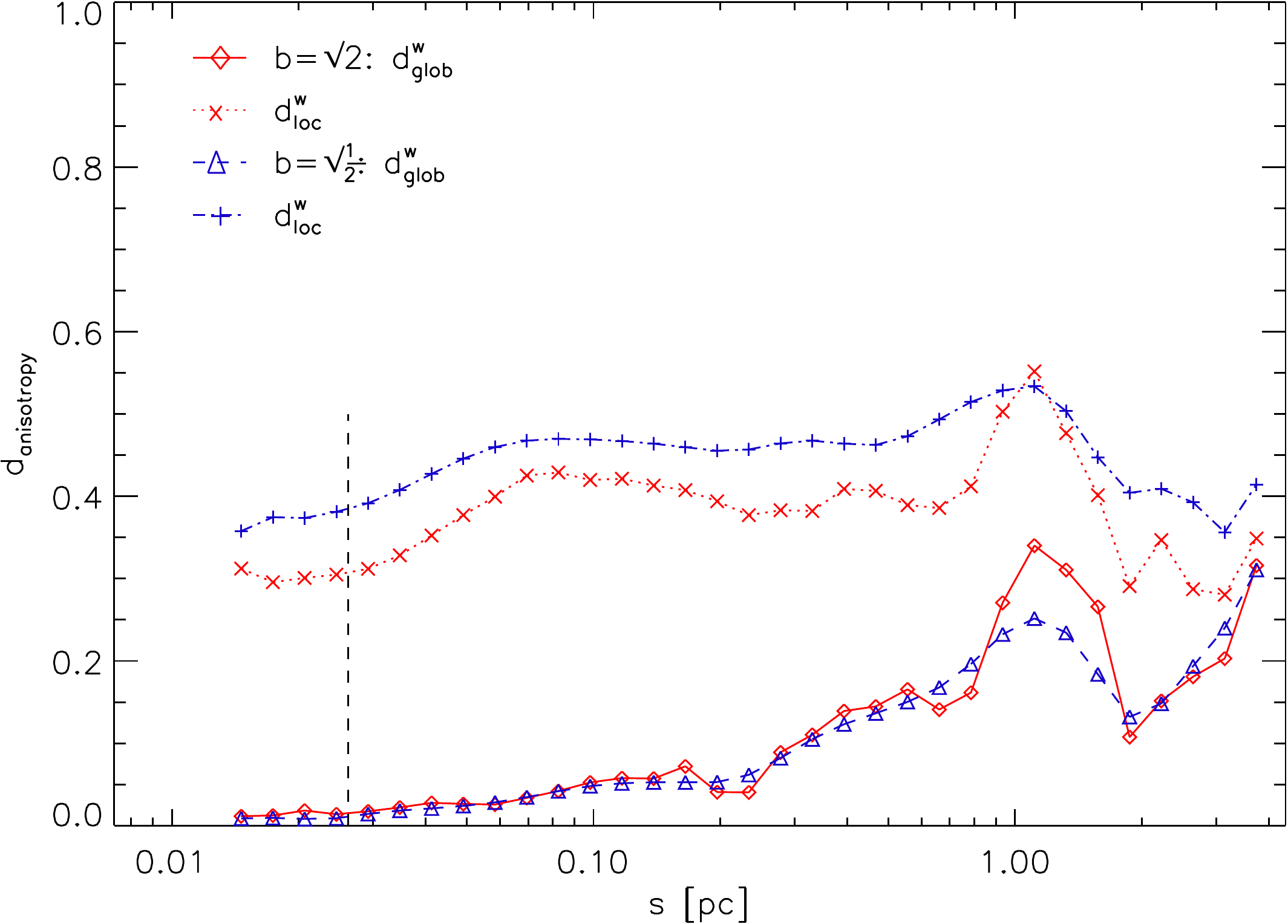}
\caption{Spectra of isotropic wavelet coefficients (top panel), angular
spectrum of anisotropic wavelet coefficients (central panel), and
local and global degrees of anisotropy (bottom panel) for the Polaris
column density map computed for filter parameters of $b=1/\sqrt{2}$
and $b=\sqrt{2}$. In the 2-D map of wavelet coefficients, the results from
the $b=\sqrt{2}$ filter are given in colors, those from the $b=1/\sqrt{2}$
filter in contours. Angles are plotted in radians.
The dashed lines marks the spatial resolution of the map of 36~arcsec
limited by the 500~$\mu$m observations.}
\label{fig:spectra-polaris}
\end{figure}

Figure~\ref{fig:spectra-polaris} shows the \changed{rescaled} spectra of the isotropic wavelet coefficients,
the angular spectrum of anisotropic wavelet coefficients, and the degrees of anisotropy
for the Polaris column density map obtained from the different filter shapes. \changed{
As we are interested in the anisotropy, angular distribution and size distribution
of the filaments we show the results for two different filter shapes again.}
The spectra of isotropic wavelet coefficients show \changed{relatively little
variation} between 0.03 and 1~pc. \changed{The $b=\sqrt{2}$-filter picks up
a weak broad peak around 0.12~pc, corresponding to a filament FWHM of about
0.045~pc.}
In this scale range the angular spectrum shows a wide distribution of weak filaments at
all angles, weakly concentrated around two contributions that are
approximately perpendicular at 0.03~pc and merge into a broad peak at about
$\varphi=40$~degrees at 0.5~pc. The resulting local degree of anisotropy is
approximately constant at 0.4, and the global degree of anisotropy is slightly
increasing due to the convergence of the angular components. \changed{As
expected from the studies in Sect.~\ref{sect_filtershape} the angular spectrum
for the $b=\sqrt{2}$ filter is better confined in spatial and angular scale.
The smaller localization parameter $b=1/\sqrt{2}$
instead leads to the detection of more small scale anisotropic structures
leading to a somewhat higher local degree of anisotropy below 1~pc.}

A single prominent structure appears at a scale of about 1.1~pc at angles between 45 and
60~degrees. It can be attributed to the `saxophone' and corresponds to a minimum
filament width of $\sigma\sub{minor}=0.18$~pc or a FWHM of about 0.4~pc.
This matches the width of the main `saxophone' filament that represents the main
anisotropy in the map \changed{at those scales}. Here, the local degree
grows to a value of 0.55, the global one to 0.35. At larger scales all coefficients drop
again. A secondary, smaller peak appears at the scale of the map size where the
emission pattern appears as an approximately horizontal strip in the map.
In contrast to Sect.~\ref{sect_filtershape}, the $b=1/\sqrt{2}$ filter detects a
somewhat larger local degree of anisotropy. This is expected for curved
filaments where the better localized $b=1/\sqrt{2}$ filter can follow
the structure of the filament while the $b=\sqrt{2}$ filter provides
higher coefficients for straight filaments.
At very small scales where no cloud structure can be resolved in the finite
resolution of the map the spectra are dominated by observational noise, visible
as an increase of the wavelet coefficients in the two upper plots. \changed{In this
small scale range we cannot expect to measure any significant structure
size, as discussed in Sect.~\ref{sec:simpleGaussian}.}

\changed{Even with the scale sensitive $b=\sqrt{2}$ filter we find no
statistically significant excess of structure corresponding to a filament
width around 0.1~pc. At all scales below about 0.4~pc}
the spectrum indicates an almost self-similar hierarchy of smaller and smaller
filaments without any prominent scale. The `saxophone' only represents
the upper end of that hierarchy. The wavelet analysis provides no
support for a universal filament width.

\begin{figure}
\centering
\includegraphics[angle=90,width=0.88\columnwidth]{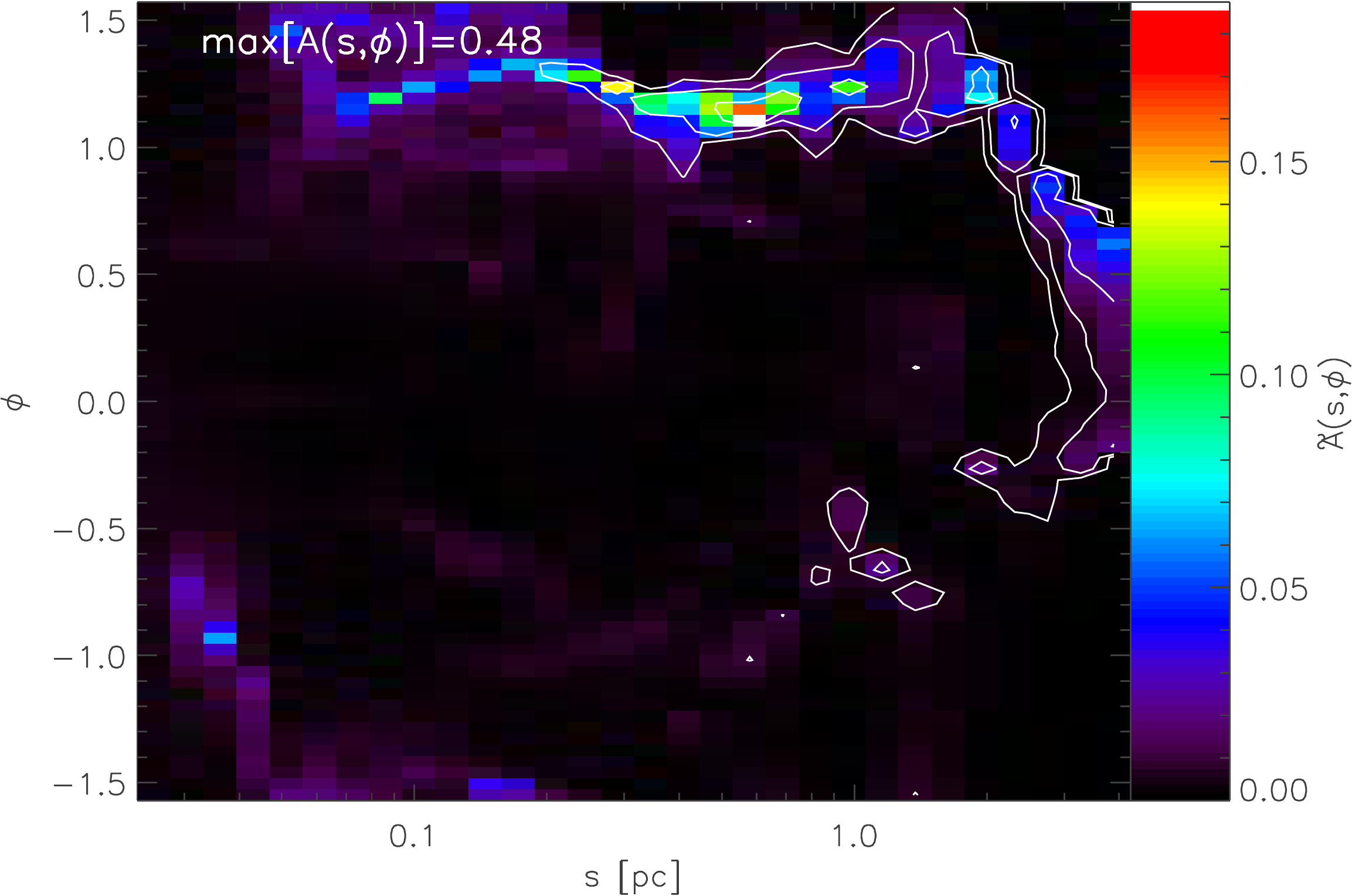}
\caption{2-D spectrum of anisotropic wavelet coefficients (contours) and
degrees of anisotropy (color) for Aquila. The contours are drawn at levels of 1/30,
1/10, and 1/3 of the peak value indicated in the top left corner of the plot.
The angles are plotted in radians.
}
\label{fig:a-spectra-aquila}
\end{figure}

To focus on the characteristic scales in the Aquila column density map
we also show the analysis results for the $b=\sqrt{2}$ filter.
Figure~\ref{fig:a-spectra-aquila} contains the two-dimensional spectrum of wavelet coefficients
$A(s,\varphi)$ and the coefficients normalized by the spectrum of isotropic modes, providing
the angular degree of anisotropy $\tilde{A}(s,\varphi)$.
The Aquila spectrum differs strongly from the Polaris spectrum. It is dominated by a single
warped filament that covers all scales from 0.05~pc to the map size.
Apart from a structure at an angle of $\phi=-0.6\mathrm{rad}=-35$~degrees,
stemming from the filaments east of the main ridge, all other structures
are statistically negligible compared to the main filament. The corresponding map of wavelet
coefficients shows a single strong peak at the location of the main ridge in the upper
center of the map.

\begin{figure}
\centering
\includegraphics[angle=90,width=0.88\columnwidth]{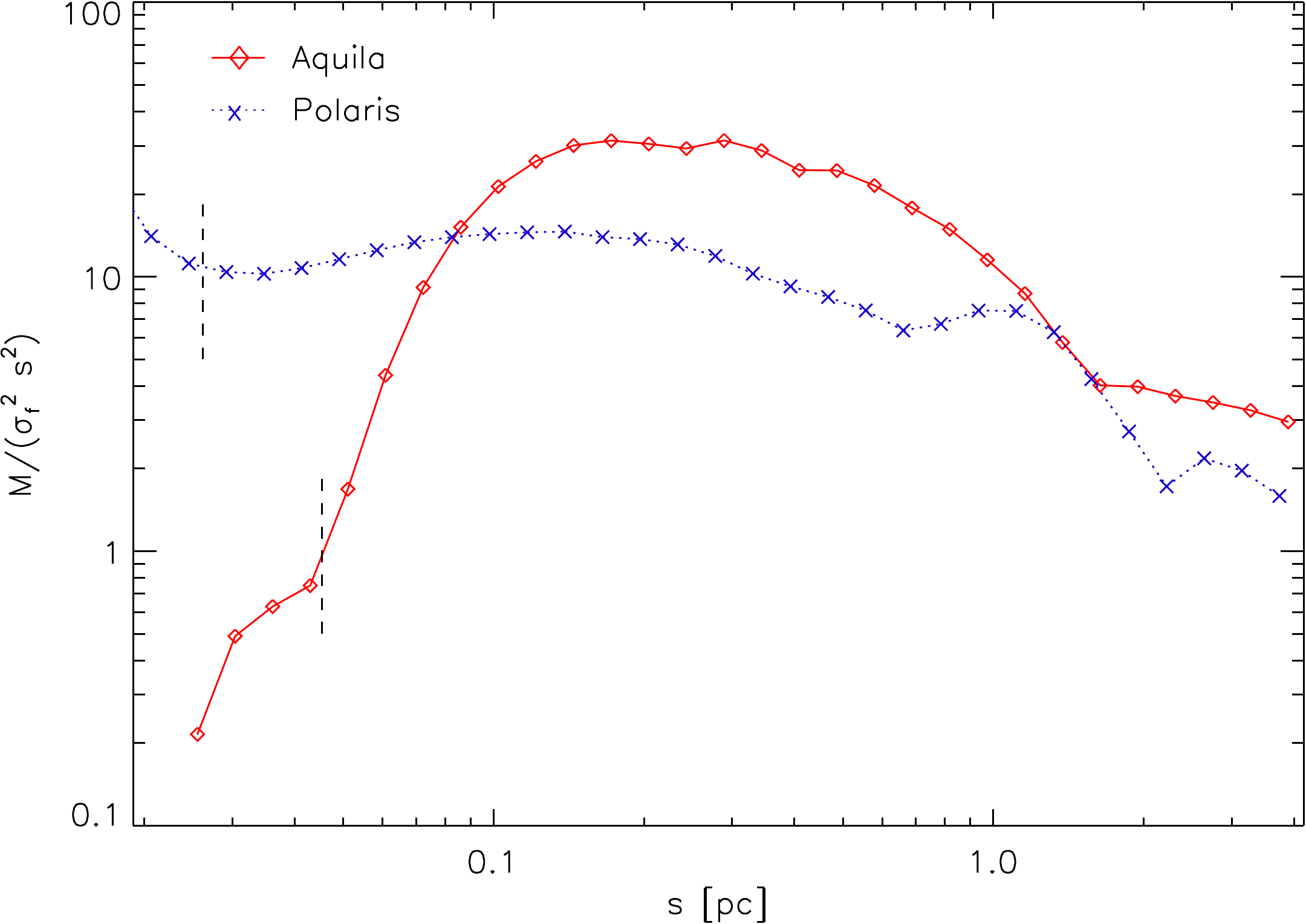}\vspace{3mm}
\includegraphics[angle=90,width=0.88\columnwidth]{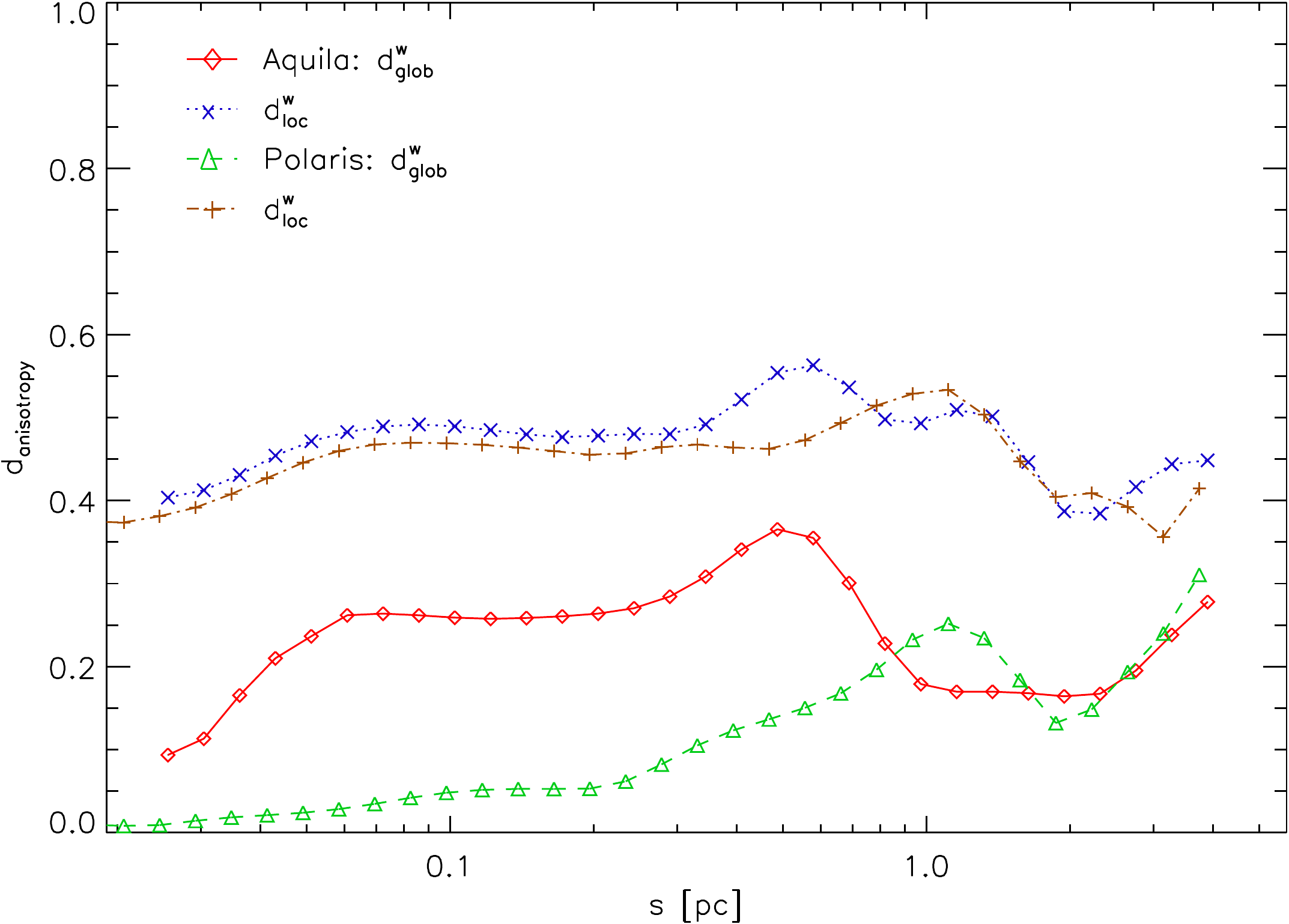}
\caption{Spectra of rescaled isotropic wavelet spectra $M^i(s)/s^2$ (top panel, $b=\sqrt{2}$) and local and global degrees of
anisotropy (bottom panel, $b=1/\sqrt{2}$) for the two maps from Fig.~\ref{fig:maps-aquila-polaris}. The dashed
lines in the panel for the spectra mark the spatial resolution in the two maps. Around that scale
and below, the structures in the map are dominated by the telescope beam and
observational noise.}
\label{fig:spectra-aquila-polaris}
\end{figure}

Figure~\ref{fig:spectra-aquila-polaris} compares the resulting spectra of the
wavelet coefficients and the degrees of anisotropy for the Aquila map with the
Polaris results. \changed{Between 0.15 and 0.7~pc, both maps show an approximately self-similar
scaling with about the same exponent $M^i(s)/s^2 \propto s^{-0.3}$ corresponding
to a  power spectral index of $\beta \approx 2.7$.  The local degree of anisotropy is approximately the
same at all scales. However, we find three main differences: The Aquila spectrum
steeply rises at small scales. At large scales it drops already above 0.8~pc
compared to 1.3~pc for Polaris. And the global degree of anisotropy is significantly
higher in Aquila.}

Part of the rise of the Aquila spectrum above the beam size may stem from
the blurring of the observable structure by the telescope beam as discussed 
in detail for the $\Delta$-variance by \citet{Bensch2001}.
The finite resolution suppresses variations at scales up to scales of a few beam
sizes. In Polaris the effect is present as well, but is compensated by
observational noise. The difference between the two spectra in this scale range is thus consistent with the significantly higher signal-to-noise ratio in Aquila
compared to Polaris.

The relative lack of structure at scales above 0.8~pc in Aquila is
\changed{accompanied by a peak in local and global degree of anisotropy
that also falls at smaller scales compared to Polaris. The two quantities
measure width and length of filaments (see Sect.~\ref{sec:simpleGaussian}).
As there is only a very broad peak in the wavelet spectrum for Aquila, we can
only use the upper edge criterion to measure the maximum filament width.
The $\hat{s}_{90\,\%}$ scale of 0.7~pc corresponds to a FWHM of 0.21~pc.
This matches approximately the width that one would measure
naively with a ruler for the cyan area around the} brightest
structure in the map (Fig.~\ref{fig:maps-aquila-polaris}).
The peaks in the degrees of anisotropy at $s=0.6$~pc for Aquila \changed{and
at 1.1~pc for Polaris correspond to a similar difference in the filament
length of a factor two. For Aquila the} upper end of the filament hierarchy
falls a factor of two below the one for Polaris, \changed{both in filament
length and width.}

Smaller filaments are preferentially aligned with the main filament, creating
a coherent structure in the $\tilde{A}(s,\varphi)$ diagram, so that
the global degree of anisotropy closely follows the local
degree of anisotropy, quite different from the Polaris behavior.
In contrast to Aquila, the Polaris map shows much weaker global anisotropy.
It shows a high filamentariness measured by the local anisotropy but
the filaments with scales below 1~pc are oriented in random directions. Around 1~pc
the `saxophone' dominates as a global structure.

\begin{figure}
\centering
\includegraphics[angle=90,width=0.86\columnwidth]{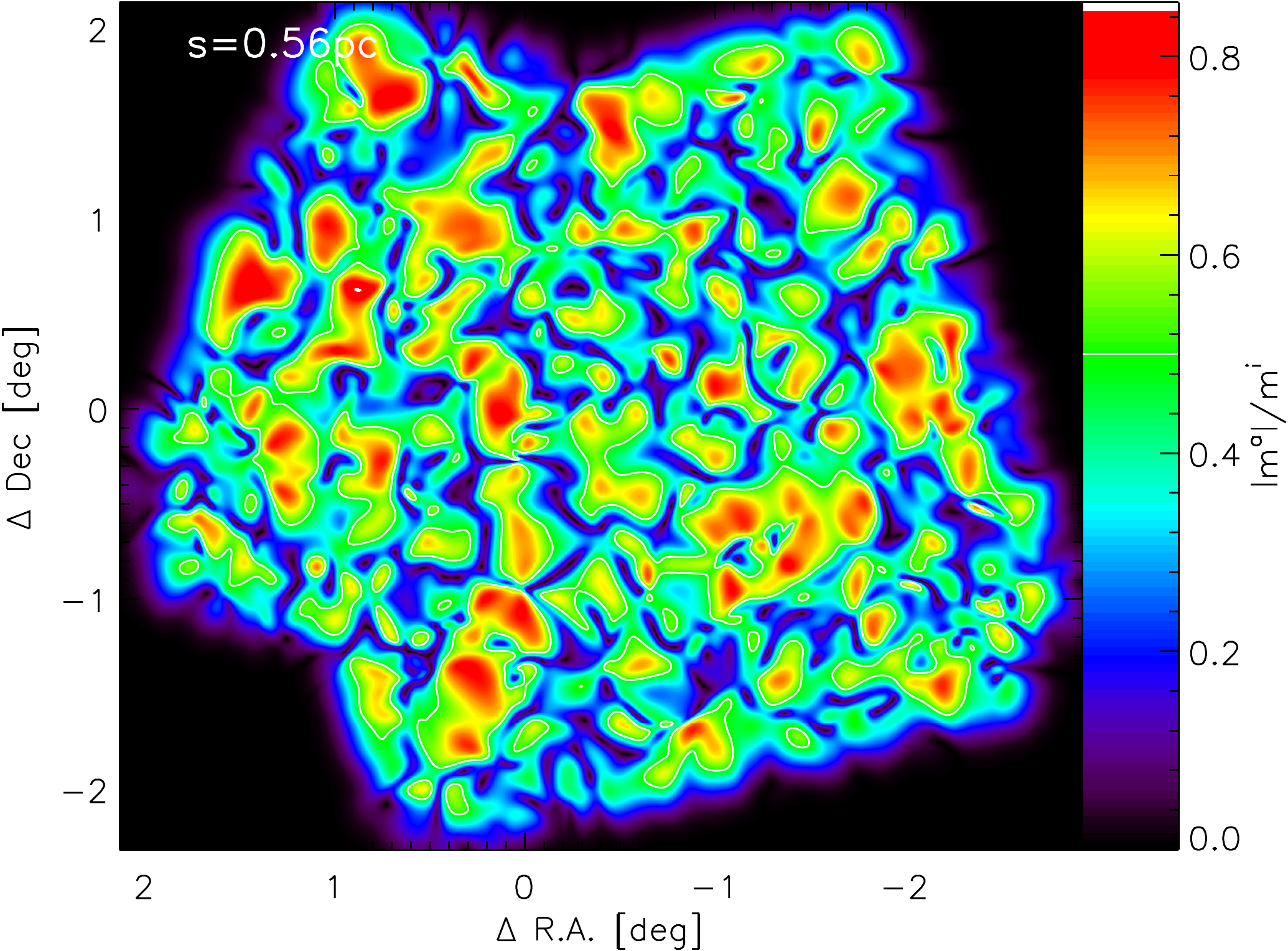}
\caption{Map of the local degree of anisotropy $|m^a(s,\vec{x})|/m^i(s,\vec{x})$ for the Polaris
column density map when using the $b=1/\sqrt{2}$-filter with a size of $s=0.56~\mathrm{pc}=0.21\deg$.
The white countour at a value of 0.5 separates regions
dominated by anisotropic fluctuations from those dominated by isotropic fluctuations. There is
no obvious correlation to the corresponding map of wavelet coefficients shown in the
lower panel of Fig.~\ref{fig:ma_polaris}.}
\label{fig:degree_polaris}
\end{figure}

\changed{We can use the degree of anisotropy to address the question whether
the cloud structure is ``dominated''  by filaments \citep{Federrath2016}). Using
a threshold of $d\sub{loc}^w=0.5$ for such a distinction, the lower panel of Fig.~\ref{fig:spectra-aquila-polaris}
shows that both Polaris and Aquila fall at approximately this limit for all
scales between 0.05~pc and 1.3~pc. Globally, the maps are neither dominated
by isotropic nor by anisotropic structures. Isotropic fluctuations are
dominant at the scale of the clouds and close to the resolution limit, anisotropic
fluctuations are slightly higher at the scales of the lengths of the two
dominant filaments in the maps, at 0.6~pc and 1.1~pc, respectively.
To study whether this global result also holds locally
Fig.~\ref{fig:degree_polaris} shows the map of the local degree of anisotropy
equivalent to Fig.~\ref{fig:ellipse_coefficients}. When we compare it with the
corresponding map of anisotropic wavelet coefficients in the lower panel of
Fig.~\ref{fig:ma_polaris}, we find no clear correlation to the amplitude of
the wavelet coefficients. Around the ``saxophone'' the degree of anisotropy is
high on average, but not across the whole structure. Low
degrees are found e.g. at its southern end. This variation of the degree continues
through the whole map. High degrees of anisotropy are usually seen at the centers
of visible filaments but low degrees occur at their ends. The same is true for
other size scales and for the corresponding Aquila maps. The local degree of
anisotropy varies strongly across the whole map. An area-weighted average is
close to the wavelet-coefficient weighted average in $d\sub{loc}^w$. A dominance
of filamentary structures is thus only seen in confined parts of the clouds.
Globally, isotropic and anisotropic fluctuations are comparable.}

\section{Gravitational stability}
\label{sec:discussion}

The spectra of isotropic and anisotropic wavelet coefficients provide us with a means
of directly comparing the relative amount of structure in isotropic and filamentary
modes. When interpreting the filaments as cylindrical structures
we can measure the relative importance of spherical and cylindrical collapse modes
by comparing the power in isotropic and anisotropic structures. As their
relative importance changes as a function of scale and position in the maps
we can try to use the anisotropic wavelet analysis to constrain the local collapse
modes.


The standard criterion for gravitational instability is the Jeans
criterion, first computed for an unbound spherical mass distribution
\citep{Jeans1902}. Equivalent criteria can be derived for various
geometries, differing from the original criterion by
prefactors in the order of unity.
\changed{The first solution for the geometry of an infinite cylinder
was provided by \citet{Stodolkiewicz1963} and \citet{Ostriker1964}.
They found a hydrostatically stable configuration for an isothermal
cylinder that is described by a Plummer density profile with an
exponent $p=4$ providing a steep boundary to the filaments
described. It is described by a mass per length of
\begin{equation}
M/l = {2 c\sub{s}^2 \over G} \;.
\end{equation}
Here, $G$ describes the gravitational constant and $c\sub{s}$ is
the sound speed in the gas.
More massive filaments will undergo radial collapse,
finally leading to fragmentation and star-formation \citep[see e.g.
discussion in][]{Anathpindika2015}.

Relatively similar spherically symmetric configurations are
described by Bonnor-Ebert spheres \citep{Ebert1955,Bonnor1956},
solutions of the hydrostatic equations for isothermal spheres
confined by some external pressure. They consist of a central part
with a relatively constant density and an outer density decay
inversely proportional to the square of the radius, similar
to Plummer profiles with $p=2$. The critical Bonnor-Ebert
sphere is the densest stable solution having a density contrast
of 14 between the center and the outer boundary and a mass of
\begin{equation}
M\sub{crit} = 4.42 {c\sub{s}^3 \over {G^{3/2} {\rho\sub{c}^{1/2}}}}
\end{equation}
\citep[e.g.][]{FischeraDopita2008}. The quantity $\rho\sub{c}$
denotes the central gas density.}

Both criteria can be translated into \changed{an approximate criterion
for the column density when treating the clumps or filaments as uniform
spheres or cylinders, ignoring the density variation within them,
and substituting the core} density $\rho\sub{c}$ by
the peak column number density
\begin{equation}
\hat{N}= 2 R \, {\rho\sub{c} \over \mu}
\end{equation}
where $R$ is the radius of the sphere or cylinder and $\mu$ the
average molecular mass. Together with the relations between density,
size, and mass of a sphere and cylinder, we can write them
as new criteria for the gravitational instability
\begin{eqnarray}
\hat{N} R &>& \left[ {1\over 2} \left({3\over \pi}\times 4.42 \right)^2\right]^{1/3} {c\sub{s}^2 \over G \mu} \nonumber\\
& \approx & 2.07 \, {c\sub{s}^2 \over G \mu} \; {\rm for\; spheres} \label{eq_stability_sphere}\\
\hat{N} R &>&  {4 \over \pi} {c\sub{s}^2 \over G \mu} \nonumber\\
& \approx & 1.27 \, {c\sub{s}^2 \over G \mu} \; {\rm for\; cylinders} \label{eq_stability_cylinder}
\end{eqnarray}
The factor ${c\sub{s}^2 / G\mu } = 0.65\times 10^{21}$~cm$^{-2} $pc for
\changed{isothermal molecular gas with a temperature of 15~K and an
average molecular mass of 2.36 atomic hydrogen masses, typical conditions
for molecular clouds \citep[e.g.][]{MacLowKlessen2005}. With
these criteria the easily measurable
product of column density and radius of the structures
allows us to estimate their stability just as a function of their temperature
independent of the difficult-to-measure density.}
We can use any measurement of the product of column density and the smallest
one-dimensional extent of a structure to characterize the gravitational stability
of the structure, \changed{both for filaments and cores.
For H$_2$ gas we only have to divide the column density $N\sub{H}$ from
Fig.~\ref{fig:maps-aquila-polaris} by a factor of two.}


\changed{As the wavelet analysis provides a way to measure column density fluctuations
as a function of their size and geometry we can interpret the wavelet coefficients
in terms of the stability analysis using the criteria from Eqs.~\ref{eq_stability_sphere} and \ref{eq_stability_cylinder}.
If we use the observed column density $N$ as the map function $f$ in Eq.~\ref{cwt}
the wavelet transform $W$ will scale as $N s^{1/2}$ because
the wavelet $\psi$ is not normalized relative to $s$ (Eq.~\ref{eq:morlet}) so
that it provides an $s^2$ contribution in the integral. We thus obtain a scaling
of the wavelet coefficients $m(s)\propto (\hat{N})^2 s$.
When multiplying the wavelet coefficients $m$ with the wavelet \changed{scale $s$ we
obtain a quantity that measures} the gravitational stability modes, $m\times s \approx (\hat{N}R)^2$.
Maps of $\sqrt{m^i\times s}$ and $\sqrt{m^a \times s}$ \changed{characterize} the
gravitational stability of structures of varying size. The isotropic coefficients
$m^i$ characterizes the total variations while we can assign $m^a$ to anisotropic
collapse modes. By comparing the coefficients $\sqrt{m^i\times s}$ and $\sqrt{m^a \times s}$
with the critical values $(\hat{N} R)\sub{crit}$ from Eqs.~\ref{eq_stability_sphere} and \ref{eq_stability_cylinder}
the wavelet coefficient maps provide information on the individual collapse modes for every point
and scale $s$.}

If a map is dominated by a single structure, we can assess its stability from the
average coefficients $M^i(s)$ and $M^a(s)$
discussed in the previous chapters. The contribution of the  wavelet coefficients
$m^i(s,\vec{x})$, $m^a(s,\vec{x})$ of an individual structure to the average
over the whole map (Eq.~\ref{w_spec}) is scaled by the area over which the
coefficients of that structure contribute. Due to the blurring nature of the
wavelets, the area is proportional to $s^2$ so that the relative contribution
of any structure to the average spectrum is given by $s^2/l\sub{map}^2$,
where $l\sub{map}$ gives the linear size of the whole map. \changed{
With the contribution of the individual structure to the total normalized wavelet
spectrum $M\sub{indiv} \propto m\times s^2$ we obtain the stability
of individual structures from $(M/s)\sub{indiv} \propto m\times s \propto (\hat{N} R)^2$.
Consequently, we can judge the gravitational stability
in an analyzed map by taking the square root of the normalized wavelet spectra $M/s$.}
For the example of an singular isothermal
sphere (Appx.~\ref{appx_tests})
$M/s$ is flat, in line with the scale-independence
of this configuration being equally stable or unstable at every scale.


For real maps we always have a hierarchy of filaments in the map.
Due to the linear nature of the wavelet analysis their contributions add up in the
wavelet spectra, so that we cannot judge the stability of any individual structure
from the global spectrum. Instead we need to select a particular contribution.
We focus here on the single structure that dominates the wavelet maps
$m^i$ and $m^a$ providing their \changed{maxima}. For the two observed maps they correspond
to points in the `saxophone' region in Polaris and in the main ridge in Aquila.
\changed{The spectra of
$\sqrt{(m^i)/s}$ and $\sqrt{m^a/s}$,} then quantify the gravitational stability
of the spherical and cylindrical modes in the main structures.

\begin{figure}
\centering
\includegraphics[angle=90,width=0.88\columnwidth]{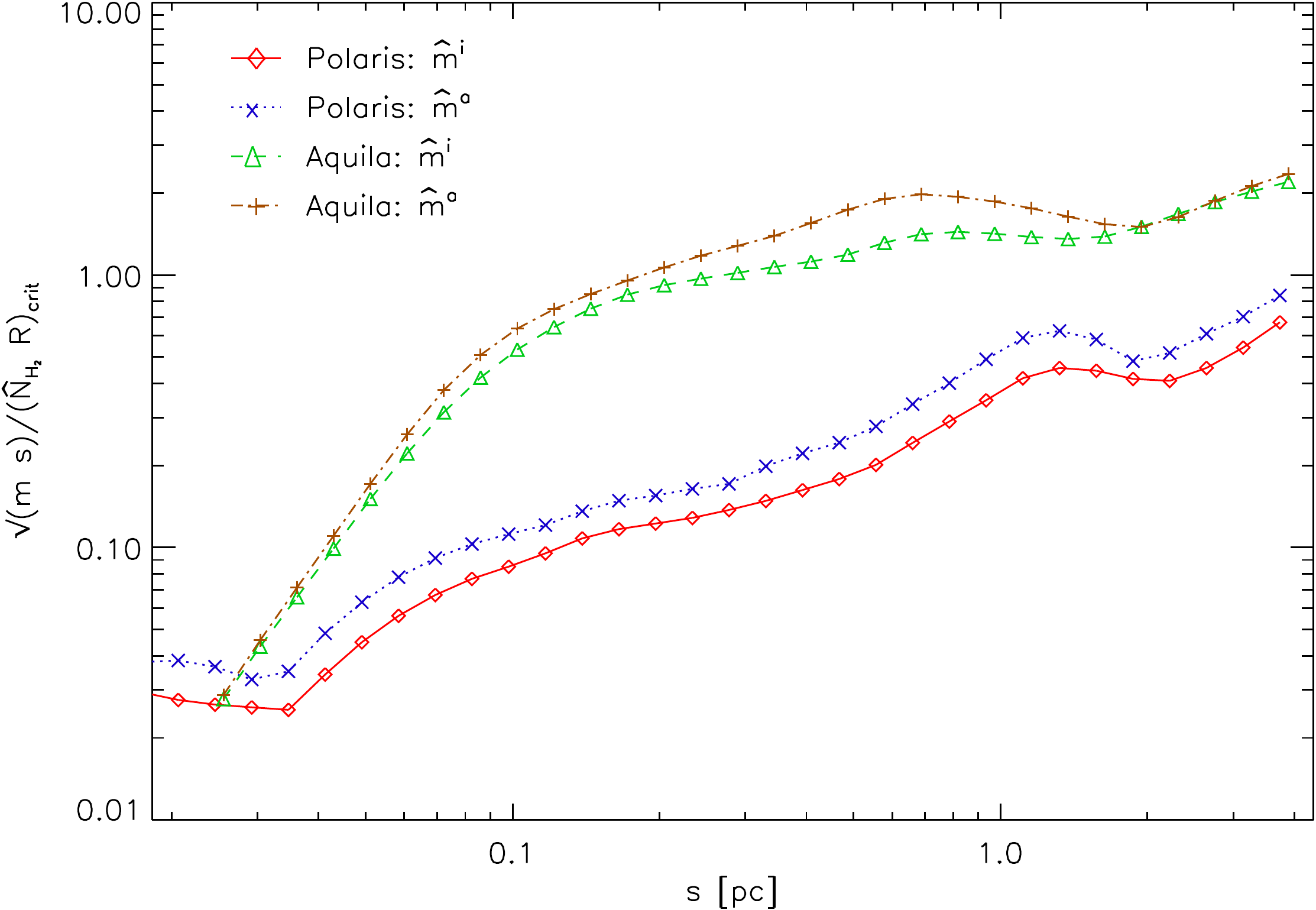}
\caption{Gravitational stability spectra for the dominating structures in the two maps
from Fig.~\ref{fig:maps-aquila-polaris}. The \changed{isotropic and anisotropic
wavelet coefficients $m^i$ and $m^a$ characterize the size dependence of the
local column density variation. Taking the size-scaled coefficients relative to
the critical values of $(\hat{N} R)\sub{crit}=1.35\times 10^{21}$~cm$^{-2} $pc
and $0.83\times 10^{21}$~cm$^{-2} $pc allows us to measure the gravitational stability
of the structures against spherical and cylindrical collapse modes. Values above
unity indicate structures that are unstable against collapse. Lower values stand
for structures that are not gravitationally bound.}}
\label{fig:stability-aquila-polaris}
\end{figure}

In Fig.~\ref{fig:stability-aquila-polaris} we show the \changed{spectra of
the maxima of the wavelet coefficients, $\sqrt{m^i/s}$ and $\sqrt{m^a/s}$,
relative to the critical values $(\hat{N} R)\sub{crit}=1.35\times 10^{21}$~cm$^{-2} $pc
and $0.83\times 10^{21}$~cm$^{-2} $pc for spherical and cylindrical collapse
modes, respectively.
To guarantee the best localization of the wavelet coefficients on the filaments
we used the $b=1/\sqrt{2}$ filter here. Inspection of the maps of $m^i$
and $m^a$ shows that the peaks in the two maps that are used to compute
the stability spectra are usually very close.
The anisotropic modes always peak at the centers of the filaments with the main peak given by the
center of the `saxophone' and the Aquila rift while the isotropic modes}
cover some broader regions around the filaments.

The absolute \changed{magnitude of the wavelet coefficients} confirms that the Polaris main filament
is subcritical for the assumed temperature of 15~K. No collapse is expected there.
In contrast the ridge in Aquila should be collapsing \changed{over a broad range
of scales.} This is in line with the known star formation activity in the two regions.
\changed{In both regions we find a clear dominance of the anisotropic instability.
Cylindrical collapse modes will dominate. The Aquila spectra show that the limit
for the cylindrical gravitational instability is reached at scales of 0.15~pc, the
spherical limit is only reached at sizes of 0.2~pc. On scales of 2~pc and more
filamentary collapse is no longer preferred over the radially symmetric one,
both are similarly unstable. The maximum of the anisotropic spectrum at
0.7--0.8~pc suggests that the structure on that size is as unstable against
filamentary collapse as the whole cloud.}

For the Polaris `saxophone' \changed{the isotropic and anisotropic spectra
just differ by an almost constant factor in agreement with the constant}
local degree of anisotropy around $d^w_{loc}\approx 0.5$ measured in the whole
Polaris map (Fig.~\ref{fig:spectra-aquila-polaris}). This indicates that we
have a uniform distribution of filamentariness throughout the whole region.
In contrast, for the Aquila map we measured about the same local
degree of anisotropy, but now find \changed{that for the most unstable structure
the dominance of the anisotropic modes is limited to scales between about 0.04 and 2~pc.
This suggests a higher filamentariness of other structures on smaller
and larger scales.} To understand the full distribution of the filamentariness across
a map it is therefore necessary to combine the spatial information with the
spectra.

\begin{figure}
\centering
\includegraphics[angle=90,width=0.88\columnwidth]{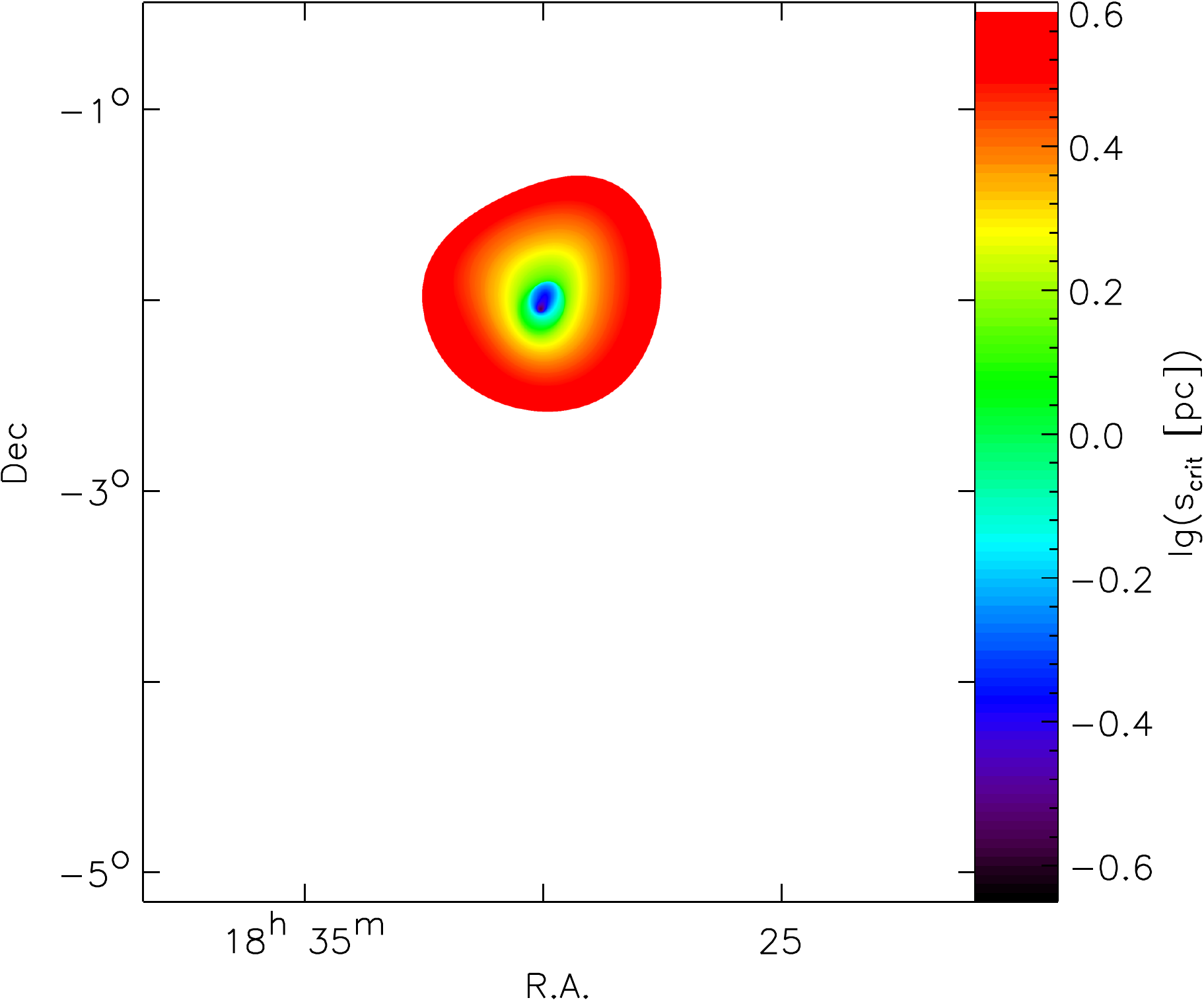}
\includegraphics[angle=90,width=0.88\columnwidth]{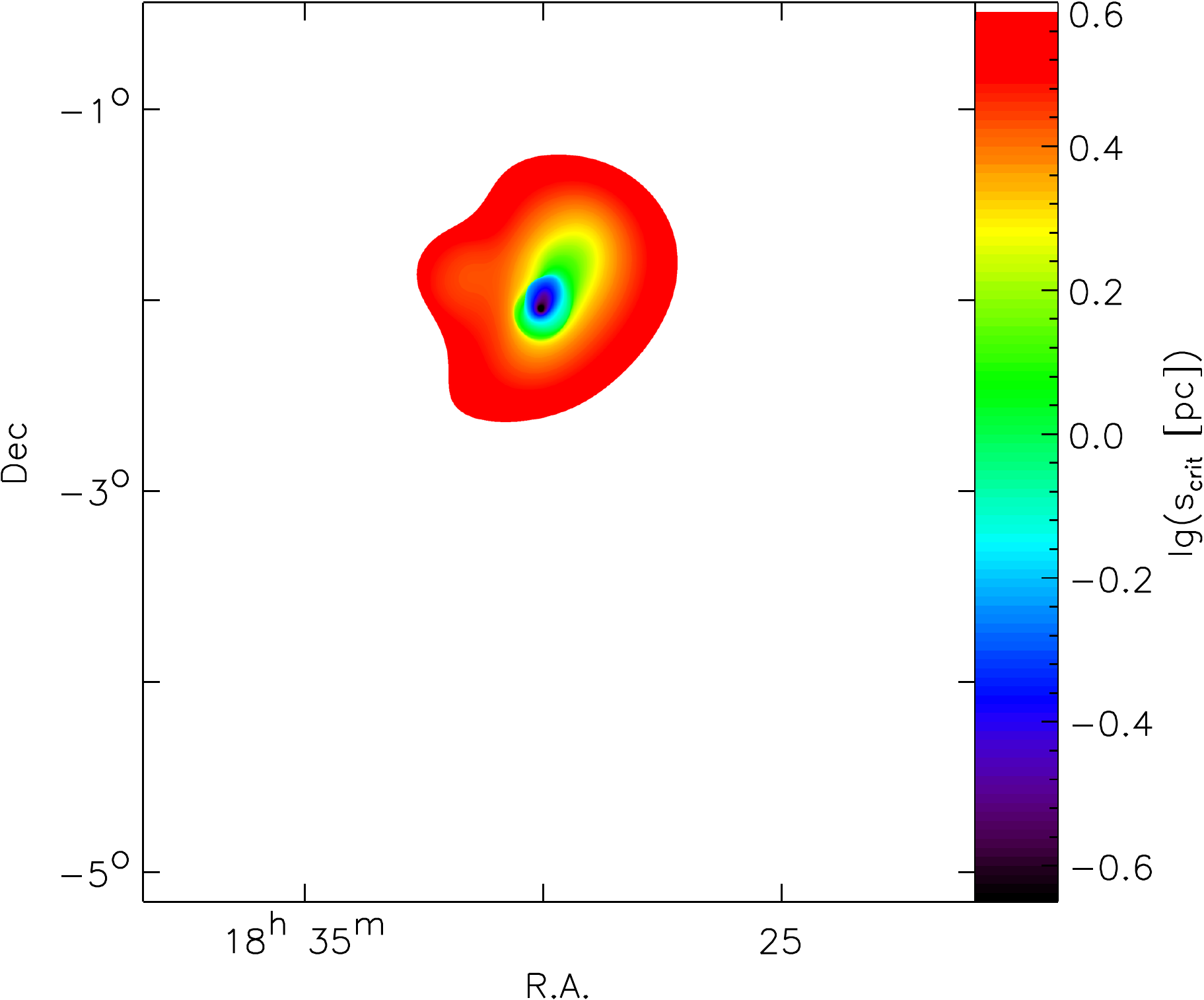}
\caption{Maps of the critical size for gravitational instability in the
Aquila column density map (Fig.~\ref{fig:maps-aquila-polaris}). The upper
plot shows the critical scale for isotropic modes, the lower plot for anisotropic
modes. White areas are regions
with too low densities for collapse. \label{fig:stability-maps-aquila}}
\end{figure}

In case of supercritical gravitational modes we can go one step further by
determining the minimum scale $s\sub{crit}$
for which the condition of \changed{supercriticality
$\hat{N} R > (\hat{N} R)\sub{crit}$ is met.} From this quantity we can
assess the size of the region affected by the characteristic
collapse mode for the density distribution around the considered point.
As the free-fall time only depends on the density, the collapse time
should also follow the pattern measured by the critical length scale.
Figure~\ref{fig:stability-maps-aquila} shows this map of critical scales
computed from the wavelet coefficients for the Aquila column densities.
White areas in the map indicate regions with too low densities for collapse.
The corresponding figure for Polaris would be white only.

First we notice the global consistency of the map values with the map
topology. The red region, showing a critical size $s\sub{crit}\approx 3$~pc has about
a \changed{radius} of 3~pc, the green region indicating $s\sub{crit}\approx 1.2$~pc has a
\changed{radius} of 1.2~pc, and so on. Fast collapse on small scales is only expected
within the main filament. \changed{The anisotropic modes cover a more elongated region
compared to the isotropic modes. The anisotropic modes also show a significant
instability for the side arm of the main rift further east. Both modes identify the most unstable
structure within the main filament, the minimum for the anisotropic modes
is slightly further south-east compared to the isotropic modes. }

Assessing the gravitational stability of structures in an observed map from
the wavelet analysis only, is of course a very crude approximation as it ignores all
information about the dynamics in the gas flows and the details of the geometry
of the individual structures decomposing them into two modes only.
However it should allow for a qualitative assessment of the behavior of
the different regions and the relative importance of cylindrical versus
spherical collapse modes.
Observations of molecular lines should try to verify a relation between
these column-density based gravitational instability modes and the actual
velocity field of infalling gas.

\section{Conclusions and outlook}

The anisotropic wavelet analysis inherits the strengths of the
$\Delta$-variance \citep{Stutzki1998, Ossenkopf2008a} but
allows us to quantify and distinguish local and global anisotropies and measure the
size of filaments \changed{(Sect.~\ref{sec:single})}. In contrast to many existing filament
finders it is not biased by typical selection effects as discussed e.g. by \citet{Panopoulou2017}.
Therefore it can provide \changed{a spectrum of filament widths} from
a data set starting from the resolution limit up to about 20\,\% of
the map size. \changed{However, the spectra of wavelet coefficients and
degrees of anisotropy result from a combination of the size spectrum
of filaments with the distribution of their aspect ratios
(Sect.~\ref{sect_fractal}). Here we concentrated on the derivation
of the size range of filaments. The derivation of the full distributions
from the slope of the spectra will be the topic of a subsequent study.
Similarly, an assignment of anisotropic structures in a map to underlying
prolate or oblate 3-D structures is only possible if we have a priori knowledge
on their aspect ratios. (Sect.~\ref{sect_3d}).}

Maps of wavelet coefficients permit to identify the spines of filamentary
structures of a particular scale \changed{(Sect.~\ref{sect_superposition}
and \ref{sect:obs})}, to quantify the angular distribution of the different filaments
\changed{(Sect.~\ref{sect_ecc_calibration})}, and to measure the local and global degree of
anisotropy and angular alignment for the structures in a map
\changed{(Sect.~\ref{sect_ecc_calibration} and \ref{sect_filtershape})}.

The application of the analysis to molecular line maps from MHD simulations in
Sect.~\ref{sect:MHD} confirms
the findings from \cite{LazarianPogosyan2004} and \cite{Burkhart2013} about the impact
of the optical depth on the size spectrum of the observable structures. With increasing
optical depth wider filaments become more prominent reflecting the broader velocity
dispersion on larger scales and the global magnetic field structure. The
observed filament sizes depend on the combination of magnetic-field dominated density-velocity
correlations with radiative transfer effects. This can be exploited by observing tracers
with different optical depth. Optically thin lines trace small-scale
filaments that are entangled with the field lines while optically thick lines
rather trace the velocity structure that inherits more of the global anisotropy
from the large-scale magnetic field. Both types of maps show filamentary
structures, but the ratio between local and global anisotropy
clearly separates them.

Applying the analysis to column density maps obtained from Herschel observations
\citep{Schneider2013} \changed{in Sect.~\ref{sect:obs}},
we find a broad range of filament widths. We do not find a universal characteristic filament width
\changed{but a approximately self-similar behavior covering at least a factor of four in Aquila
and a factor of ten in Polaris. The self-similar range can be somewhat smaller when assuming an intrinsic $p=2$ Plummer profile.
The main filament in Polaris, having a width of
0.4~pc, is accompanied by many smaller filaments. They} have
no preferred direction so that a high degree of local anisotropy is not leading
to a global anisotropy. In contrast in the Aquila map \changed{the main filament,
having a total width of 0.2~pc, dominates the statistics. However, it also splits up
into a hierarchy of associated subfilaments that are closely related in position and
orientation.} Consequently the Aquila map shows a significant
global anisotropy. \changed{Globally the degree of anisotropy in both maps falls
at about 0.5, indicating no clear dominance of either isotropic or anisotropic
fluctuations. Locally, the situation may be different.}

\changed{For individual structures} we can measure the relative importance of
spherical and cylindrical collapse modes by comparing the power in isotropic and
anisotropic \changed{fluctuations (Sect.~\ref{sec:discussion}).
By translating the gravitational stability criteria for isothermal gas
in a Bonnor-Ebert sphere or a hydrostatic cylinder into a product of central column
density and size we obtain an equivalence to the isotropic or anisotropic wavelet
coefficients. The application to the observed column density maps shows that
in both regions that we studied} the cylindrical modes dominate \changed{for
the largest fluctuations in the maps. In Polaris
this dominance is scale independent, in Aquila the spherical modes are comparable
to the cylindrical ones at scales below 0.05 and above 1.5~pc. The amplitude of
the gravitational modes is much higher in Aquila compared to Polaris.
For a gas temperature of 15~K} the higher column density in Aquila leads to mode amplitudes
above the limit for gravitational stability. They should lead to collapse at all scales
above 0.15~pc. All resolved structures in the Polaris Flare are gravitationally
stable. Measuring the critical spatial scale for a gravitational mode should allow us to
estimate the size of a region affected by a particular collapse mode. Fast collapse
is expected around the main Aquila rift, but larger regions should contribute to mass
accretion.

This pattern asks for an observational verification. Molecular line measurements
should trace the change of the velocity structure across that region to test the
correlation between the gravitational modes computed from the column density and actual
infall motions.

To constrain the scope of this paper we have limited ourselves to the two-dimensional
case here. However, it is essential to study the relation between three-dimensional
filamentary density structures and their projection in observable two-dimensional maps.
Therefore, we plan to extend the method to three dimensions in a subsequent paper.

\begin{acknowledgements}
This project was financed through DFG project number Os 177/2-2. We thank
Tigran Arshakian for taking a major initiative in the collaboration and
many useful discussions. We are grateful to Blakesley Burkhart for providing
us with the MHD simulations analyzed in Sect.~\ref{sect:MHD} \changed{and
Nicola Schneider for providing us with the column density maps analyzed in Sect.~\ref{sect:obs}.
We thank an anonymous referee for an extremely detailed and constructive report
on the first version of the manuscript.
}
\end{acknowledgements}


\clearpage

\begin{appendix}
\section{Analytical expressions for single Gaussian ellipses}
\label{appx_an}

For the special test case consisting of a single Gaussian elliptical clump
we can provide an analytic solution of the wavelet analysis that can serve
as a reference.
We define the Gaussian ellipse as
\begin{equation}\label{ex1}
  g(x,y)=(a c)^{-1/2} e^{-\frac{1}{2} \left(\frac{x^2}{a^2}+\frac{y^2}{c^2}\right)},
\end{equation}
where the prefactor provides a normalized integral of $g(x,y)^2$. \changed{To keep the equations shorter we use a notation here that deviates from the main text in using directly the letters $a$ and $c$ for the standard deviations of the main axes, not $\sigma_a$ and $\sigma_c$.}

%
%
%


The wavelet transform of $g(x,y)$ gives
\begin{strip}
\begin{equation}
  W(s,\varphi,\bx)=\frac{2 \pi \sqrt{c}  \exp \left(-\frac{2 \pi ^2 b^2 c^2 \sin ^2(\varphi ) \left(2 a^2+b^2 s^2\right)+2 \pi ^2 a^2 b^2 \cos ^2(\varphi ) \left(b^2 s^2+2 c^2\right)-2 i \pi
   b^2 s y \sin (\varphi ) \left(2 a^2+b^2 s^2\right)+2 a^2 y^2+2 i \pi  b^2 s x \cos (\varphi ) \left(b^2 s^2+2 c^2\right)+b^2 s^2 x^2+b^2 s^2 y^2+2 c^2
   x^2}{\left(2 a^2+b^2 s^2\right) \left(b^2 s^2+2 c^2\right)}\right)}{\sqrt{a s \left(\frac{1}{a^2}+\frac{2}{b^2 s^2}\right) \left(b^2 s^2+2 c^2\right)}},
  \end{equation}
where we neglect the constant $\exp(-\pi^2b^2)$ in Eq. (\ref{eq:morlet}). Its contribution to the result is less than 1\% even for the smallest $b$ used here. The resulting maps of isotropic $m^i$ and anisotropic $m^a$ are given by
\begin{eqnarray}
   m^i(s,x,y)&=&\frac{4 \pi ^2 a b^2 c s \times I_0\left(\frac{2 b^4 (c^2-a^2) \pi ^2 s^2}{\left(2 a^2+b^2 s^2\right) \left(2 c^2+b^2 s^2\right)}\right) \exp \left(-\frac{2 \left(a^2
   \left(\pi ^2 b^4 s^2+4 \pi ^2 b^2 c^2+2 y^2\right)+\pi ^2 b^4 c^2 s^2+b^2 s^2 \left(x^2+y^2\right)+2 c^2 x^2\right)}{\left(2 a^2+b^2 s^2\right) \left(b^2 s^2+2
   c^2\right)}\right)}{\left(2 a^2+b^2 s^2\right) \left(b^2 s^2+2 c^2\right)},\\
   m^a(s,x,y)&=&\frac{4 \pi ^2 a b^2 c s \times I_1\left(\frac{2 b^4 \left(c^2-a^2\right) \pi ^2 s^2}{\left(2 a^2+b^2 s^2\right) \left(2 c^2+b^2 s^2\right)}\right) \exp \left(-\frac{2
   \left(a^2 \left(\pi ^2 b^4 s^2+4 \pi ^2 b^2 c^2+2 y^2\right)+\pi ^2 b^4 c^2 s^2+b^2 s^2 \left(x^2+y^2\right)+2 c^2 x^2\right)}{\left(2 a^2+b^2 s^2\right) \left(b^2
   s^2+2 c^2\right)}\right)}{\left(2 a^2+b^2 s^2\right) \left(b^2 s^2+2 c^2\right)},
   \end{eqnarray}
where $I_0$ and $I_1$ denote the Bessel-$I$ functions of the zeroth and first order. Integration over the $(x,y)$ - plane gives the isotropic and anisotropic spectra
\begin{eqnarray}
   M^i(s)&=&\frac{2 \pi ^3 a b^2 c s \times \exp \left(-\frac{2 \pi ^2 b^2 \left(a^2 \left(b^2 s^2+4 c^2\right)+b^2 c^2 s^2\right)}{\left(2 a^2+b^2 s^2\right) \left(b^2 s^2+2
   c^2\right)}\right) I_0\left(\frac{2 b^4 \left(c^2-a^2\right) \pi ^2 s^2}{\left(2 a^2+b^2 s^2\right) \left(2 c^2+b^2 s^2\right)}\right)}{\sqrt{\left(2 a^2+b^2 s^2\right)
   \left(b^2 s^2+2 c^2\right)}},\\
    M^a(s)&=&\frac{2 \pi ^3 a b^2 c s \times \exp \left(-\frac{2 \pi ^2 b^2 \left(a^2 \left(b^2 s^2+4 c^2\right)+b^2 c^2 s^2\right)}{\left(2 a^2+b^2 s^2\right) \left(b^2 s^2+2
   c^2\right)}\right) I_1\left(\frac{2 b^4 \left(c^2-a^2\right) \pi ^2 s^2}{\left(2 a^2+b^2 s^2\right) \left(2 c^2+b^2 s^2\right)}\right)}{\sqrt{\left(2 a^2+b^2
   s^2\right) \left(b^2 s^2+2 c^2\right)}}.
   \end{eqnarray}
\end{strip}

\begin{figure}
\centering
\includegraphics[angle=90,width=0.88\columnwidth]{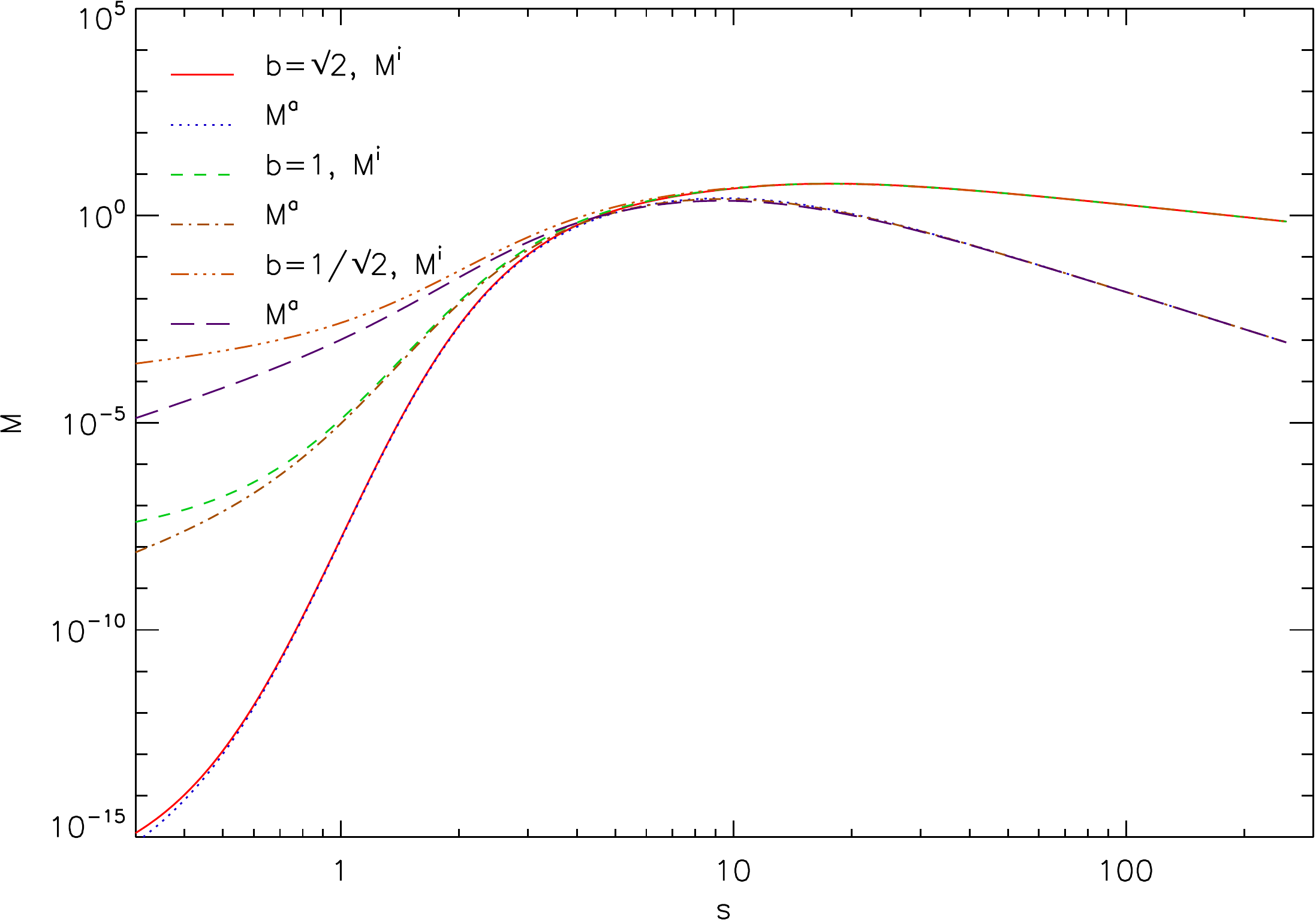}
\includegraphics[angle=90,width=0.88\columnwidth]{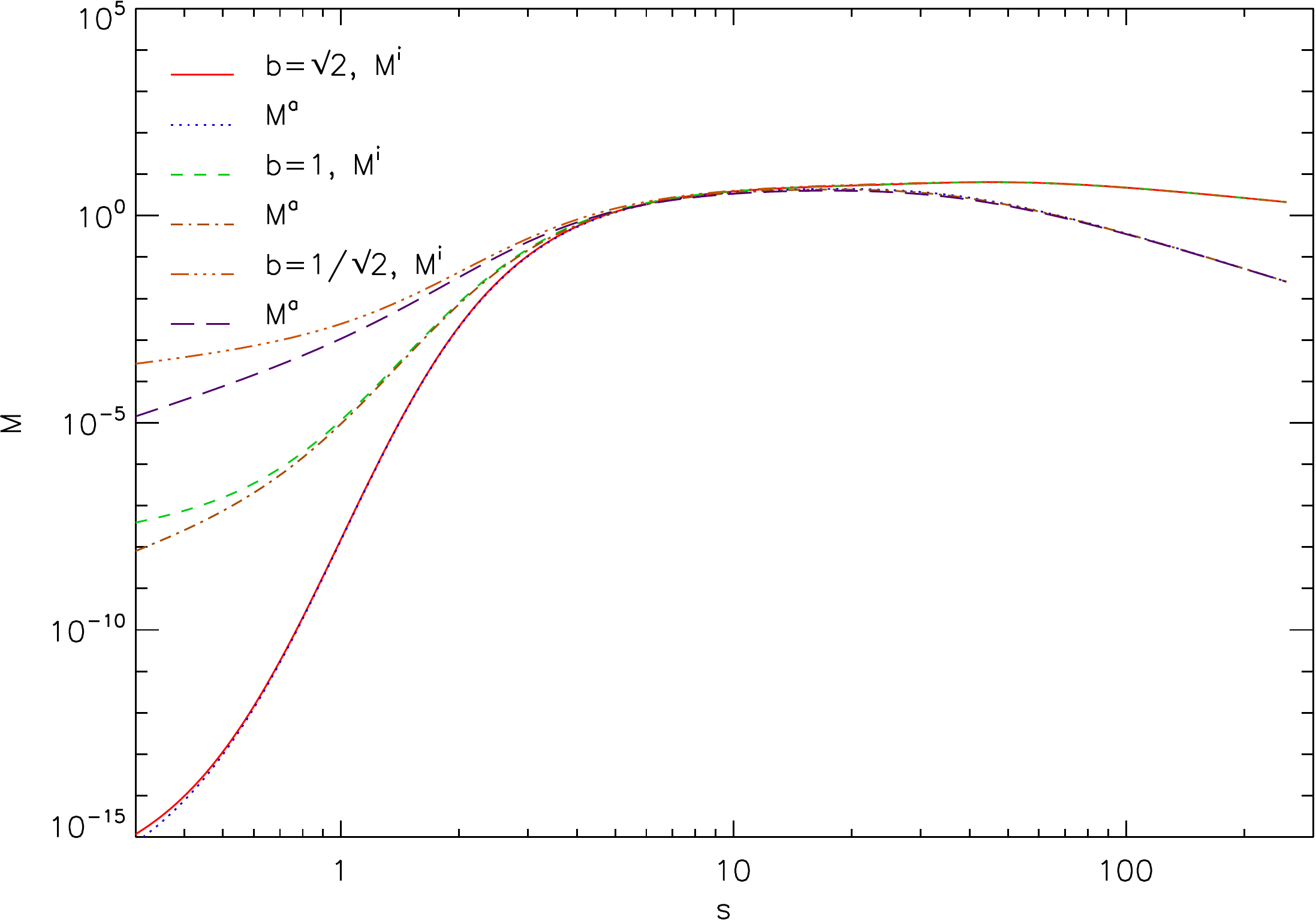}
\caption{Wavelet spectra from the analytic wavelet convolution of elliptical
Gaussians with \changed{axes of $1 \times 3$ pixels (top panel) and $1 \times 9$ ~pixels} (bottom panel), respectively.
\label{fig:gaussian-spectra}}
\end{figure}
In Fig.~\ref{fig:gaussian-spectra} we show the spectra of isotropic and
anisotropic coefficients for two ellipses with $a=1, c=3$ and $a=1, c=9$ computed
for the three different values of the localization parameter discussed in the
paper. The $b$ parameter only affects the small scales. At large scales the
curves for the three $b$ values fall on top of each other. In contrast the
extension of the larger main axis of the ellipses does not change the spectra
at small scales, but extends the peaks towards longer scales.

The ratio of the anisotropic and isotropic spectra provides the degrees of anisotropy
\begin{equation}\label{ex7}
  d^w_{loc}(s)=\frac{I_1\left(\frac{2 b^4 \left(c^2-a^2\right) \pi ^2 s^2}{\left(2 a^2+b^2 s^2\right) \left(2 c^2+b^2 s^2\right)}\right)}{I_0\left(\frac{2 b^4 \left(c^2-a^2\right) \pi ^2
   s^2}{\left(2 a^2+b^2 s^2\right) \left(2 c^2+b^2 s^2\right)}\right)}
\end{equation}

\begin{figure}
\centering
\includegraphics[angle=90,width=0.88\columnwidth]{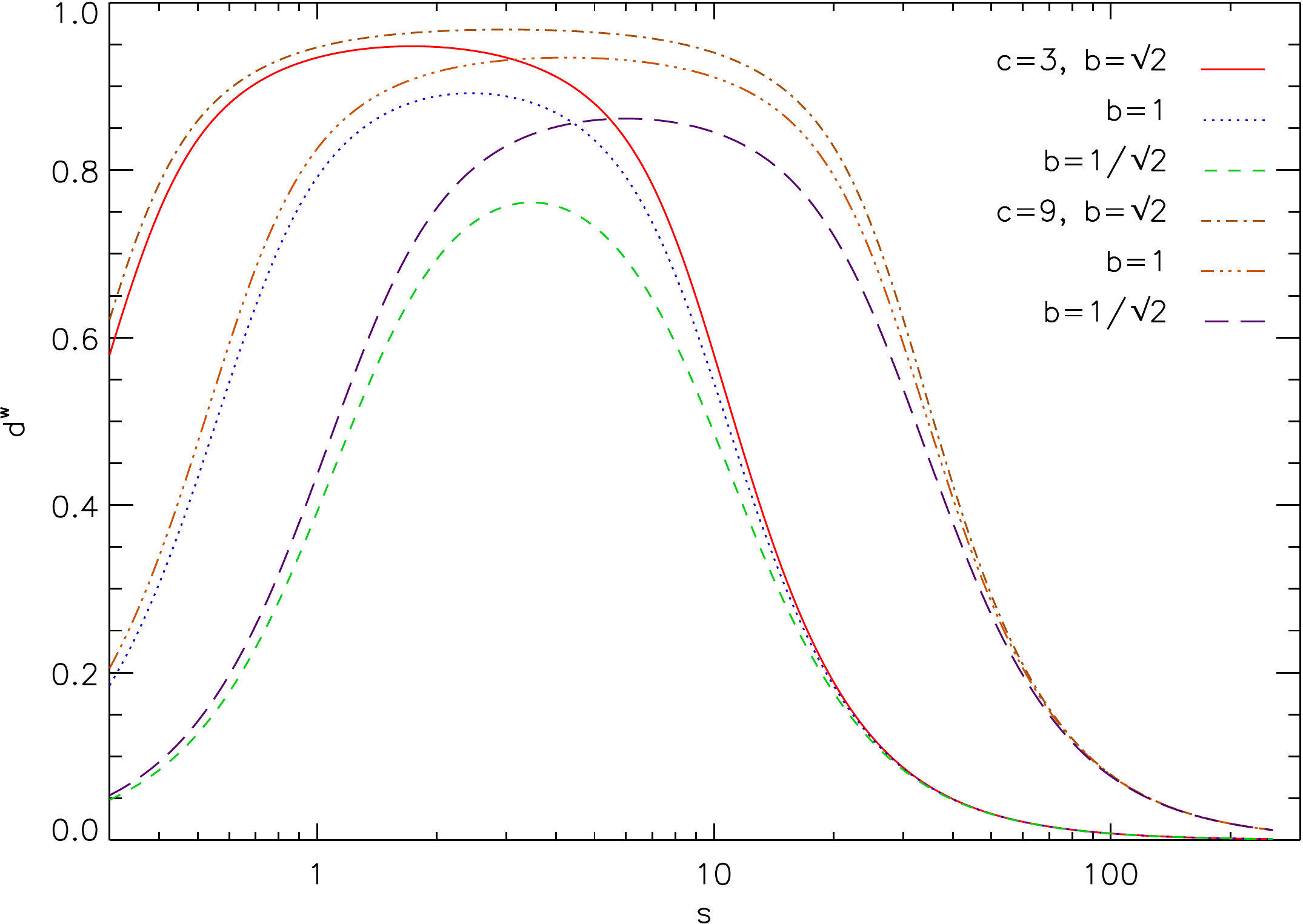}
\caption{Degree of local and global anisotropy from the analytic wavelet
convolution of elliptical Gaussians with \changed{axes of $1 \times 3$ pixels and $1 \times 9$ ~pixels}.
\label{fig:gaussian-degrees}}
\end{figure}
Since $m^a(s,x,y)$ is real and positive, local and global anisotropy agree $d^w_{loc}(s)=d^w_{glob}(s)$.
Figure~\ref{fig:gaussian-degrees} shows those ratios for the spectra
from Fig.~\ref{fig:gaussian-spectra}. We see that the large-scale wing of the
anisotropy spectra is only determined by the larger major axis of the ellipses --
the wings are shifted exactly by the factor of three. In contrast, the
small-scale wing and the resulting width of the plateau with high anisotropies
is mainly determined by the localization parameter $b$. A $b=\sqrt{2}$ filter creates
a plateau that is much wider than the factor of three or nine between the
two major axes of the ellipses.

\changed{The maximum of the local degree of anisotropy $d^w_{loc}(s)$ at the
peak scale $s^*$ depends only on the aspect ratio $\gamma=c/a$ and is given
by}
\begin{equation}\label{ex8}
  d^w_{loc}(s^*)={I_1\left(\frac{\pi^2 b^2 \left(\gamma-1\right)}{\left(\gamma+1\right)}\right)}\Bigg/
	{I_0\left(\frac{\pi^2 b^2 \left(\gamma-1\right)}{\left(\gamma+1\right)}\right)}\,.
\end{equation}

The location of the peak of the degree of anisotropy falls at $s^*=\sqrt{2 a c}/b$.
For the peaks of the spectra of wavelet coefficients we cannot give an analytic expression,
but they are approximated by simple fitting functions in Sect.~\ref{sec:simpleGaussian}.

\section{Comparison to the $\Delta$-variance}
\label{sect_comp_deltavar}

\changed{
Our approach presented in Sect.~\ref{sec:theory} is similar to the
$\Delta$-variance. Both methods make use of wavelet functions
for filtering of spatial structures. However,} the wavelet used
for the $\Delta$-variance is purely real and has an isotropic,
slightly different shape.
Because of a normalization by $s^{-2}$ in the equivalent of Eq.~\ref{cwt} for
the $\Delta$-variance definition, the $\Delta$-variance spectra are
steeper by $s^{-1}$ than the wavelet spectra computed here. For a direct comparison
we \changed{have to rescale the $\Delta$-variance spectra by a multiplication with
$s$ and a corresponding normalization factor to get the scale-dependent energy density
or divide the wavelet coefficients from Eq.~\ref{w_parts} by $s$ to get the
scale-dependent variance.}

\citet{Ossenkopf2008a} calibrated the scale dependence of the
$\Delta$-variance by the systematic analysis of test structures with
well defined size scales, comparing the width of individual
emission peaks in the map with the scale of the $\Delta$-variance
peak. It provided a relation between filter size $s$ and scale of
the highest sensitivity of the $\Delta$-variance.
For Gaussian circular structures \citet{ArshakianOssenkopf2016} showed that
the $\Delta$-variance spectra have a prominent peak at
$s_\Delta(\sigma_\Delta^2)\sub{max}=4\sigma$ of the Gaussians.

\changed{When using the amplitude rescaling $M^i(s)/s^2$ for the wavelet
coefficients introduced in Sect.~\ref{sec:theory} the $\Delta$-variance
has to be rescaled by $1/s$ to obtain the same dimensionality.
The analysis of the peak of the rescaled
$\Delta$-variance for circular Gaussian structures, equivalent to
Appx.~\ref{appx_an}, provides a shift to $s_\Delta(\sigma_\Delta^2/s)\sub{max}=2.4\sigma$.
As the peak of the corresponding wavelet spectra, $M^i(s)/s^2$, falls
at $4.8\sigma$ we find that a simple shift of the $\Delta$-variance lag scale
by the factor 2.0 provides a match of the peak scales of the $\Delta$-variance
spectra and the isotropic wavelet spectra.

\begin{figure}
\centering
\includegraphics[angle=90,width=0.88\columnwidth]{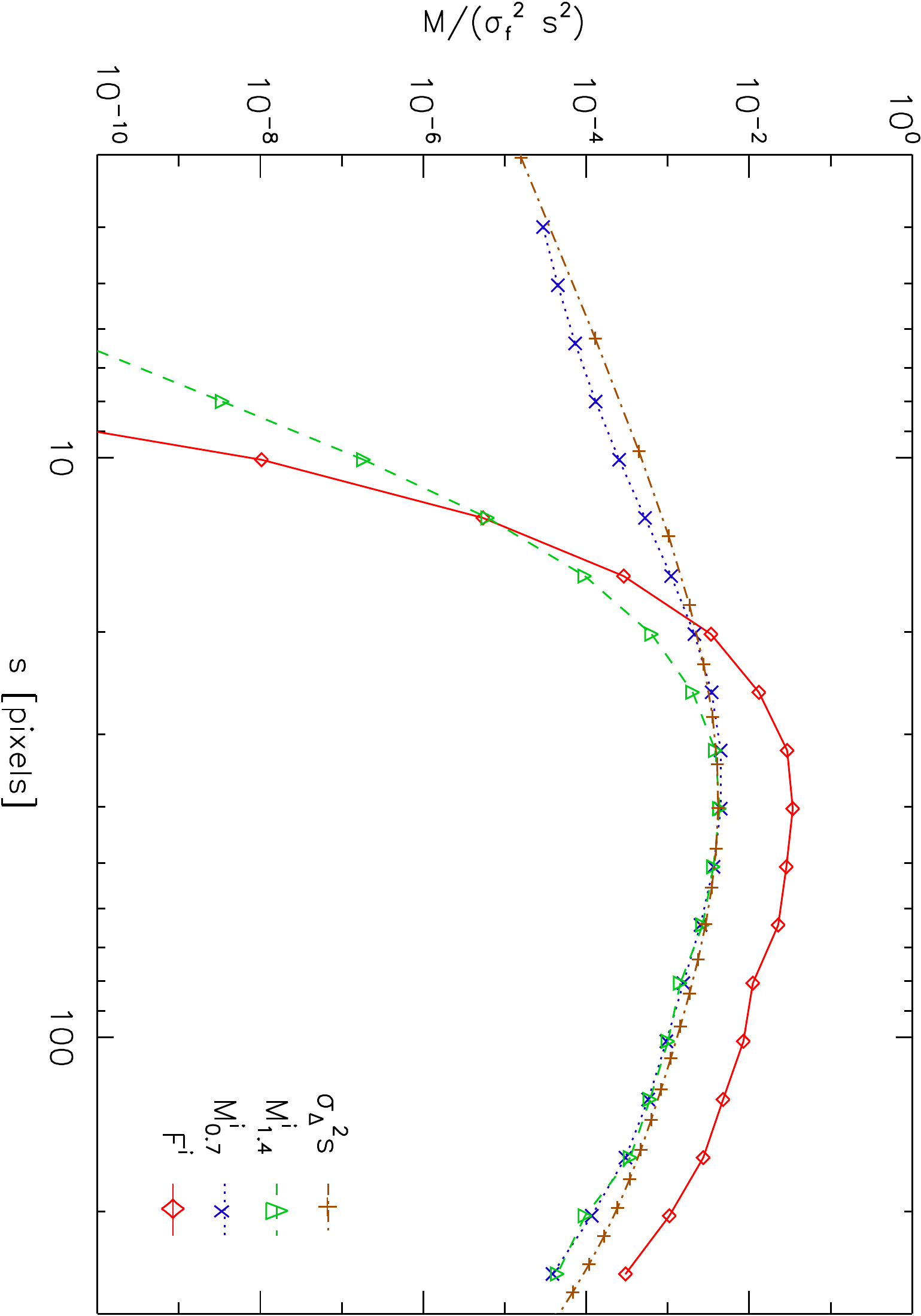}\vspace{3mm}
\includegraphics[angle=90,width=0.88\columnwidth]{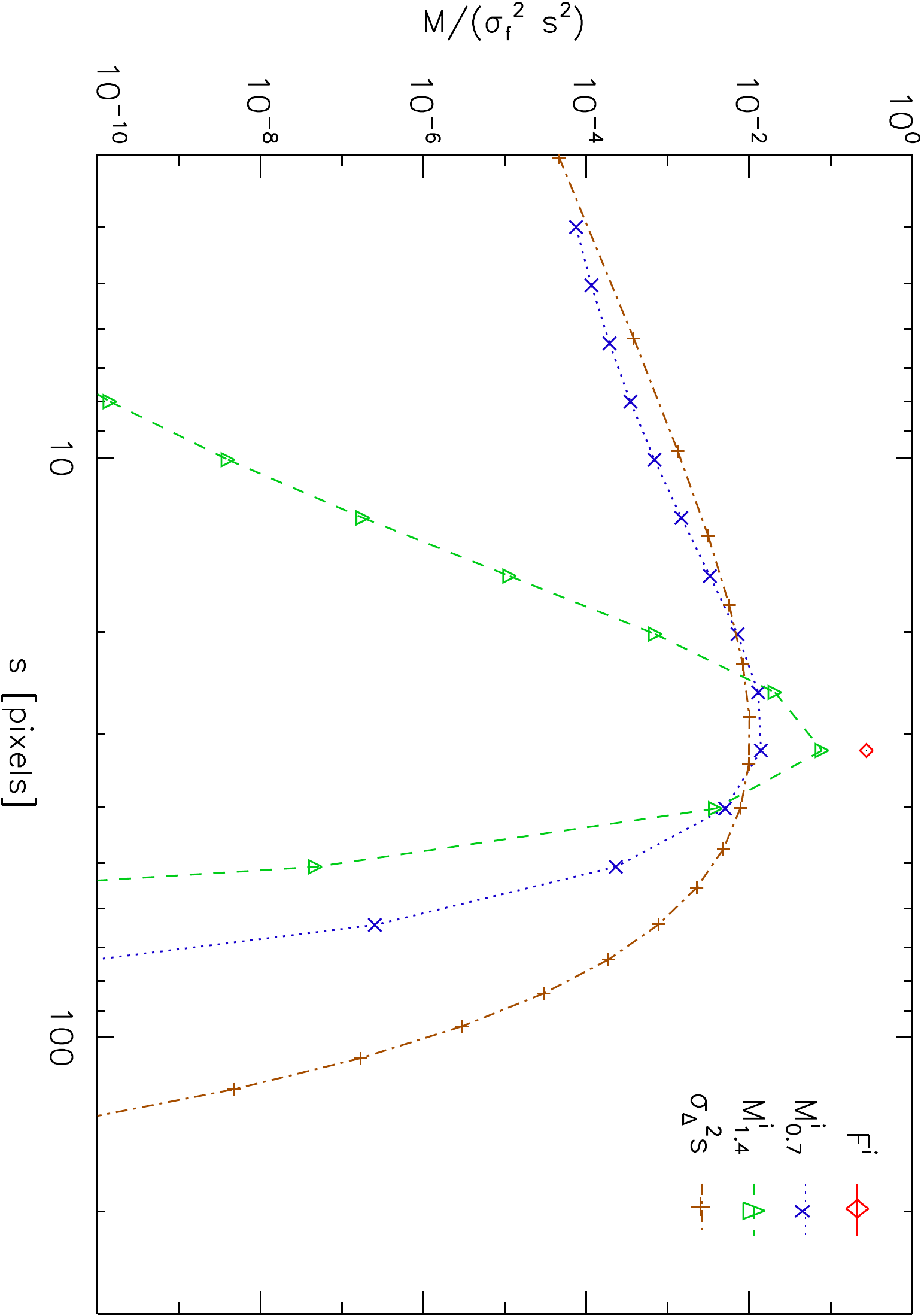}
\caption{Comparison of the $\Delta$-variance spectra with the isotropic wavelet
and Fourier spectra for the test structures from Figs.~\ref{fig:clumps} and
\ref{fig_stripes}, i.e. a map containing 10 Gaussian clumps with a standard
deviation of $\sigma=8$~pixels (top) an a map containing
a sinusoidal stripe pattern with a period of $p=32$~pixels (bottom). All spectra
are plotted using the amplitude rescaling, $M/s^2$, that always guarantees a peak.
The wavelet spectra were computed for localization parameters $b=\sqrt{2}$ and
$b=1/\sqrt{2}$.}
\label{fig:deltavar}
\end{figure}

In Fig.~\ref{fig:deltavar} we compare the \changed{rescaled} $\Delta$-variance spectra with
the wavelet and Fourier spectra for the isotropic example from Fig.~\ref{fig:clumps}
and the anisotropic example from Fig.~\ref{fig_stripes}.
Here and in all the spectra in Appx.~\ref{appx_tests} we have shifted the $\Delta$-variance
spectra by the constant factor of 2.0 to allow for a direct
comparison to the wavelet spectra, using the $s$ parameter as a uniform scale.
Both for the isotropic test case but also for the extreme example of the
most anisotropic structure, given by the one-dimensional sinusoidal
pattern (Fig.~\ref{fig_stripes}), we obtain a match of the
peak positions in the spectra by this $\Delta$-variance shift.}
It confirms that both methods provide an independent,
uniform, and consistent assignment between wavelet scale and measured
structure size. \changed{When comparing the spectra, one only has to
use the appropriate scale definition. Here, we stick to the wavelet scale
defined in Sect.~\ref{sec:theory} that has a peak at 4.8 times the
standard deviation of isotropic Gaussian structures in spectra of
amplitude per scale for all methods compared throughout the paper.}

\changed{For the scale-free cases of fractional
Brownian motion structures
(see e.g. Fig.~\ref{fig:appx_std1} and Fig.~\ref{fig:appx_fbm})
there is always} a good match of the isotropic spectra. For all
cases with pronounced size scales the $\Delta$-variance spectrum shows
a similar, \changed{but wider peak than} the isotropic wavelet and
Fourier spectra. \changed{As known from the sampling theorem and
discussed in Sect.~\ref{sect_filtershape} the scale sensitivity
decreases with increasing localization of the filtering. The Fourier
spectra, without any localization filtering, show the sharpest
peaks around prominent scales. Our anisotropic wavelet spectra
show broader peaks with an increasing width for decreasing
localization parameter $b$. The $\Delta$-variance peaks finally
are always the broadest among the compared methods. In terms of
scale sensitivity} our new anisotropic wavelet filter is thus
superior to the original $\Delta$-variance wavelet as shown already
for another type of isotropic wavelet by \citet{Arevalo2012}.

\section{Comparison with structure-function-based anisotropy}
\label{sect_comp_sf}

\cite{EsquivelLazarian2011} also proposed a very generic approach
to characterize the scale dependent isotropy or anisotropy in maps.
An extension to position-position-velocity cubes was proposed by
\citet{Burkhart2014}.

The method is based on the structure function
\begin{equation}
S(\vec{r}) = \left\langle \left[ f(\vec{x}) - f (\vec{x} + \vec{r})\right]^2 \right\rangle_{\vec{x}}.
\end{equation}
Inspecting the contours of the structure function in the $\vec{r}$-plane
the anisotropies in the map are quantified by the deviation of these
contours from a round shape. As all anisotropies characterized by
\cite{EsquivelLazarian2011} were aligned with the $y$-direction,
it was sufficient to measure the values of the structure function
$x$ and $y$-direction, i.e. at $x=|\vec{r}|$ and $y=|\vec{r}|$ and
take their ratio as a function of the scale $|\vec{r}|$. Dividing the smaller by the
larger value provided the degree of isotropy\changed{, $d\sub{iso}^{\rm sf}$,} going from zero for
extremely anisotropic structures to unity for isotropic structures.

Generalizing the approach, we can use the minimum and maximum
value of the structure function at any point on the circle of constant $|\vec{r}|$
being independent from a known direction of anisotropy to define the
structure-function-based degree of isotropy. As the structure function averages
over the whole map, this measure can only characterize the global degree of
isotropy in a map.  For a direct comparison to our degree of anisotropy, not isotropy,
we plot $1-d\sub{iso}^{\rm sf}$ as a measure of anisotropy.

\begin{figure}
\centering
\includegraphics[angle=90,width=0.88\columnwidth]{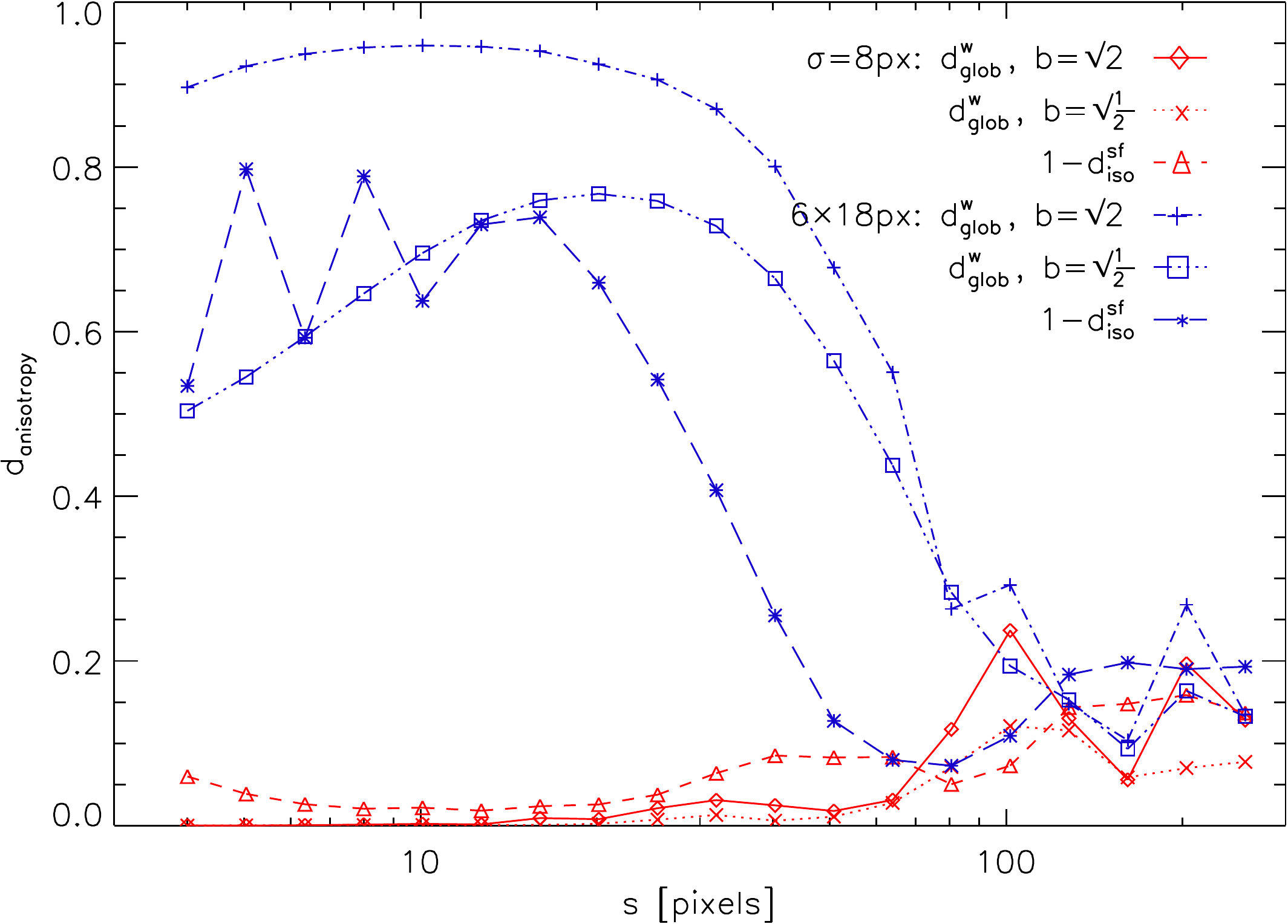}
\caption{Comparison of the global degree of anisotropy measured through the
wavelet analysis with the degree of isotropy measured through the
structure function for two maps containing spherical clumps ($\sigma=8$~pixels,
Fig.~\ref{fig:clumps}) and elliptical clumps ($\sigma=6x18$~pixels) all aligned
at 45~degrees (Fig.~\ref{fig:aligned_ellipses}).}
\label{fig:structure_function}
\end{figure}

Figure~\ref{fig:structure_function} compares the output of the structure function
with the wavelet-based global degree of anisotropy from the two filter shapes with
$b=\sqrt{2}$ and $b=1/\sqrt{2}$ for the simple examples of a random arrangement of
circular clumps and a superposition of aligned elliptic clumps (see Sects.
\ref{sect_superposition}, \ref{sect_ecc_calibration}, and Appx.~\ref{appx_tests}).
All methods agree on
the isotropy of the spherical clumps on small scales and the small random
anisotropy on large scales. For the aligned clumps, the structure function shows
a moderate anisotropy at small scales similar to the value obtained from the
less elongated filter with  $b=1/\sqrt{2}$. However, the anisotropy shows no
increase at very small scales and drops already at shorter scales than in the
wavelet spectrum.

\begin{figure}
\centering
\includegraphics[angle=90,width=0.88\columnwidth]{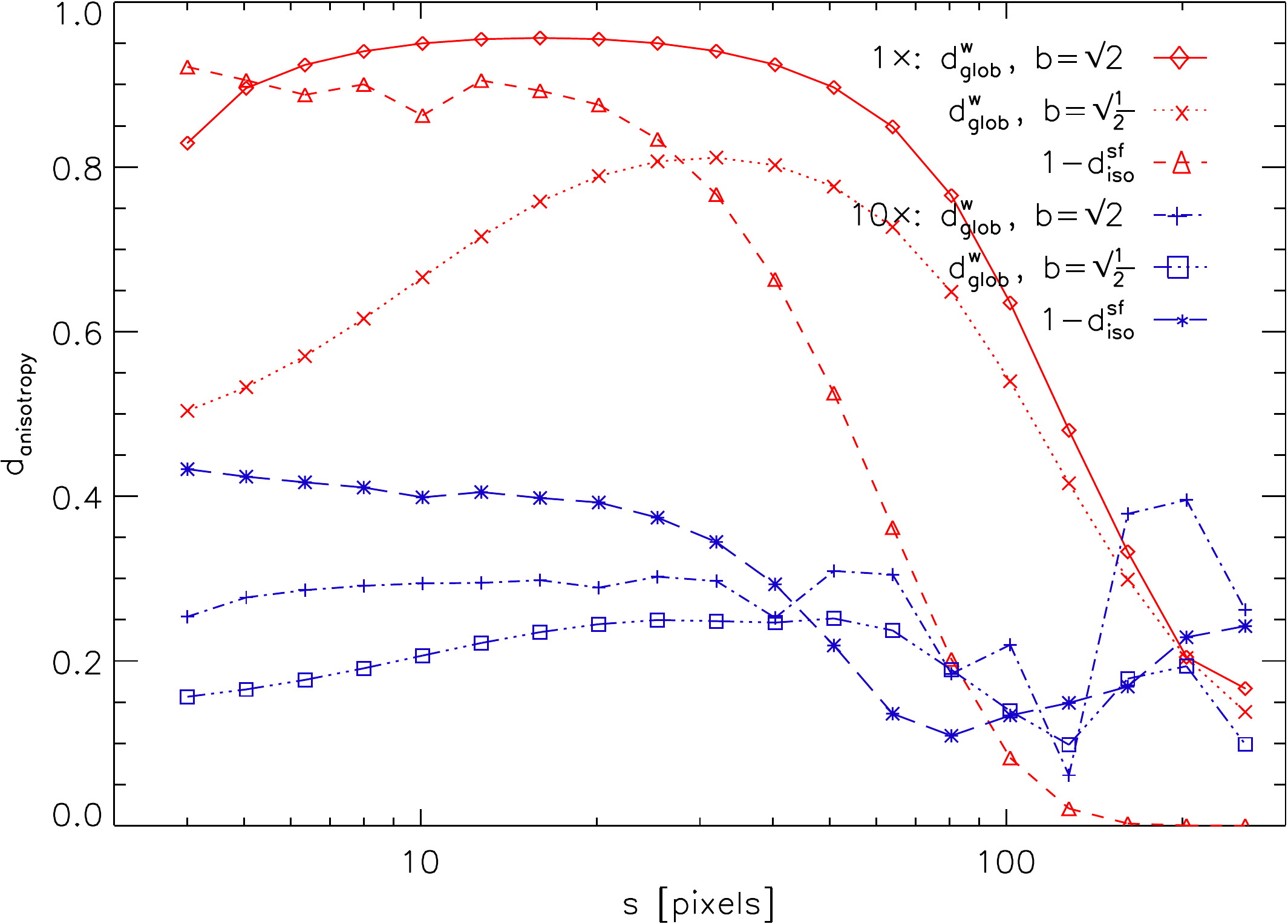}
\caption{Comparison of the global degree of anisotropy measured through the
wavelet analysis with the degree of isotropy measured through the
structure function for two maps containing either a single or ten
randomly distributed and oriented elliptic clumps with an axes
ratio of 4:1 ($\sigma=8\times 32$~pixels, see Fig.~\ref{fig:compare_sizes}).}
\label{fig:structure_function_multiple}
\end{figure}

To \changed{test} whether this scale shift results from the destruction of the global
anisotropy by the superposition of many clumps, we repeat the test with the
varying number of ellipses from Figs.~\ref{fig:compare_sizes} and
\ref{fig:comparefilterizes} \changed{and include a calculation} of the structure function.
Figure~\ref{fig:structure_function_multiple} compares the output of the structure function
with the wavelet-based global degree of anisotropy. The single clump  results
in a high global anisotropy seen through all measures and the relative
shift in the scales is the same as in Fig.~\ref{fig:structure_function} excluding
superposition effects. Consequently, the two methods only differ in a scale
calibration. However, even when further increasing the size of the ellipses,
the anisotropy seen through the structure function remains at the same high
level at small scales preventing any detection of the minimum filament size
through that method.

For the superposition of multiple clumps we find similar values for the remaining
global anisotropy at small scales. The structure function provides the highest values
being least sensitive to superposition effects as it naturally ignores
\changed{low map values in contrast to the wavelet analysis that always considers
variations.} \changed{It is an appropriate method to measure the global
anisotropy when we recalibrate the scales by a factor of two relative to
each other, that means we can shift the structure function based degree of anisotropy
to two times higher scales to apply the scaling relations from Sect.~\ref{sec:simpleGaussian}.}
When looking for global anisotropies, the wavelet analysis
has the advantage of being sensitive to the minimum size of the structures
while the structure function is less affected by superposition effects.
However, the structure function can only characterize the global anisotropy.
For an unbiased analysis of the local filamentariness in a map, the wavelet
analysis is the only approach.

\section{Understanding the degree of anisotropy for clump ensembles}
\label{appx_anisotropy}

\begin{figure}
\centering
\includegraphics[angle=90,width=0.8\columnwidth]{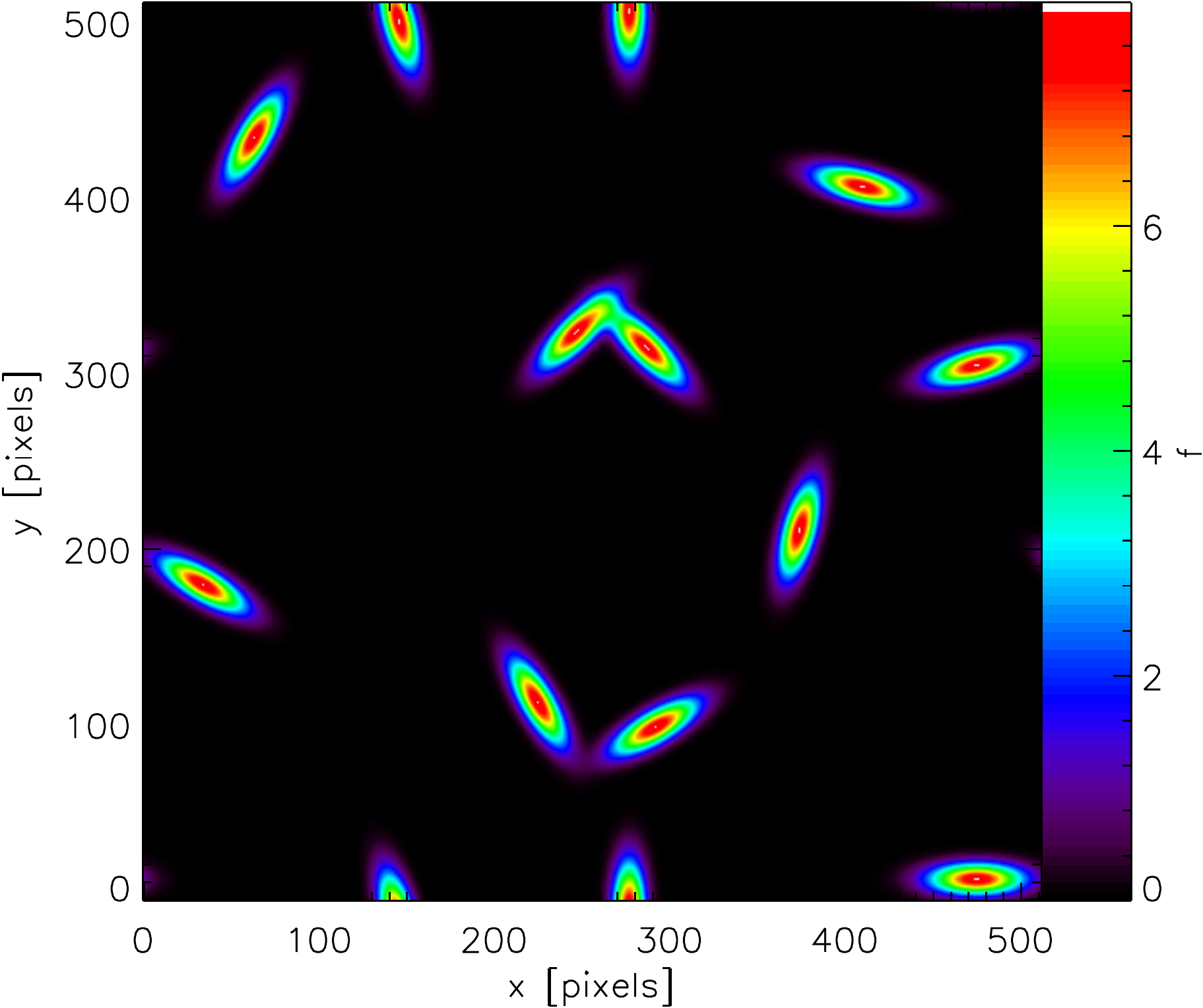}\vspace{3mm}
\includegraphics[angle=90,width=0.87\columnwidth]{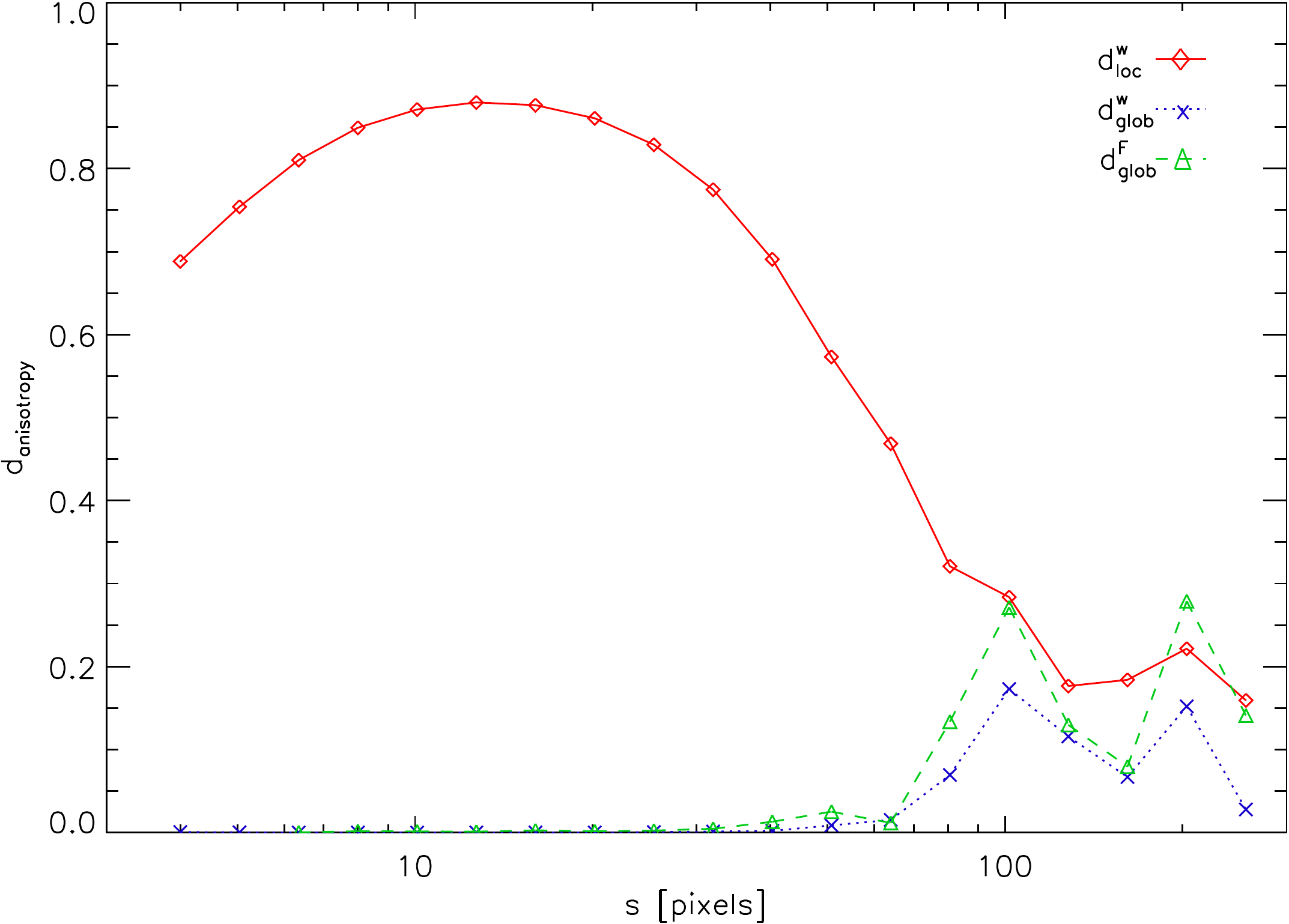}
\caption{Anisotropic wavelet analysis of a map containing
elliptic Gaussian clumps with standard deviations of 6 and 18~pixels in
orthogonal directions providing an aspect ratio of 3:1 and
having random orientations. The upper panel shows the original map,
the lower plot the degrees of anisotropy.}
\label{fig:ellipses}
\end{figure}

\changed{To illustrate how the degree of anisotropy changes for an ensemble of
clumps relative to the degree measured for individual clumps in
Sect.~\ref{sec:simpleGaussian} we give some examples for superpositions
of anisotropic clumps here, similar to the case of isotropic clumps
in Fig.~\ref{fig:clumps}, but now including the additional degree of freedom
of their relative orientation.
Figure~\ref{fig:ellipses} shows the wavelet analysis for an ensemble
of ten anisotropic} Gaussian clumps. The clumps have the same
aspect ratio of 1:3 and a uniform distribution of orientations. The spectra of wavelet and
Fourier coefficients are similar to those of the individual clumps
(Fig.~\ref{fig:singleGaussian}) so that we do not show them here.
The main difference is seen in the local \changed{and global} degree of anisotropy.

\begin{figure}
\centering
\includegraphics[angle=90,width=0.8\columnwidth]{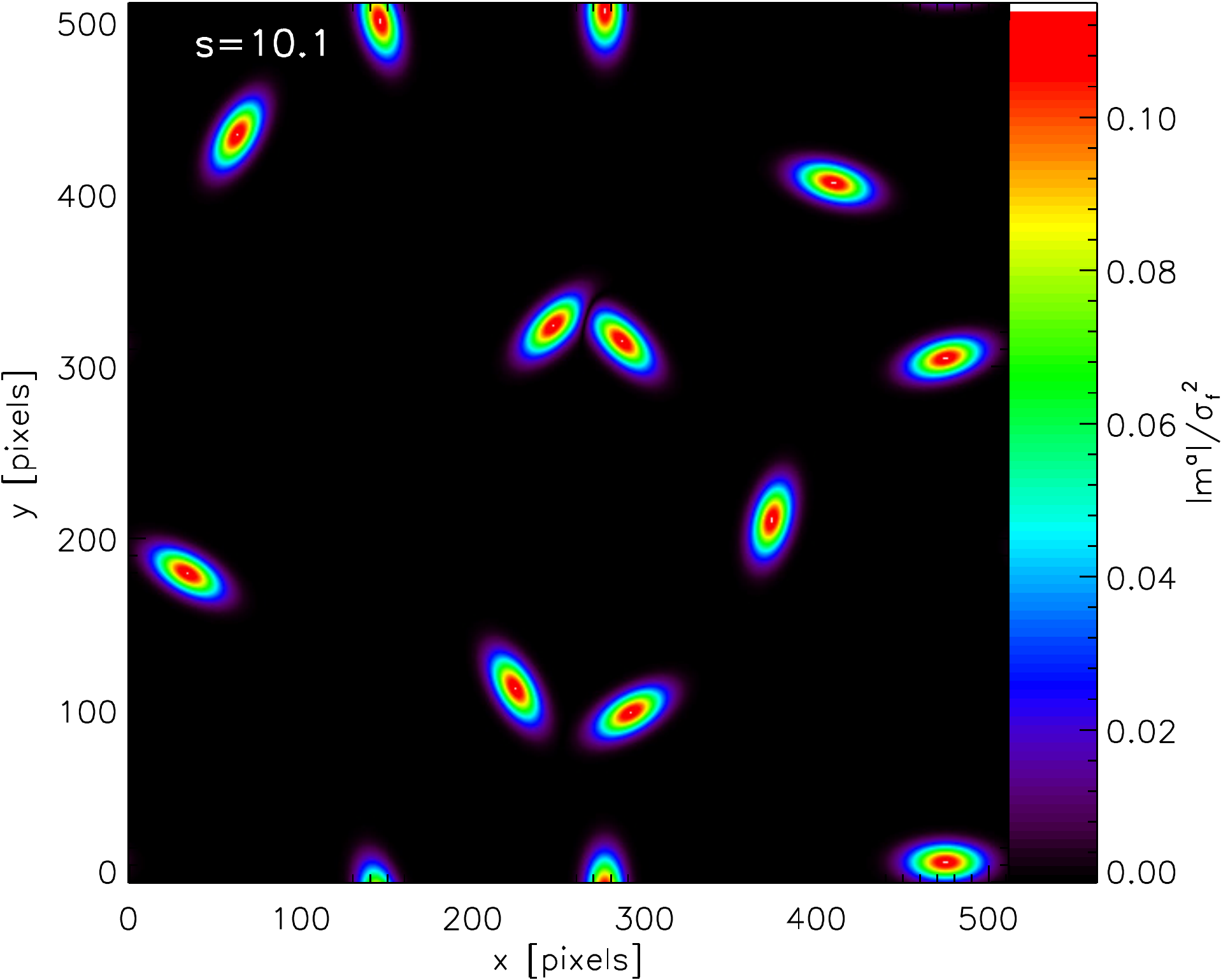}\vspace{3mm}
\includegraphics[angle=90,width=0.8\columnwidth]{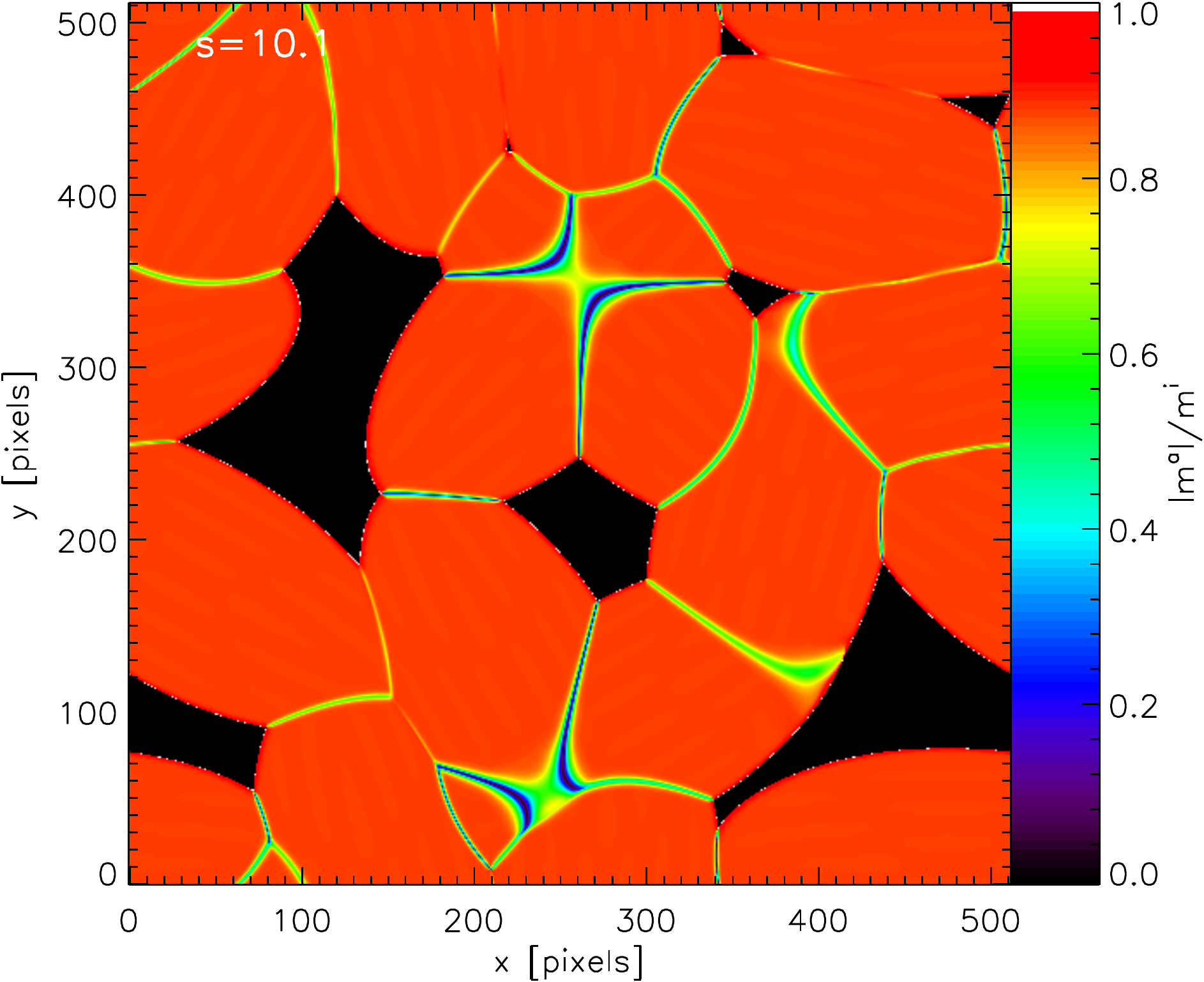}
\caption{Map of anisotropic wavelet coefficients (top) and map of the
ratio of anisotropic to isotropic wavelet coefficients (bottom) for the
structure of elliptic Gaussian clumps from Fig.~\ref{fig:ellipses}
computed for a wavelet size $s$ of 10.1~pixels.}
\label{fig:ellipse_coefficients}
\end{figure}

The shape of the spectrum of local anisotropies can be understood when
looking at a map of anisotropies defined as $|m^a(s,\vec{x})|/m^i(s,\vec{x})$.
This quantity is not useful to characterize the general anisotropy
in a map because it is dominated by contributions from very small
$m^i(s,\vec{x})$ far away from any real structure in the map, but it
can be used to visualize the spatial location of the different contributions.
Fig.~\ref{fig:ellipse_coefficients} shows the map of anisotropic
wavelet coefficients used to compute the degrees of anisotropy and
the map of the ratio of anisotropic to isotropic wavelet coefficients
for the structure from Fig.~\ref{fig:ellipses}
when applying a filter size $s=10$~pixels. The anisotropic wavelet
coefficients peak at the locations of the individual clumps but they
have a less elongated structure being dominated by the gradients
at the sharper boundaries of the individual clumps. Around the clumps
we find identical values for the isotropic
and anisotropic wavelet coefficients providing a ratio of unity.
Lower values occur at the boundaries where
two clumps approach each other so that the overall structure becomes less
anisotropic. This is in particular prominent for the two groups of clumps at about
$(x,y)=(250,350)$ and (250,50), where the approximately orthogonal direction
of the two clumps leads to a reduced local degree of anisotropy. When
going to larger filter sizes, the effect of the relative arrangement of
the different clumps becomes more and more prominent relative to the
anisotropy of the individual clumps so that the overall degree of anisotropy
$d^w_{loc}(s)$ decreases as seen in Fig.~\ref{fig:ellipses}. At smaller filter sizes
the anisotropic filter does not fully match the structure of the ellipses
so that the local degree of anisotropy $|m^a(s,\vec{x})|/m^i(s,\vec{x})$
falls somewhat below unity explaining the decrease of $d^w_{loc}(s)$
towards very small filter sizes.

\begin{figure}
\centering
\includegraphics[angle=90,width=0.87\columnwidth]{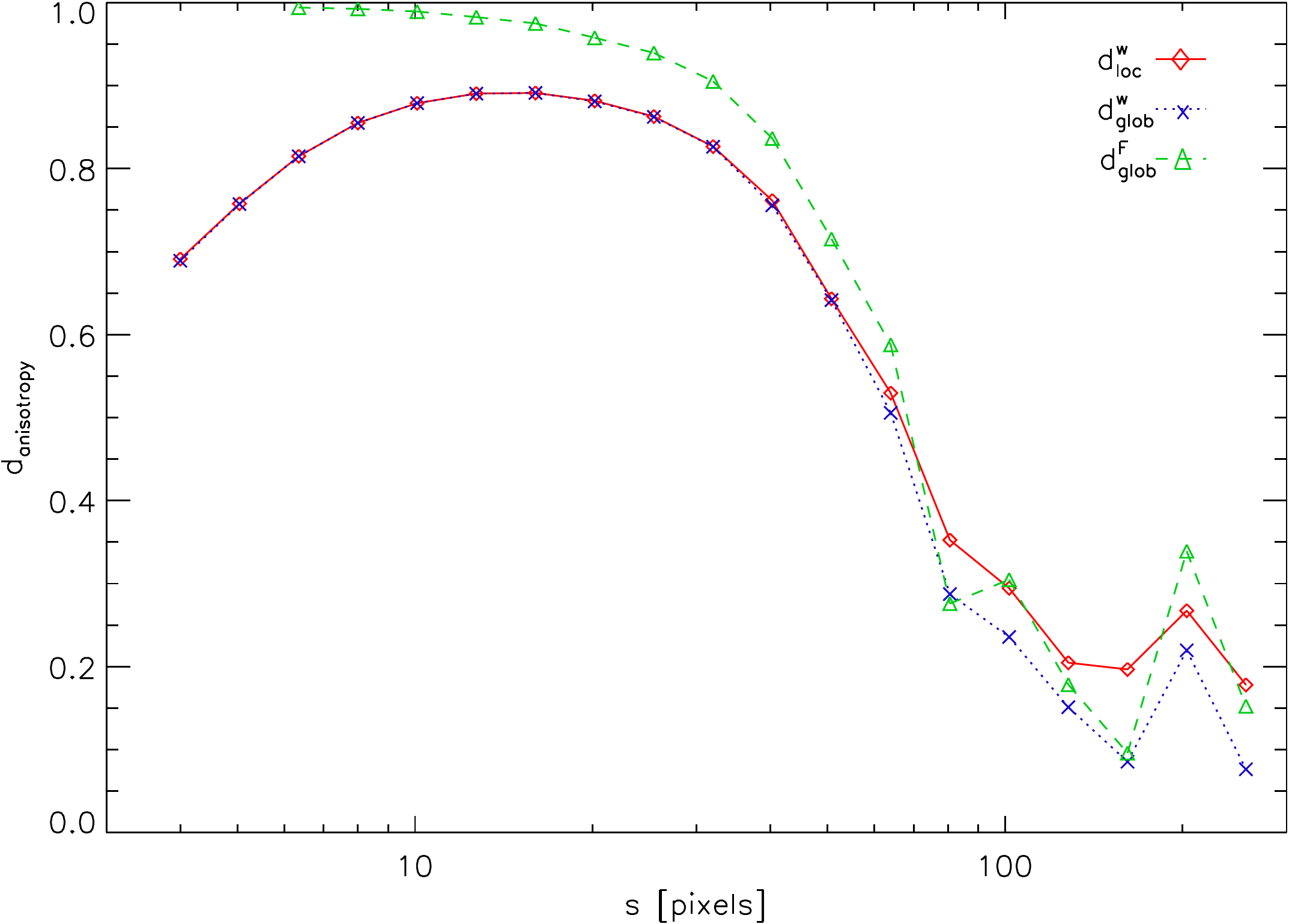}
\caption{Degrees of anisotropy for a map containing the elliptic
Gaussian clumps from Fig.~\ref{fig:ellipses}, but all aligned in the
same 45~degrees orientation.}
\label{fig:aligned_ellipses}
\end{figure}

For Fig.~\ref{fig:aligned_ellipses} we changed the configuration from
Fig.~\ref{fig:ellipses} by introducing a global anisotropy.
All clumps are aligned in the same direction with their major axis.
The resulting spectra of wavelet coefficients and isotropic Fourier
coefficients are practically the same as those
from the unaligned ellipses in Fig.~\ref{fig:ellipses} and also the
local degree of anisotropy is the same as in Fig.~\ref{fig:ellipses}
but the wavelet-based degree of global anisotropy now matches
the local anisotropy spectrum.  The degree of global anisotropy
measured through the Fourier spectrum follows the one measured
through the wavelet coefficients at large and intermediate scales
but remains high at scales below the clumps size. Combining local and global
degree of anisotropy allows us to characterize both the anisotropy of
individual structures and their mutual alignment.

\section{Analysis of test structures}
\label{appx_tests}

Here we present the standard analysis plots for a subset
of test structures that we processed to measure the sensitivity of the
different tools to different aspects of the maps. \changed{We only
show data sets that are not described by the analytical formulae
in Appx.~\ref{appx_an} or where superposition effects are small
enough that they still closely follow those formulae.}
The figures presented here always show the analyzed
map in the first panel,  the angular distribution of the
anisotropic wavelet coefficients $A(s,\varphi)$ in the
second panel, the spectrum of isotropic wavelet, $\Delta$-variance
and Fourier coefficients  in the third panel, and the corresponding
spectra of the degrees of anisotropy in the fourth panel.

\changed{In contrast to the plots up to Sect.~\ref{sect_fractal}
that used the $b=1$ filter we show the results for two different
filters.}
The plots always contain the results for the analysis with
a broad Gaussian kernel, $b=\sqrt{2}$, indicated by the
superscript "1.4", and a narrow kernel, $b=1/\sqrt{2}$,
indicated by the superscript "0.7". In the second panel,
the \changed{angular spectrum for} the $b=1/\sqrt{2}$ filter
is plotted in colors and for the $b=\sqrt{2}$ filter in
contours.

\begin{figure}
\centering
\includegraphics[angle=90,width=0.8\columnwidth]{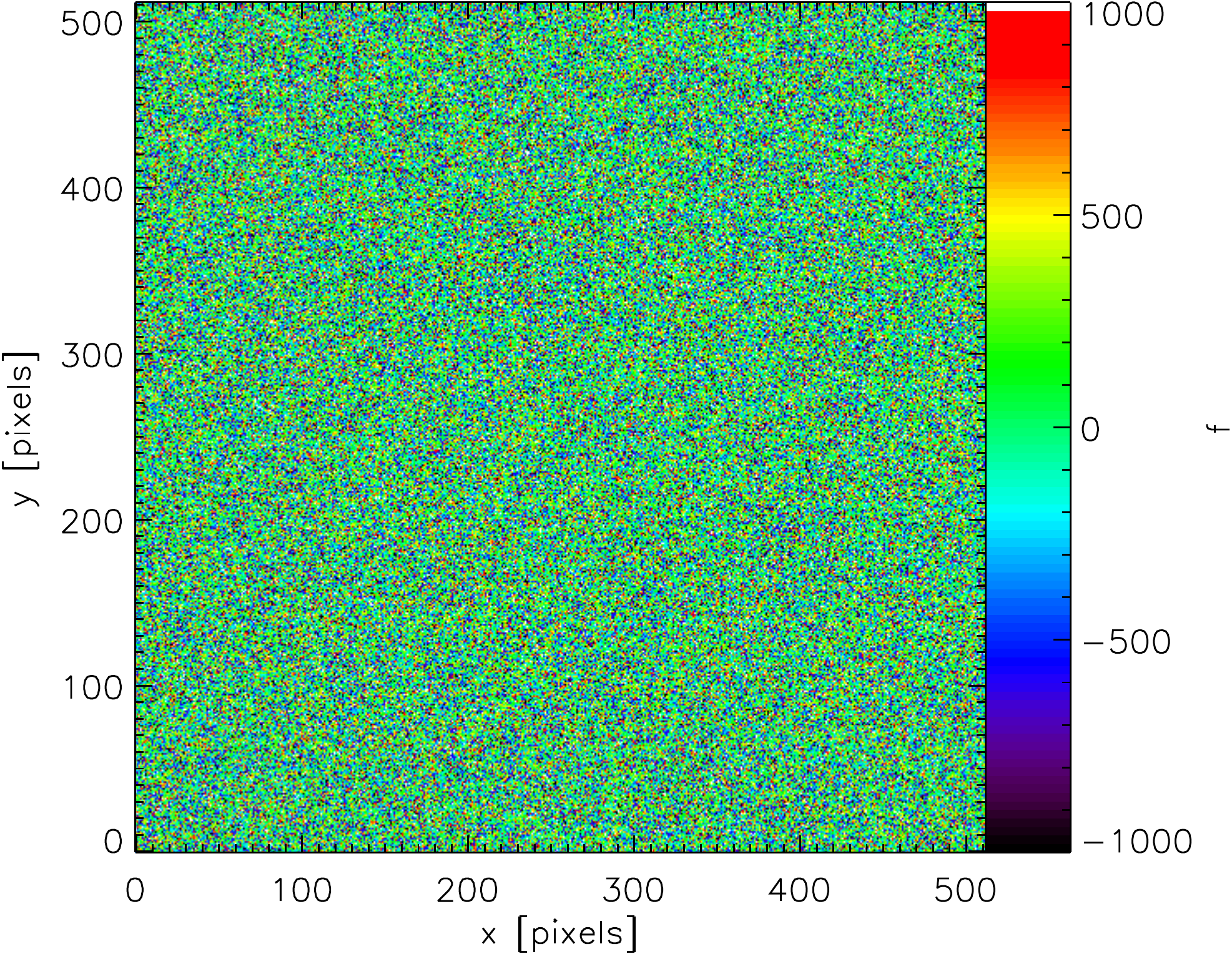}\vspace{3mm}
\includegraphics[angle=90,width=0.9\columnwidth]{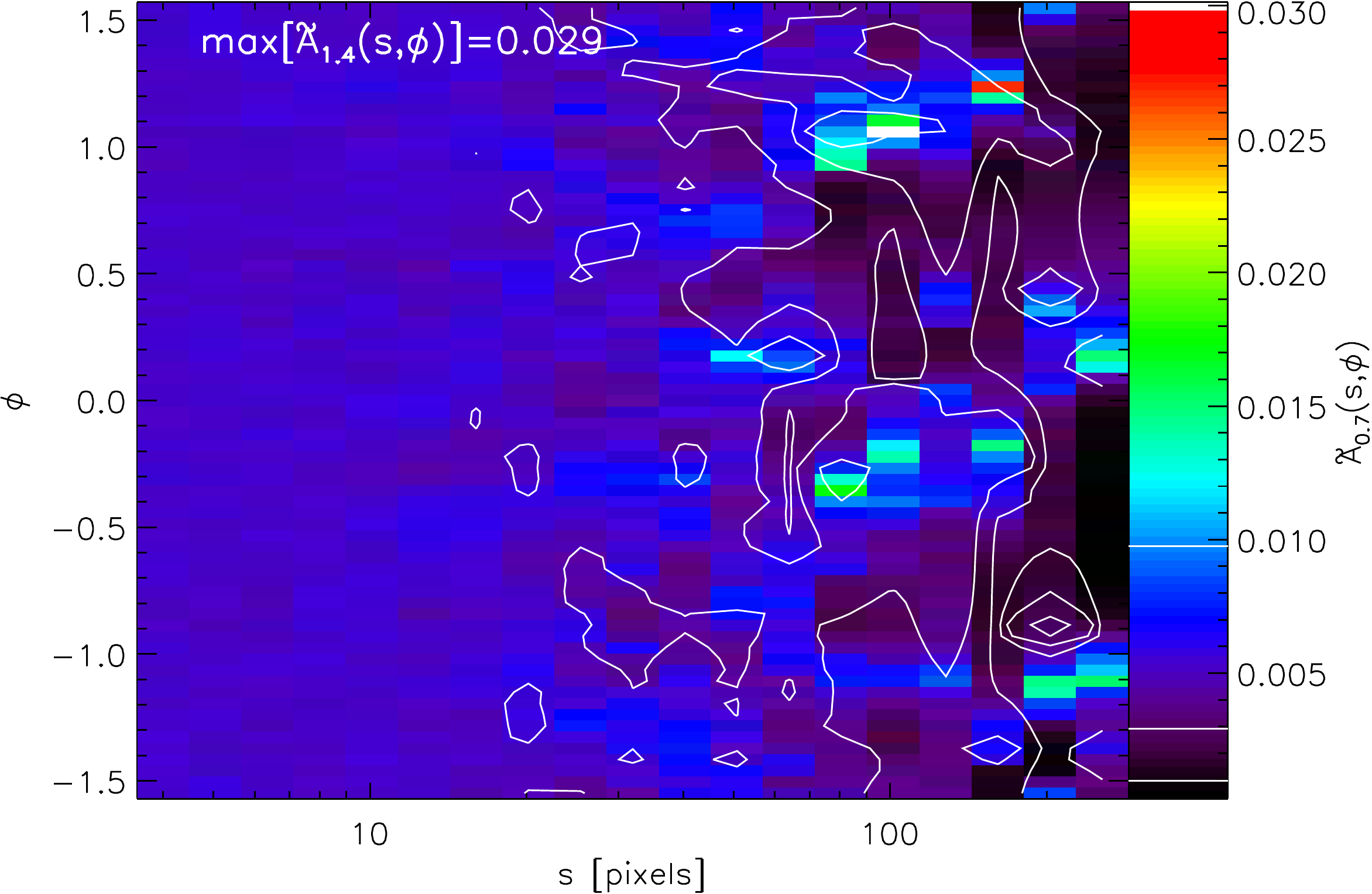}\vspace{3mm}
\includegraphics[angle=90,width=0.88\columnwidth]{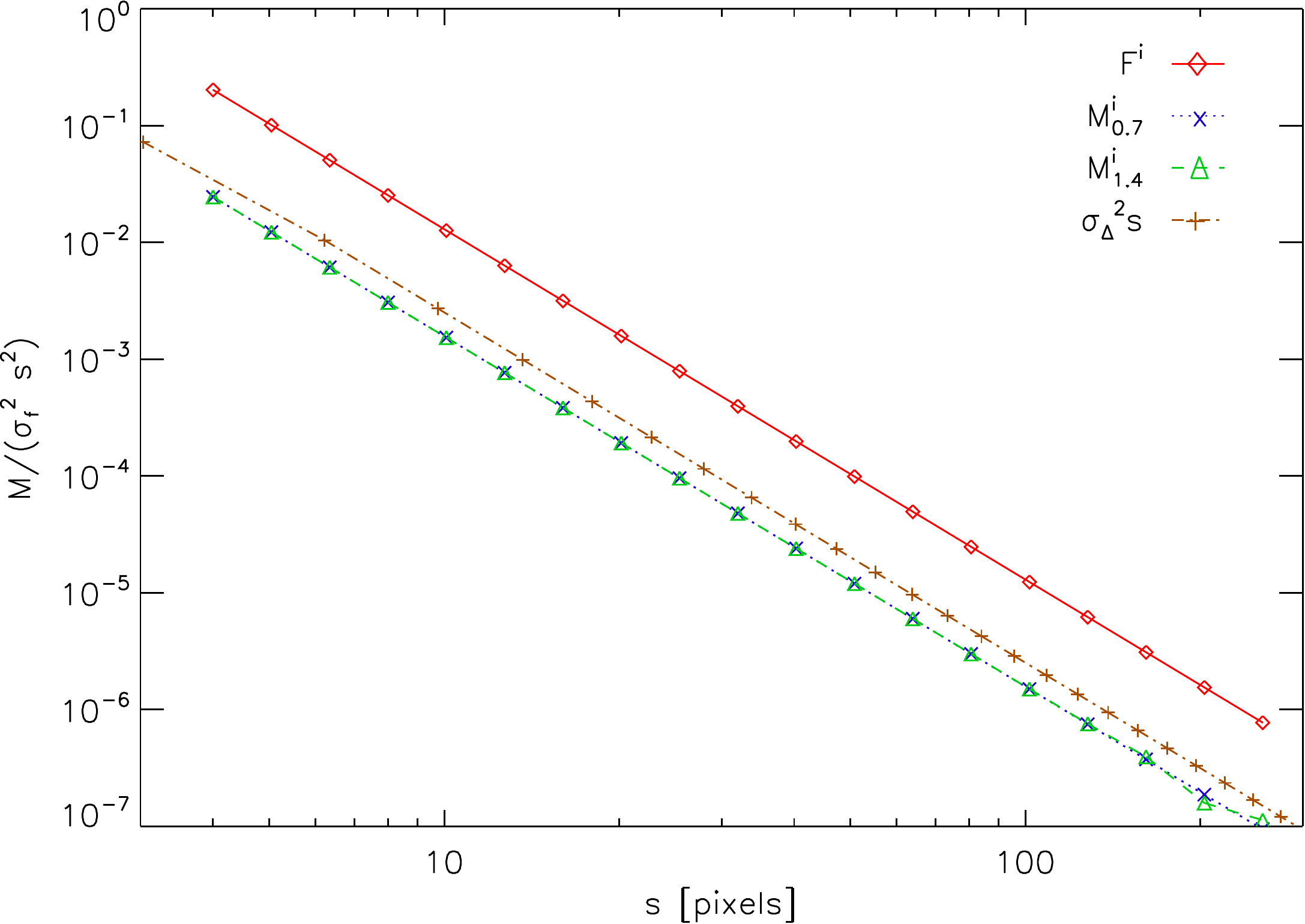}\vspace{3mm}
\includegraphics[angle=90,width=0.88\columnwidth]{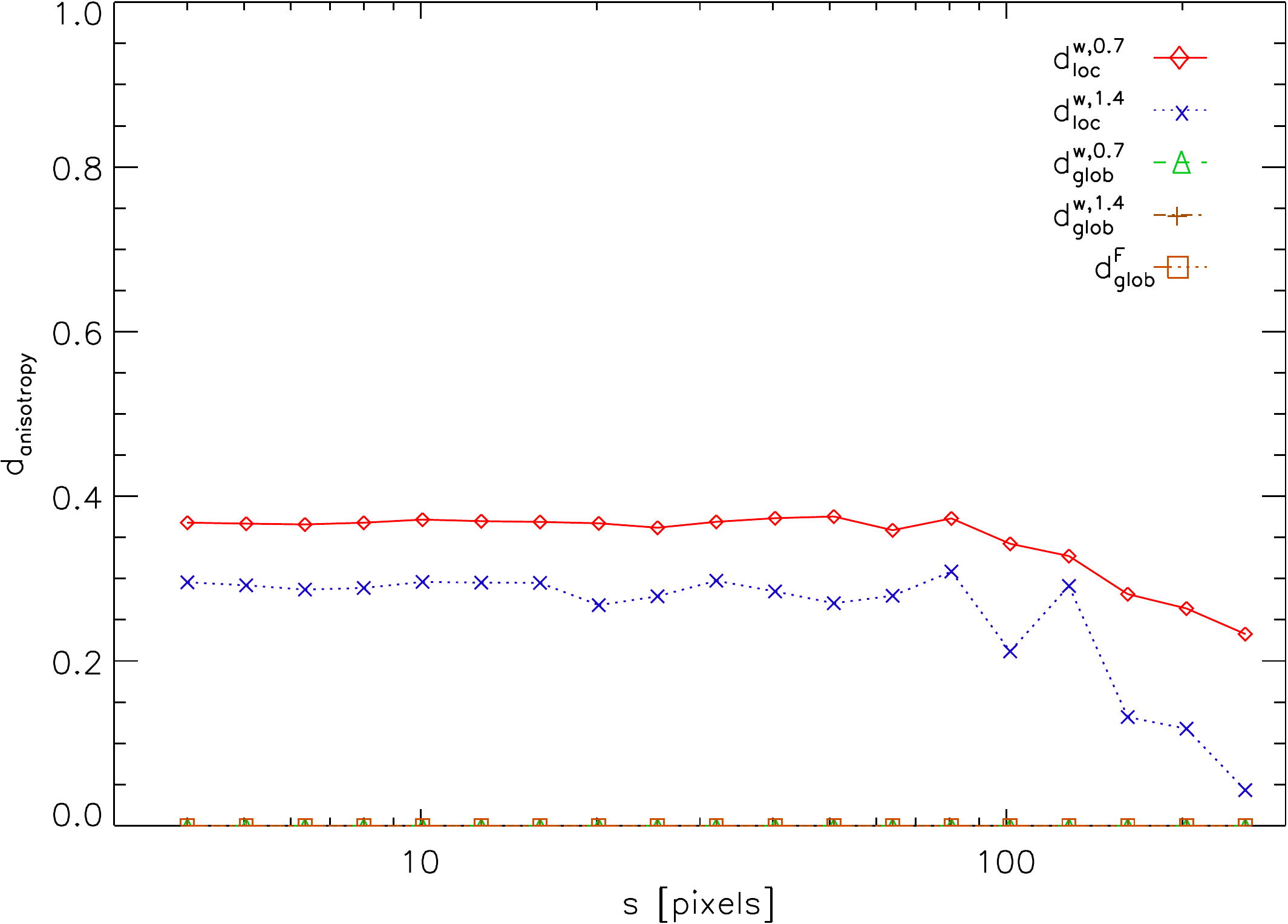}
\caption{Anisotropic wavelet analysis of a white noise map. }
\label{fig:appx_std1}
\end{figure}

\begin{figure}
\centering
\includegraphics[angle=90,width=0.8\columnwidth]{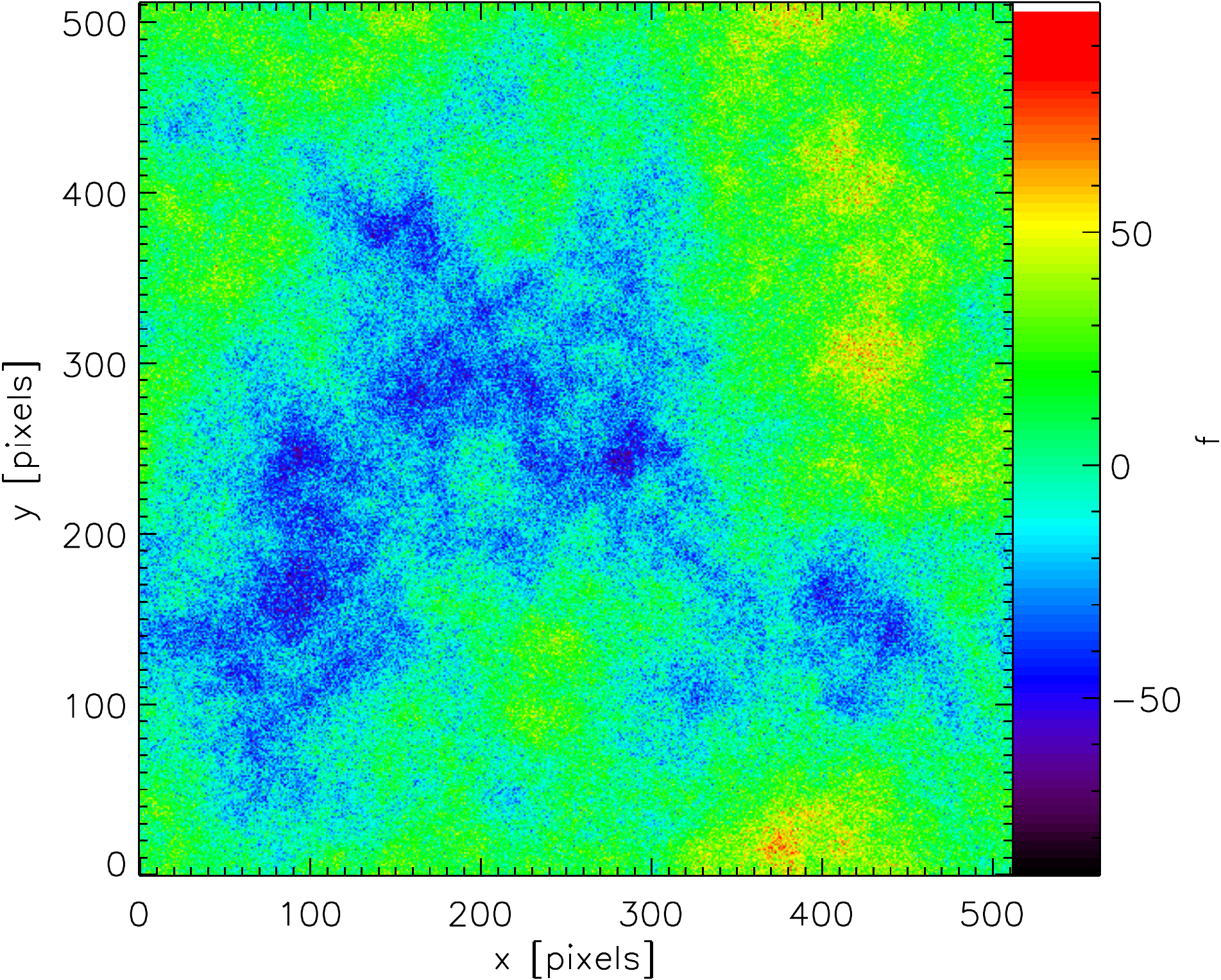}\vspace{3mm}
\includegraphics[angle=90,width=0.9\columnwidth]{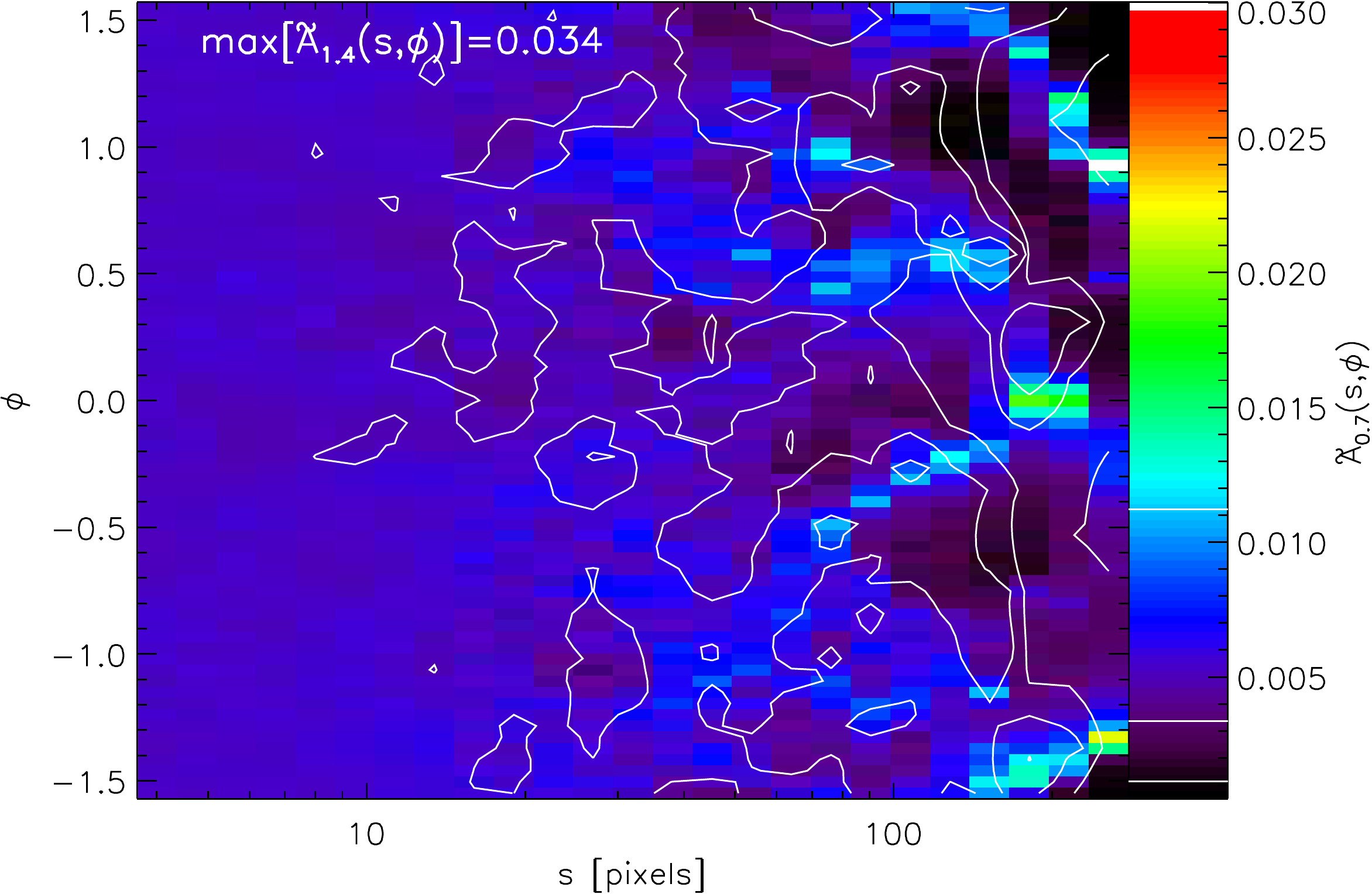}\vspace{3mm}
\includegraphics[angle=90,width=0.88\columnwidth]{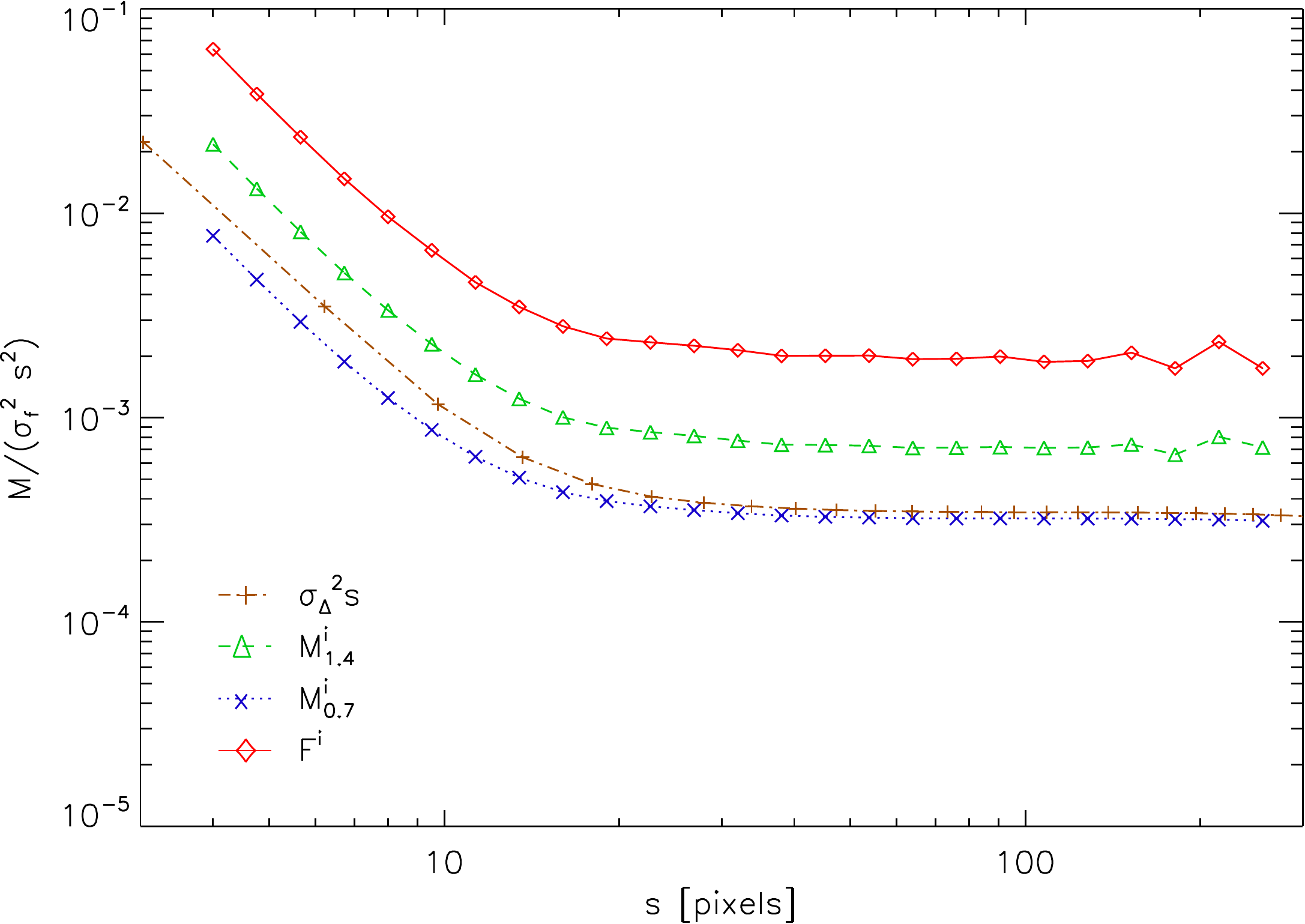}\vspace{3mm}
\includegraphics[angle=90,width=0.88\columnwidth]{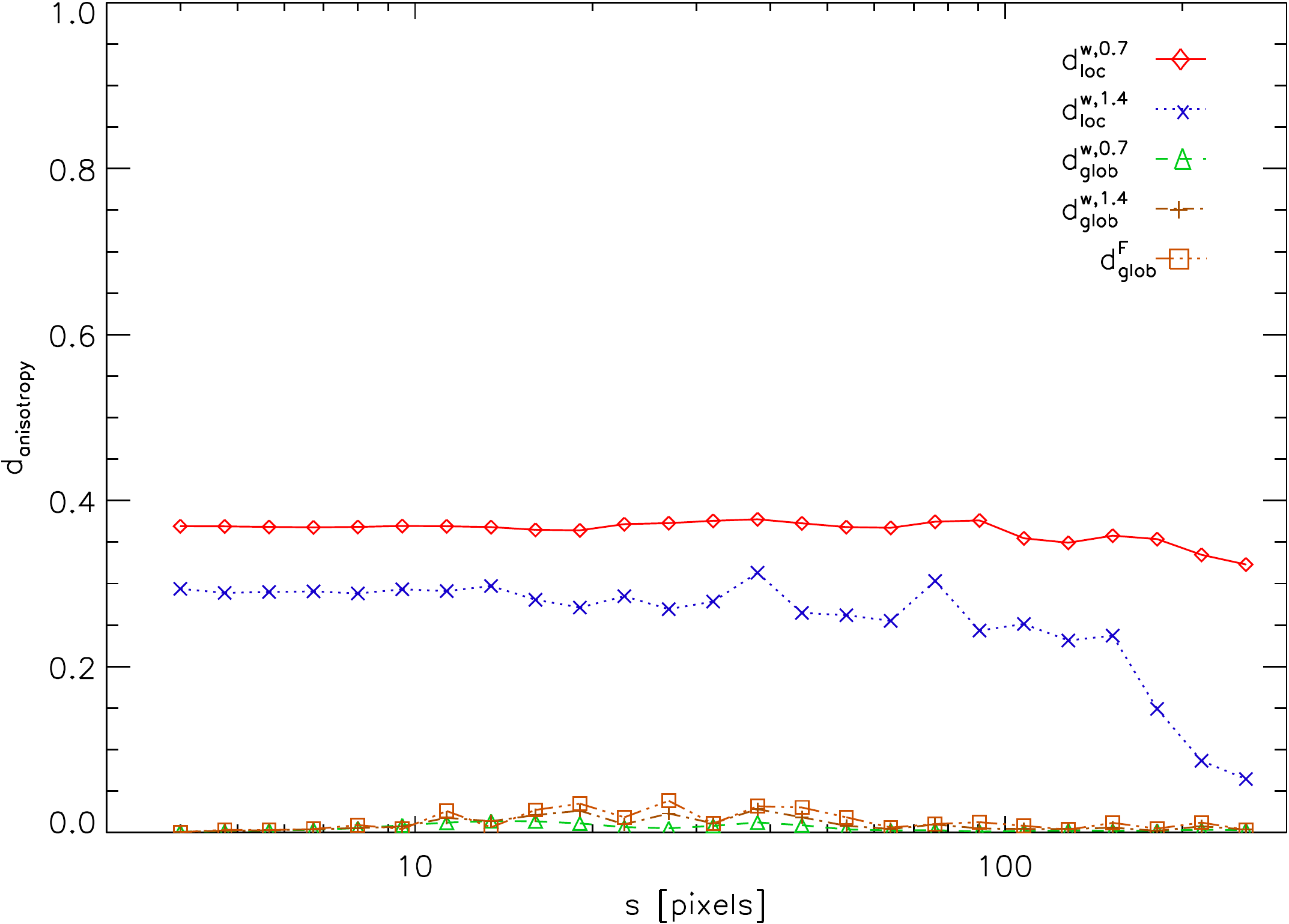}
\caption{Anisotropic wavelet analysis of a map containing a fractional Brownian motion structure \changed{with a spectral index $\beta=-3$}
superimposed by Gaussian noise (S/N$=5$).}
\label{fig:appx_fbm}
\end{figure}

Figure~\ref{fig:appx_std1} shows the results for a Gaussian
distribution of white noise and Fig.~\ref{fig:appx_fbm} the more general
case of a fractional Brownian motion structure (see Sect.~\ref{sec:noisemaps})
with a power spectral index of 3.0 corresponding to a box-car fractal dimension
of 2.5 or a Hurst index of $H=0.5$, superimposed by Gaussian
white noise with a signal to noise ratio S/N$=5$.
For the white noise, the {\changed{renormalized Fourier,
$\Delta$-variance,} and wavelet spectra \changed{show a $s^{-3}$
dependence} following a $s^{-1}$ dependence of the original spectra
given by the normalization of the Fourier and wavelet coefficients. White
noise has a two-dimensional Fourier spectrum $\hat{f}(k_x,k_y)$
that is independent of the wavenumbers. The radial integration
in the $\vec{k}$-plane provides the energy density spectrum $F(k) \propto k$
so that the resulting energy densities
per filter size drop like $s^{-1}$. The spectra of anisotropic coefficients
also fall like $s^{-1}$ providing an almost constant degree
of anisotropy. \changed{At scales above 30-50~pixels multiple
random anisotropies are visible in the angular spectra with no
preferred directions.}
When summing over different orientations of the anisotropies,
as done in the sum of the total values of the anisotropic wavelet
coefficients in the local degree of anisotropy, all accidental
local anisotropies are caught so that we obtain a relatively
constant ratio between anisotropic and isotropic coefficients
of about 0.30 \changed{for $b=\sqrt{2}$ and 0.37 for $b=1/\sqrt{2}$,
respectively,} providing a constant degree of local anisotropy
(Eq.~\ref{eq:dwloc}) with these values for all scales.
When only considering global anisotropies, as measured by the
Fourier coefficients or the sum of anisotropic wavelet coefficients
including their phase (Eq.~\ref{eq:dwglob} and \ref{eq_da}),
we obtain vanishing global degrees of anisotropy.
The same result is obtained for noise structures with a
different noise spectral index. As long as they have
random phases, the local degree of anisotropy remains at
\changed{a value around 0.3} while the global degree of anisotropy
vanishes for all scales. The
only difference in these structures is the different spectral slope
of the \changed{renormalized} wavelet and Fourier coefficients
\changed given by} the spectral index of the noise \changed{as $\beta -3$}.

\begin{figure}
\centering
\includegraphics[angle=90,width=0.8\columnwidth]{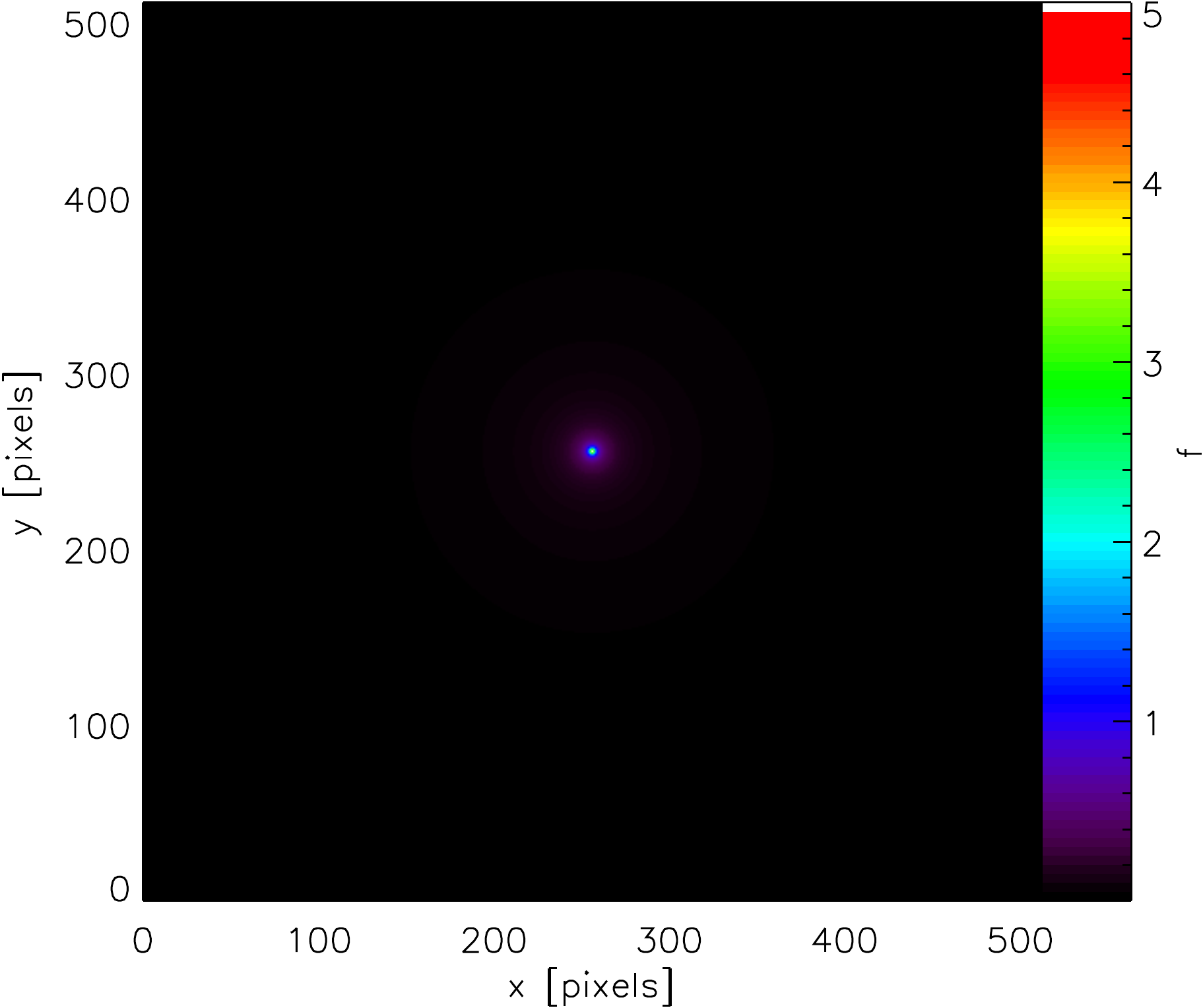}\vspace{3mm}
\includegraphics[angle=90,width=0.9\columnwidth]{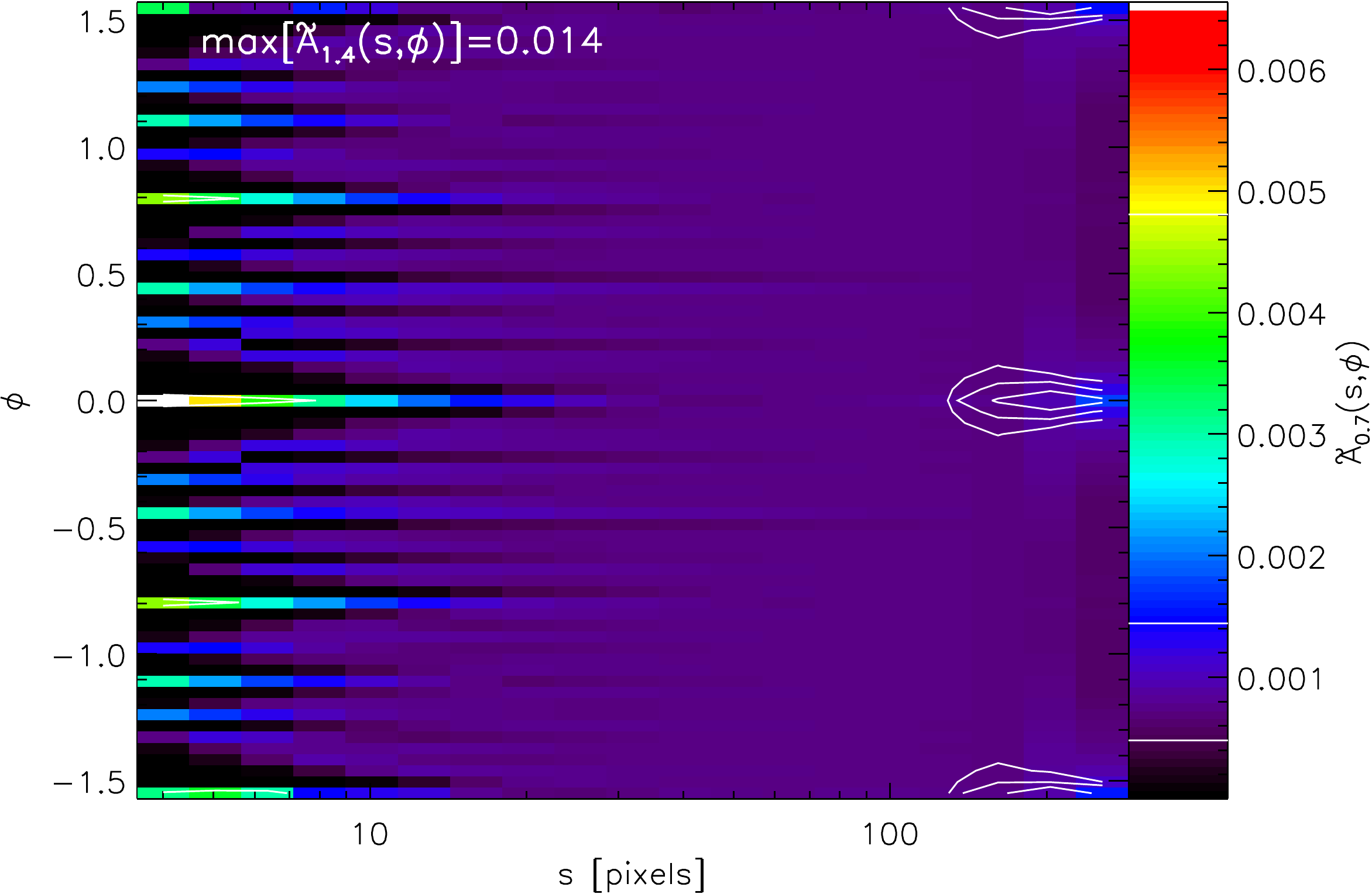}\vspace{3mm}
\includegraphics[angle=90,width=0.88\columnwidth]{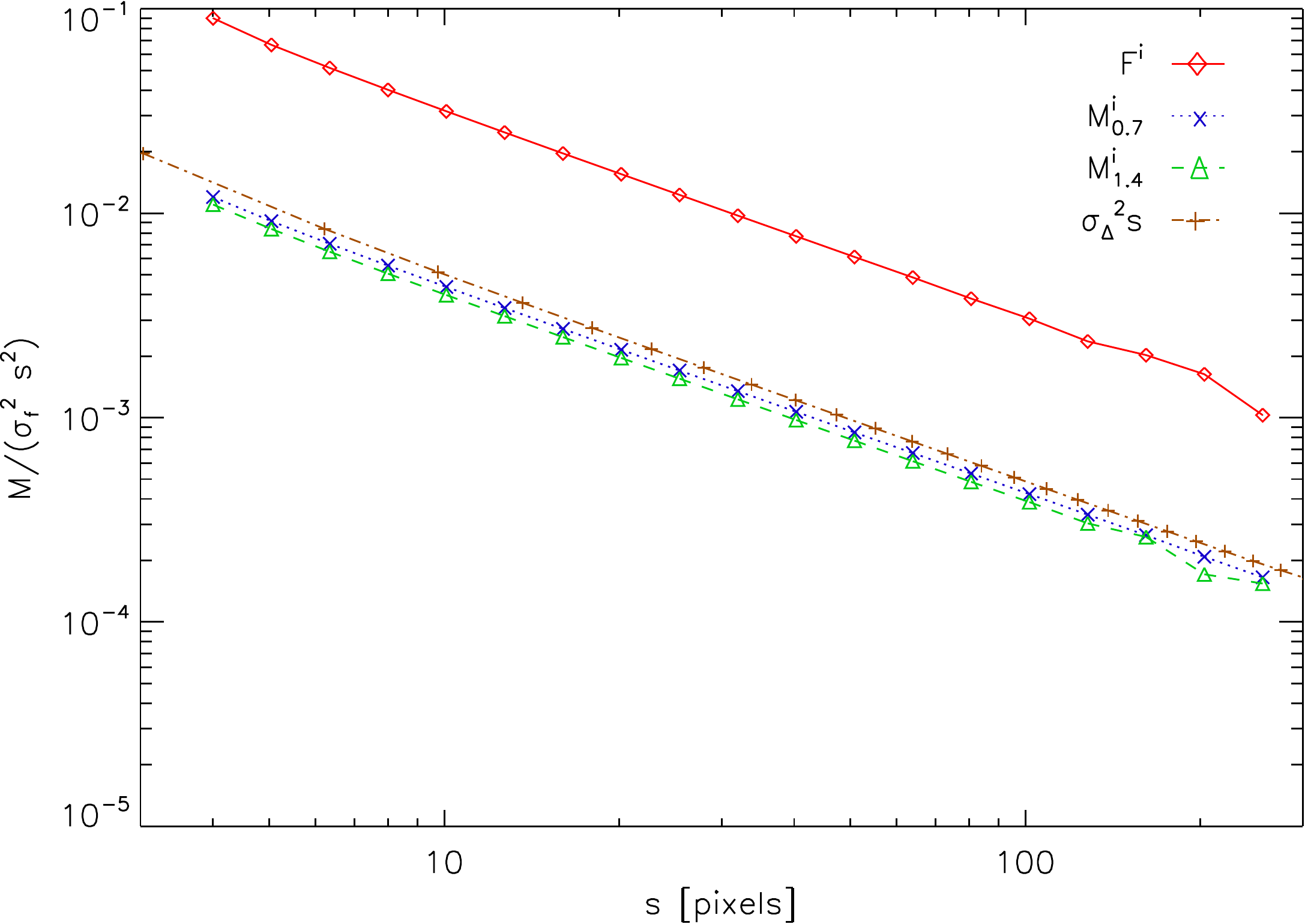}\vspace{3mm}
\includegraphics[angle=90,width=0.88\columnwidth]{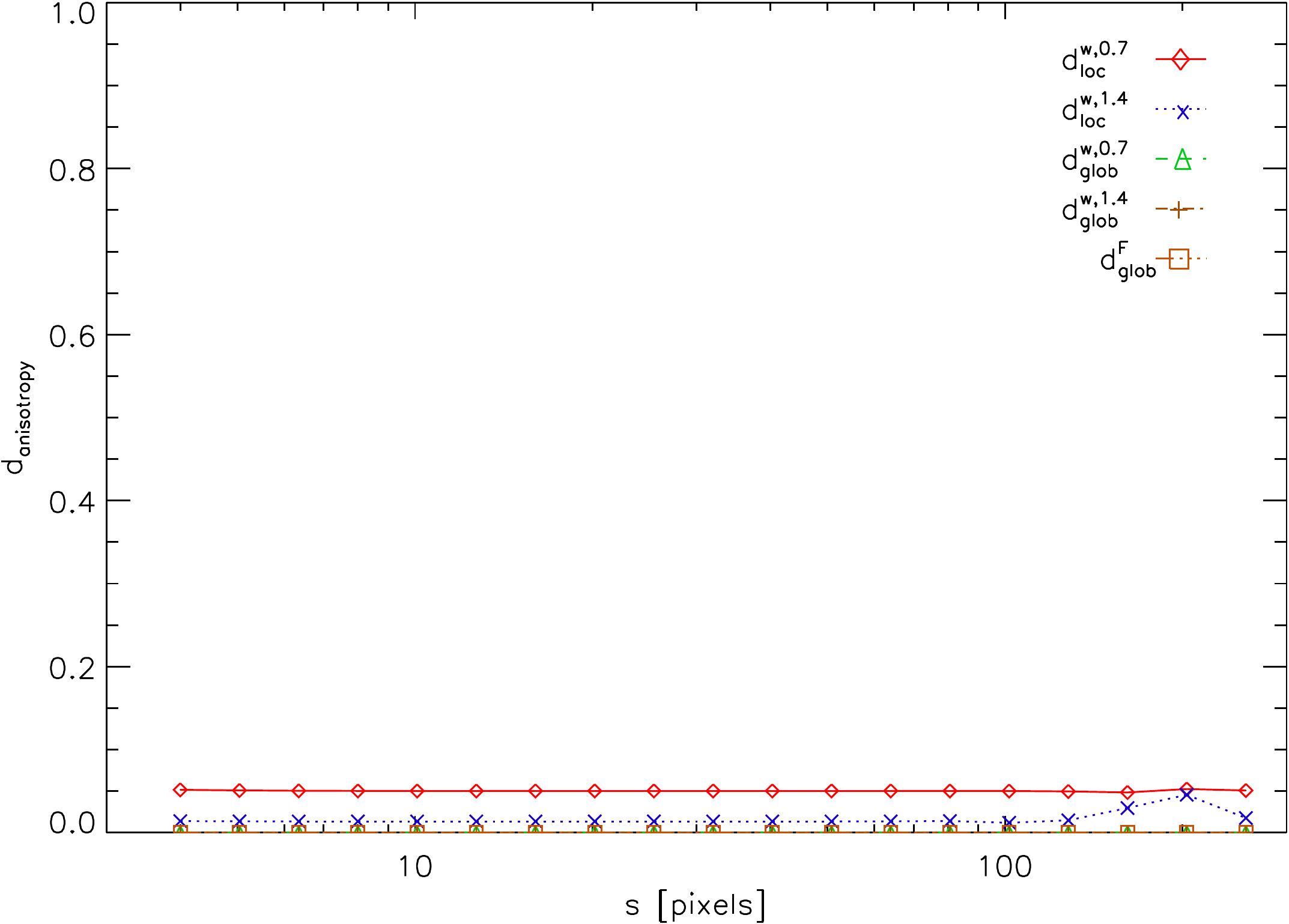}
\caption{Anisotropic wavelet analysis of a map containing the projection of a sphere
with a $r^{-2}$ density profile.}
\label{fig:appx_sis}
\end{figure}

Figure~\ref{fig:appx_sis} shows the
results for the projection of a singular isothermal sphere
given by a three-dimensional radial profile $n\propto r^{-2}$.
To avoid the numerical problems of the singularity in the center,
the infinite value in the central pixel has been replaced by a
value of 4.0 times the density of the closest surrounding pixels.
The \changed{angular distribution shows some artifacts at a low
level.} The somewhat stripy structure seen in $A(s,\varphi)$ at
small scales stems from the gridding in the map that enhances
\changed{all periodicities in a two-dimensional square lattice.
The enhancements in $\tilde{A}(s,\varphi)$ at angles of 0 and $\pi/2$
stem from the truncation of the sphere by the square map. Those
distortions are only visible in the angular spectra. In the radially
integrated spectra of wavelet coefficients and the degree of anisotropy
they are not visible. The renormalized spectra follow a strict $s^{-1}$
dependence like $\Delta$-variance and the power spectrum. The degree
of anisotropy is vanishing.}

\begin{figure}
\centering
\includegraphics[angle=90,width=0.8\columnwidth]{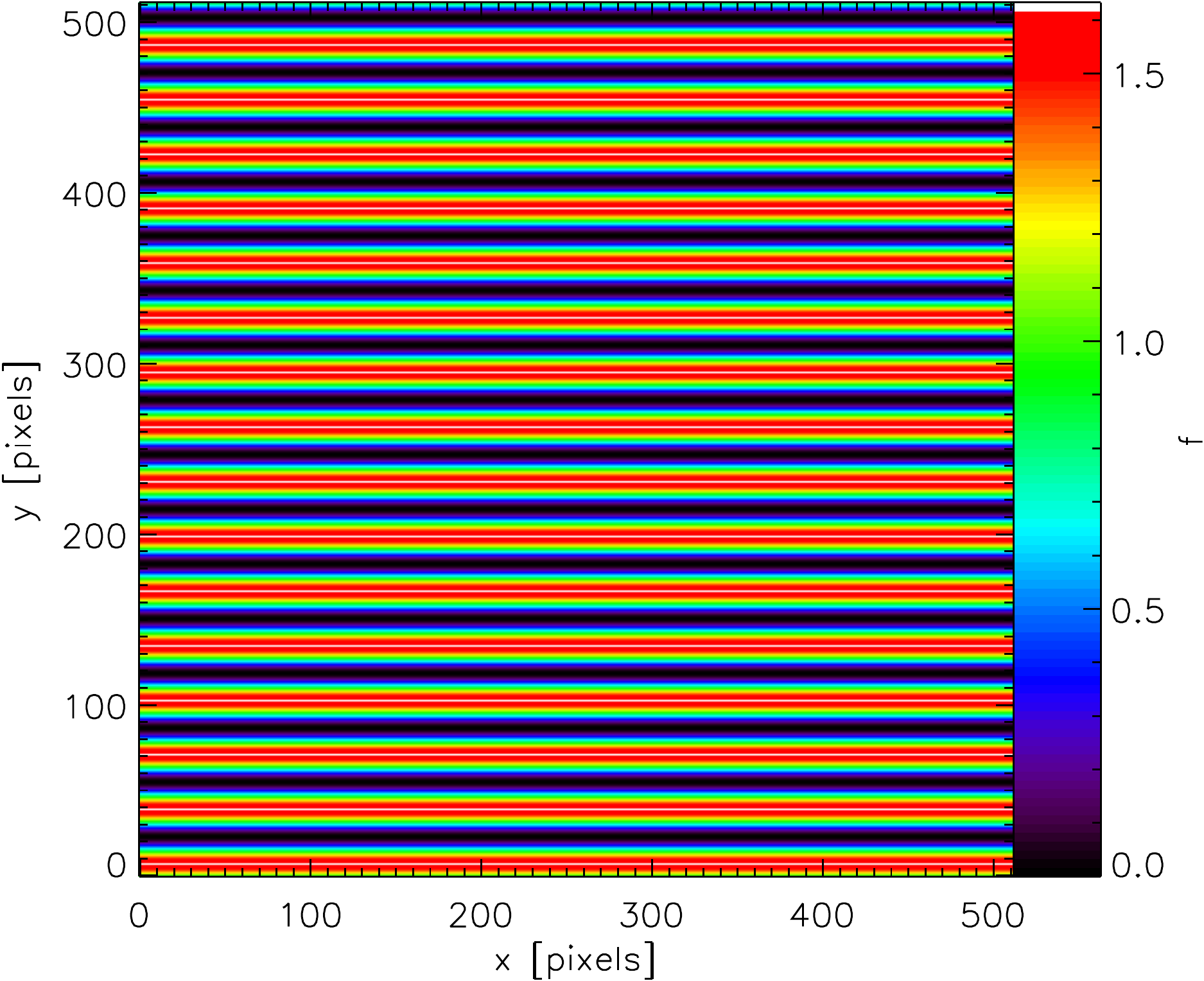}\vspace{3mm}
\includegraphics[angle=90,width=0.9\columnwidth]{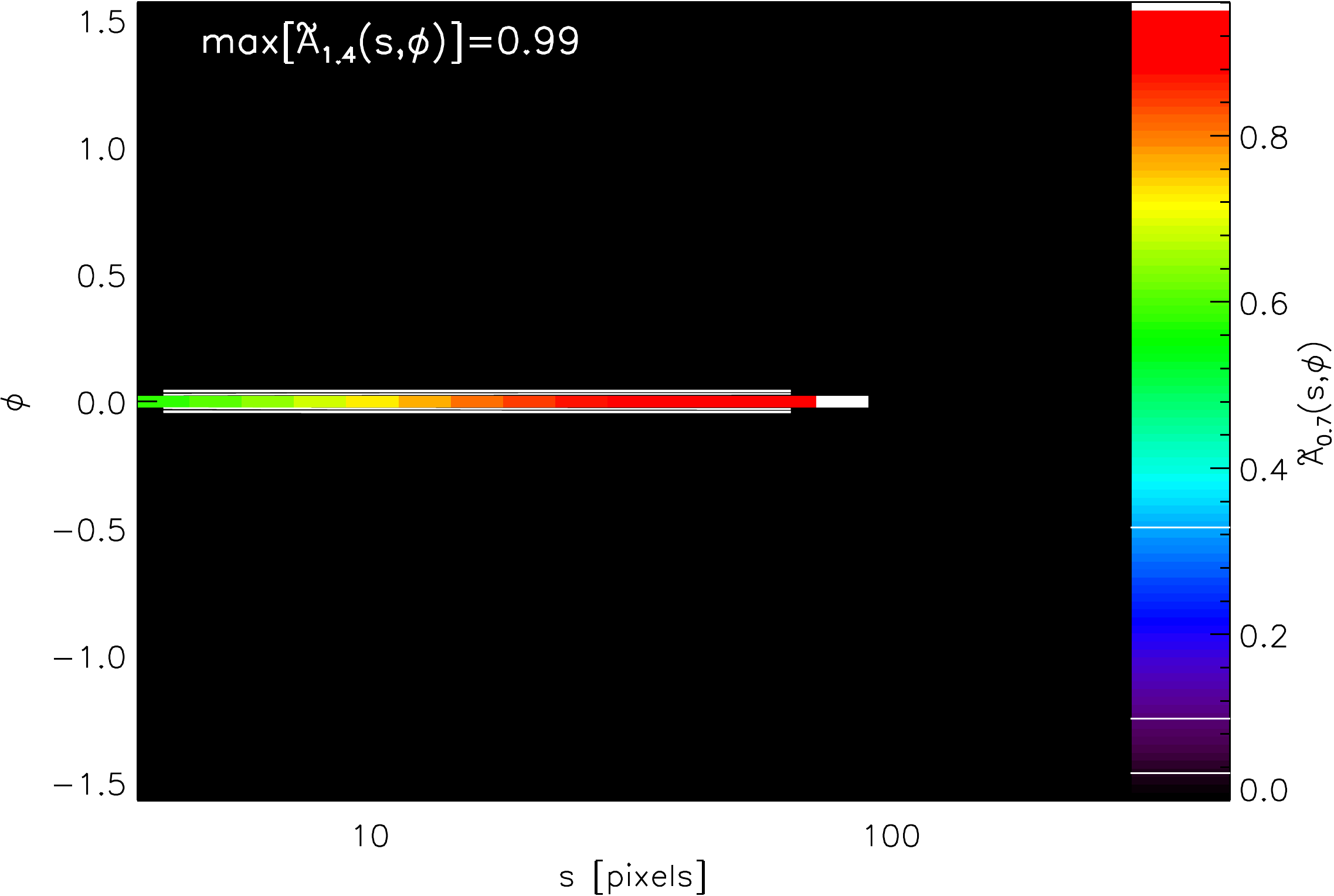}\vspace{3mm}
\includegraphics[angle=90,width=0.88\columnwidth]{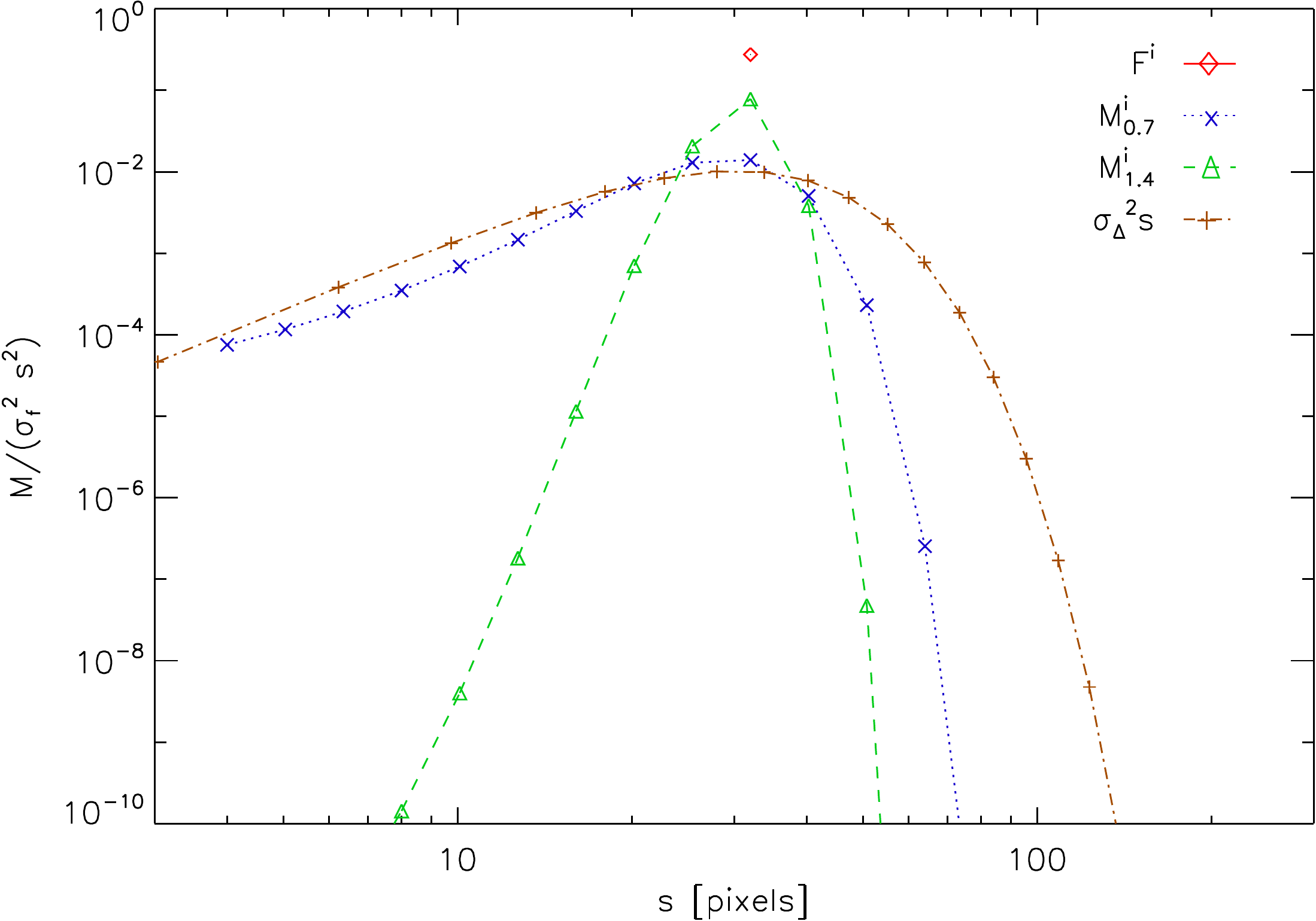}\vspace{3mm}
\includegraphics[angle=90,width=0.88\columnwidth]{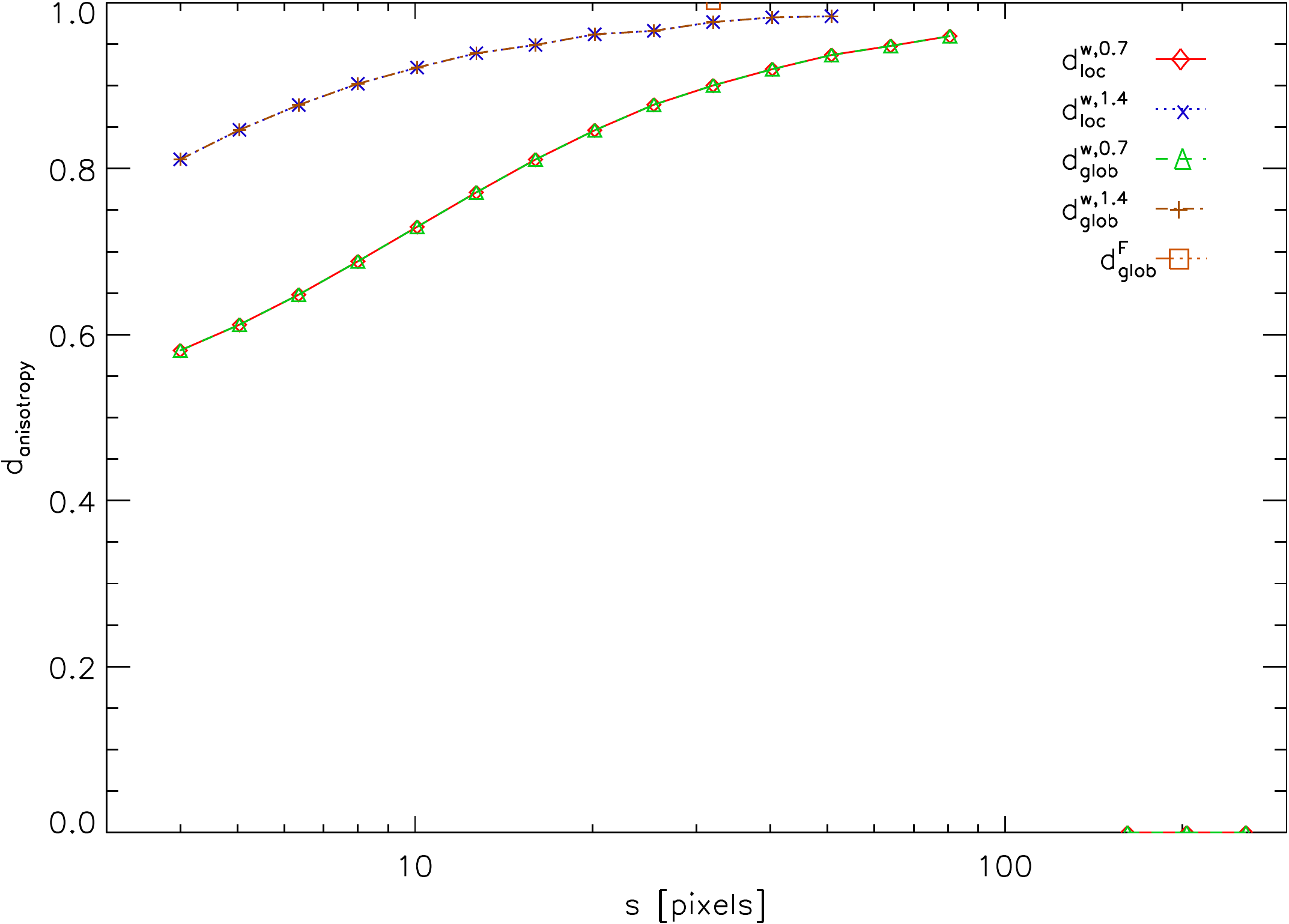}
\caption{Anisotropic wavelet analysis of a map containing
a sinusoidal stripe pattern with $k_y=16$, i.e. a period of $p=32$~pixels.}
\label{fig:appx_std2}
\end{figure}

\begin{figure}
\centering
\includegraphics[angle=90,width=0.8\columnwidth]{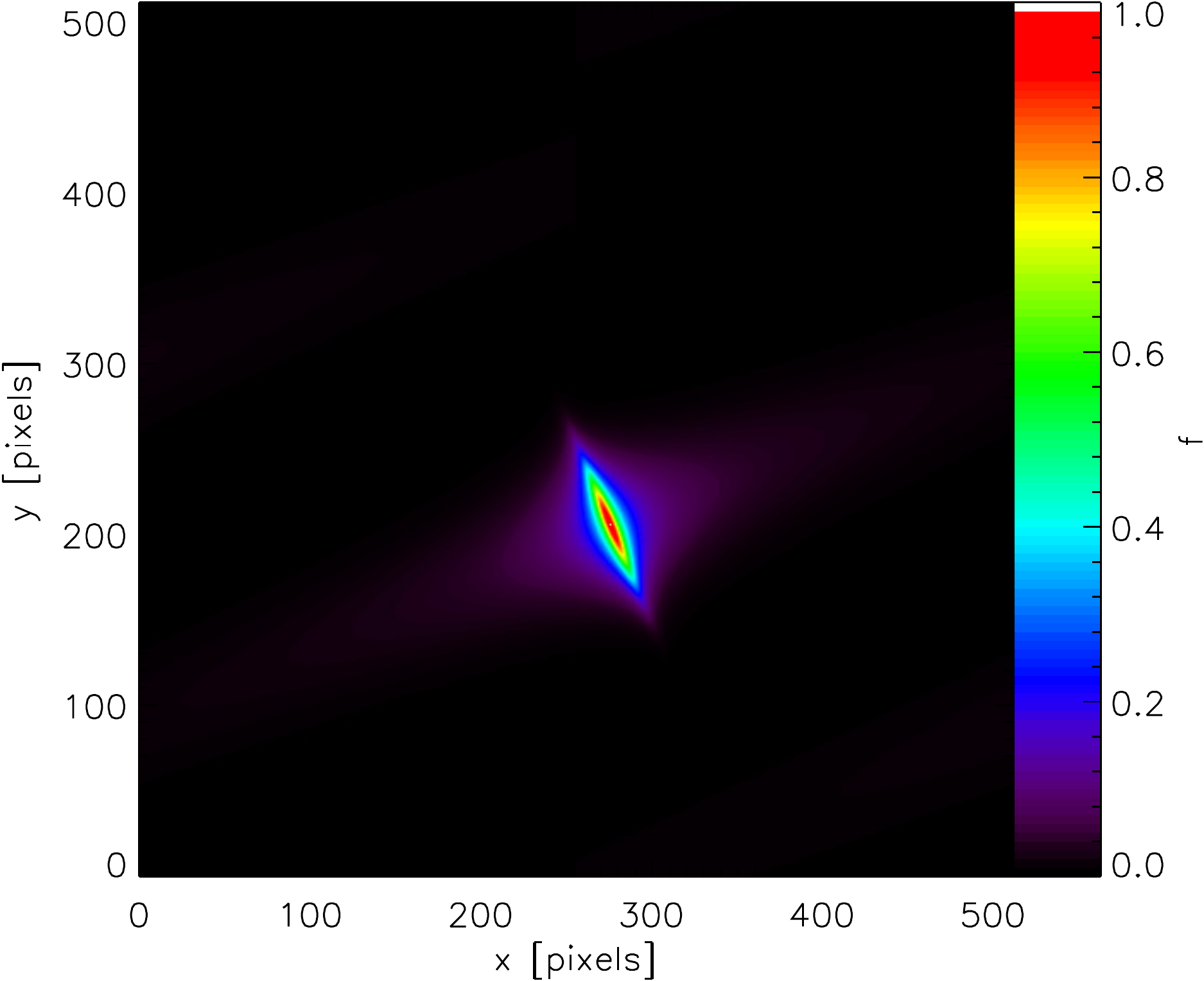}\vspace{3mm}
\includegraphics[angle=90,width=0.9\columnwidth]{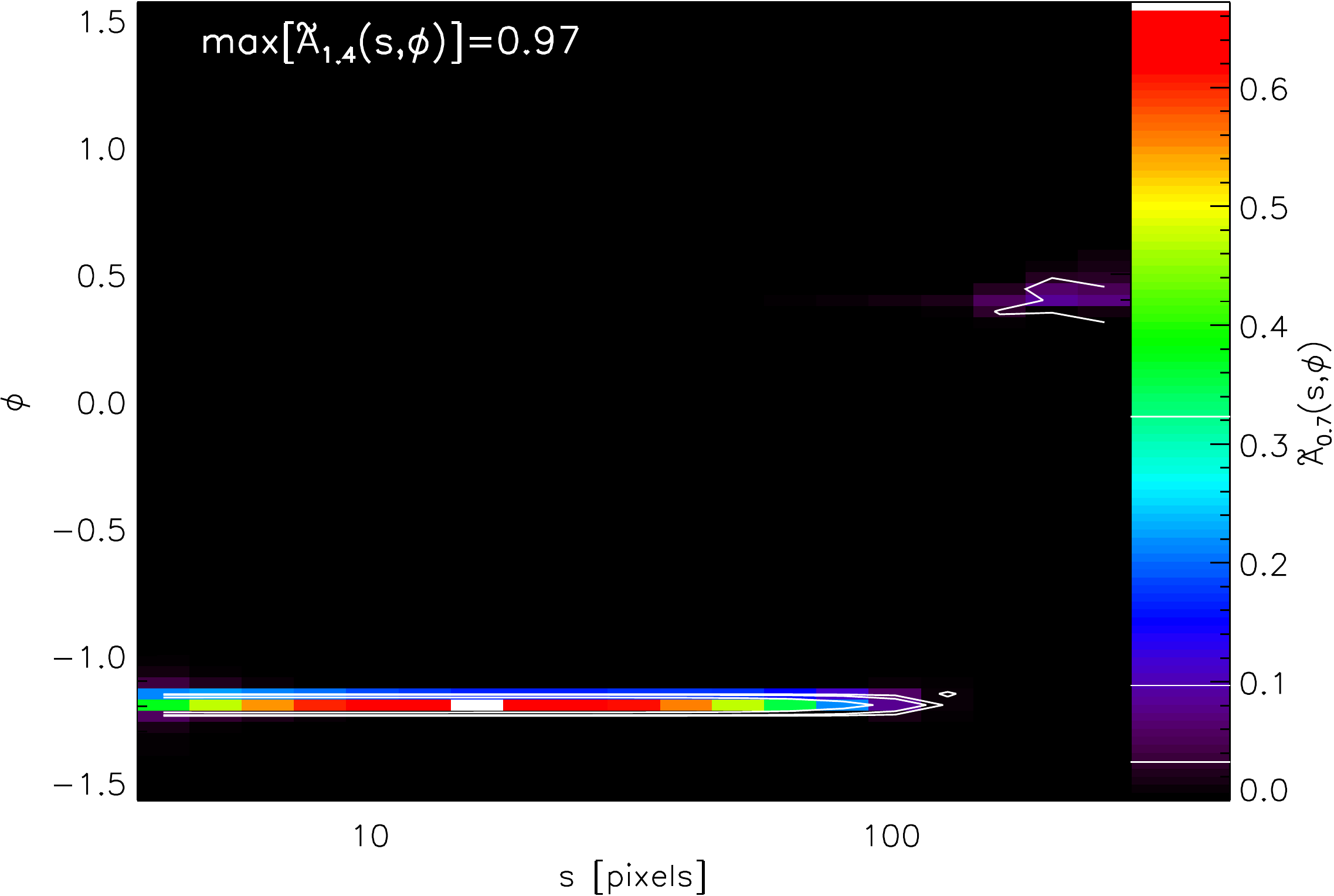}\vspace{3mm}
\includegraphics[angle=90,width=0.88\columnwidth]{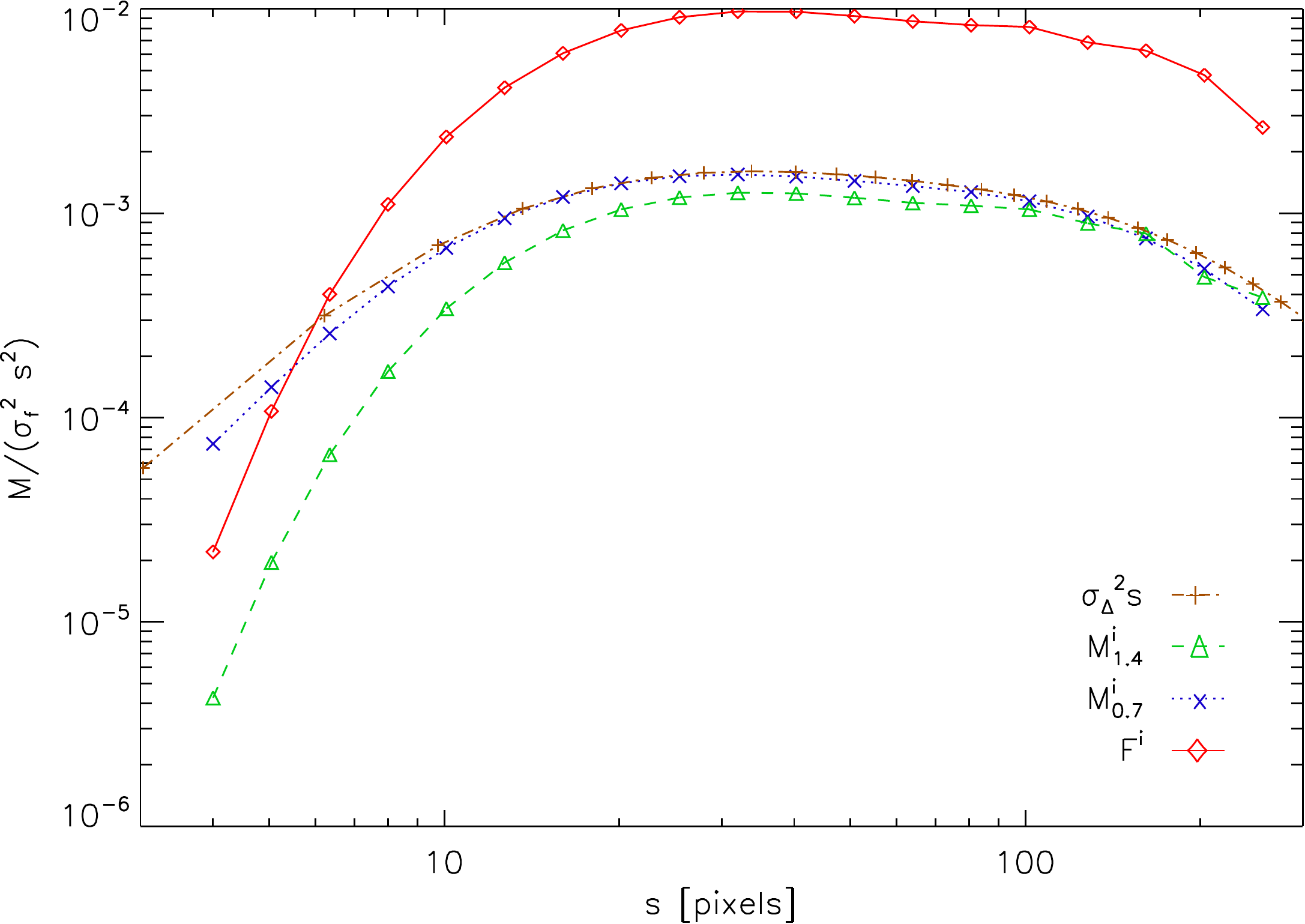}\vspace{3mm}
\includegraphics[angle=90,width=0.88\columnwidth]{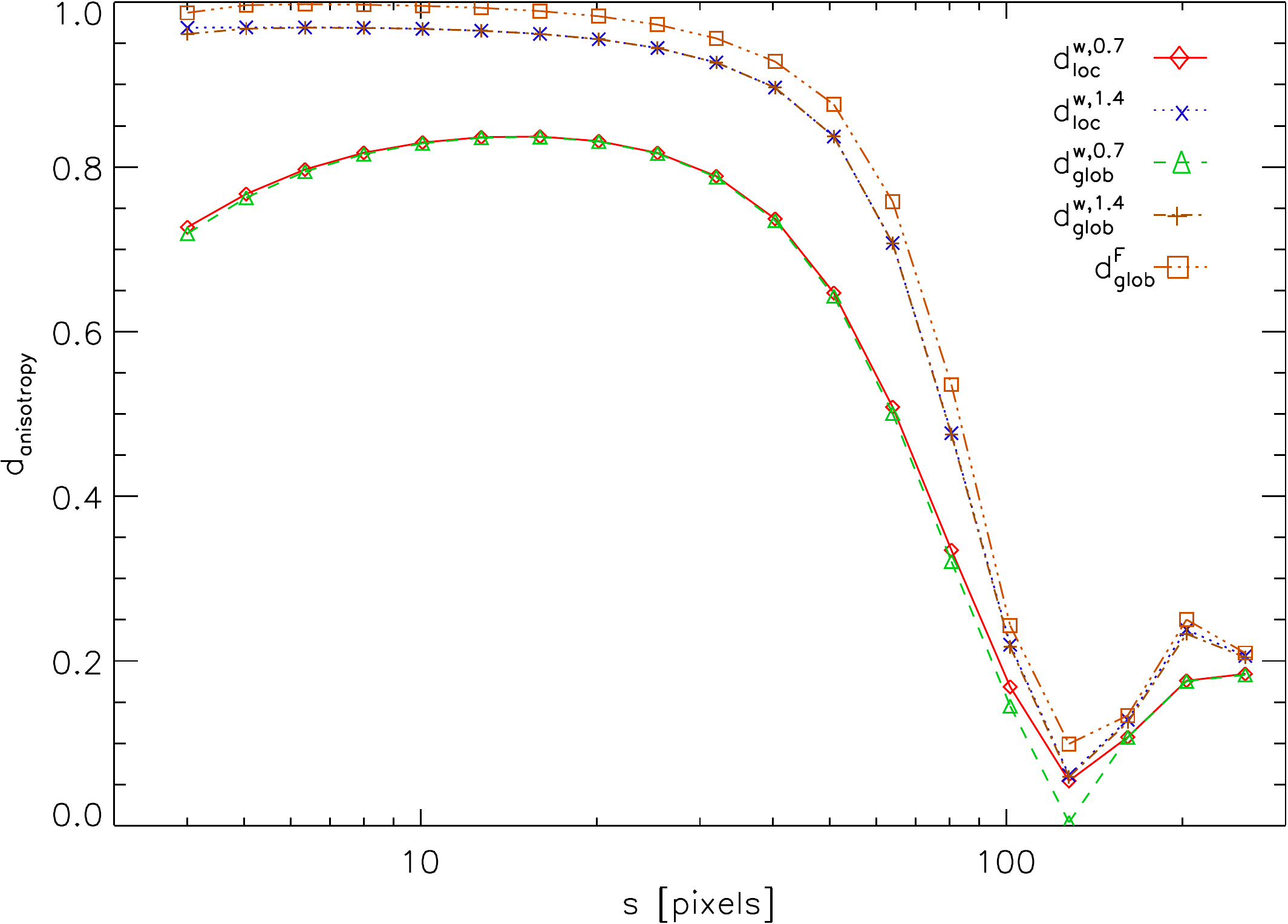}
\caption{Anisotropic wavelet analysis of a map containing an asymmetric $p=2$ Plummer profile
with $R_{a}=3.2$~pixels and $\sigma_{c}=26$~pixels.}
\label{fig:appx_plummer_long}
\end{figure}

\begin{figure}
\centering
\includegraphics[angle=90,width=0.8\columnwidth]{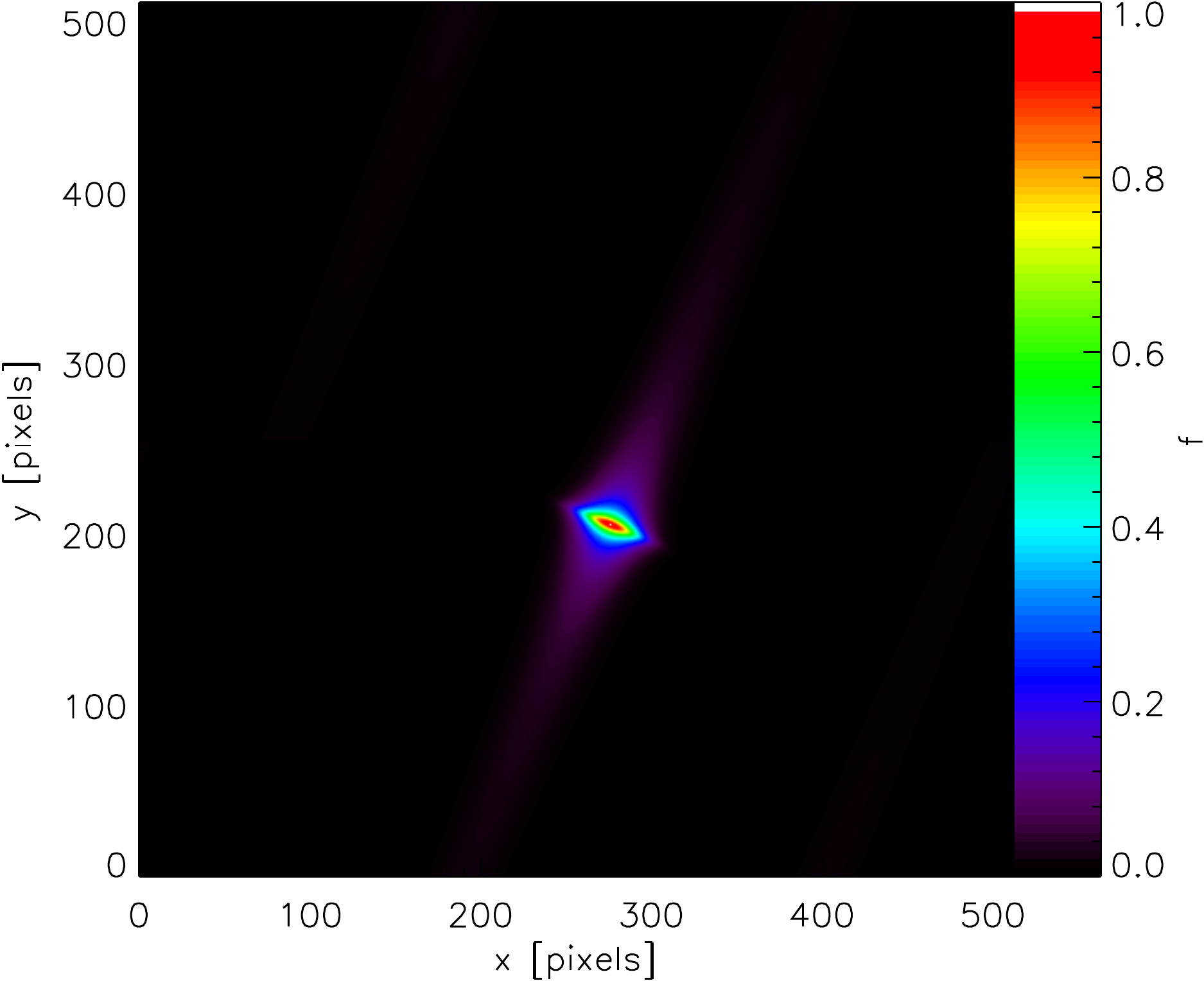}\vspace{3mm}
\includegraphics[angle=90,width=0.9\columnwidth]{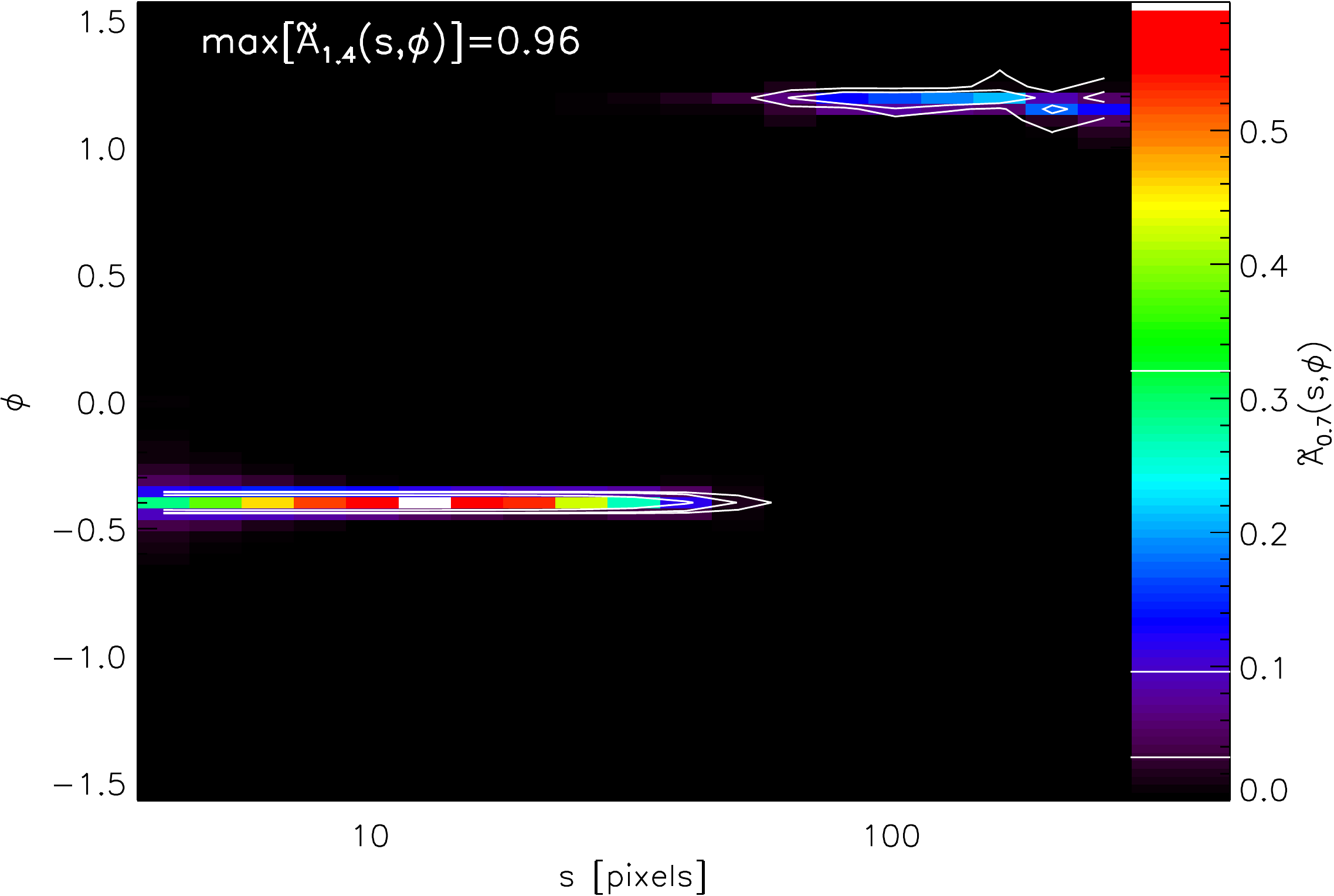}\vspace{3mm}
\includegraphics[angle=90,width=0.88\columnwidth]{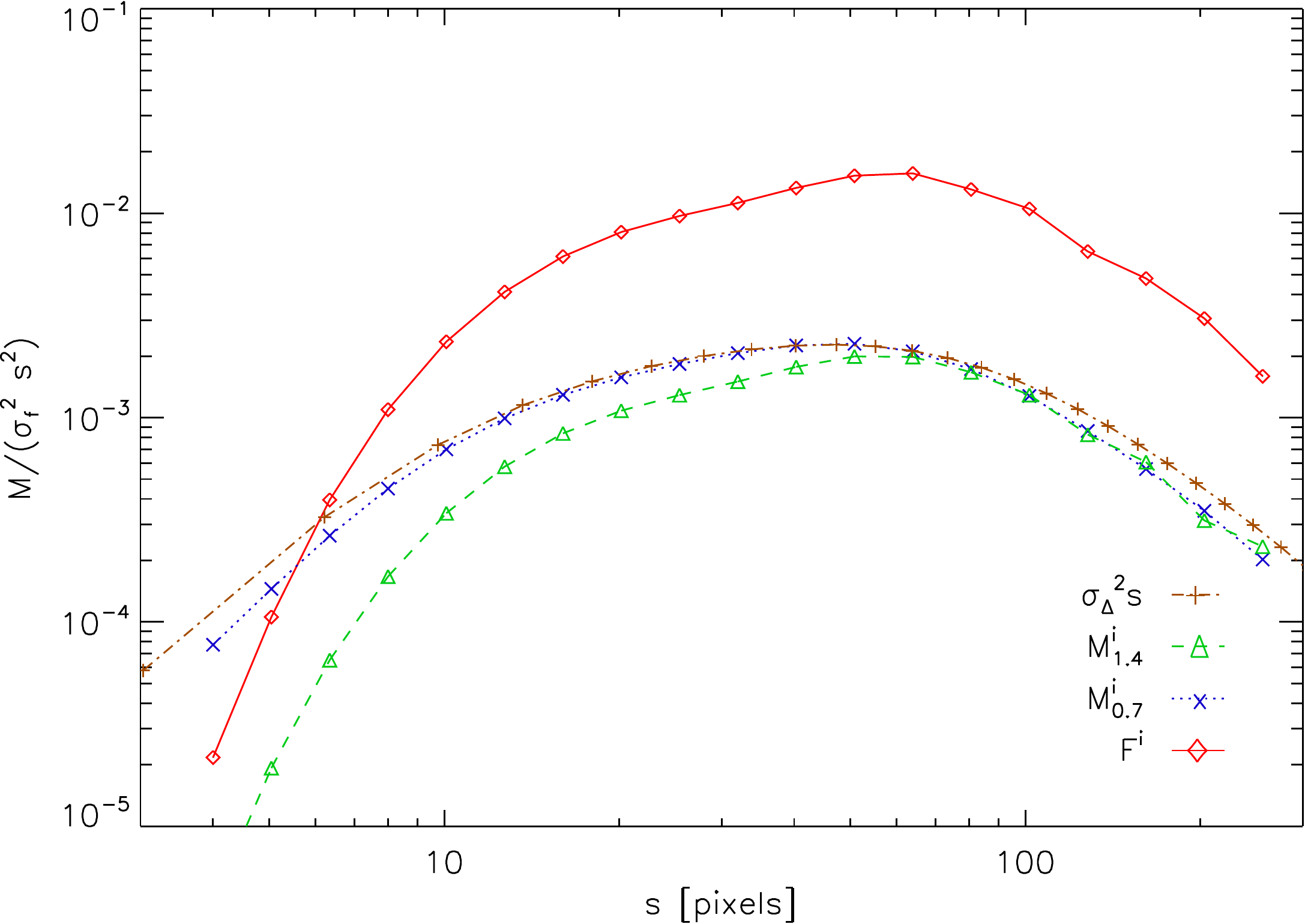}\vspace{3mm}
\includegraphics[angle=90,width=0.88\columnwidth]{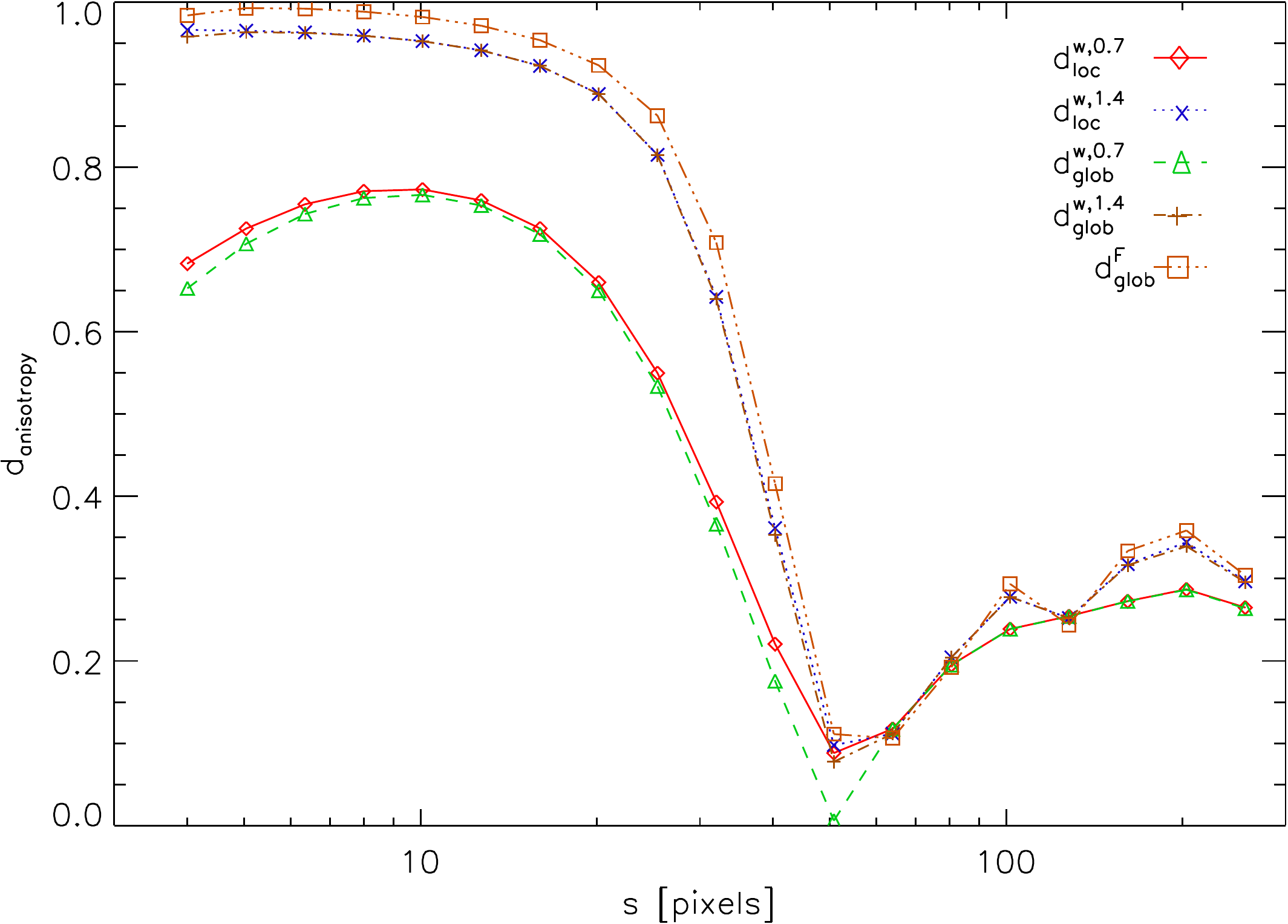}
\caption{Anisotropic wavelet analysis of a map containing an asymmetric $p=2$ Plummer profile
with $R_{a}=3.2$~pixels and $\sigma_{c}=13$~pixels.}
\label{fig:appx_plummer_short}
\end{figure}

Figure~\ref{fig:appx_std2} contains the results for the map containing
a sinusoidal stripe pattern with a period of $p=32$~pixels
\changed{discussed in Sect.~\ref{sec:sinus}. The Fourier spectrum measures
the exact period. As discussed in Sect.~\ref{sect_filtershape} the
filter with the wide localization parameter $b=\sqrt{2}$ also traces
this pronounced scale very sharply while the filter with the
narrower $b=1/\sqrt{2}$ shows a broader peak, almost as wide
as the $\Delta$-variance. Nevertheless all four approaches show a
peak at the scale reflecting the periodicity of the sinusoidal pattern.
As the sinusoidal structure is the only
structure in the map local and global degrees of anisotropy agree.
Because the power spectrum vanishes at all scales different from
$s=32$~pixels it can only measure the degree of anisotropy there,
providing a value of unity. The $b=\sqrt{2}$ provides a value close
to unity at this scale but falls off to 0.8 at small scales. The
smaller localization parameter gives lower values of 0.9 at the
peak at 0.6 at small scales. The angle of the pattern is accurately
measured by both filters.

Figures~\ref{fig:appx_plummer_long} and \ref{fig:appx_plummer_short}
show the analysis for two elliptical Plummer profiles, as discussed
in Sect.~\ref{sect_plummer}. With a ratio $\sigma_c/R_a=8$ the
Plummer ellipse in Fig.~\ref{fig:appx_plummer_long} falls into
the asymptotic regime of long filaments in Fig.~\ref{fig:plummer-plateau}
where the peak of the wavelet spectra is independent of the filament
length. The Plummer profile in Fig.~\ref{fig:appx_plummer_short}
has a smaller axis ratio $\sigma_c/R_a=4$ falling into the regime
where the scale calibration from Sect.~\ref{sect_plummer} has the
largest error. Qualitatively, all types of spectra show the same
behavior. The renormalized wavelet spectra of the long Plummer profiles
show a steep increase at scales below 20 ~pixels, a power law trend
at intermediate scales and a drop beyond 200~pixels. The peak parameters
match the fitting formula from Sect.~\ref{sect_plummer}. The $b$-dependence
is small. For the
short Plummer profile the power-law behavior is steeper than in
Figs.~\ref{fig:appx_plummer_long} and \ref{fig:plummer_spectra},
it starts already at smaller scales, in spite of the same profile
width, and it extends only to about 70~pixels, more than a factor
of two below the upper end in Fig.~\ref{fig:appx_plummer_long}. Obviously,
here the wavelet spectrum is no longer dominated by the filament
width only. In the degrees of anisotropy, we find a
shift of the upper edge of the plateau by exactly the factor of two
that is the difference in the lengths of the two profiles, confirming
the length-dependence of the spectrum of the degree of anisotropy.
The reason for the difference in the wavelet spectra can be seen from the maps showing the
structures and the angular spectra. At small scales the center
of the Plummer profile is the dominating structure. The angular spectra
peak at the angle of the main filament axis $\sigma_c$. At
larger scales the shallow decay of the Plummer profile creates a
structure that is oriented rather perpendicular, in the direction of
$R_a$. For short filaments this provides already a significant
contribution at intermediate scales. However,
the visual impression of the structure in Fig.~\ref{fig:appx_plummer_short}
also proves that the restriction of the fitting formula (Eq.~\ref{eq_plummer2D}) from
Sect.~\ref{sect_plummer} to longer axes ratios is no problem for
the characterization of real filaments as one would hardly
describe structures with $\sigma_c/R_a<8$ as significantly elongated
structures.  }

\begin{figure}
\centering
\includegraphics[angle=90,width=0.8\columnwidth]{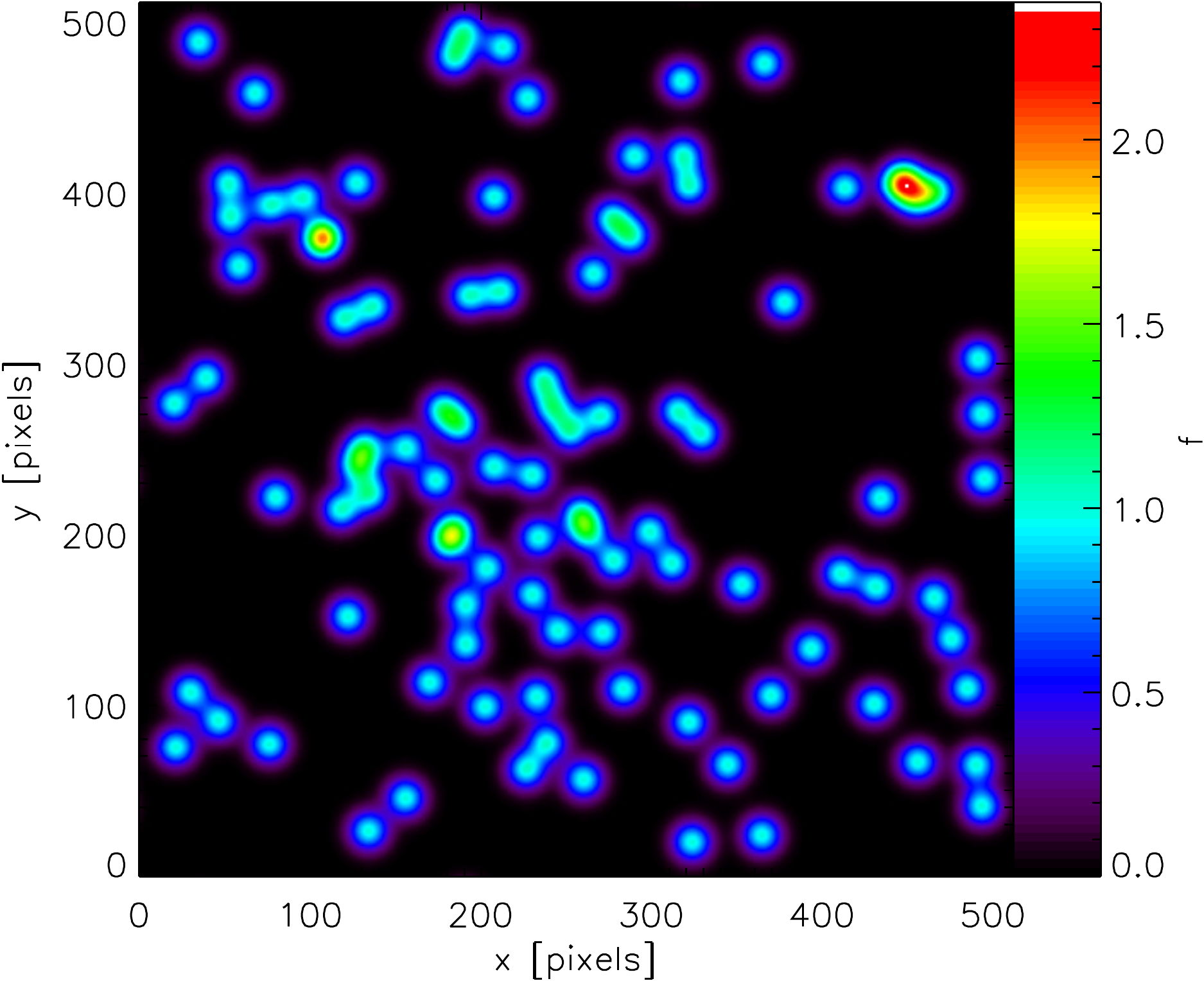}\vspace{3mm}
\includegraphics[angle=90,width=0.9\columnwidth]{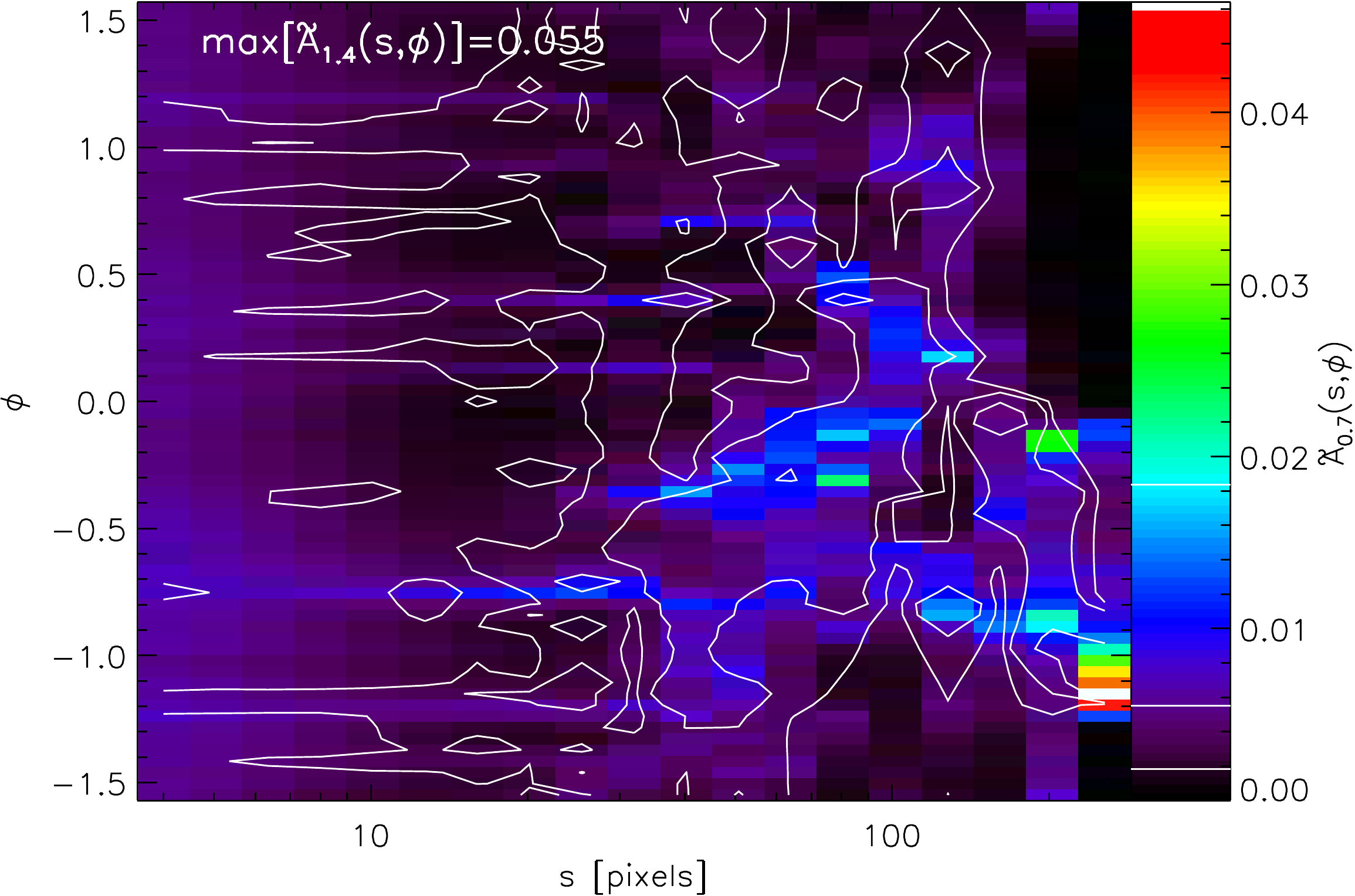}\vspace{3mm}
\includegraphics[angle=90,width=0.88\columnwidth]{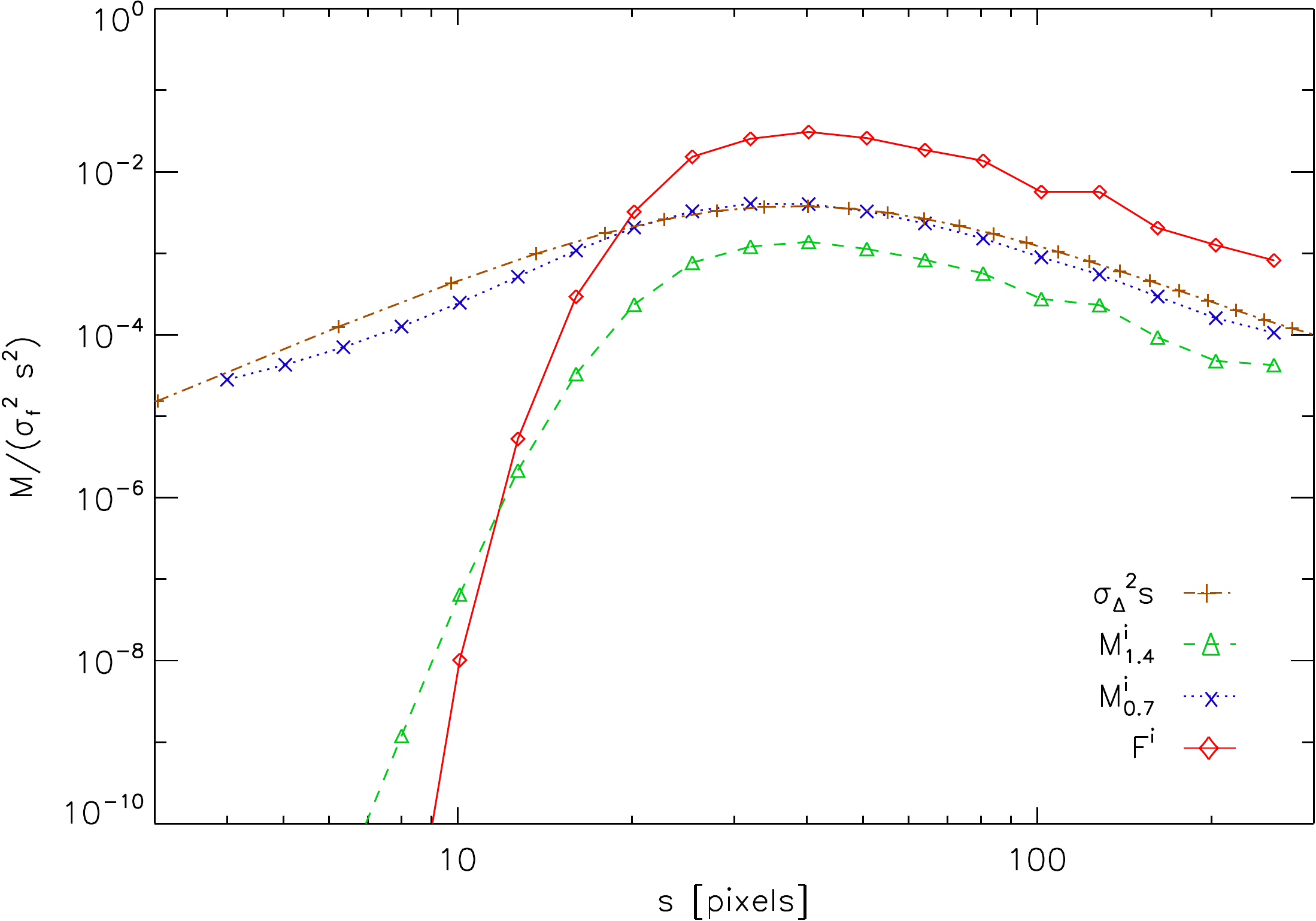}\vspace{3mm}
\includegraphics[angle=90,width=0.88\columnwidth]{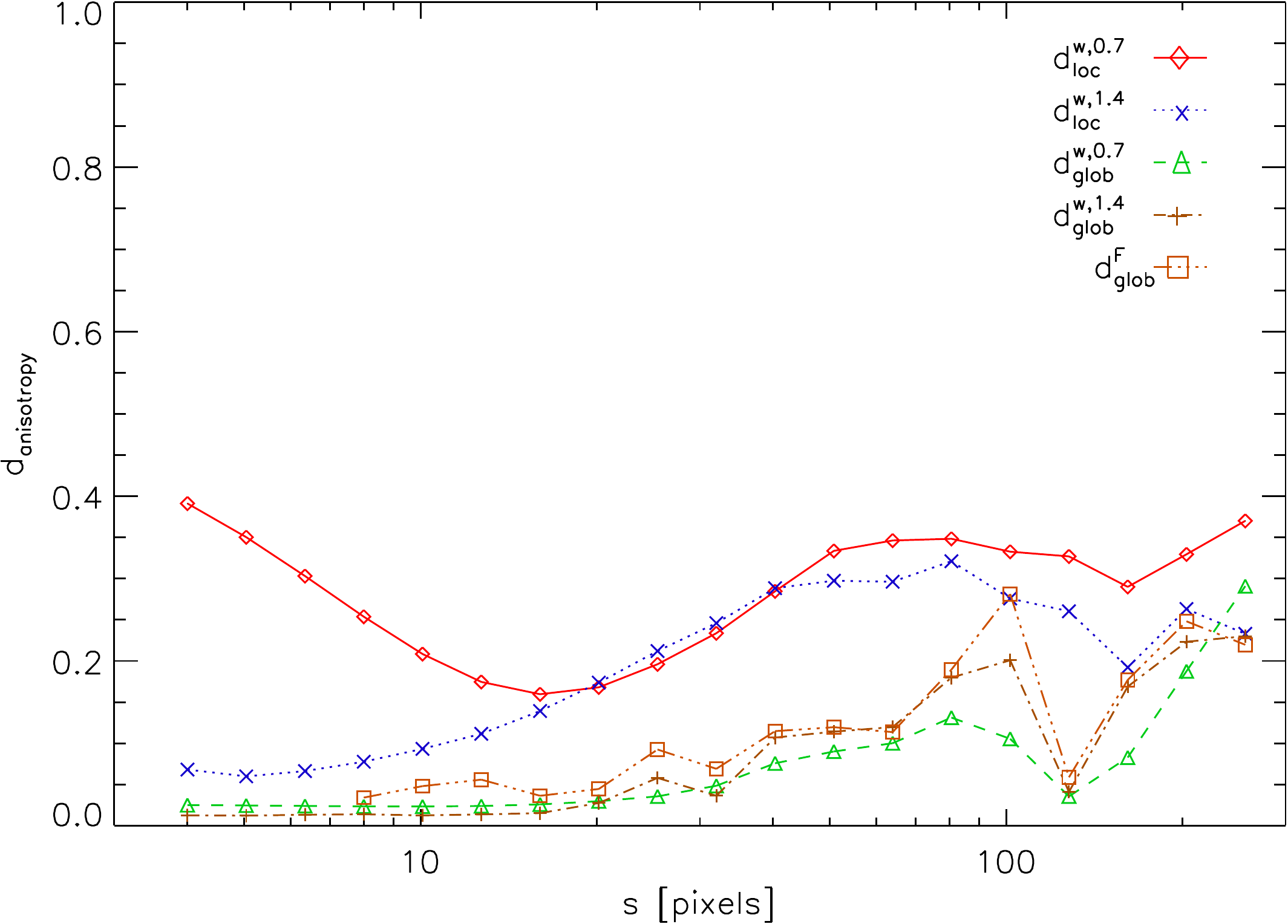}
\caption{Anisotropic wavelet analysis of a map containing a superposition of 100 randomly
placed spherical clumps with $\sigma=8$~pixels.}
\label{fig:appx_many_spheres}
\end{figure}

\begin{figure}
\centering
\includegraphics[angle=90,width=0.8\columnwidth]{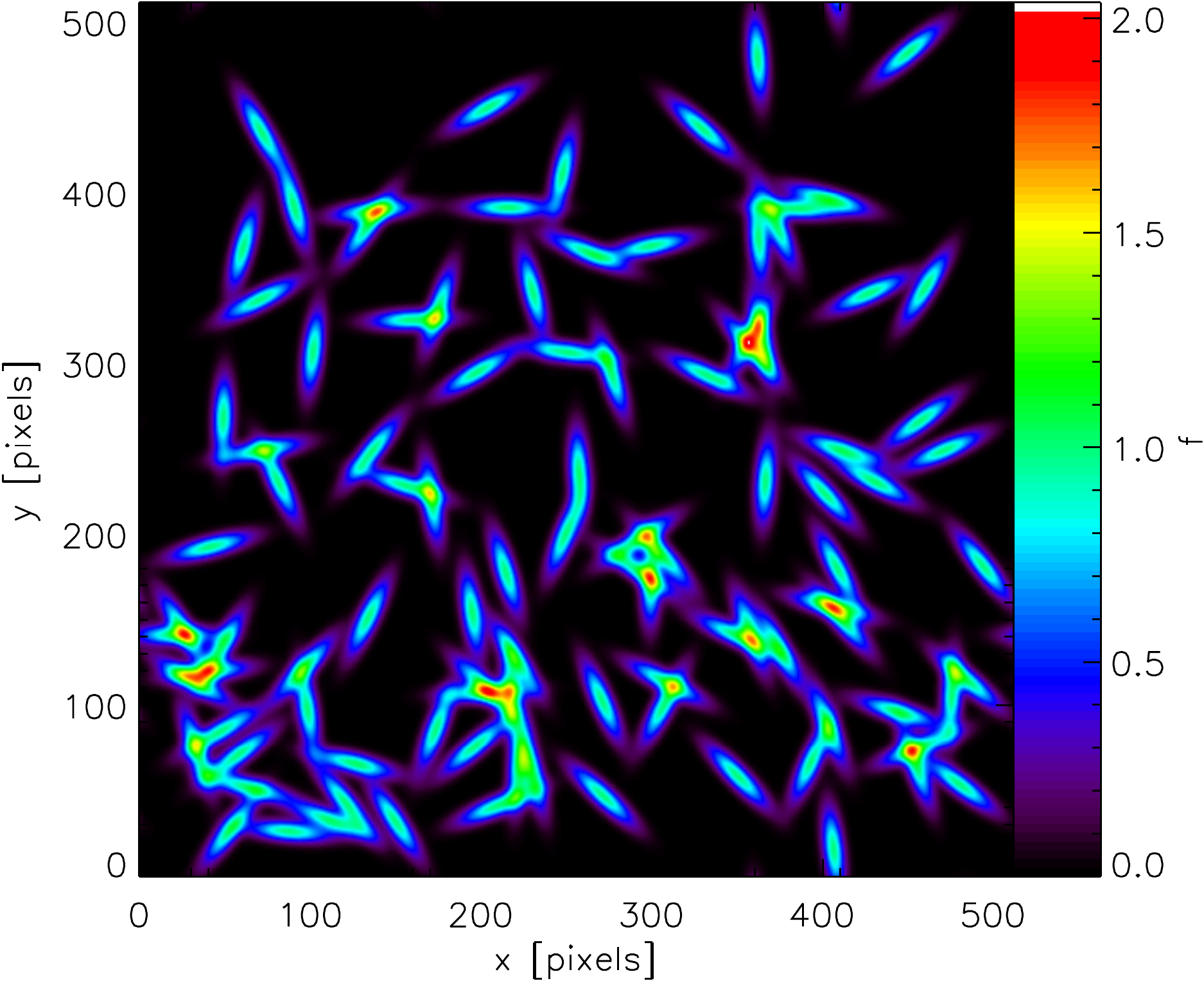}\vspace{3mm}
\includegraphics[angle=90,width=0.9\columnwidth]{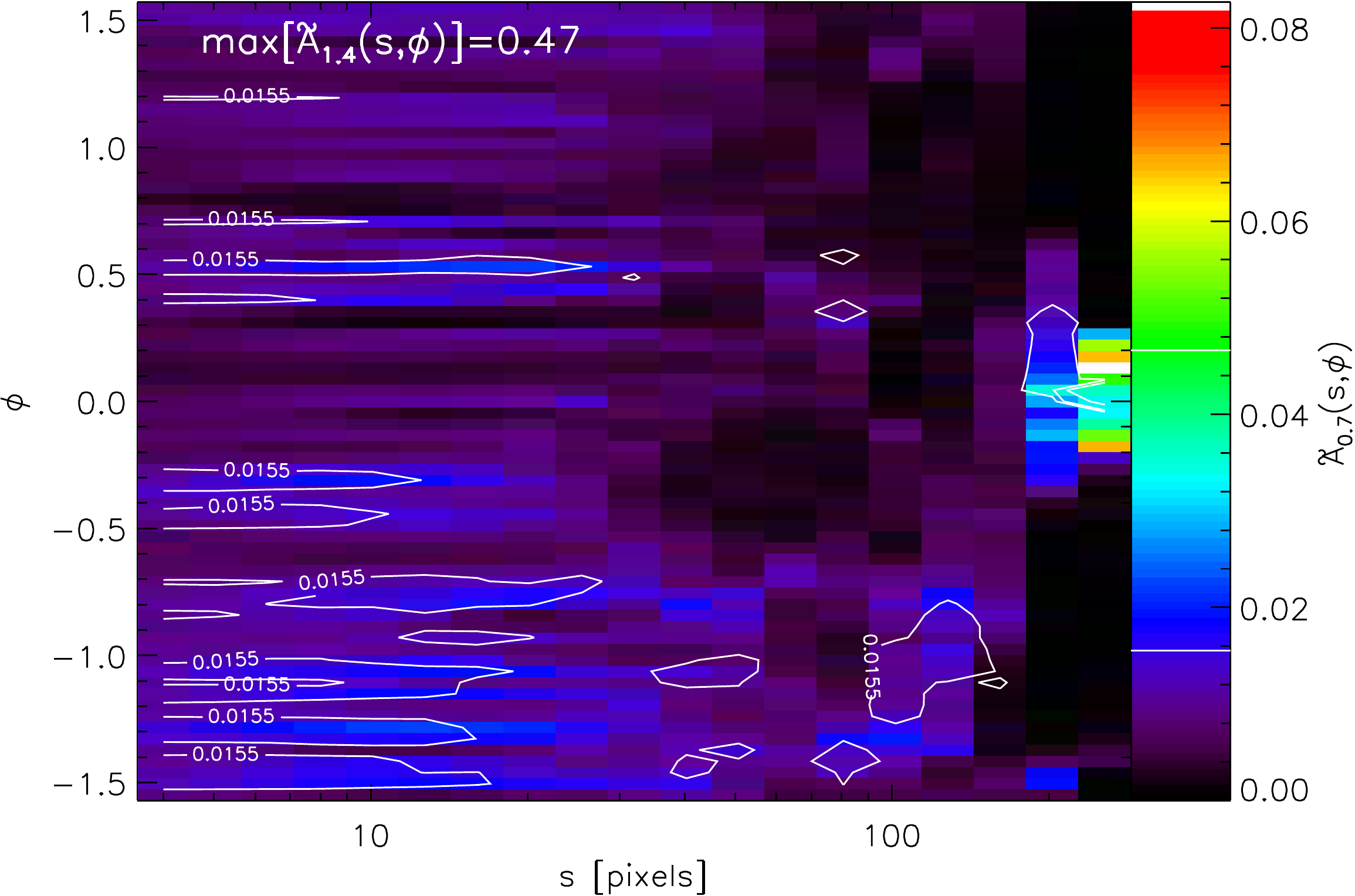}\vspace{3mm}
\includegraphics[angle=90,width=0.88\columnwidth]{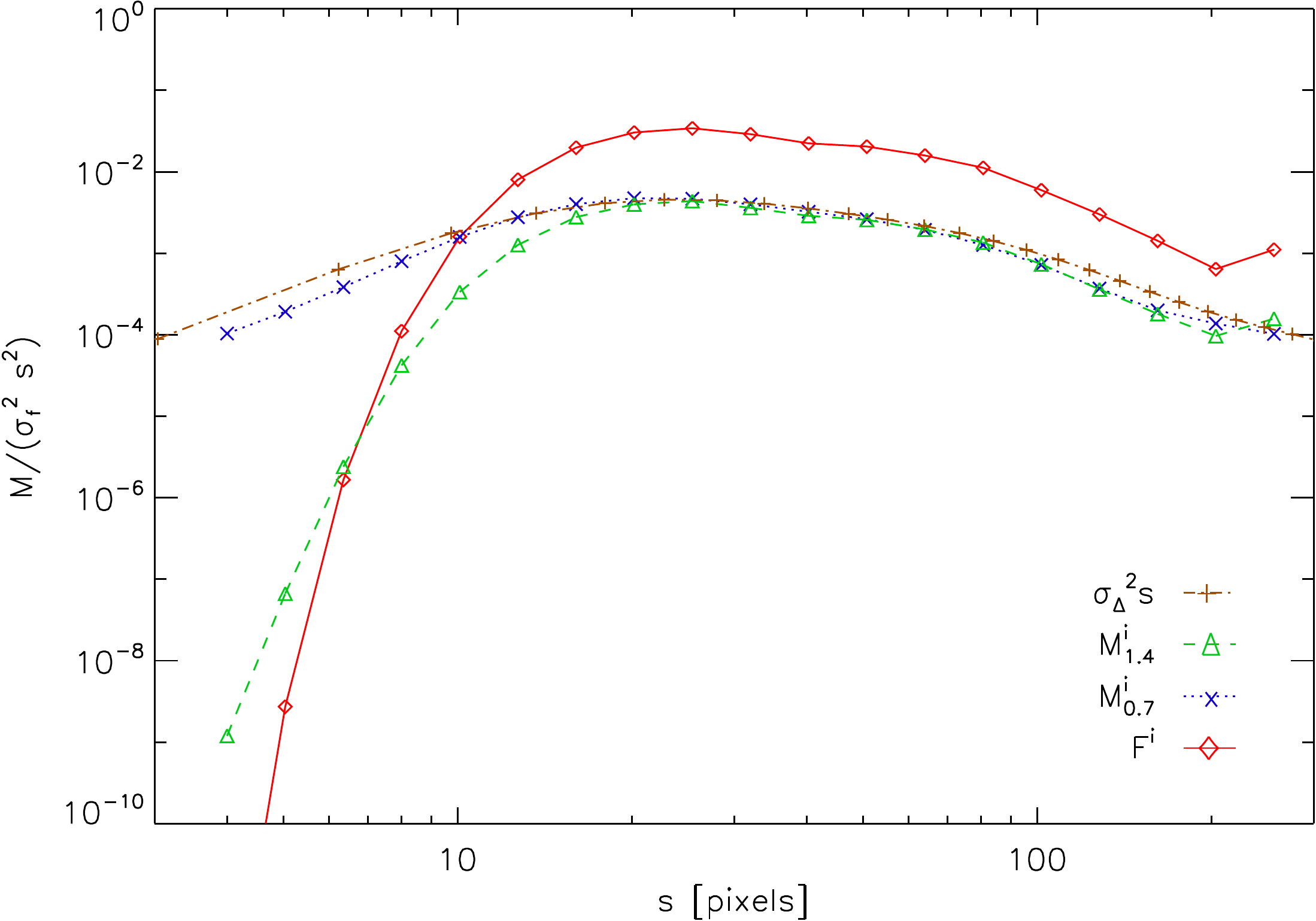}\vspace{3mm}
\includegraphics[angle=90,width=0.88\columnwidth]{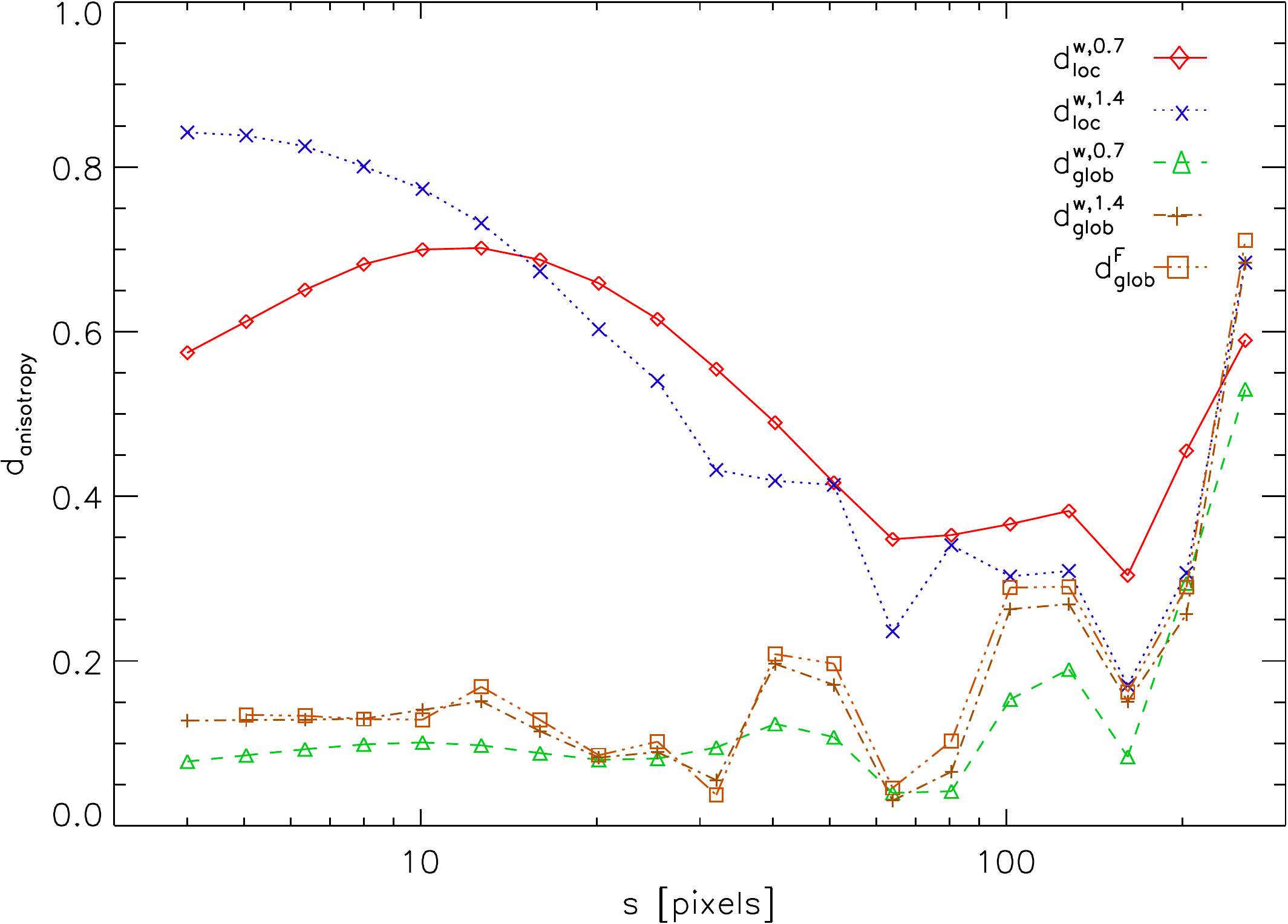}
\caption{Analysis of a map containing a superposition of 100 randomly
placed and oriented Gaussian elliptical clumps with $\sigma_{a}=4$~pixels and $\sigma_{c}=16$~pixels.}
\label{fig:appx_many_ellipses}
\end{figure}

\begin{figure}
\centering
\includegraphics[angle=90,width=0.8\columnwidth]{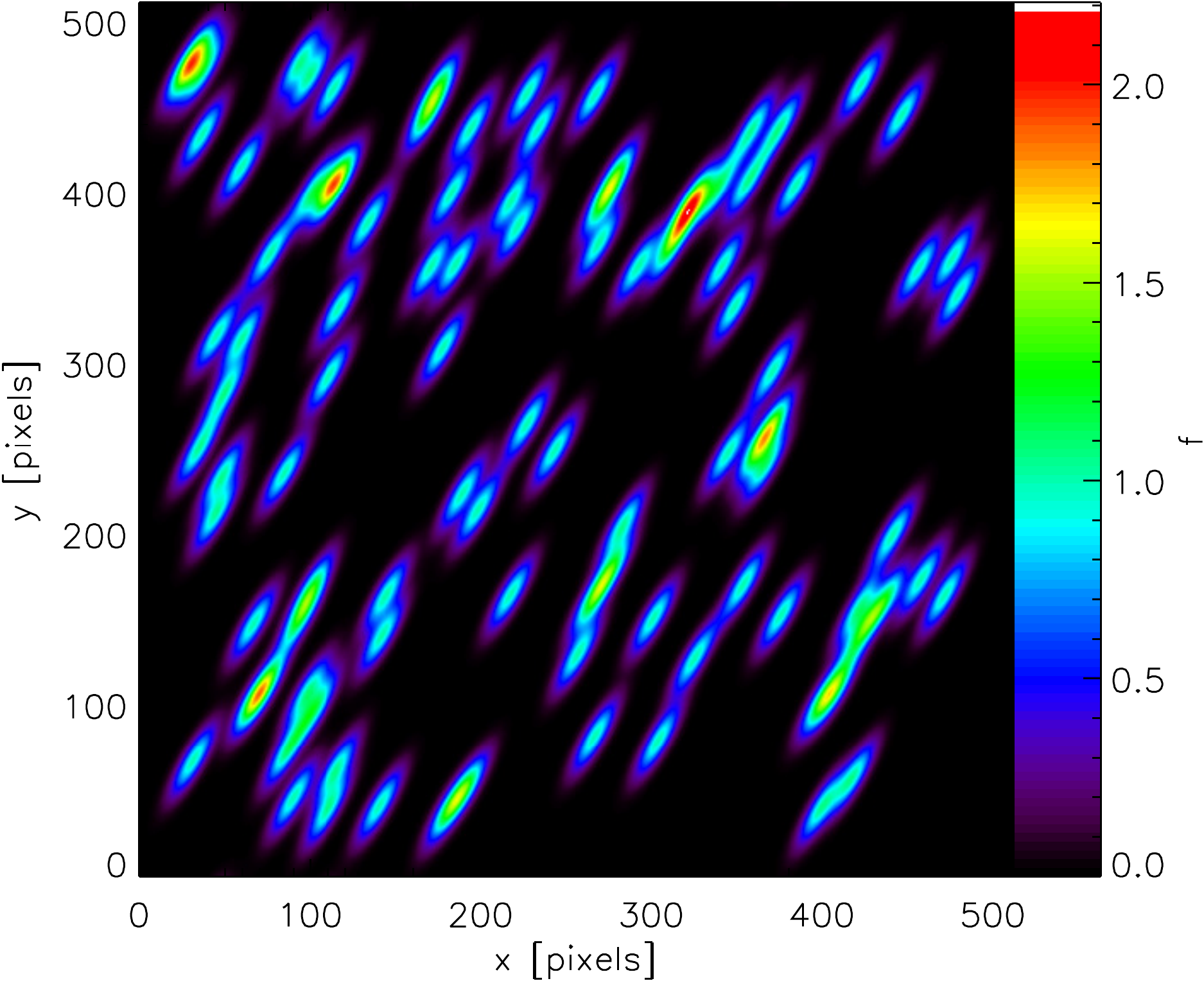}\vspace{3mm}
\includegraphics[angle=90,width=0.9\columnwidth]{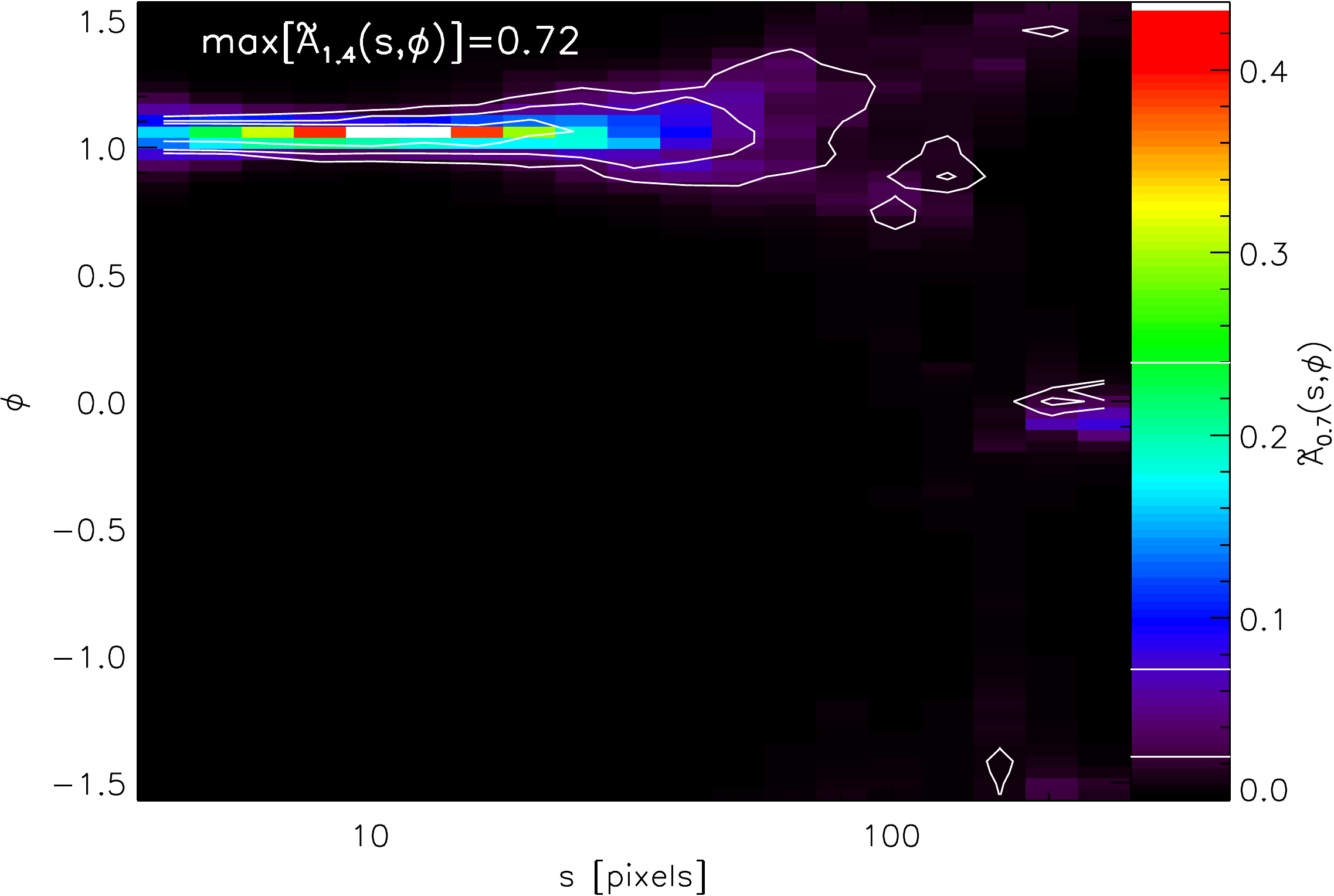}\vspace{3mm}
\includegraphics[angle=90,width=0.88\columnwidth]{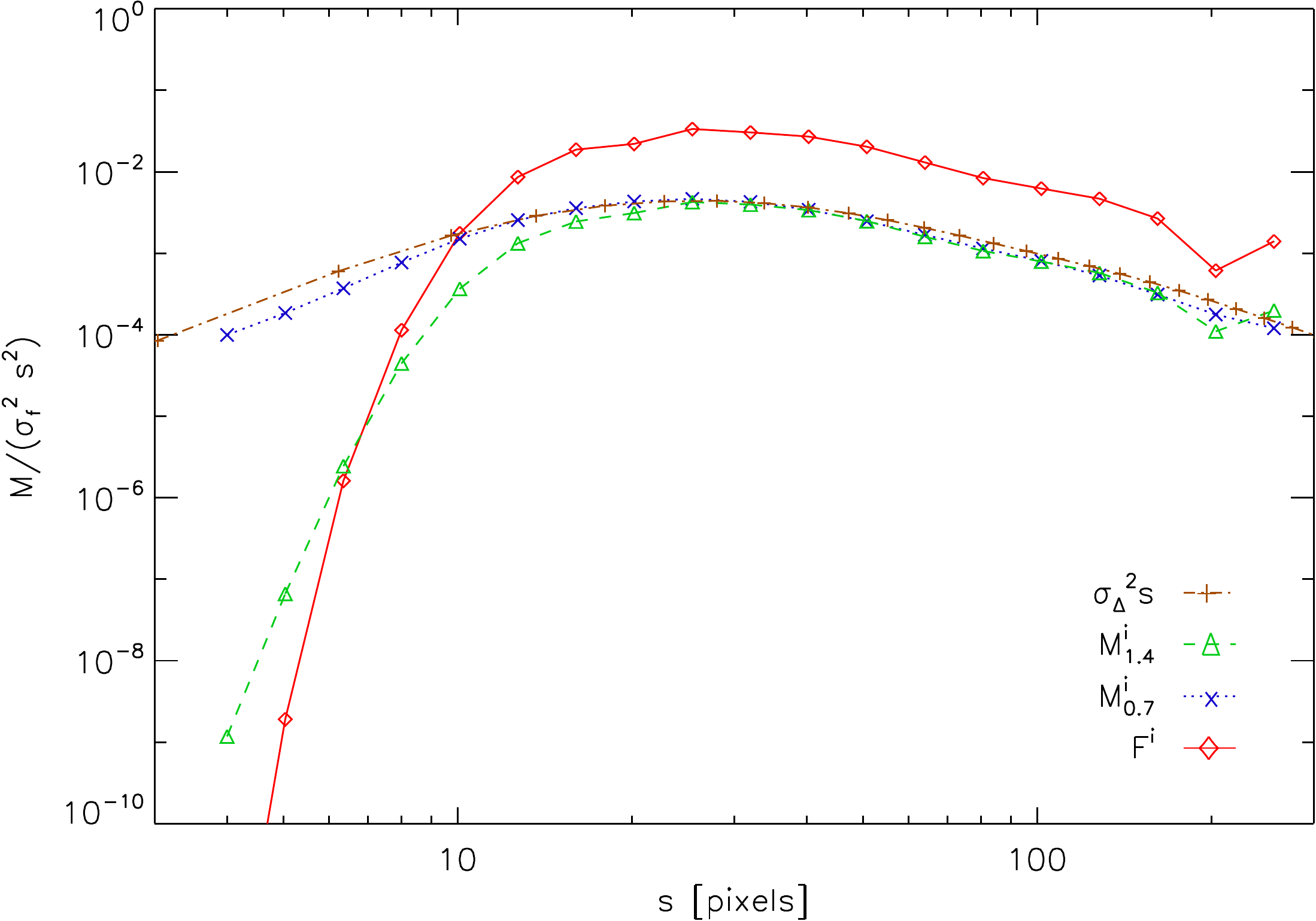}\vspace{3mm}
\includegraphics[angle=90,width=0.88\columnwidth]{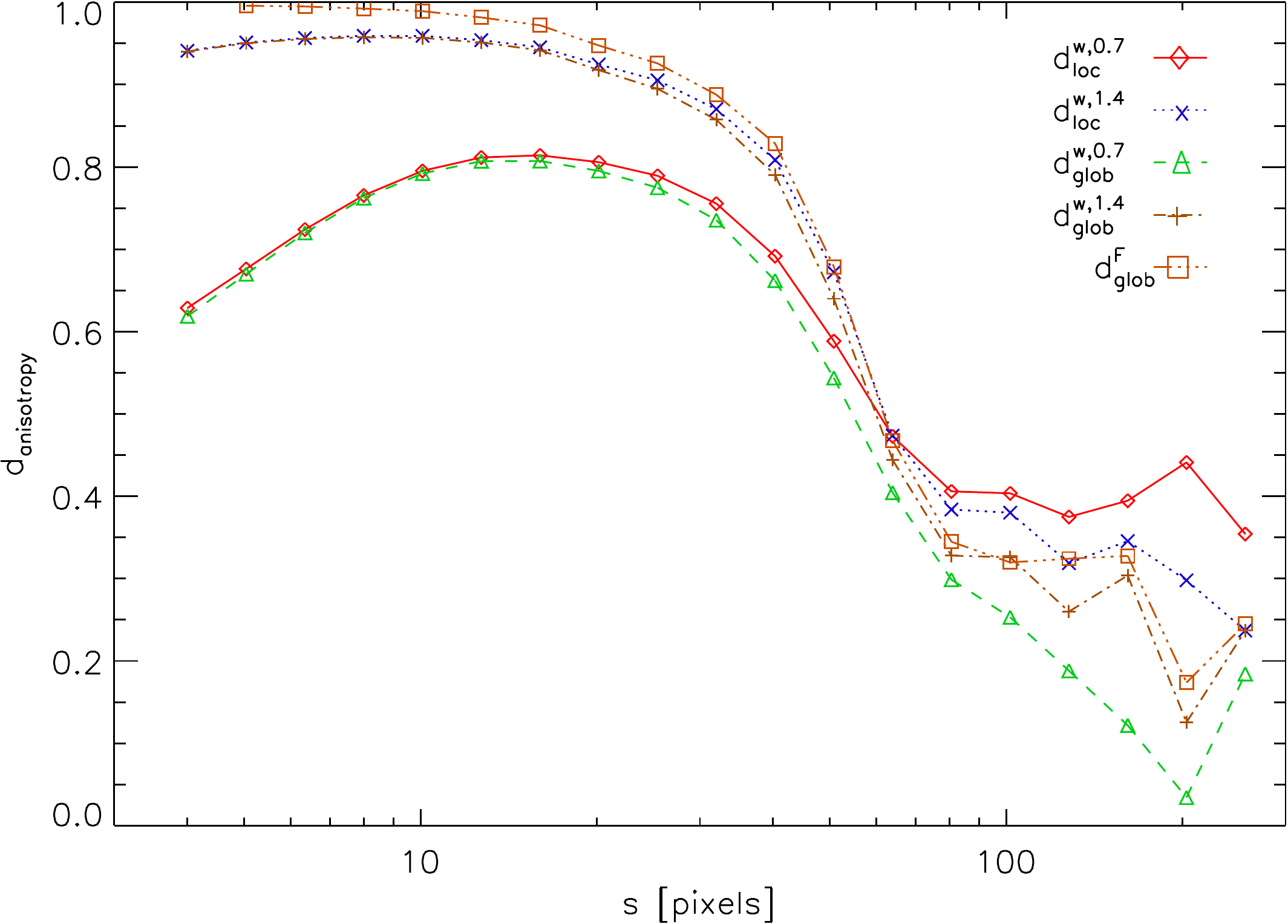}
\caption{Analysis of a map containing a superposition of 100 randomly
placed Gaussian elliptical clumps with $\sigma=4\times 16$~pixels all oriented at an angle of 60~degrees.}
\label{fig:appx_many_aligned_ellipses}
\end{figure}

\begin{figure}
\centering
\includegraphics[angle=90,width=0.8\columnwidth]{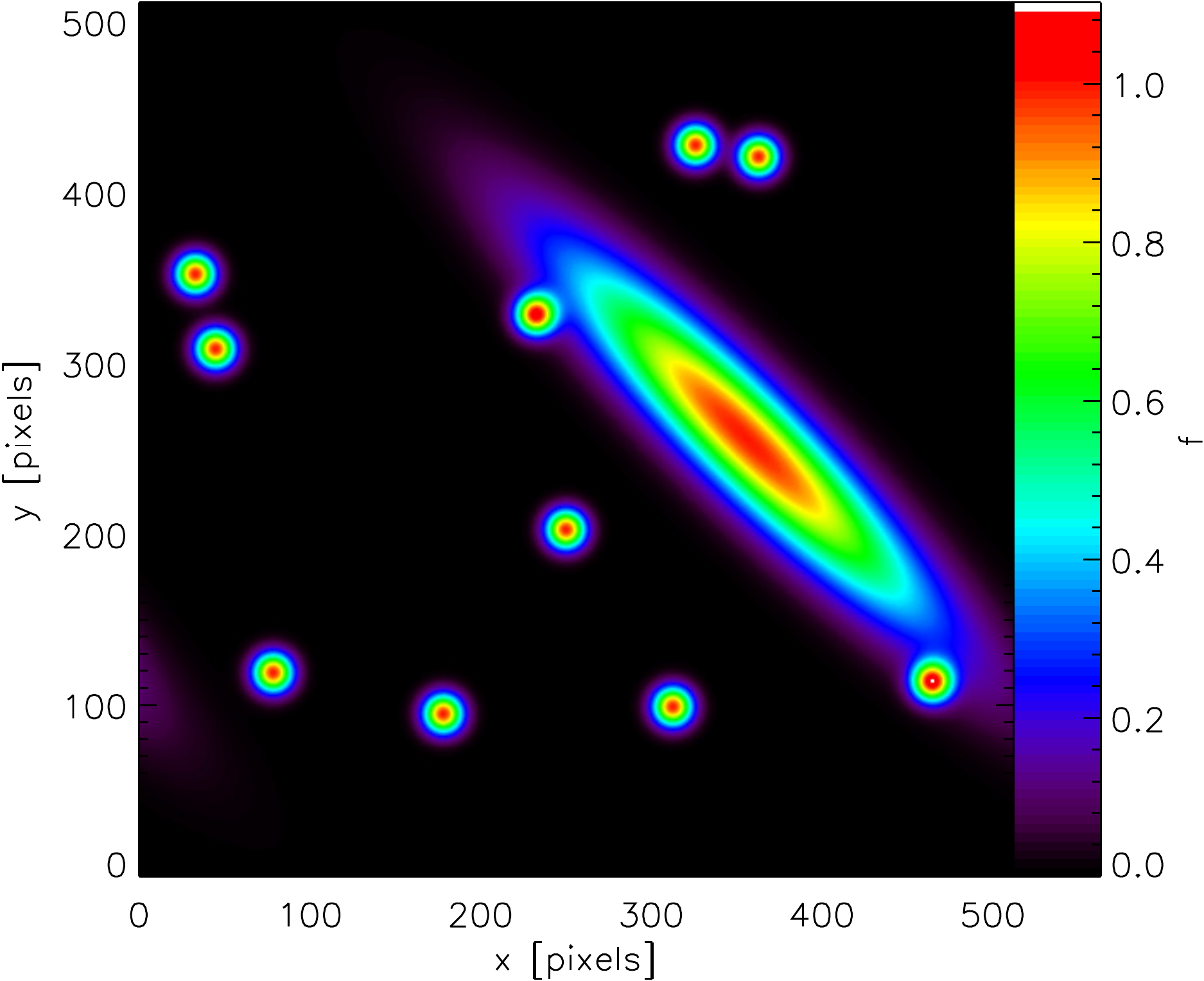}\vspace{3mm}
\includegraphics[angle=90,width=0.9\columnwidth]{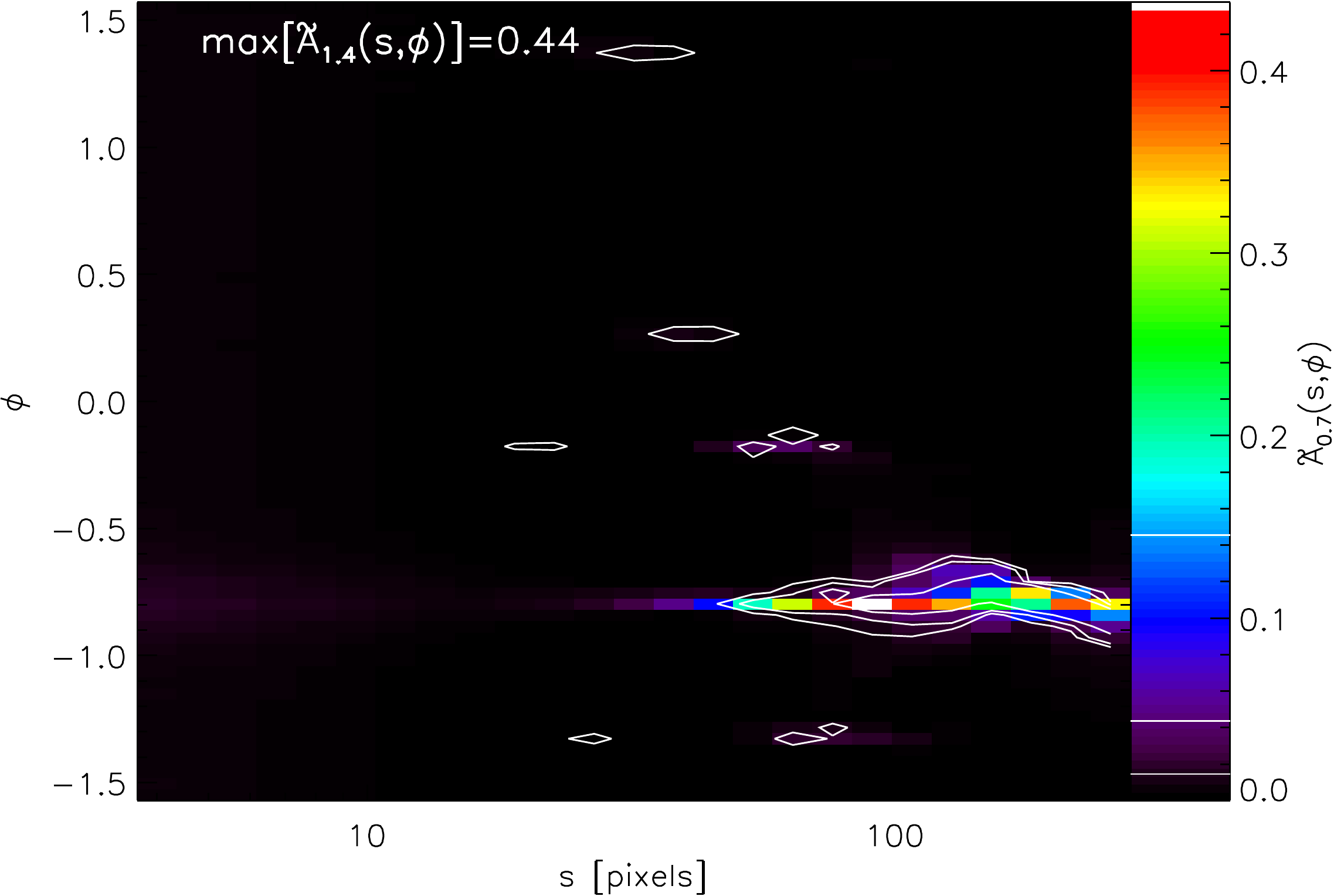}\vspace{3mm}
\includegraphics[angle=90,width=0.88\columnwidth]{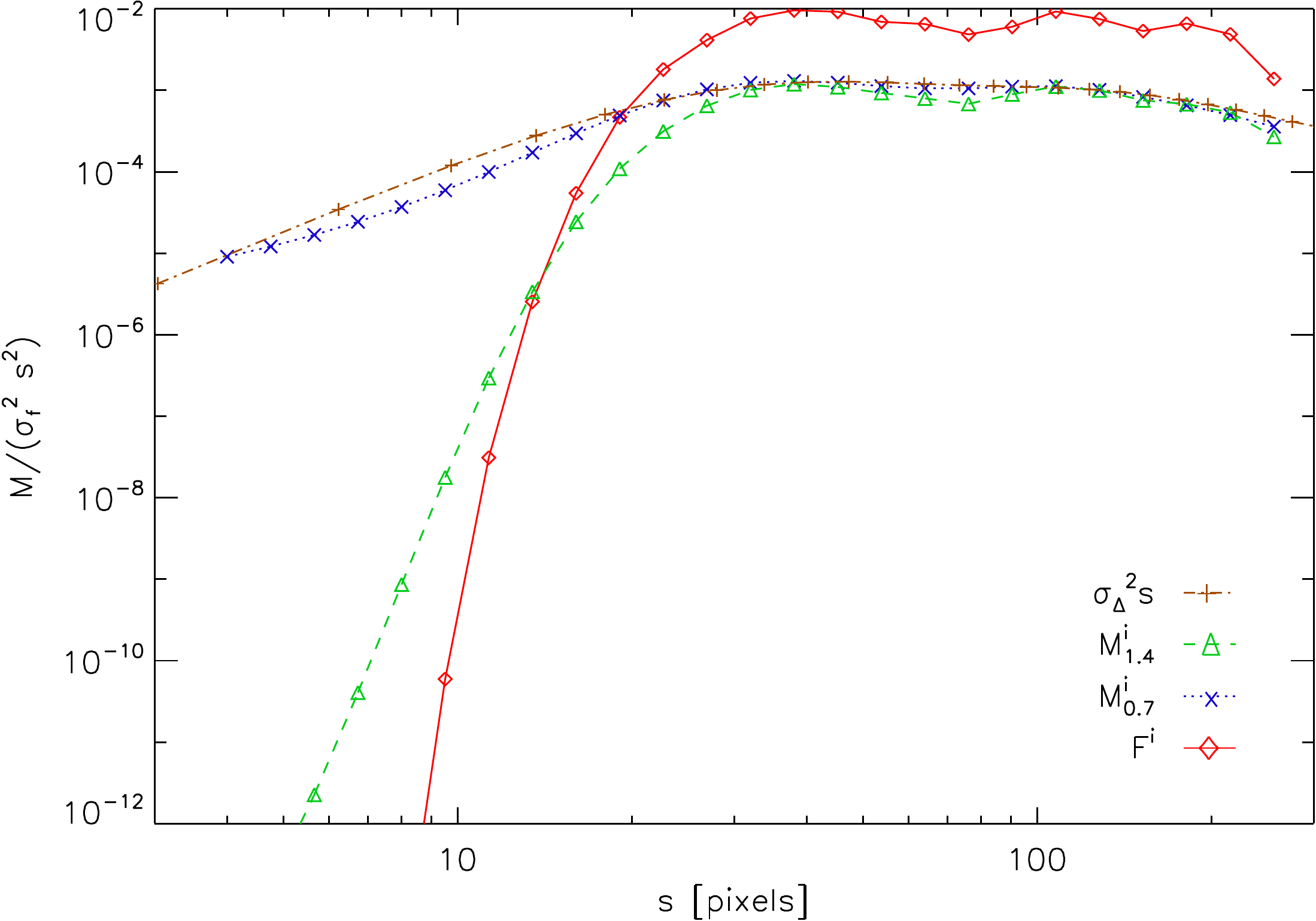}\vspace{3mm}
\includegraphics[angle=90,width=0.88\columnwidth]{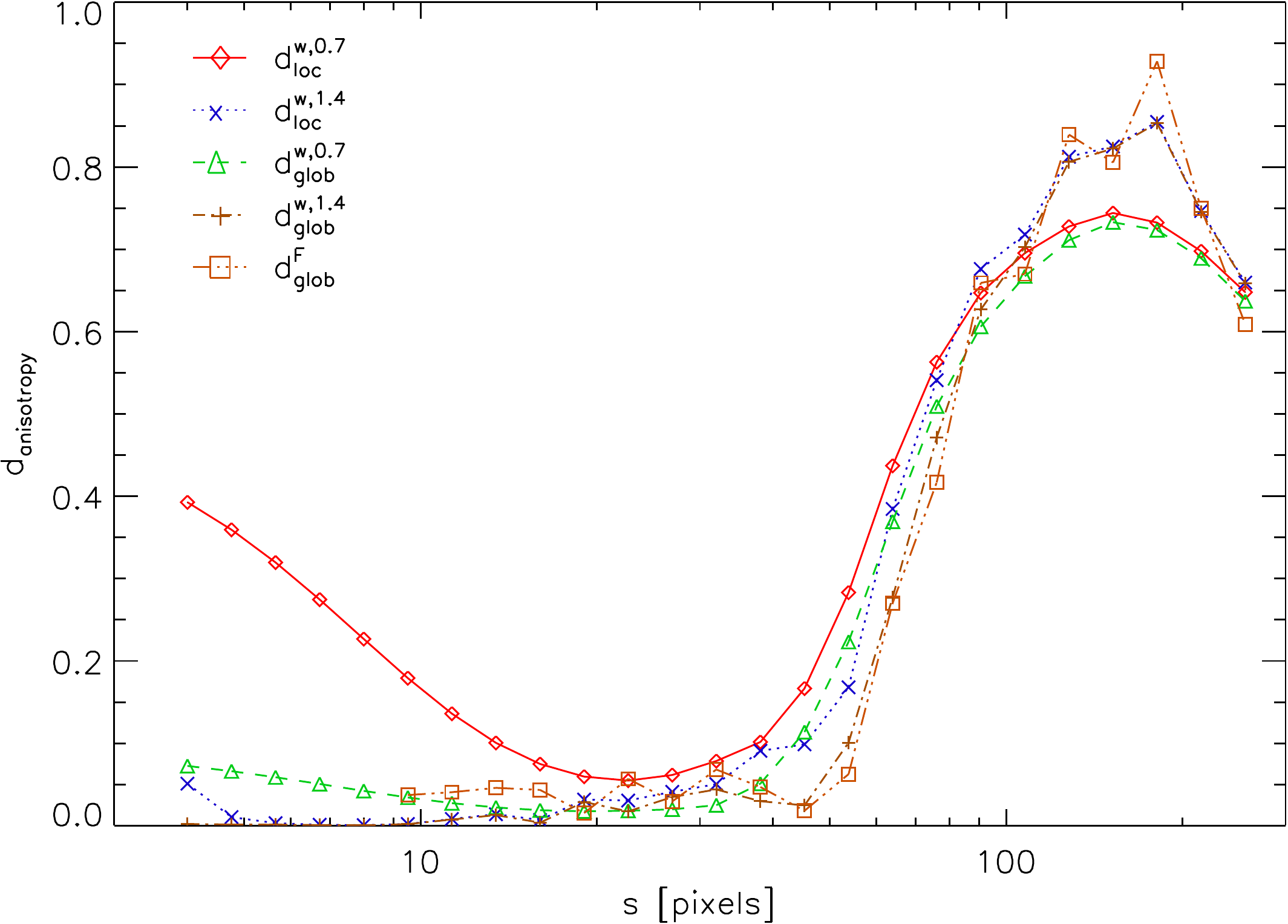}
\caption{Analysis for a superposition of a big elliptic
clump with $\sigma=20\times100$~pixels with 10 small spherical clumps with $\sigma=8$~pixels. }
\label{fig:appx_small_isotropic}
\end{figure}

\changed{
As examples with strong superposition effects Figs.~\ref{fig:appx_many_spheres},
\ref{fig:appx_many_ellipses}, and \ref{fig:appx_many_aligned_ellipses} show
ensembles of 100 Gaussian clumps each. In Fig.~\ref{fig:appx_many_spheres}
we used isotropic clumps with $\sigma_a=\sigma_c=8$~pixels, in
Fig.~\ref{fig:appx_many_ellipses} randomly oriented elliptical clumps
with standard deviations of $\sigma=4\times 16$~pixels, and in
Fig.~\ref{fig:appx_many_aligned_ellipses} the same type of elliptical clumps
all oriented at an angle of 60~degrees. The maps show that the superposition
of the individual clumps creates in the order of ten larger scale structures
per map. In Fourier, wavelet, and $\Delta$-variance
spectra we find no significant difference to the appearance of individual clumps
as discussed in Sect.~\ref{sec:simpleGaussian}. There the superposition
effects seem to be negligible. The angular spectra show few prominent structures
but the degrees of anisotropy are noticeably changed. In all cases the
superposition of clumps creates a local degree of anisotropy of about 0.4 at
intermediate and large scales, beyond the characteristic scales of the individual clumps.
For the randomly oriented clumps, the superposition noticeably lowers the local
degree of anisotropy at the small scales that are characteristic for the clumps.
For the ensemble of aligned clumps this effect does not happen. Here, only
intermediate and large scales are affected. The angular distribution is only
slightly widened there when using the $b=\sqrt{2}$ filter. In case of a
strong global anisotropy, superposition effects are thus almost negligible
in the anisotropic wavelet analysis.

Figure~\ref{fig:appx_small_isotropic} finally
uses a superposition of one a big ellipsoidal clump with standard deviations of
20 and 100~pixels, oriented at -45 degrees, with a set of ten smaller isotropic
clumps. The small clumps have a radius of $\sigma=8$~pixels and random positions.
The configuration is thus similar to Fig.~\ref{fig:small_anisotropic}. The
$b=\sqrt{2}$ wavelet and the Fourier spectra also show a small dip separating
the contributions from large and small clumps. This dip is smeared out when
using the smaller localization parameter, $b=1/\sqrt{2}$, or the $\Delta$-variance.
Although the small clumps do not contribute individually to the degree of anisotropy
they still distort the anisotropy spectrum of the large clump suppressing it
at scales between 30 and 60~pixels. In agreement with Sect.~\ref{sect_filtershape}
the effect is minimized when using a small localization parameter. For a
large parameter $b=\sqrt{2}$ the superposition even tends to warp the
angular spectrum of the large clump.
}
\clearpage

\end{appendix}
\end{document}